%% file: Tutorial_revtex.tex
\documentclass[onecolumn,superscriptaddress,final]{revtex4-2}

\usepackage{lineno}
\usepackage{graphicx}

\usepackage{amsmath}
\usepackage{amssymb}
\usepackage{graphicx}
\usepackage{afterpage}
\usepackage{amsfonts}
\usepackage{graphicx}
\usepackage{braket}
\usepackage{textcomp}
\usepackage{bm}
\usepackage{bbm}
\usepackage{color}
\usepackage{bbold}
\usepackage{hyperref}

\newcommand{\E}{\ensuremath{{\cal E}}}
\newcommand{\dvt}{\ensuremath{\frac{\text{d}}{\text{d}t}}}
\linenumbers 

\begin{document}
	\nolinenumbers

\title{Fundamentals of heterodyne wave mixing spectroscopy: a tutorial}

\author{Daniel~Groll}
\email{daniel.groll@uni-muenster.de}
\affiliation{Institute of Solid State Theory, University of M{\"u}nster, 48149 M{\"u}nster, Germany}

\author{Thilo~Hahn}
\affiliation{Institute of Solid State Theory, University of M{\"u}nster, 48149 M{\"u}nster, Germany}

\author{Pawe\l{}~Machnikowski}
\affiliation{Institute of Theoretical Physics, Wroc\l{}aw University of Science and Technology, 50-370 Wroc\l{}aw, Poland}

\author{Tilmann~Kuhn}
\affiliation{Institute of Solid State Theory, University of M{\"u}nster, 48149 M{\"u}nster, Germany}

\author{Jacek~Kasprzak}
\email{jacek.kasprzak@cnrs.fr}
\affiliation{Institut N{\'e}el, CNRS, Universit{\'e} Grenoble Alpes, Grenoble INP, 38000 Grenoble, France}
\affiliation{Japanese-French Laboratory for Semiconductor Physics and Technology (J-FAST), CNRS, Universit{\'e} Grenoble Alpes, Grenoble INP, University of Tsukuba, 1-1-1 Tennodai, Tsukuba, Ibaraki 305-8573, Japan}
\affiliation{Institute of Experimental Physics, Faculty of Physics, University of Warsaw, 02-093 Warsaw, Poland}

\author{Daniel~Wigger}
\email{d.wigger@uni-muenster.de}
\affiliation{Department of Physics, University of M{\"u}nster, 48149 M{\"u}nster, Germany}

%



\begin{abstract}
This tutorial provides a joint theoretical and experimental overview of heterodyne wave mixing spectroscopy, focusing mainly on four-wave mixing (FWM). This powerful and versatile time-resolved nonlinear optical spectroscopy technique enables the investigation of individual localized single photon emitters, as well as microscopy of extended samples, e.g., two-dimensional transition metal dichalcogenides. Starting with the fundamental theory of optically driven two-level systems, we motivate the utility of wave mixing spectroscopy via a discussion on homogeneous and inhomogeneous linewidths which can be independently measured using FWM. We then provide a detailed overview of the heterodyne wave mixing setup operated by one of the authors (JK) at Institut N{\'e}el in Grenoble, supported by theoretical modeling of the signal detection process. Throughout the paper we elaborate on important benefits of heterodyne wave mixing spectroscopy, e.g., background-free detection, measurement of the full signal field including amplitude and phase, and investigation of coupling mechanisms in few-level systems. Within the context of the latter point we discuss the significance of two-dimensional (2D) FWM spectra. This tutorial is dedicated to students, young researchers, as well as experts in the field of nonlinear spectroscopy in general and FWM in particular. It explains the fundamental concepts and building blocks required to operate a heterodyne wave mixing experiment both from the experimental and theoretical side. This joint approach is helpful for theoreticians who want to accurately and quantitatively model wave mixing signals, as well as for experimentalists who aim to interpret their recorded data.
\end{abstract}
\maketitle

\tableofcontents

\clearpage

\input{1_Introduction}

\input{2_Theory}

\input{3_Theory_2}

\input{4_Experiment}

\input{5_Experiment_Theory}

\input{6_nls}

\input{7_Conclusions}



~\\[1cm]

\noindent\textbf{Funding:} German Federal Ministry of Education and Research: Research Group Linkage Program of the Alexander von Humboldt Foundation.\\
National Science Centre, Poland, grant number 2023/51/B/ST3/01710.

~\newline
\noindent\textbf{Acknowledgments:} We thank Iris Niehues and Anton Plonka for proofreading. We thank the groups of Jon Finley (TU Munich), Wojciech Pacuski (University of Warsaw), Stephan Reitzenstein (TU Berlin), Arne Ludwig (Ruhr Uni Bochum), Marek Potemski (University of Warsaw), and Takashi Taniguchi and Kenji Watanabe (NIMS Tsukuba) for providing samples that were used in previous experimental studies resulting in data that was used in this tutorial.

~\newline
\noindent\textbf{Disclosures:} The authors declare no conflicts of interest.


\newpage
\begin{appendix}
	
	\input{A_Appendix}
\end{appendix}
~\\[1cm]
\bibliography{bib_FWM}

\end{document}

%% file: 1_Introduction.tex
\section{Introduction and overview}
\subsection{Linear and nonlinear optical spectroscopy}\label{sec:intro}
Driving a material out of equilibrium using an external light source, e.g., a laser, leads to a response of the material in the form of a macroscopic polarization $\bm{P}$. For weak excitations the polarization can be approximated as a linear function in the external electric field $\bm{E}$, which for homogeneous and isotropic media reads
\begin{equation}
	\bm{P}(t)\approx \bm{P}^{(1)}(t)=\epsilon_0\int\limits_{-\infty}^t\text{d}t'\chi^{(1)}(t-t')\bm{E}(t')
\end{equation}
with $\chi^{(1)}$ denoting the linear susceptibility of the material and $\epsilon_0$ being the vacuum permittivity. The linear susceptibility already contains valuable information on (linear) absorption, reflection and refraction properties, as well as propagation of light through the material~\cite{griffiths2005introduction,klingshirn2012semiconductor}. There are however also details of the material under investigation which remain elusive in the linear response regime, most notably coupling mechanisms between different optically active transitions~\cite{kasprzak2011coherent,smallwood2018multidimensional}, as well as different sources of line-broadening~\cite{cho1994fifth,langbein2010coherent,cundiff2012optical}. These are contained in higher order nonlinear susceptibilities $\chi^{(n>1)}$, which relate the external field $\bm{E}$ with the nonlinear parts of the macroscopic polarization $\bm{P}^{(n>1)}=\mathcal{O}\left(\bm{E}^{(n>1)}\right)$~\cite{maker1965study,shen1984principles,mukamel1995principles,klingshirn2012semiconductor}, with the total macroscopic polarization being
\begin{equation}
	\bm{P}=\sum\limits_{n=1}^{\infty}\bm{P}^{(n)}\,.
\end{equation}
Investigating the response of a sample to multiple laser pulses naturally yields information on these nonlinear susceptibilities while at the same time providing access to the underlying dynamics of optically excited charge carriers by tuning the delay between the pulses (see Chap.~\ref{sec:theory_2} for details). 

\begin{figure}[b]
	\centering
	\includegraphics[width = 0.55\textwidth]{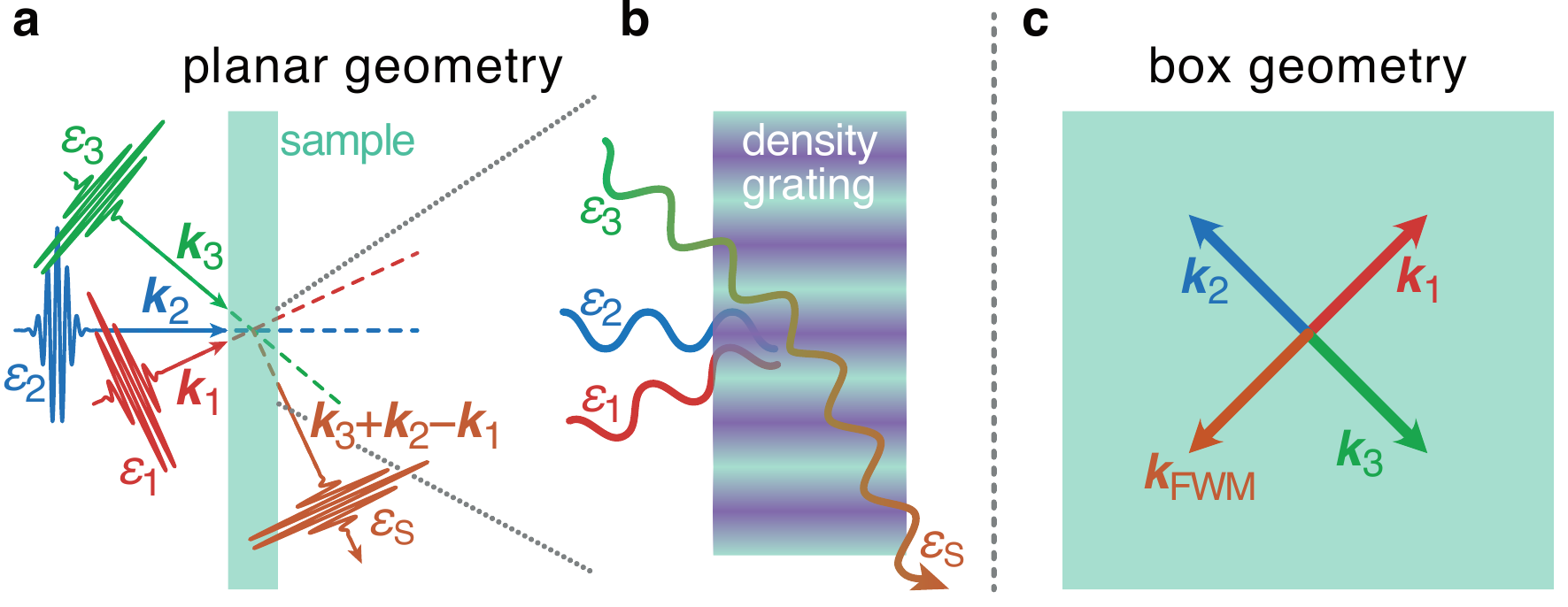}
	\caption{Schematic picture of the angle-resolved four-wave mixing experiment. (a) Excitation pulses traveling with different wave vectors ${\bm k}_{1,2,3}$, i.e., directions. The signal is emitted in the nonlinear mixing direction ${\bm k}_{\rm FWM}={\bm k}_3+{\bm k}_2-{\bm k}_1$. (b) Zoom-in illustrating the generation of the transient grating by pulses $\E_1$ and $\E_2$ which diffracts pulse $\E_3$ generating the signal~$\E_{\rm S}$. (c) Schematic of the box geometry.}
	\label{fig:WM_scheme}
\end{figure}

This is the realm of ultrafast nonlinear spectroscopy methods which have a long-standing history in semiconductor optics and have been routinely applied for example to study exciton dynamics in quantum well structures~\cite{gobel1990quantum,kim1992giant,chemla2001many}. In the traditional setup of wave mixing spectroscopy laser pulse excitation and signal read-out are carried out under different angles with respect to the sample normal as schematically shown in Fig.~\ref{fig:WM_scheme}(a)~\cite{mukamel1995principles,shah2013ultrafast}. Here, the excitation fields impinge with wave vectors ${\bm k}_{1,2,3}$ and the general four-wave mixing (FWM) signal is detected in the direction ${\bm k}_3+{\bm k}_2-{\bm k}_1$. Often instead of this transmission geometry an equivalent reflection geometry is used. We directly see that the detection direction does not agree with any excitation direction, which means that -- in contrast to linear or pump-probe spectroscopy~\cite{shah2013ultrafast,hilton2011ultrafast} -- the FWM signal is in principle background free.

The fundamental mechanism that creates this signal in an unorthodox direction is highlighted in the zoom-in in Fig.~\ref{fig:WM_scheme}(b)~\cite{rossi2002theory}. Considering a semiconductor sample and assuming that each pulse only acts weakly in the first order of the electric field (see Sec.~\ref{sec:FWM_2LS} for details), the first arriving pulse $\E_1$ (red) is absorbed in the sample creating exciton coherences and the second pulse $\E_2$ (blue) converts these coherences into exciton occupations. While the details of this explanation might appear rather abstract at this point, more context will be provided in the following chapters. Because the two fields arrive under different angles the created coherences and $\E_2$ have a spatially-dependent phase relation, which depends on the wave vector difference ${\bm k}_2-{\bm k}_1$. Consequently, the generated exciton density (violet) forms a grating with the corresponding periodicity. The last pulse $\E_3$ (green) now scatters from this density grating which results in the signal propagating in the phase-matching direction ${\bm k}_\text{FWM}={\bm k}_3+{\bm k}_2-{\bm k}_1$. To suppress undesired contributions from scattered light the most commonly used excitation and detection configuration is the box geometry~\cite{li2013unraveling,nardin2015multi}. There the ${\bm k}_i$ and the signal ${\bm k}_{\rm FWM}$ are three-dimensional vectors and their in-plane components (with respect to the sample surface) form a square as depicted in Fig.~\ref{fig:WM_scheme}(c). 

From the interference-grating argument we see that a reliable signal generation requires a spatial extension of the density grating over many periods, typically beyond 10~\textmu m~\cite{scarpelli2017resonantly}. This strongly limits the lateral resolution of the traditional wave mixing spectroscopy method and calls for a sufficiently large and homogeneous sample.

The focus of semiconductor optics shifted from extended systems to point-like systems, e.g., in the form of semiconductor quantum dots, in the early 2000s~\cite{woggon1997optical,jacak2013quantum,gywat2010spins,tartakovskii2012quantum}. The reason is their potential use as single photon emitters or qubits for quantum technological applications~\cite{arakawa2020progress, garcia2021semiconductor}. While the traditional angle-resolved FWM technique allows to study spatially spread-out quantum dot ensembles, e.g., their inhomogeneous broadening~\cite{borri2001ultralong}, accessing single quantum dots requires rethinking of the wave mixing process.

A breakthrough solution for this issue has been achieved by converting the direction-information encoded in the wave vectors (see Fig.~\ref{fig:WM_scheme}) into the microwave frequency domain~\cite{langbein2006heterodyne,langbein2010coherent} (explained in Sec.~\ref{sec:exp}). Consequently, all excitation fields can travel in the same spatial mode and the previously required density grating in real space is converted into a phase-modulation in time. This allows to significantly improve the spatial resolution of the FWM technique because the operation with diffraction limited pulses becomes feasible. In this way it is not only possible to perform spectroscopy on single quantum dots~\cite{langbein2005microscopic}, but also high-resolution micro-spectroscopy images of extended systems can be taken. Recently, this development has proven to be very useful in the latest trends in semiconductor materials research, which is massively focusing on atomically thin crystals, e.g., in the form of transition metal dichalcogenide (TMDC) monolayers and their heterostructures~\cite{boule2020coherent,rodek2023controlled,polczynska2023coherent}.

\subsection{Brief review on applications of four-wave mixing spectroscopy}
Without giving a full review of the past and current progress in the field of wave mixing spectroscopy, in the following we will shortly highlight some outstanding works in applying this technique to various material platforms with different dimensionalities of the charge carrier confinement. Starting from zero-dimensional systems, atomic vapors have been studied extensively~\cite{vaughan2007coherently,dai2012two,li2013unraveling,lomsadze2017frequency}. For molecular systems it is worth mentioning works on quantum coherent energy transfer and perspectives on quantum technology applications~\cite{collini2013spectroscopic, collini20212d}, on photosynthesis~\cite{brixner2005two, collini2010coherently, schlau2011two}, on polaritons~\cite{duan2021isolating}, and on dye molecules in liquids and polymers~\cite{yajima1979spatial,weiner1985three}. In the realm of solid state nano-systems, wave mixing spectroscopy has recently been applied to color centers in diamond~\cite{day2022coherent}, which are quasi zero-dimensional atomic defects. The investigation of artificial atoms in the form of quantum dots and their ensembles has a longer-lasting tradition and is now routinely performed~\cite{borri2001ultralong, moody2011exciton,kasprzak2010up,wigger2018rabi,liu2019non,groll2020four,grisard2022multiple}. 

While traditionally two-dimensional semiconductors in the form of quantum wells have been studied widely~\cite{miller1983degenerate,schultheis1985photon,gobel1990quantum,leo1991subpicosecond,kim1992giant,bartels1995chi,kim1998femtosecond, phillips1999coherent,vaughan2007coherently,cundiff2009optical}, within the past decade wave mixing spectroscopy has also been applied very successfully to other two-dimensional material platforms: quasi-two-dimensional nanoplatelets~\cite{naeem2015giant}, perovskites~\cite{titze2019ultrafast,nguyen2023phonon}, and novel layered two-dimensional semiconductors in particular in the form of TMDCs~\cite{hao2017neutral, hao2017trion, policht2021dissecting, purz2022imaging, hahn2022destructive, muir2022interactions, huang2023quantum, rodek2023controlled}. Pioneering works on wave mixing spectroscopy have also investigated three-dimensional thin crystal films~\cite{masumoto1983optical,schultheis1986picosecond}. Finally, more exotic excitations like plasmon-polaritons~\cite{pres2023detection}, magneto-excitons~\cite{mapara2021multidimensional} and low-energy transitions in cuprates~\cite{novelli2020persistent} can be accessed via wave mixing spectroscopy.

A recently published book also gives a wide overview on the topic~\cite{li2023optical}.

\subsection{Overview: How to read this tutorial}
This tutorial is dedicated to students, young researchers, and everyone who wants to learn about nonlinear spectroscopy in general and FWM in particular. With this scope in mind, it covers a wide range from basic theoretical concepts and detailed descriptions of the experimental setup to advanced methods for the analysis of the experimental signals, which can be applied to various physical systems. Depending on the personal background and the specific interest of the reader a focus on specific chapters is possible in a first reading.

To lay the foundation for a more in-depth discussion of FWM spectroscopy in later chapters, in Chap.~\ref{sec:theory} we present the basic theory of the optically driven two-level system (2LS). This chapter can be skipped by readers who are familiar with concepts like the Bloch vector, the optical Bloch equations and linear absorption spectra. The physics behind time-resolved spectrosocopy and its theoretical description are then introduced in Chap.~\ref{sec:theory_2} using the example of ensembles of 2LSs. Finally, in Sec.~\ref{sec:FWM_2LS} we explain the concept of FWM spectroscopy, applying the theory established so far.

After the previous theoretical description of FWM spectroscopy, Chap.~\ref{sec:exp} is dedicated to the experiment. In Sec.~\ref{sec:setup} we explain the individual components of the heterodyne FWM setup at Institut N\'eel in Grenoble, France operated by one of the authors (JK). We continue with a step-by-step description of the data acquisition and processing in Sec.~\ref{sec:FWM_data}. A deeper understanding might require some theoretical context from the following Chap.~\ref{sec:exp_theo} and we provide according references where useful. Afterwards in Sec.~\ref{sec:2d_spectra} we finally discuss the concept of two-dimensional (2D) FWM spectra and compare experimental data measured on different excitonic 2LSs with the theory established in Chaps.~\ref{sec:theory} and \ref{sec:theory_2}.

In Chap.~\ref{sec:exp_theo} we provide theoretical modeling of the heterodyne FWM signal detection process in the setup from Chap.~\ref{sec:exp}. The presented treatment can serve as a blueprint that can be adapted to the description of several other spectroscopy methods, e.g., time-dependent photoluminescence, resonance fluorescence, and coherent control. Here we highlight important subtleties of the FWM data acquisition and processing described in Sec.~\ref{sec:FWM_data}. This chapter may be skipped if the reader is only interested in the experimental implementation or the general FWM concept.

In Chap.~\ref{sec:nls} we give a brief overview of how one can extend modeling of FWM signals from 2LSs to more general $N$-level systems. In this context we introduce the notion of double-sided Feynman diagrams and apply them explicitly to the case of different three-level systems. This then allows us to understand more nuanced 2D FWM spectra measured on different quantum dot structures. We finally briefly discuss higher order $N$-wave mixing signals as a natural extension of FWM spectroscopy. 

We conclude this tutorial in Chap.~\ref{sec:conclusions} by presenting a short summary of the benefits of heterodyne wave mixing spectroscopy, which have been elaborated throughout the different chapters.


%% file: 2_Theory.tex
\section{Basics on light-matter coupling: the optically driven two-level system}\label{sec:theory}
The simplest example of a quantum mechanical system is that of a two-level system (2LS). If the transition between the two energy levels can be driven by optical means we speak of an optically driven 2LS. In terms of light-matter interaction, this is the most basic model, considering a single bright transition, which can be used to describe, e.g., atoms~\cite{cohen2022photons,cohen1998atom}, quantum dots~\cite{schmitt1987theory,krummheuer2002theory,stufler2005quantum,hohenester2007quantum,ramsay2010damping,tartakovskii2012quantum,groll2020four}, optically active defect centers~\cite{duke1965phonon,mahan2000many,aharonovich2011diamond,wigger2019phonon,preuss2022resonant,fischer2023combining}, or excitons in two-dimensional semiconductors under certain conditions~\cite{schultheis1985photon,wegener1990line,steinbach1999electron,rodek2021local,hahn2022destructive}. This model is also important for understanding central features of FWM spectroscopy, as it is the simplest system in which FWM can occur~\cite{schultheis1985photon,wegener1990line,langbein2010coherent}. With this in mind, we start in this Chap.~\ref{sec:theory} by introducing fundamental properties of optically driven 2LSs, before discussing their FWM signal in detail in the following Chap.~\ref{sec:theory_2}.
\subsection{Dipole and rotating wave approximation}\label{sec:dipole_RWA}
We consider a 2LS consisting of a ground state $\ket{g}$ and an excited state $\ket{x}$ with a transition energy $\hbar\omega_x$. It is common practice to set the ground state energy to zero, which does not change any of the physical results, since only energy differences appear in the equations of motion that we derive in the following. The Hamiltonian describing the free dynamics of the 2LS is thus given by
\begin{equation}\label{eq:H_0}
	H_0=\hbar\omega_x\ket{x}\bra{x}\,.
\end{equation}
As a concrete example the ground state could describe a quantum dot without excited charge carriers, while the excited state could describe a quantum dot containing an electron-hole pair, i.e., a quantum dot exciton~\cite{schmitt1987theory,krummheuer2002theory,stufler2005quantum,hohenester2007quantum,ramsay2010damping,tartakovskii2012quantum,groll2020four}. Such electron hole pairs can be excited by optical means, e.g., with a laser. We focus here on nanoscopic systems where the spatial extent of the charge carrier wavefunction is typically much smaller than the wavelength $\lambda \approx 2\pi c/\omega_x$ of the light that is used for the excitation ($c$: speed of light in the medium). We can then assume that the optical field does not depend on the spatial coordinate across the nanosystem and describe the light matter interaction in terms of the dipole approximation with the interaction Hamiltonian~\cite{cohen1998atom,cohen2022photons}
\begin{equation}\label{eq:dipole_approx}
	H_I(t)=-\bm{E}(t)\cdot \bm{d}\,.
\end{equation}
Here, $\bm{d}$ is the so-called dipole operator, which for a single electron would be given by $\bm{d}=-e\bm{r}$ with $-e$ being the electron charge and $\bm{r}$ being its position operator. In the case of multiple charge carriers the total dipole operator is a sum of single particle ones~\cite{cohen2022photons}. The optical driving is described semi-classically via the classical electric field $\bm{E}(t)$ of the light. It is usually a good approximation to model the light field in this fashion when considering excitation with lasers~\cite{cohen1998atom}. The total Hamiltonian of the optically driven 2LS is thus given by
\begin{equation}\label{eq:H_2LS}
	H_\mathrm{2LS}(t)=H_0+H_I(t)=\hbar\omega_x\ket{x}\bra{x} -\bm{E}(t)\cdot \bm{d} = \hbar\omega_x\ket{x}\bra{x} -\sum_{i,j=g,x}\ket{i}\bra{i}\bm{E}(t)\cdot \bm{d}\ket{j}\bra{j}\,,
\end{equation}
where in the last step we used the completeness relation of the 2LS
\begin{equation}
	\mathbb{1}_\mathrm{2LS}=\sum_{i=g,x}\ket{i}\bra{i}=\ket{g}\bra{g}+\ket{x}\bra{x}\,.\label{eq:id_2LS}
\end{equation}
The dynamics of the optically driven 2LS are described by the time-dependent Schrödinger equation
\begin{equation}
	i\hbar\frac{\text{d}}{\text{d}t}\ket{\psi(t)}=H_\mathrm{2LS}(t)\ket{\psi(t)}\,,\label{eq:schroedinger_2LS}
\end{equation}
which can be formally solved by introducing the unitary time evolution operator~\cite{breuer2002theory}
\begin{equation}
	U_\mathrm{2LS}(t,t_0)=\hat{T}\exp\left[-\frac{i}{\hbar}\int\limits_{t_0}^t\text{d}\tau\,H_\mathrm{2LS}(\tau)\right]\,.\label{eq:U_2LS}
\end{equation}
Here, $\hat{T}$ denotes the time-ordering operator and the time-evolution operator has the defining properties
\begin{equation}
	U_\mathrm{2LS}(t,t_0)\ket{\psi(t_0)}=\ket{\psi(t)}\,,\quad i\hbar\frac{\text{d}}{\text{d}t}U_\mathrm{2LS}(t,t_0)=H_\mathrm{2LS}(t)U_\mathrm{2LS}(t,t_0)\,,\quad U_\mathrm{2LS}(t_0,t_0)=\mathbb{1}_\mathrm{2LS}\,.\label{eq:properties_U}
\end{equation}
It is often very practical to rewrite the full time-evolution operator in terms of the free time evolution operator
\begin{equation}
	U_0(t,t_0)=\exp\left[-\frac{i}{\hbar}H_0(t-t_0)\right]\label{eq:U_0}
\end{equation}
and the interaction picture time-evolution operator
\begin{equation}
	U_I^I(t,t_0)=\hat{T}\exp\left[-\frac{i}{\hbar}\int\limits_{t_0}^t\text{d}\tau\,H_I^I(\tau)\right]\,,\qquad H_I^I(\tau)=U_0^{\dagger}(\tau,t_0)H_I(\tau)U_0(\tau,t_0)\label{eq:U_I^I}
\end{equation}
as
\begin{equation}
	U_\mathrm{2LS}(t,t_0)=U_0(t,t_0)U_I^I(t,t_0)\,.\label{eq:dyson}
\end{equation}
Note that $U_0$ and $U_I^I$ have the same structure as $U_\mathrm{2LS}$ in Eq.~\eqref{eq:U_2LS} but with $H_\mathrm{2LS}$ replaced by $H_0$ and $H_I^I$, respectively. In the case of $U_0$, the corresponding operator $H_0$ does not depend on time, such that the time-ordering operator $\hat{T}$ can be dropped and the integral yields the factor $(t-t_0)$ in Eq.~\eqref{eq:U_0}. In this way, $U_0$ and $U_I^I$ have properties analog to those of $U_\mathrm{2LS}$ described in Eq.~\eqref{eq:properties_U}, which can be used to prove Eq.~\eqref{eq:dyson}.

The separation of the full time evolution into the free time evolution and the impact of optical driving in Eq.~\eqref{eq:dyson} can now be used to justify an important approximation to the full Hamiltonian in Eq.~\eqref{eq:H_2LS}. To do so, we first have to calculate the interaction Hamiltonian in the interaction picture, which is given by
\begin{align}
	H_I^I(t)=&-\bm{E}(t)\ket{g}\bra{g}\bm{d}\ket{g}\bra{g}-\bm{E}(t)\ket{x}\bra{x}\bm{d}\ket{x}\bra{x}\notag\\
	&-\bm{E}(t)e^{i\omega_x(t-t_0)}\ket{x}\bra{x}\bm{d}\ket{g}\bra{g}-\bm{E}(t)e^{-i\omega_x(t-t_0)}\ket{g}\bra{g}\bm{d}\ket{x}\bra{x}\,.\label{eq:H_I^I}
\end{align}
The external light field will usually be separated into a slow envelope $\bm{E}_0(t)$ and a fast carrier wave with the carrier frequency $\omega_l$ as
\begin{equation}
	\bm{E}(t)=\bm{E}_0(t)\cos(\omega_l t-\phi)\,,\label{eq:E(t)=E_0(t)}
\end{equation}
where $\phi$ is the carrier-envelope phase. When considering excitations close to the transition energy of the 2LS, i.e., with a detuning $\delta=\omega_l-\omega_x$ whose absolute values are small compared to both $\omega_l$ and $\omega_x$, we can usually apply the so-called rotating wave approximation (RWA)~\cite{cohen1998atom} by noting that $H_I^I(t)$ in Eq.~\eqref{eq:H_I^I} contains slowly varying terms $\sim \exp(\pm i\delta t)$ and fast oscillating terms $\sim \exp(\pm i\omega_l t)$, $\exp[\pm i (\omega_l+\omega_x)t]$. We can then drop the fast oscillating terms, assuming that they average out during the 2LS's optically induced time evolution, yielding the interaction Hamiltonian in the RWA
\begin{equation}\label{eq:H_I_RWA}
	H_{I,\text{RWA}}(t)=-\frac{1}{2}\bm{E}_0(t)\bra{x}\bm{d}\ket{g}e^{-i\omega_l t+i\phi} \ket{x}\bra{g} +h.c.\,,
\end{equation}
where we went back to the original picture from the interaction picture and used that the dipole operator is hermitian $\bra{x}\bm{d}\ket{g}^*=\bra{g}\bm{d}\ket{x}$. The term "$+h.c.$" is a shorthand for the expression hermitian conjugate, i.e., $A+A^{\dagger}=A+h.c.$

Using Eq.~\eqref{eq:U_I^I} we can now investigate this "averaging out" of the fast oscillating terms more thoroughly and discuss under which conditions the RWA is justified. To this aim we have to remember that any time-evolution operator $U(t,t_0)$ of the form in Eq.~\eqref{eq:U_2LS} has the semigroup property $U(t,t_1)U(t_1,t_0)=U(t,t_0)$, i.e., instead of propagating from $t_0$ to $t$ directly, we can propagate to an intermediate time $t_1$ followed by a propagation to $t$. We can do this arbitrarily often, such that we can slice the time evolution from $t_0$ to $t$ into $N$ parts, e.g.,
\begin{equation}
	U_I^I(t,t_0)=U_I^I(t,t-\epsilon)U_I^I(t-\epsilon,t-2\epsilon)...U_I^I(t_0+\epsilon,t_0)\,,\quad \epsilon=(t-t_0)/N\,.\label{eq:U_I^I_slice}
\end{equation}
If the length $\epsilon$ of one time slice is short enough, we can approximate the interaction picture time evolution by first order perturbation theory
\begin{equation}
	U_I^I(t'+\epsilon,t')=\hat{T}\exp\left[-\frac{i}{\hbar}\int\limits_{t'}^{t'+\epsilon}\text{d}\tau\,H_I^I(\tau)\right]\approx 1-\frac{i}{\hbar}\int\limits_{t'}^{t'+\epsilon}\text{d}\tau\,H_I^I(\tau)\,.\label{eq:U_I^I_linear}
\end{equation}
This approximation works better, the shorter the time slice. The maximum length of the time slice, for which this is a good approximation, is dictated by the strength of the interaction described by $H_I^I$. The weaker the interaction, the larger $\epsilon$ can be. If the interaction strength, as described by the matrix elements of $\bm{E}_0(t)\cdot\bm{d}$, when inserting Eq.~\eqref{eq:E(t)=E_0(t)} in Eq.~\eqref{eq:H_I^I}, is sufficiently weak, we can choose $\epsilon$ in Eq.~\eqref{eq:U_I^I_linear} much larger than the 2LS period $2\pi/\omega_x$ or the laser period $2\pi/\omega_l$. If then the envelope $\bm{E}_0(t)$ is sufficiently slowly varying such that it is nearly constant on the interval of length $\epsilon$, we can state that all terms in $H_I^I(\tau)$ that are $\sim \exp(\pm i\omega_l \tau)$, $\exp[\pm i (\omega_l+\omega_x)\tau]$ can be neglected to a good approximation, as they average out in the integral in Eq.~\eqref{eq:U_I^I_linear}. We can then replace $H_I^I(\tau)$ by its RWA version $U_0^{\dagger}(\tau,t_0)H_{I,\text{RWA}}(\tau)U_0(\tau,t_0)$ in all slices in Eq.~\eqref{eq:U_I^I_slice}, if the approximation holds for all slices separately. Doing this for all slices implies that we can also simply replace $H_I^I(\tau)$ by $U_0^{\dagger}(\tau,t_0)H_{I,\text{RWA}}(\tau)U_0(\tau,t_0)$ in Eq.~\eqref{eq:U_I^I}, or equivalently approximate the total 2LS Hamiltonian as
\begin{equation}
	H_\mathrm{2LS}(t)\approx H_\mathrm{2LS,RWA}(t)=H_0+H_{I,\text{RWA}}(t)=\hbar\omega_x\ket{x}\bra{x}+\frac{\mathcal{A}(t)}{2}\ket{x}\bra{g}+\frac{\mathcal{A}^*(t)}{2}\ket{g}\bra{x}\,,\label{eq:H_2LS_RWA}
\end{equation}
where we introduce the shorthand notation
\begin{equation}
	\mathcal{A}(t)=\mathcal{A}_0(t)e^{-i\omega_l t}=-\bm{E}_0(t)\bra{x}\bm{d}\ket{g}e^{-i\omega_l t+i\phi}\,,\label{eq:def_eff_field}
\end{equation}
which describes the effective complex amplitude of the laser when interacting with the 2LS. From the previous derivation we can state that there are three requirements to the validity of the RWA:\\[-5mm]
\begin{itemize}
\item[(i)] The laser envelope $\bm{E}_0(t)$ has to be nearly constant on a laser period $2\pi/\omega_l$,\\[-5mm]
\item[(ii)] the laser driving has to be nearly resonant with $|\delta|=|\omega_l-\omega_x|\ll\omega_l,\omega_x$, and\\[-5mm]
\item[(iii)] the laser driving cannot be too strong, requiring the matrix elements of $\bm{E}_0(t)\cdot\bm{d}$ to be much smaller than the 2LS energy $\hbar\omega_x$.\\[-5mm]
\end{itemize}
FWM experiments can be designed such that these requirements are met (see Chap.~\ref{sec:exp}). In the case that the physical system under consideration has a well defined parity, such that the diagonal matrix elements $\bra{g}\bm{d}\ket{g}$ and $\bra{x}\bm{d}\ket{x}$ vanish anyway, requirement (ii) can be relaxed to $|\delta|\ll\omega_l+\omega_x$~\cite{cohen1998atom,cohen2022photons}.

In the following we will always use the 2LS Hamiltonian in Eq.~\eqref{eq:H_2LS_RWA} and drop the index "RWA" in all expressions like $H_{I,\mathrm{RWA}}$ understanding that the RWA is already applied. While the interaction picture with respect to $H_0=\hbar\omega_x\ket{x}\bra{x}$ was only used to clarify the requirements for the RWA in the previous discussion, we will often use this interaction picture to simplify either analytical or numerical calculations. The Schrödinger picture dynamics of the 2LS in Eq.~\eqref{eq:schroedinger_2LS} contain fast components due to the large energies $\hbar\omega_l$ and $\hbar\omega_x$, while the corresponding interaction picture equation of $\ket{\psi_I(t)}=U_0^{\dagger}(t,t_0)\ket{\psi(t)}=U_0^{\dagger}(t,t_0)U_\mathrm{2LS}(t,t_0)\ket{\psi(t_0)}=U_I^I(t,t_0)\ket{\psi_I(t_0)}$
\begin{equation}
	i\hbar\frac{\text{d}}{\text{d}t}\ket{\psi_I(t)}=i\hbar\frac{\text{d}}{\text{d}t}U_I^I(t,t_0)\ket{\psi_I(t_0)}=H_{I}^I(t)U_I^I(t,t_0)\ket{\psi_I(t_0)}=H_{I}^I(t)\ket{\psi_I(t)}\label{eq:schroedinger_interaction}
\end{equation} 
contains only the comparably small energies $\hbar\delta$ and $\mathcal{A}_0(t)$ describing the detuning and the laser interaction strength. If we had not applied the RWA, there would still be large energy scales like $\hbar(\omega_l+\omega_x)$ contained in the interaction picture Schrödinger equation. Typical values of $\hbar\omega_l$ are on the order of $\sim$eV in semiconductor systems like quantum dots~\cite{stufler2005quantum,ramsay2010damping,tartakovskii2012quantum,groll2020four}, corresponding to a timescale of about 1~fs, while typical relevant values of the detuning $\delta$ are on the order of meV. Therefore, calculating the dynamics of Eq.~\eqref{eq:schroedinger_interaction} numerically (with RWA applied) requires much less effort than calculating the dynamics of Eq.~\eqref{eq:schroedinger_2LS} numerically (without RWA applied), while yielding virtually the same information, if the RWA is applicable. This interaction picture representation of the dynamics is usually referred to as the frame rotating with the frequency $\omega_x$, a term which will be explained in the following Sec.~\ref{sec:bloch}. One could also consider alternative interaction pictures, e.g., separating $H_\mathrm{2LS}(t)$ into a free part $\tilde{H}_0=\hbar\omega_l\ket{x}\bra{x}$ and an interaction part $\tilde{H}_I(t)=H_\mathrm{2LS}(t)-\tilde{H}_0$. One would then speak of the frame rotating with the laser frequency $\omega_l$ in an analog fashion~\cite{cohen1998atom}.
\subsection{Optical Bloch equations and Bloch sphere}\label{sec:bloch}
The dynamics of the optically driven 2LS are often presented in terms of its excited state occupation $n_x(t)=\braket{\ket{x}\bra{x}}(t)$ and its polarization $p(t)=\braket{\ket{g}\bra{x}}(t)$. The dynamics of such operator expectation values $\braket{A}(t)=\bra{\psi(t)}A\ket{\psi(t)}$ can be determined via the Ehrenfest theorem~\cite{breuer2002theory}
\begin{equation}
	\frac{\text{d}}{\text{d}t}\braket{A}(t)=\frac{1}{i\hbar}\braket{\left[A, H_\mathrm{2LS}(t)\right]}(t)
\end{equation}
for not explicitly time-dependent operators $A$. Here, $\left[A,B\right]=AB-BA$ denotes the commutator between operators. Using the 2LS Hamiltonian in Eq.~\eqref{eq:H_2LS_RWA}, we can derive the so-called optical Bloch equations~\cite{haken1979light, allen1987optical,cohen1998atom}
\begin{subequations}\label{eq:bloch_eq}
\begin{align}
	\dvt n_x(t)&=-\frac{i}{\hbar}\left[\frac{\mathcal{A}(t)}{2}p^*(t)-\frac{\mathcal{A}^*(t)}{2}p(t)\right]=\frac{1}{\hbar}\text{Im}\left[\mathcal{A}(t)p^*(t)\right]\,,\\
	\dvt p(t)&=-i\omega_x p(t)-\frac{i}{\hbar}\frac{\mathcal{A}(t)}{2}\left[1-2n_x(t)\right]\,,
\end{align}
\end{subequations}
where we used that $n_g=\braket{\ket{g}\bra{g}}=1-\braket{\ket{x}\bra{x}}=1-n_x$ due to Eq.~\eqref{eq:id_2LS} and that $\braket{\ket{g}\bra{x}}^*=\braket{\ket{x}\bra{g}}$. We can rewrite this equation in terms of the Bloch vector~\cite{allen1987optical,cohen1998atom,haroche2006exploring}
\begin{equation}
	\bm{v}= \left(\begin{matrix}2\text{Re}(p)\\ 2\text{Im}(p)\\\Delta n \end{matrix}\right)\,,\label{eq:bloch_vector}
\end{equation}
where $\Delta n=2n_x-1=n_x-(1-n_x)=n_x-n_g=\braket{\ket{x}\bra{x}}-\braket{\ket{g}\bra{g}}$ is the so-called inversion of the 2LS, which takes on values between $-1$ and $1$ for $n_x=0$ ($\ket{x}$ unoccupied) and $n_x=1$ ($\ket{x}$ fully occupied), respectively. In terms of this vector the optical Bloch equations read
\begin{equation}
	\dvt \bm{v}(t)=\bm{v}(t)\times\bm{R}(t)\,,\qquad \bm{R}(t)=\frac{1}{\hbar}\left(\begin{matrix}\text{Re}\left[\mathcal{A}(t)\right]\\\text{Im}\left[\mathcal{A}(t)\right]\\\hbar\omega_x\end{matrix}\right)\,.\label{eq:bloch_eq_vector}
\end{equation}
Note that these optical Bloch equations are formally equivalent to the Bloch equations for spins in an external magnetic field~\cite{bloch1946nuclear}.

\begin{figure}[t]
	\centering
	\includegraphics[width = 0.72\textwidth]{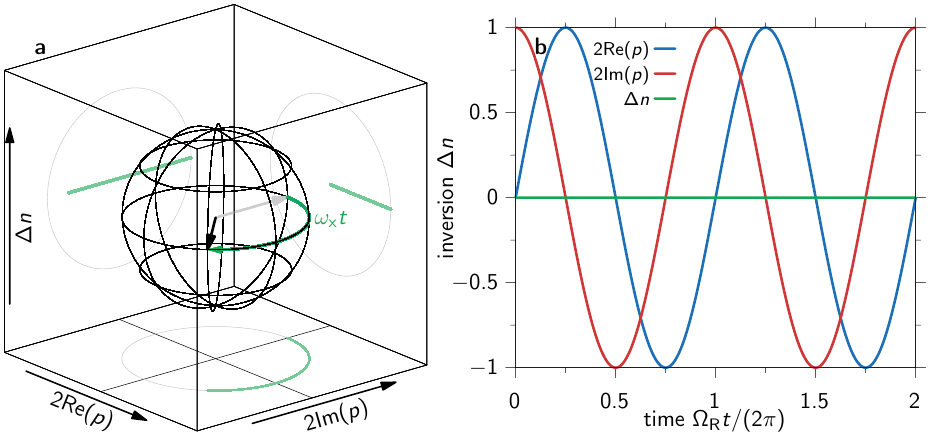}
	\caption{Free dynamics ($\mathcal{A}=0$) of the 2LS [see Eq.~\eqref{eq:2LS_free_dyn}]. (a) Free rotation of a Bloch vector for the initial superposition $\ket{\psi(t=0)}=(\ket{g}+i\ket{x})/\sqrt{2}$ (grey arrow) with the state $\ket{\psi(t)}$ as black arrow. Trajectory $\ket{\psi(0)}\rightarrow\ket{\psi(t)}$ in green. (b)~Components of the Bloch vector $\bm{v}$ in Eq.~\eqref{eq:bloch_vector} for the free dynamics in Eq.~\eqref{eq:2LS_free_dyn} with the inital state from (a).}
	\label{fig:Bloch_rot}
\end{figure}

The infinitesimal time evolution of the Bloch vector from time $t$ to time $t+\epsilon$ is thus given by an infinitesimal rotation around the axis defined by $\bm{R}(t)$. Importantly, this automatically implies that the length of the Bloch vector is conserved for all times since
\begin{equation}
	\dvt |\bm{v}(t)|^2=2\bm{v}(t)\cdot \dvt \bm{v}(t)= 2\bm{v}(t)\cdot\left[\bm{v}(t)\times\bm{R}(t)\right]=0\,.
\end{equation}
For a 2LS in the ground state $\ket{\psi(t_0)}=\ket{g}$ we obviously have $p(t_0)=\braket{\psi(t_0)|g}\braket{x|\psi(t_0)}=0$ and $n_x(t_0)=0$ analog, which implies $\bm{v}(t_0)=(0,0,-1)^T$. This means that the Bloch vector has unit length at time $t_0$ and thus unit length at all times for the optically driven 2LS. While this fact no longer holds when considering dissipation, for the moment it motivates us to introduces the so-called Bloch sphere~\cite{allen1987optical,haroche2006exploring}, a sphere in a vector space spanned by the axes defined in Eq.~\eqref{eq:bloch_vector}, whose surface is given by the condition $|\bm{v}|=1$. This sphere is shown in Fig.~\ref{fig:Bloch_rot}~(a). Its south pole at $\bm{v}=(0,0,-1)^T$ represents the ground state $\ket{g}$ of the 2LS with $n_x=0,\ p=0$, while its north pole at $\bm{v}=(0,0,1)^T$ represents the excited state $\ket{x}$ with $n_x=1,\ p=0$. The axis connecting south and north pole has the obvious property $p=0$, while at the equator the polarization takes on its maximum value of $|p|=1/2$ and the occupation is then $n_x=1/2$.

In the absence of optical driving with $\mathcal{A}=0$, the optical Bloch equations~\eqref{eq:bloch_eq} yield the solutions
\begin{equation}
	n_x(t)=n_x(t_0)\,,\qquad p(t)=e^{-i\omega_x(t-t_0)}p(t_0)\,\quad\text{for}\quad\mathcal{A}=0\,.\label{eq:2LS_free_dyn}
\end{equation}
In other words: the occupation of the excited state is conserved in absence of optically induced transitions, while the polarization rotates with the eigenfrequency $\omega_x$. In terms of the Bloch vector representation in Eq.~\eqref{eq:bloch_eq_vector} this clearly implies that any point on the Bloch sphere, representing the current state of the 2LS, rotates around the $z$-axis with the frequency $\omega_x$ as depicted in Fig.~\ref{fig:Bloch_rot}. We can of course transform into the co-rotating reference frame, defining
\begin{equation}
	\bar{p}^{(x)}(t)=e^{i\omega_xt}p(t)\,,\quad\bar{n}^{(x)}_x(t)=n_x(t)\,,\quad \bar{\bm{v}}^{(x)}(t)=\left(\begin{matrix}2\text{Re}\left[\bar{p}^{(x)}(t)\right]\\ 2\text{Im}\left[\bar{p}^{(x)}(t)\right]\\2\bar{n}^{(x)}_x(t)-1 \end{matrix}\right)\,.
\end{equation}
In terms of these new variables, the optical Bloch equations~\eqref{eq:bloch_eq} read
\begin{subequations}
	\begin{align}
		\dvt \bar{n}^{(x)}_x(t)&=\frac{1}{\hbar}\text{Im}\left[\mathcal{A}_0(t)e^{-i\delta t}\bar{p}^{(x)*}(t)\right]\,,\\
		\dvt \bar{p}^{(x)}(t)&=-\frac{i}{\hbar}\frac{\mathcal{A}_0(t)}{2}e^{-i\delta t}\left[1-2\bar{n}^{(x)}_x(t)\right]\,,
	\end{align}
		or equivalently
	\begin{align}
		\dvt \bar{\bm{v}}^{(x)}(t)&=\bar{\bm{v}}^{(x)}(t)\times \bar{\bm{R}}^{(x)}(t)\,,\qquad \bar{\bm{R}}^{(x)}(t)=\frac{1}{\hbar}\left(\begin{matrix}\text{Re}\left[\mathcal{A}_0(t)e^{-i\delta t}\right]\\\text{Im}\left[\mathcal{A}_0(t)e^{-i\delta t}\right]\\0\end{matrix}\right)\,,
	\end{align}
\end{subequations}
where $\delta=\omega_l-\omega_x$ is again the detuning between laser excitation and 2LS transition. In this co-rotating frame, both the occupation and the polarization are stationary in the absence of optical driving $\mathcal{A}_0=0$, such that points on the Bloch sphere do not rotate around the $z$-axis anymore. This co-rotating frame is of course nothing else than the interaction picture with respect to $H_0$ discussed at the end of the previous Sec.~\ref{sec:dipole_RWA}, justifying the name \textit{rotating frame} given there. This can be seen from (setting $t_0=0$)
\begin{align}
	\braket{\psi_I(t)|g}\braket{x|\psi_I(t)}&=\bra{\psi(t)}U_0(t,0)\ket{g}\bra{x}U_0^{\dagger}(t,0)\ket{\psi(t)}=e^{i\omega_xt}\braket{\psi(t)|g}\braket{x|\psi(t)}\notag\\
	&=e^{i\omega_xt}p(t)=\bar{p}^{(x)}(t)
\end{align}
and analog for the occupation. Alternatively, we can introduce the frame rotating with the laser frequency $\omega_l$, which corresponds to the interaction picture with respect to $\tilde{H}_0=\hbar\omega_l\ket{x}\bra{x}$
\begin{subequations}\label{eq:bloch_rot_laser}
\begin{align}
	\bar{p}^{(l)}(t)&=e^{i\omega_lt}p(t)\,,\quad\bar{n}^{(l)}_x(t)=n_x(t)\,,\quad \bar{\bm{v}}^{(l)}(t)=\left(\begin{matrix}2\text{Re}\left[\bar{p}^{(l)}(t)\right]\\ 2\text{Im}\left[\bar{p}^{(l)}(t)\right]\\2\bar{n}^{(l)}_x(t)-1 \end{matrix}\right)\,,\\
	\dvt\bar{\bm{v}}^{(l)}(t)&=\bar{\bm{v}}^{(l)}(t)\times \bar{\bm{R}}^{(l)}(t)\,,\qquad \bar{\bm{R}}^{(l)}(t)=\frac{1}{\hbar}\left(\begin{matrix}\text{Re}\left[\mathcal{A}_0(t)\right]\\\text{Im}\left[\mathcal{A}_0(t)\right]\\-\hbar\delta\end{matrix}\right)\,.
\end{align}
\end{subequations}
The virtue of this reference frame is that the rotation axis $\bar{R}^{(l)}$ only depends on time via the envelope $\mathcal{A}_0(t)$ of the exciting light field and becomes stationary for continuous wave excitation with $\mathcal{A}_0(t)=$ const.

Finally we note that the Bloch equations~\eqref{eq:bloch_eq} have a stationary solution with $\dvt n_x=\dvt p=0$, independent of whether there is optical driving or not. This is given by the case $n_x=1/2$ and $p=0$ and therefore describes the center of the Bloch sphere in Fig.~\ref{fig:Bloch_rot}~(a). With the dynamics of the optically driven 2LS described so far, this point cannot be reached purely by optical driving since it does not correspond to a unit length Bloch vector. We will later see that such a situation, where the 2LS becomes transparent, i.e., its dynamics cannot be influenced by the light field anymore and become stationary, can occur when dissipation is included in the description of the 2LS~\cite{forstner2003phonon,kruegel2005role}.
\subsection{Different forms of excitations}
\subsubsection{Continuous wave excitation}\label{sec:cw_nodiss}
\begin{figure}[h]
	\centering
	\includegraphics[width = 0.72\textwidth]{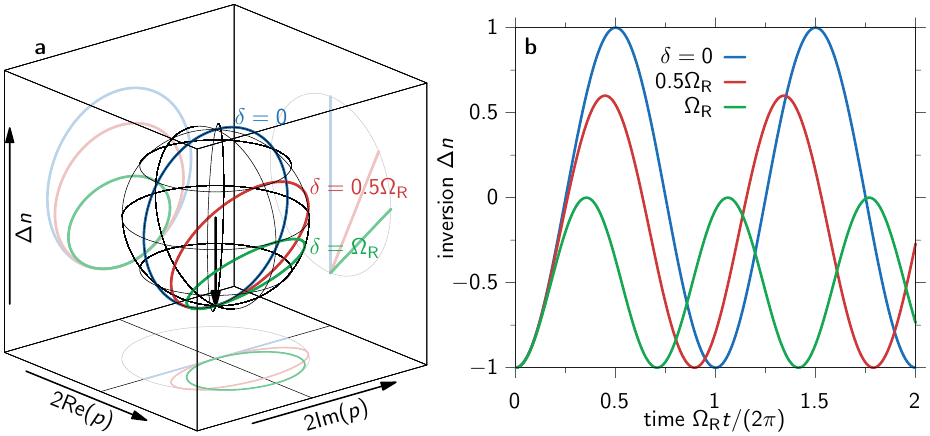}
	\caption{Rabi oscillations due to cw excitation in the frame rotating with the laser frequency $\omega_l$ [see Eqs.~\eqref{eq:bloch_rot_laser} and \eqref{eq:laser_cw}]. (a)~Bloch vector dynamics for different detunings $\delta$ with colored lines marking the trajectories (blue: $\delta=0$, red: $\delta=0.5\Omega_\mathrm{R}$, green: $\delta=\Omega_\mathrm{R}$) and a 2LS starting in the ground state (black arrow). (b) Dynamics of the inversion $\Delta n$ for different detunings with colors corresponding to the trajectories in (a).}
	\label{fig:Bloch_Rabi}
\end{figure}

The simplest form of excitation is given by continuous wave (cw) excitation with
\begin{equation}
	\mathcal{A}_0(t)=\hbar \Omega_\mathrm{R} e^{i\phi}\label{eq:laser_cw}\,,
\end{equation}
where $\phi$ is an arbitrary but constant phase and $\Omega_\mathrm{R}\geq 0$ denotes the (resonant) Rabi frequency, i.e., the amplitude of the optical excitation, which drives the 2LS continuously with constant intensity. In the frame rotating with the laser frequency~$\omega_l$ in Eqs.~\eqref{eq:bloch_rot_laser}, this implies that the axis of rotation for the Bloch vector is constant and given by
\begin{equation}
	\bar{\bm{R}}^{(l)}=\left(\begin{matrix}\Omega_\mathrm{R}\cos(\phi)\\\Omega_\mathrm{R}\sin(\phi)\\-\delta\end{matrix}\right)\,.\label{eq:cw_axis}
\end{equation}
A constant axis of rotation implies periodic dynamics and the frequency of this rotation is given by the length of the rotation vector $\bar{\bm{R}}^{(l)}$
\begin{equation}
	|\bar{\bm{R}}^{(l)}|=\sqrt{\Omega_\mathrm{R}^2+\delta^2}\,.\label{eq:cw_axis_length}
\end{equation}
Figure~\ref{fig:Bloch_Rabi} shows these periodic dynamics of the 2LS for an initial ground state $\ket{g}$ and different values of the detuning with (a) showing the trajectory on the Bloch sphere (colored lines) and (b) showing the dynamics of the inversion $\Delta n$ [colored lines corresponding to trajectories in (a)].

In the case of vanishing detuning $\delta=0$ (blue) the axis of rotation in Eq.~\eqref{eq:cw_axis} lies in the x-y-plane, such that the trajectory in (a) runs through the north pole, if it starts in the south pole. In other words: for vanishing detuning the excitation is most efficient and we can reach maximum inversion $\Delta n=1$ (the state $\ket{x}$), as shown in (b). When increasing the detuning $\delta$ (red and green lines in Fig.~\ref{fig:Bloch_Rabi}), the axis of rotation tilts stronger towards the z-direction with increasing $\delta$. This leads to trajectories whose minimum distance to the north pole increases with detuning, i.e., we cannot excite the 2LS as effectively anymore, since the laser frequency $\omega_l$ is not in resonance with the 2LS frequency $\omega_x$. This implies that the maximum inversion decreases with increasing detuning, as can be seen in Fig.~\ref{fig:Bloch_Rabi}~(b).
Finally, the frequency of the so-called Rabi oscillations~\cite{allen1987optical,cohen1998atom,forstner2003phonon,kruegel2005role,wigger2018rabi} in Fig.~\ref{fig:Bloch_Rabi} (b) increases with the detuning, as can already be seen from the length of the rotation vector in Eq.~\eqref{eq:cw_axis_length}, which is equivalent to the rotation frequency.

Note that due to the spherical symmetry of the Bloch sphere, we can rotate in the x-y-plane to effectively choose $\phi=0$, which has been done in Fig~\ref{fig:Bloch_Rabi}. This corresponds to a unitary transformation on the 2LS Hilbert space with the operator $U(\phi)=e^{i\phi\ket{x}\bra{x}}$, such that the polarization transforms as $p=\braket{\psi|g}\braket{x|\psi}\rightarrow \bra{\psi}U(\phi)^{\dagger}\ket{g}\bra{x}U(\phi)\ket{\psi}=e^{i\phi}p$, while the occupation remains unchanged. This effectively removes the constant phase $\phi$ in Eq.~\eqref{eq:laser_cw} from the Bloch equations~\eqref{eq:bloch_eq}. The rotation vector in Eq.~\eqref{eq:cw_axis} then reads $\bar{\bm{R}}^{(l)}=(\Omega_\mathrm{R},0,-\delta)^T$ in this rotated frame of reference. As long as we are dealing with a single excitation, even if it is pulsed and not cw (see following Sec.~\ref{sec:2LS_pulsed}), we can always remove an overall constant phase in this fashion. Note however, that this does not imply that we can remove all relative phases when dealing with multiple pulses. Indeed such relative phases will become important in later chapters on wave mixing. In the present case though, removing the overall phase allows us to simplify the calculation of the solution of the Bloch equations~\eqref{eq:bloch_rot_laser} in the frame rotating with the laser frequency, which for cw excitation from Eq.~\eqref{eq:laser_cw} and phase $\phi=0$ read
\begin{equation}
	\frac{\text{d}}{\text{d}t}\left(\begin{matrix}2\text{Re}\left[\bar{p}^{(l)}(t)\right]\\ 2\text{Im}\left[\bar{p}^{(l)}(t)\right]\\\Delta\bar{n}^{(l)}(t) \end{matrix}\right)=\left(\begin{matrix}-\delta 2\text{Im}\left[\bar{p}^{(l)}(t)\right]\\ \delta 2\text{Re}\left[\bar{p}^{(l)}(t)\right]+\Omega_\mathrm{R}\Delta\bar{n}^{(l)}(t)\\-\Omega_\mathrm{R} 2\text{Im}\left[\bar{p}^{(l)}(t)\right] \end{matrix}\right)
\end{equation}
with $\Delta\bar{n}^{(l)}=2\bar{n}^{(l)}_x-1$ denoting the inversion in this rotating frame. Calculating the second derivative of $2\text{Im}\left[\bar{p}^{(l)}(t)\right]$ yields a harmonic oscillator differential equation with the frequency $|\bar{\bm{R}}^{(l)}|$ given in Eq.~\eqref{eq:cw_axis_length}. The solution can then be inserted in the equations of motion for $2\text{Re}\left[\bar{p}^{(l)}(t)\right]$ and $\Delta\bar{n}^{(l)}(t)$ from above which are solved by simple integration. Considering the initial conditions $\bar{p}^{(l)}(t=0)=0$ and $\Delta\bar{n}^{(l)}(t=0)=-1$ for $\ket{\psi(t=0)}=\ket{g}$, as well as $|\bar{\bm{v}}^{(l)}|=1$, we obtain the Bloch vector dynamics for cw excitation
\begin{equation}
	\bar{\bm{v}}^{(l)}(t)=\left(\begin{matrix}2\text{Re}\left[\bar{p}^{(l)}(t)\right]\\ 2\text{Im}\left[\bar{p}^{(l)}(t)\right]\\\Delta\bar{n}^{(l)}(t) \end{matrix}\right)=\left(\begin{matrix}\frac{\delta\Omega_\mathrm{R}}{\Omega_\mathrm{R}^2+\delta^2}\left[1-\cos\left(\sqrt{\Omega_\mathrm{R}^2+\delta^2}t\right)\right]\\-\frac{\Omega_\mathrm{R}}{\sqrt{\Omega_\mathrm{R}^2+\delta^2}}\sin\left(\sqrt{\Omega_\mathrm{R}^2+\delta^2}t\right)\\\frac{\Omega_\mathrm{R}^2}{\Omega_\mathrm{R}^2+\delta^2}\left[1-\cos\left(\sqrt{\Omega_\mathrm{R}^2+\delta^2}t\right)\right]-1 \end{matrix}\right)
\end{equation}
which provides the solutions shown in Fig.~\ref{fig:Bloch_Rabi}, i.e., Rabi oscillations.
\subsubsection{Pulsed excitation}\label{sec:2LS_pulsed}
\begin{figure}[t]
	\centering
	\includegraphics[width = 0.55\textwidth]{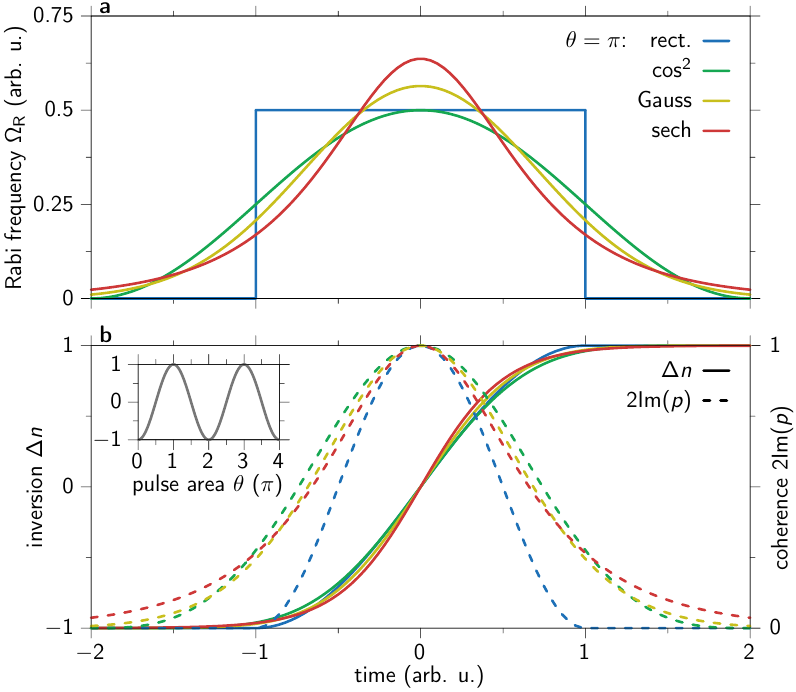}
	\caption{Impact of the pulse shape for a full inversion with $\theta=\pi$ and resonant excitation with $\delta=0$. (a) Pulse dynamics for different pulse shapes and identical pulse areas $\theta=\pi$. (b) Corresponding inversion (solid) and polarization (dashed) dynamics. Inset: Rabi rotations, i.e., value of the inversion $\Delta n$ after a pulse with varying pulse area $\theta$ for an initial value of $\Delta n=-1$.}
	\label{fig:Rabi_pulses}
\end{figure}

We can also consider a situation, where the optical driving is switched on and off again, such that the pulse envelope $\mathcal{A}_0(t)$ is truly time-dependent in contrast to cw excitation in Eq.~\eqref{eq:laser_cw}. As discussed previously, we can ignore any global constant phase and define
\begin{equation}
	\mathcal{A}_0(t)=\hbar\Omega_\mathrm{R}(t)
\end{equation}
with $\Omega_\mathrm{R}(t)\in\mathbb{R}$ now being the so-called time-dependent Rabi frequency. This time-dependence now implies that the rotation vector
\begin{equation}
	\bar{\bm{R}}^{(l)}(t)=\left(\begin{matrix}\Omega_\mathrm{R}(t)\\0\\-\delta\end{matrix}\right)\label{eq:rotation_pulse}
\end{equation}
is time-dependent. However, for the special case of vanishing detuning $\delta=0$, it always points in the x-direction, but the instantaneous frequency of rotation changes. This implies that the system undergoes a rotation similar to the case of cw excitation with vanishing detuning [blue curve in Fig.~\ref{fig:Bloch_Rabi} (a)], however it starts at some initial time, for which $\Omega_\mathrm{R}(t\leq t_1)=0$ and ends after the pulse has reached $\Omega_\mathrm{R}(t\geq t_2>t_1)=0$ again. Since the instantaneous angular frequency is given by $\Omega_\mathrm{R}(t)$, the total angle that the trajectory of the Bloch vector covers on the Bloch sphere is given by
\begin{equation}
	\theta=\left|\,\int\limits_{-\infty}^{\infty}\text{d}t\,\Omega_\mathrm{R}(t)\right|\,,\qquad \delta=0\,.\label{eq:pulse_area}
\end{equation}
This angle is called the pulse area~\cite{mccall1969self,allen1987optical,ramsay2010damping}. We can derive this result also by considering the interaction picture time evolution operator in Eq.~\eqref{eq:U_I^I}. For vanishing detuning and real $\mathcal{A}_0(t)$ the interaction Hamiltonian in the interaction picture has the form [see Eq.~\eqref{eq:H_2LS_RWA}]
\begin{equation}
	H_I^I(t)=\frac{\mathcal{A}_0(t)}{2}(\ket{x}\bra{g}+\ket{g}\bra{x})\label{eq:H_I^I_res}
\end{equation}
and therefore commutes with itself at different times, such that we can drop the time-ordering operator $\hat{T}$ in Eq.~\eqref{eq:U_I^I}, leading to
\begin{equation}
	U_I^I(\infty,-\infty)=\exp\left[-\frac{i}{\hbar}\int\limits_{-\infty}^{\infty}\text{d}\tau\,H_I^I(\tau)\right]=\exp\left[-\frac{i}{2}(\ket{x}\bra{g}+\ket{g}\bra{x})\int\limits_{-\infty}^{\infty}\text{d}\tau\,\Omega_\mathrm{R}(\tau)\right]\label{eq:U_I_I_pulse}\,.
\end{equation}
Using $(\ket{x}\bra{g}+\ket{g}\bra{x})^2=\mathbb{1}_{\rm 2LS}$ and the series expansion of the exponential function, we can separate this expression into sine and cosine terms, obtaining
\begin{equation}
	U_I^I(\infty,-\infty)=\cos\left(\frac{\theta}{2}\right)\mathbb{1}_\mathrm{2LS} - i\sin\left(\frac{\theta}{2}\right)(\ket{x}\bra{g}+\ket{g}\bra{x})\,,
\end{equation}
where we assumed that the integral over $\Omega_\mathrm{R}(t)$ is positive semidefinite and thus equal to the pulse area $\theta$. Note that this is not a restriction, since we can always assume that we removed any global phase, including an overall minus sign, via rotation of the Bloch sphere. The interaction picture corresponds to the frame rotating with the 2LS frequency $\omega_x$, which for vanishing detuning $\delta=\omega_l-\omega_x=0$ considered here coincides with the frame rotating with the laser frequency, which has been used before to derive Eq.~\eqref{eq:pulse_area}. In this frame, the full time evolution of the 2LS that is being represented by the Bloch vector is exactly given by $\ket{\psi_I(\infty)}=U_I^I(\infty,-\infty)\ket{\psi_I(-\infty)}$. When considering the specific example of $\ket{\psi_I(-\infty)}=\ket{g}$, we can quickly see that the pulse area $\theta$ that appears in this alternative derivation clearly corresponds to the angle that the trajectory covers throughout the pulse
\begin{equation}
	U_I^I(\infty,-\infty)\ket{g}=\cos\left(\frac{\theta}{2}\right)\ket{g}-i\sin\left(\frac{\theta}{2}\right)\ket{x}\,.\label{eq:U_I^I_pa_theorem}
\end{equation}
For an angle of $\theta=\pi$ we end up in the excited state, covering exactly half of the blue trajectory in Fig.~\ref{fig:Bloch_Rabi}. For an angle of $\theta=2\pi$ we are back in the ground state. The result that the pulse area $\theta$ as defined in Eq.~\eqref{eq:pulse_area} describes the angle that is covered by the trajectory on the Bloch sphere, independent of the specific pulse shape, is known as the pulse area theorem~\cite{allen1987optical}. One will often encounter terms like "$\pi$-pulse" in the literature, implying that the intensity of a pulse is chosen in a way that $\theta=\pi$, leading to full inversion in the case of vanishing detuning~\cite{holthaus1994generalized, stievater2001rabi, wigger2018rabi}. Figure~\ref{fig:Rabi_pulses} shows examples of different shapes of $\pi$-pulses in (a) with corresponding dynamics of the inversion $\Delta n$ and the polarization $p$ in (b) for resonant excitation. In all cases the system clearly reaches full inversion with $\Delta n=1$ after starting in the ground state with $\Delta n=-1$, irrespective of the specific pulse shape that is used. The inset in Fig.~\ref{fig:Rabi_pulses}~(b) shows the inversion $\Delta n$ after excitation from the ground state $\ket{g}$ ($\Delta n=-1$) as a function of the pulse area $\theta$ of the exciting laser pulse. We find sinusoidal oscillations as a function of the pulse area [see Eq.~\eqref{eq:U_I^I_pa_theorem}]. These oscillations with respect to the parameter $\theta$ are called Rabi \textit{rotations}~\cite{zrenner2002coherent, ramsay2010damping, mccutcheon2010quantum, grisard2022multiple}, in contrast to the temporally-resolved Rabi \textit{oscillations} in the case of cw excitation in Fig.~\ref{fig:Bloch_Rabi}~(b).

Note that for non-vanishing detuning we cannot find such a simple result as in Eqs.~\eqref{eq:pulse_area} and \eqref{eq:U_I^I_pa_theorem}. As in the derivation above, we can use two perspectives to understand this: (i) The rotation axis in Eq.~\eqref{eq:rotation_pulse} changes direction with time, if the detuning is non-vanishing. At each instant, the change of the Bloch vector can be understood as a rotation around the z-axis with fixed frequency $\delta$ and around the x-axis with varying frequency $\Omega_\mathrm{R}(t)$. Since two rotations do not commute in three dimensions, this makes it hard to keep track of the Bloch vector and to make any simple analytical prediction on where it will land after the pulse. In contrast to this, the rotation axis does not change direction for vanishing detuning or in the case of cw excitation in Eq.~\eqref{eq:cw_axis}, such that in these cases we could make simple analytical predictions on the dynamics of the Bloch vector. (ii) For non-vanishing detuning the interaction Hamiltonian in the interaction picture has the form
\begin{equation}
	H_I^I(t)=\frac{\mathcal{A}_0(t)}{2}\left(e^{-i\delta t}\ket{x}\bra{g}+e^{i\delta t}\ket{g}\bra{x}\right)
\end{equation}
and does not commute with itself at different times in contrast to the resonant case in Eq.~\eqref{eq:H_I^I_res}. This implies that we cannot ignore the time-ordering operator $\hat{T}$ in the time evolution operator $U_I^I$ from Eq.~\eqref{eq:U_I^I}, which makes it generally hard to find a simple analytical solution to the time evolution of the 2LS.
\subsubsection{Ultrashort pulse limit}\label{sec:delta_pulses}
A very important special case where analytical solutions are still possible is that of ultrashort pulsed excitation, modeled via $\delta$-pulses with
\begin{equation}
	\mathcal{A}(t)=\hbar\theta e^{i\phi} \delta(t-t_0)\,,\label{eq:delta_pulse}
\end{equation}
where $\theta\geq 0$ is the pulse area, $\phi$ is an arbitrary phase and $\delta(t)$ denotes the $\delta$-function. Note that such pulses are infinitely narrow in time and therefore infinitely broad in frequency space. This means that their carrier frequency is arbitrary, implying that we can (or better should) always interpret them as resonant excitations with vanishing detuning $\delta=0$. This hints towards the fact that the pulse area theorem always holds for ultrashort pulses. We can easily see this, when calculating the full time evolution operator of the 2LS around the time of excitation
\begin{align}
	\lim\limits_{\epsilon\rightarrow 0}U_\mathrm{2LS}(t_0+\epsilon, t_0-\epsilon)&=\lim\limits_{\epsilon\rightarrow 0}\exp\left\lbrace-\frac{i}{\hbar}\int\limits_{t_0-\epsilon}^{t_0+\epsilon}\text{d}\tau\,\left[H_0+H_I(\tau)\right]\right\rbrace \notag\\
	&=\exp\left[-\frac{i}{2}\theta\left(\ket{x}\bra{g} e^{i\phi}+\ket{g}\bra{x} e^{-i\phi}\right)\right]\,,\label{eq:U_P_deriv}
\end{align}
where we dropped the contribution from $H_0$ in the limit $\epsilon\rightarrow 0$ and used Eq.~\eqref{eq:H_2LS_RWA} together with Eq.~\eqref{eq:delta_pulse}. This time evolution operator for the ultrashort pulse excitation at time $t_0$ is of the same form as the interaction picture time evolution operator for an arbitrary pulse in Eq.~\eqref{eq:U_I_I_pulse} (apart from the additional phase). We can evaluate the exponential in an analogous fashion to obtain
\begin{equation}
	U_P(\theta,\phi)=\lim\limits_{\epsilon\rightarrow 0}U_\mathrm{2LS}(t_0+\epsilon, t_0-\epsilon)=\cos\left(\frac{\theta}{2}\right)\mathbb{1}_\mathrm{2LS}-i\sin\left(\frac{\theta}{2}\right)\left(\ket{x}\bra{g}e^{i\phi}+\ket{g}\bra{x}e^{-i\phi}\right)\,,\label{eq:U_P}
\end{equation}
which describes how a single $\delta$-pulse at time $t_0$ transforms the 2LS. At any other time the 2LS Hamiltonian is simply given by $H_0$, implying that the system evolves freely, apart from the impact of the pulse at time $t_0$. In terms of the excited state occupation $n_x$ and polarization $p$, this pulse transformation reads
\begin{subequations}\label{eq:delta_pulse_transform}
\begin{align}
	p_+&=\bra{x}U_P\ket{\psi_-}\bra{\psi_-}U_P^{\dagger}\ket{g}=\cos^2\left(\frac{\theta}{2}\right) p_-+\frac{i}{2}e^{i\phi}\sin(\theta) (2n_{x-}-1)+\sin^2\left(\frac{\theta}{2}\right)e^{2i\phi}p_-^*\,,\\
	n_{x+}&=\bra{x}U_P\ket{\psi_-}\bra{\psi_-}U_P^{\dagger}\ket{x}=n_{x-}+\sin^2\left(\frac{\theta}{2}\right)(1-2n_{x-})-\sin(\theta) \text{Im}\left(e^{-i\phi}p_-\right)\,,
\end{align}
\end{subequations}
where $\alpha_\pm=\lim\limits_{\epsilon\rightarrow 0} \alpha(t_0\pm\epsilon)$ with $\alpha=p,n_x,\psi$ denote quantities right after/before the ultrashort pulse. In terms of the Bloch vectors right after/before the pulse, we obtain the transformation
\begin{align}
	{\bm v}_+ = \begin{pmatrix}
		2\cos^2(\phi)\sin^2\left(\frac{\theta}{2}\right) + \cos(\theta) & \sin(2\phi) \sin^2\left(\frac{\theta}{2}\right) & -\sin(\phi)\sin(\theta) \\
		\sin(2\phi)\sin^2\left(\frac{\theta}{2}\right)	& 2\sin^2(\phi)\sin^2\left(\frac{\theta}{2}\right) + \cos(\theta) & \cos(\phi)\sin(\theta) \\
		\sin(\phi)\sin(\theta) & -\cos(\phi)\sin(\theta) & \cos(\theta)
	\end{pmatrix}
	{\bm v}_-\,,
\end{align}
which describes a general rotation in three dimensions around the axis $\bm{R}=(-\cos(\phi),-\sin(\phi),0)^T$ with a total angle~$\theta$~\cite{morawiec2003orientations}.

\begin{figure}[t]
	\centering
	\includegraphics[width = 0.34\textwidth]{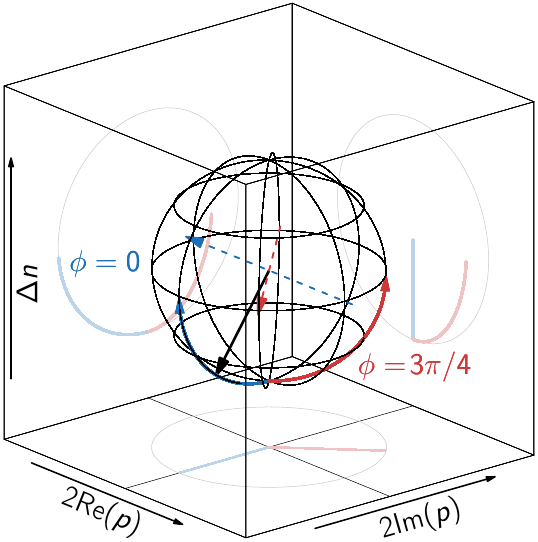}
	\caption{Examples of Bloch vector rotations induced by ultrashort pulses with pulse areas of $\theta = \pi/2$ and pulse phases of $\phi=0$ (blue) and $\phi=3\pi/4$ (red). The curved colored arrows indicate the rotation of the Bloch vector around the rotation axes (dashed arrows) in the respective color.}
	\label{fig:pi_2}
\end{figure}

Figure~\ref{fig:pi_2} illustrates the action of a single $\delta$-pulse with pulse area $\theta=\pi/2$. Both displayed examples rotate the Bloch vector from the ground state $\left| g\right>$, i.e., the south pole of the Bloch sphere, into an equal superposition of $\ket{g}$ and $\ket{x}$ on the equator of the Bloch sphere. The final orientation depends on the pulse phase $\phi$, i.e., the rotation axis $\bm{R}$ (dashed arrows). There, the case of $\phi=0$ (blue) leads to a rotation around the $-2$Re$(p)$ axis, while in the $\phi=3\pi/4$ case (red) the Bloch vector ${\bm v}$ rotates around a diagonal in the 2Re$(p)$-2Im$(p)$ plane.

If we recall the requirements for the validity of the RWA in Sec.~\ref{sec:dipole_RWA}, we see that one necessary ingredient was a slowly varying pulse envelope, which is strictly speaking not satisfied for $\delta$-pulses. We can however interpret these pulses as a mathematical limiting case~\cite{vagov2002electron}, e.g., of a Gaussian pulse of width $\sigma\gg 2\pi/\omega_l$. If the optical driving is sufficiently weak, the RWA is clearly satisfied, as long as the detuning $\delta$ is not too large. The detuning on the other hand quantifies the internal timescale of the 2LS. An ultrashort pulse is now given by the case $\sigma\ll\delta^{-1}$, i.e., the pulse is much quicker than any internal timescale of the optically driven 2LS. If the RWA is fulfilled for such a Gaussian pulse, we can approximate this Gaussian pulse by a $\delta$-pulse \textit{after} applying the RWA. This approximation then only requires $\sigma\ll\delta^{-1}$, while the RWA is assumed to still be correct in any physical realization of the $\delta$-pulse, which of course has a finite temporal width. In this sense an ultrashort $\delta$-pulse is a mathematical idealization, which can be used to strongly simplify analytical calculations even for systems in which the RWA has already been applied.
\subsection{Linear absorption spectrum}\label{sec:abs_spec}
Now that we have established some fundamental results on optical driving of a 2LS, we can investigate our first spectroscopic signal. Before dealing with FWM spectroscopy in a later section however, we want to start with a simpler optical signal: the linear absorption spectrum~\cite{schmitt1987theory,mahan2000many,krummheuer2002theory}. As is well known from classical electrodynamics in matter~\cite{griffiths2005introduction}, the linear absorption spectrum in extended media is directly related to the imaginary part of the linear susceptibility. We will derive here a similar relation also for single 2LSs. To do so, we make the following assumptions: (i) The 2LS is excited by a laser pulse of arbitrary length but small pulse area, i.e., weak integrated intensity, such that we can perform perturbation theory in orders of the light field. (ii) The occupation $n_x$ after the excitation is measured via the detection of a photon that is emitted after spontaneous decay of the excited state $\ket{x}\rightarrow\ket{g}$. While until now technically speaking we did not include such a dissipative process in our model and the occupation $n_x$ is stationary after optical excitation [see Eq.~\eqref{eq:2LS_free_dyn}], it is clear that a 2LS that can be excited by a laser pulse also couples to the continuum of surrounding light modes into which it can decay spontaneously~\cite{breuer2002theory}. We will discuss the details of such dissipative processes in the following Sec.~\ref{sec:theory_dissipation}. Right now we simply use our knowledge that such a process has to happen in any physical realization of an optically driven 2LS to argue that $n_x(\infty)$ is our observable for measuring how well the 2LS absorbed a weak laser pulse of known carrier frequency. This kind of spectroscopy method is usually referred to as photoluminescence excitation~\cite{yu2010fundamentals, wigger2019phonon,preuss2022resonant}.

We consider the 2LS initially in its ground state $\ket{g}=0$, i.e., $p(-\infty)=n_x(-\infty)=0$. From the discussion on the free time evolution in Eq.~\eqref{eq:2LS_free_dyn} we know that the system remains in this state until it is disturbed by the laser pulse. From the Bloch equations~\eqref{eq:bloch_eq} we can then conclude that the lowest order contribution to the polarization is linear in the light field $\mathcal{A}$, making the occupation $n_x$ at least quadratic in $\mathcal{A}$. Linearization of the Bloch equations~\eqref{eq:bloch_eq} yields
\begin{subequations}\label{eq:bloch_eq_linear}
	\begin{align}
		\label{eq:n_x_lin}\dvt n_x(t)&=\frac{1}{\hbar}\text{Im}\left[\mathcal{A}(t)p^*(t)\right]\,,\\
		\dvt p(t)&\approx-i\omega_x p(t)-\frac{i}{\hbar}\frac{\mathcal{A}(t)}{2}\,,
	\end{align}
\end{subequations}
where we dropped $n_x$ in the differential equation for $p$. We can solve the linearized equation for $p$ by integration
\begin{equation}
	p(t)=-\frac{i}{2\hbar}\int\limits_{-\infty}^{t}\text{d}\tau\,\mathcal{A}(\tau)e^{-i\omega_x(t-\tau)}=-(\mathcal{A}*\chi)(t)\,,\label{eq:2LS_p_linear}
\end{equation}
which we have written as a convolution between the effective laser field $\mathcal{A}$ and the linear susceptibility of the 2LS~\cite{mahan2000many,krummheuer2002theory,wigger2019phonon}
\begin{equation}
	\chi(t)=\frac{i}{2\hbar}e^{-i\omega_x t}\Theta(t)\,,\label{eq:chi_nodiss}
\end{equation}
where $\Theta(t)$ is the Heaviside function, which vanishes for $t<0$ and takes on unit value for $t\geq 0$. This function makes sure that the upper integration limit is automatically set to $t$ in the integral in Eq.~\eqref{eq:2LS_p_linear}. Note that we defined the effective laser field $\mathcal{A}(t)$ in Eq.~\eqref{eq:def_eff_field} with an additional minus sign for notational convenience in the interaction Hamiltonian $H_I(t)$, such that $-\mathcal{A}\sim \bra{x}\bm{d}\ket{g}\cdot\bm{E}$ actually describes the effective field acting on the 2LS, which is accounted for by the overall sign in Eq.~\eqref{eq:2LS_p_linear}. We can understand the linear susceptibility $\chi(t)$ as the response of the 2LS to an ultrashort optical excitation at $t=0$ with $\mathcal{A}(t)=-\mathcal{A}_0\delta(t)$. When inserted into Eq.~\eqref{eq:2LS_p_linear}, this yields $p(t)=\mathcal{A}_0\chi(t)$.

We can now use these results to determine the occupation after the pulse by integrating Eq.~\eqref{eq:n_x_lin}
\begin{equation}
	n_x(\infty)=\frac{1}{\hbar}\text{Im}\left[\int\limits_{-\infty}^{\infty}\text{d}t\,\mathcal{A}(t)p^*(t)\right]=\frac{1}{\hbar}\text{Im}\left[\int\limits_{-\infty}^{\infty}\text{d}t\int\limits_{-\infty}^{\infty}\text{d}\tau\,\mathcal{A}^*(t)\chi(t-\tau)\mathcal{A}(\tau)\right]\,,
\end{equation}
where we used $\text{Im}(z)=-\text{Im}(z^*)$. This result can be brought into a more intuitive form by Fourier transformation via 
\begin{equation}
	f(t)=\frac{1}{2\pi}\int\limits_{-\infty}^{\infty}\text{d}\omega\,\tilde{f}(\omega)e^{-i\omega t}\,,\qquad f=\mathcal{A},\chi
\end{equation}
to obtain
\begin{align}
		n_x(\infty)&=\frac{1}{\hbar (2\pi)^3}\text{Im}\left[\,\int\limits_{\mathbb{R}^5}\text{d}t\text{d}\tau\text{d}\omega_1\text{d}\omega_2\text{d}\omega_3\,\tilde{\mathcal{A}}^*(\omega_1)\tilde{\chi}(\omega_2)\tilde{\mathcal{A}}(\omega_3)e^{i\omega_1 t}e^{-i\omega_3\tau}e^{-i\omega_2(t-\tau)}\right]\notag\\
		&=\frac{1}{2\pi\hbar}\text{Im}\left[\,\int\limits_{\mathbb{R}^3}\text{d}\omega_1\text{d}\omega_2\text{d}\omega_3\,\tilde{\mathcal{A}}^*(\omega_1)\tilde{\chi}(\omega_2)\tilde{\mathcal{A}}(\omega_3)\delta(\omega_1-\omega_2)\delta(\omega_2-\omega_3)\right]\notag\\
		&=\frac{1}{2\pi\hbar}\int\limits_{-\infty}^{\infty}\text{d}\omega\,|\tilde{\mathcal{A}}(\omega)|^2\text{Im}\left[\tilde{\chi}(\omega)\right]\label{eq:2LS_n_x_absorb}\,,
\end{align}
where we used that the integral over $\exp(i\omega t)\text{d}t$ yields $2\pi\delta(\omega)$, i.e., it is a representation of the $\delta$-function. We see that the occupation after optical excitation is given by the overlap integral between the intensity spectrum of the pulse $|\tilde{\mathcal{A}}(\omega)|^2$ and the imaginary part of the Fourier transform of the linear susceptibility $\text{Im}\left[\tilde{\chi}(\omega)\right]$. In a similar way to the imaginary part of the linear susceptibility being responsible for absorption in classical electrodynamics in matter~\cite{griffiths2005introduction}, it is here providing the strength of absorption at a frequency $\omega$, which is then weighted by the intensity of the incoming light at that frequency $|\tilde{\mathcal{A}}(\omega)|^2$ and summed over all frequencies to give a measure for the overall absorption strength of the driven 2LS. The term $\text{Im}\left[\tilde{\chi}(\omega)\right]$ is obviously an inherent property of the 2LS as it does not relate to the external excitation. Since we considered weak excitation, this motivates us to call
\begin{align}
	\text{Im}\left[\tilde{\chi}(\omega)\right]&=\text{Im}\left[\int\limits_{-\infty}^{\infty}\text{d}t\,\chi(t)e^{i\omega t}\right]=\frac{1}{2\hbar}\text{Re}\left[\int\limits_0^{\infty}\text{d}t\,e^{i(\omega-\omega_x) t}\right]\notag\\
	&=\frac{1}{4\hbar}\left[\int\limits_0^{\infty}\text{d}t\,e^{i(\omega -\omega_x)t}+\int\limits_0^{\infty}\text{d}t\,e^{-i(\omega-\omega_x) t}\right]\notag\\
	&=\frac{1}{4\hbar}\int\limits_{-\infty}^{\infty}\text{d}t\,e^{i(\omega -\omega_x)t}=\frac{\pi}{2\hbar}\delta(\omega-\omega_x)\label{eq:lin_abs_nodiss}
\end{align}
the linear absorption spectrum of the 2LS~\cite{mahan2000many,krummheuer2002theory,wigger2019phonon}. It is proportional to a $\delta$-function which has its peak at the resonance frequency $\omega_x$ of the 2LS, i.e., the 2LS absorbs exclusively at $\omega=\omega_x$. This is nothing else than a representation of energy conservation, which has to be strictly satisfied here since we calculated the occupation at $t\rightarrow\infty$ as a measure for absorption. Using Eq.~\eqref{eq:2LS_n_x_absorb} it is obvious that the occupation is determined by how strong the 2LS is driven at its resonance frequency, i.e., $n_x(\infty)\sim |\tilde{\mathcal{A}}(\omega_x)|^2$. 
\subsection{Dissipation in the optically driven 2LS}\label{sec:theory_dissipation}
So far we have considered an ideal dissipation-less optically driven 2LS, i.e., the described dynamics were purely unitary and could be calculated in terms of the time evolution operator $U_\mathrm{2LS}$ in Eq.~\eqref{eq:U_2LS}. Any physical realization of an optically driven 2LS, e.g., an exciton in a quantum dot or a color center in a solid, will however be subject to dissipative processes induced by its environment~\cite{breuer2002theory,krummheuer2002theory,forstner2003phonon,preuss2022resonant}. In the following we will consider the two most important types of dissipation for the 2LS: spontaneous decay (possibly radiative) and pure dephasing.
\subsubsection{Density matrix formalism}
First of all we have to reformulate our mathematical description of the 2LS. While we used pure states described by ket-vectors $\ket{\psi}$ so far, it is not possible to apply this description when including dissipation. In such cases the density matrix formalism is required, which is a natural extension of quantum mechanics to mixed states~\cite{breuer2002theory}. A density matrix
\begin{equation}
	\rho=\sum_i P_i\ket{\psi_i}\bra{\psi_i}\label{eq:def_rho}
\end{equation}
describes a situation where we do not know with certainty, in which quantum mechanical state $\ket{\psi_i}$ the system is in, such that we can only give classical probabilities $P_i$ for the possible quantum states. Note that the states $\ket{\psi_i}$ do not need to be orthogonal in this representation. This classical uncertainty can for example be introduced when a small system couples to a large environment whose complicated dynamics are not tracked, such that we lose information on the total system. Nonetheless, we can still give probabilities for the states that the small system can be in. In such a situation it is also typical to encounter dissipation, such that density matrices are ideal for describing dissipative or open quantum systems~\cite{breuer2002theory}. The probabilities $P_i$ are (i)~real, (ii)~non-negative, and (iii)~add up to one, which imposes the three requirements of (i) hermiticity, (ii) positive semi-definiteness, and (iii) normalization
\begin{subequations}\label{eq:prop_rho}
\begin{align}
	\text{(i)}&\quad \rho^{\dagger}=\rho\,,\\
	\text{(ii)}&\quad \bra{\psi}\rho\ket{\psi}\geq 0\quad\forall\ket{\psi}\,,\\
	\text{(iii)}&\quad \text{Tr}(\rho)=1\,
\end{align}
\end{subequations}
on arbitrary density matrices $\rho$, where $\text{Tr}$ denotes the operator trace, i.e., the sum over the diagonal in its matrix representation.

Calculating mean values (expectation values) of operators with the density matrix works via a straightforward extension of the pure state case. In any pure state $\ket{\psi_i}$ the mean value of the operator $A$ is given by $\bra{\psi_i}A\ket{\psi_i}$. Since we only know with a certain probability $P_i$ that the system is in the state $\ket{\psi_i}$ we have to take the additional average with respect to these classical probabilities to obtain the mean value
\begin{align}
	\braket{A}=\sum_i P_i\bra{\psi_i}A\ket{\psi_i}&=\sum_n\sum_i P_i\bra{\psi_i}A\ket{\phi_n}\braket{\phi_n|\psi_i}\notag\\
		&=\sum_n \bra{\phi_n}\sum_i\ket{\psi_i}P_i\bra{\psi_i}A\ket{\phi_n}=\text{Tr}(\rho A)\,,\label{eq:average_density}
\end{align}
where we used an arbitrary orthonormal basis $\left\lbrace \ket{\phi_n}\right\rbrace$ and inserted the corresponding completeness relation $\mathbb{1}=\sum_n\ket{\phi_n}\bra{\phi_n}$, as well as the definition of the trace
\begin{equation}
	\text{Tr}(B)=\sum_n\bra{\phi_n}B\ket{\phi_n}\,.
\end{equation}

We can represent the density matrix, which is an operator on the considered Hilbert space, as an actual matrix in a chosen basis. If we do this for the 2LS in the basis given by $\ket{g}$ and $\ket{x}$, we retrieve the $2\times 2$ matrix
\begin{equation}
	\rho\hat{=}\left(\begin{matrix}\bra{g}\rho\ket{g}&\bra{g}\rho\ket{x}\\\bra{x}\rho\ket{g}&\bra{x}\rho\ket{x}\end{matrix}\right)=\left(\begin{matrix}\braket{\ket{g}\bra{g}}&\braket{\ket{x}\bra{g}}\\\braket{\ket{g}\bra{x}}&\braket{\ket{x}\bra{x}}\end{matrix}\right)\,,
\end{equation}
where in the last step we wrote the matrix elements as mean values of projection and transition operators, using Eq.~\eqref{eq:average_density}. We see that the matrix elements of the density matrix in this basis are given by the quantities used in Sec.~\ref{sec:bloch} in order to calculate the Bloch Eqs.~\eqref{eq:bloch_eq}, but now generalized to the case of mixed states, i.e., the ground state occupation $n_g=\braket{\ket{g}\bra{g}}$, the excited state occupation $n_x=\braket{\ket{x}\bra{x}}$, and the polarization $p=\braket{\ket{g}\bra{x}}$. The properties of the density matrix in Eqs.~\eqref{eq:prop_rho} ensure that (i) $n_g, n_x \in\mathbb{R}$, $\braket{\ket{x}\bra{g}}^*=\braket{\ket{g}\bra{x}}=p$, (ii) $n_g, n_x\geq 0$, and (iii) $n_g+n_x=1$. With this we can write the 2LS density matrix as
\begin{equation}
	\rho\hat{=}\left(\begin{matrix}n_g&p^*\\p&n_x\end{matrix}\right)=\left(\begin{matrix}1-n_x&p^*\\p&n_x\end{matrix}\right)\,,\label{eq:rho_2LS}
\end{equation}
showing that the polarization and excited state occupation completely define any mixed state of the 2LS.

Note that the length of the Bloch vector defined in Eq.~\eqref{eq:bloch_vector} is given by
\begin{equation}
	|\bm{v}|=\sqrt{4|p|^2+(2n_x-1)^2}\,.
\end{equation}
In the pure state case this length was always unity. In the general mixed state case we can use the positive semi-definiteness of the density matrix [see Eq.~\eqref{eq:prop_rho}(ii)] to argue that its length has to lie between 0 and 1. This positive semi-definiteness implies that all eigenvalues of the density matrix are positive, while the normalization $\text{Tr}(\rho)=1$ implies that they cannot exceed unity. We can now calculate the eigenvalues $\lambda$ of the general 2LS density matrix in Eq.~\eqref{eq:rho_2LS}, which yields
\begin{equation}
	\lambda_\pm=\frac{1}{2}\pm\sqrt{\frac{1}{4}+|p|^2+n_x(n_x-1)}=\frac{1}{2}\pm\frac{1}{2}|\bm{v}|\in[0,1]\,.
\end{equation}
Since these eigenvalues have to lie between $0$ and $1$, the Bloch vector length $|\bm{v}|$ also lies between $0$ and $1$, i.e., the Bloch vector is confined to the volume within the Bloch sphere (including its surface). We thus get the restriction
\begin{equation}\label{eq:restriction_p_n}
	0\leq 4|p|^2+(2n_x-1)^2 \leq 1
\end{equation}
when it comes to the possible values of occupation $n_x$ and polarization $p$. We have to keep this in mind when introducing dissipation to the system. 

Having defined the density matrix, we can now investigate its equation of motion in detail. In the case of purely unitary dynamics there is no information loss in the system and the classical probabilities $P_i$ for the different quantum states do not change. This implies that the time evolution of the density matrix in Eq.~\eqref{eq:def_rho} is determined by the time evolution of the ket and bra, which evolve via the time evolution operator $U_\mathrm{2LS}(t,t_0)$ from Eq.~\eqref{eq:U_2LS} in this case of a 2LS, i.e.,
\begin{equation}
	\rho(t)=U^{}_\mathrm{2LS}(t,t_0)\rho(t_0)U^{\dagger}_\mathrm{2LS}(t,t_0)\label{eq:U_rho_U}\,,
\end{equation}
where the time evolution operator on the left propagates the ket, while its adjoint on the right propagates the bra in Eq.~\eqref{eq:def_rho}. Calculating the time derivative of this expression, using the (general) properties of the time evolution operator in Eq.~\eqref{eq:properties_U} yields the von Neumann equation for our optically driven 2LS
\begin{equation}\label{eq:von_neumann}
	\dvt \rho(t)=-\frac{i}{\hbar}\left[H_\mathrm{2LS}(t),\rho(t)\right]\,,
\end{equation}
which is the generalization of the Schrödinger equation to density matrices~\cite{breuer2002theory}. This means that this result is still fully equivalent to the pure state description used earlier, i.e., from this equation we can derive the same Bloch equations~\eqref{eq:bloch_eq}.

At this point it is useful to discuss the terms \textit{occupation} and \textit{coherence}, as they are widely used in the literature on spectroscopy~\cite{demtroder2014laser1,demtroder2015laser2,cundiff2008coherent,smallwood2018multidimensional}. These are the diagonal and off-diagonal elements of the density matrix, respectively. However, since matrix elements depend on the chosen basis, we have to specify this as well. A useful choice is the energy eigenbasis of the system under investigation, e.g., the basis spanned by $\ket{g}$ and $\ket{x}$ in the case of a 2LS. In this sense, $n_g$ and $n_x$ are the occupations of the states $\ket{g}$ and $\ket{x}$, respectively. The polarization $p$ is then the coherence between $\ket{g}$ and $\ket{x}$. Since there are no other independent off-diagonal matrix elements, the terms polarization and coherence are often used interchangeably in the context of 2LSs. Choosing the energy eigenbasis as the basis with respect to which we define occupations and coherences implies that the free evolution of occupations is stationary. Coherences however oscillate in time with the energy difference of the two basis states used for calculating the off-diagonal matrix element [see Eq.~\eqref{eq:2LS_free_dyn}].
\subsubsection{Bloch equations with dissipation}
The simplest way to add dissipation to the system is by using phenomenological damping rates, changing the Bloch equations~\eqref{eq:bloch_eq} to
\begin{subequations}\label{eq:bloch_eq_diss}
	\begin{align}
		\dvt n_x(t)&=\frac{1}{\hbar}\text{Im}\left[\mathcal{A}(t)p^*(t)\right]-\gamma_\mathrm{xd}n_x(t)\,,\\
		\dvt p(t)&=-i\omega_x p(t)-\frac{i}{\hbar}\frac{\mathcal{A}(t)}{2}\left[1-2n_x(t)\right]-\gamma p(t)\label{eq:p_diss}\,.
	\end{align}
\end{subequations}
Here we have introduced two separate rates $\gamma_\mathrm{xd}$ and $\gamma$ for the occupation and polarization, respectively. If we naively just introduced a decay rate for the occupation (setting $\gamma=0$), in the absence of optical driving ($\mathcal{A}=0$) we would always end up with $n_x=0$, independent of the specific initial value of the polarization $p$. As a concrete example, consider an equal superposition $\ket{\psi}=(\ket{g}+\ket{x})/\sqrt{2}$, e.g., prepared by a $\pi/2$-pulse. Then the Bloch vector initially points to the equator of the Bloch sphere, as in Fig.~\ref{fig:pi_2}, i.e., $|p|=n_x=1/2$. If we had $\gamma=0$ and a non-vanishing decay of the excited state $\gamma_\mathrm{xd}>0$, we would end up with $|p|=1/2$ and $n_x=0$, which is a point outside of the Bloch sphere, which violates the restriction derived in Eq.~\eqref{eq:restriction_p_n}. From this we get a first hint that an excited state decay with the rate $\gamma_\mathrm{xd}>0$ has to imply a polarization decay with a rate $\gamma>0$. 

Note that we consider 2LSs with transition energies in the optical regime $\hbar\omega_x\sim1$~eV. Even at room temperature with $k_BT\sim 25$~meV we can therefore safely assume that the system thermalizes to its ground state with $n_x=0$, which is the final state of Eqs.~\eqref{eq:bloch_eq_diss} without optical driving. In systems with transition energies on the order of $k_BT$, e.g., spin systems in magnetic fields~\cite{bloch1946nuclear}, the Bloch equations would have to account for thermalization to a finite occupation of the excited state $n_x\neq 0$. In the extreme limit of $k_BT\gg\hbar\omega_x$ the final state would be $n_x=0.5$, $p=0$, i.e., the center of the Bloch sphere, instead of the south pole with $n_x=0$ as in Eqs.~\eqref{eq:bloch_eq_diss}.

To establish a more direct relation between the rates $\gamma_\mathrm{xd}$ and $\gamma$ we investigate how the equation of motion for the Bloch vector in Eq.~\eqref{eq:bloch_eq_vector} is impacted by these, yielding
\begin{equation}
	\dvt\bm{v}=\bm{v}(t)\times\bm{R}(t)-\underline{\Gamma}\bm{v}(t)-\bm{\gamma}_0\,,\qquad \underline{\Gamma}=\left(\begin{matrix}\gamma&0&0\\0&\gamma&0\\0&0&\gamma_\mathrm{xd}\end{matrix}\right)\,,\qquad \bm{\gamma}_0=\left(\begin{matrix} 0\\0\\\gamma_\mathrm{xd}\end{matrix}\right)\,,
\end{equation}
with $\bm{R}(t)$ being the same rotation vector as in Eq.~\eqref{eq:bloch_eq_vector}. The matrix $\underline{\Gamma}$ and the vector $\bm{\gamma}_0$ combined describe the impact of dissipation. The vector $\bm{\gamma}_0$ appears here since dissipation of course also has to preserve the trace of the density matrix [see Eq.~\eqref{eq:def_rho}(iii)]. While $n_x$ is damped with the rate $\gamma_\mathrm{xd}$, $n_g$ in fact increases with the rate $\gamma_\mathrm{xd}$ to ensure $n_g+n_x=1$ for all times. This implies that the inversion $\Delta n=n_x-n_g=2n_x-1$ itself, i.e., the third component of the Bloch vector, cannot be damped only via the matrix $\underline{\Gamma}$. The length of the Bloch vector now evolves as
\begin{equation}
	\dvt |\bm{v}(t)|^2=2\bm{v}(t)\cdot\dvt\bm{v}(t)=-2\bm{v}(t)\cdot[\underline{\Gamma}\bm{v}(t)]-2\bm{v}(t)\cdot\bm{\gamma}_0\,.
\end{equation}
Assuming that we start in a pure state with $|\bm{v}|=1$ anywhere on the surface of the Bloch sphere, the length of the Bloch vector must not be increased by dissipation, otherwise we do not deal with a well-defined mixed state anymore. This implies that for every vector on the surface of the Bloch sphere we have
\begin{align}
	-2\bm{v}\cdot (\underline{\Gamma}\bm{v})-2\bm{v}\cdot\bm{\gamma}_0&\leq 0\,,\qquad |\bm{v}|=1\,\notag\\
	\Leftrightarrow -2\Delta n (\gamma_\mathrm{xd}-\gamma)\Delta n -2\gamma|\bm{v}|^2-2\Delta n\gamma_\mathrm{xd}&\leq 0\notag\\
	\Leftrightarrow 2\gamma_\mathrm{xd}(-\Delta n-\Delta n^2)&\leq 2\gamma(1-\Delta n^2)\notag\\
	\Leftrightarrow \gamma_\mathrm{xd}\frac{\Delta n^2+\Delta n}{\Delta n^2-1}&\leq\gamma\,,
\end{align}
where in the last step we used that $\Delta n\in[-1,1]$ and thus $1-\Delta n^2\geq 0$. The left-hand side has a well-defined maximum at $\Delta n=-1$ and tends to $-\infty$ for $\Delta n\rightarrow 1$, such that
\begin{equation}
	\lim\limits_{\Delta n\rightarrow -1} \gamma_\mathrm{xd}\frac{\Delta n^2+\Delta n}{\Delta n^2-1}=\frac{\gamma_\mathrm{xd}}{2}\leq\gamma\,.
\end{equation}
This implies that the polarization has to decay with at least half the rate of the excited state decay. The polarization, i.e., the coherence of the 2LS, is connected to the phase relation between ground and excited state. Such a decay of the polarization is therefore also called decoherence or dephasing. While the excited state decay could for example stem from a process in which a photon is spontaneously emitted, which changes the occupation of the 2LS, there can also be mechanisms where the coupling of a 2LS to an environment does not change its occupation, e.g., because no matching frequencies $\sim\omega_x$ are provided by the environment to induce a transition $\ket{x}\rightarrow\ket{g}$. These situations can however still lead to dephasing, which we then call pure dephasing~\cite{muljarov2004dephasing,machnikowski2006change,groll2021controlling}. We can quantify any additional decay of the polarization, beyond what is induced by the excited state decay (xd), via the so-called pure dephasing (pd) rate $\gamma_\mathrm{pd}$ as
\begin{equation}\label{eq:def_gamma}
	\gamma=\frac{\gamma_\mathrm{xd}+\gamma_\mathrm{pd}}{2}\,.
\end{equation}
In the literature on optically active 2LSs one will often encounter, with varying notation and convention~\cite{haken1979light, allen1987optical,mukamel1995principles, shah2013ultrafast}, the relaxation times $T_1$, describing the lifetime of the excited state, and $T_2$, describing the decay of the polarization. Depending on the convention being used, these can be related to the rates in our notation via $\gamma_\mathrm{xd}=1/T_1$ and $\gamma=1/T_2$. In addition a pure dephasing time $T_2'$ can be defined via~\cite{shah2013ultrafast}
\begin{equation}
	\frac{2}{T_2}=\frac{2}{T_2'}+\frac{1}{T_1}\qquad\Leftrightarrow\qquad T_2'=\frac{2}{\gamma_\mathrm{pd}}\,.
\end{equation}
\subsubsection{Lindblad equation for the Bloch equations with dissipation}
As is commonly done in the literature~\cite{breuer2002theory,haroche2006exploring,roy2011influence,kasprzak2013coherence,groll2020four,wigger2021resonance}, we can also rephrase the dissipation rates in Eqs.~\eqref{eq:bloch_eq_diss} in terms of an equation of motion for the density matrix, modifying the von~Neumann equation~\eqref{eq:von_neumann}. This leads to the so-called Lindblad equation which can in principle be derived or motivated from underlying microscopic theories~\cite{breuer2002theory}. Dissipative processes therein are described by so-called Lindblad dissipators
\begin{equation}\label{eq:D_L}
	\mathcal{D}_L(\rho)=\gamma_L\left(L\rho L^{\dagger}-\frac{1}{2}\left\lbrace L^{\dagger}L,\rho\right\rbrace\right)\,,
\end{equation}
where $\lbrace A,B \rbrace=AB+BA$ is the anticommutator of two operators. Here, $\gamma_L$ is the rate associated with the dissipative process in which a transition $\rho\rightarrow L\rho L^{\dagger}$ is induced by the operator $L$.

Excited state decay with the rate $\gamma_\mathrm{xd}$ is modeled via $L=\ket{g}\bra{x}$, i.e., excited state decay leads to a transition from the excited state to the ground state. This yields the dissipator~\cite{breuer2002theory,kasprzak2013coherence,groll2020four}
\begin{equation}
	\mathcal{D}_\mathrm{xd}(\rho)=\gamma_\mathrm{xd}\left(\ket{g}\bra{x}\rho\ket{x}\bra{g}-\frac{1}{2}\left\lbrace\ket{x}\bra{x},\rho\right\rbrace\right)\,.
\end{equation}
Adding this to the right-hand side of Eq.~\eqref{eq:von_neumann} modifies the Bloch equations without dissipation~\eqref{eq:bloch_eq}. The additional terms can be calculated via
\begin{align}
	\left.\dvt \rho(t)\right|_\mathrm{xd}\hat{=}	\left.\left(\begin{matrix}\dvt n_g&\dvt p^*\\\dvt p&\dvt n_x\end{matrix}\right)\right|_\mathrm{xd}&=\left(\begin{matrix}\bra{g}\mathcal{D}_\mathrm{xd}(\rho)\ket{g}&\bra{g}\mathcal{D}_\mathrm{xd}(\rho)\ket{x}\\\bra{x}\mathcal{D}_\mathrm{xd}(\rho)\ket{g}&\bra{x}\mathcal{D}_\mathrm{xd}(\rho)\ket{x}\end{matrix}\right)\notag\\
	&=\gamma_\mathrm{xd}\left(\begin{matrix}n_x&-p^*/2\\-p/2&-n_x\end{matrix}\right)\hat{=}\mathcal{D}_\mathrm{xd}(\rho)\,.\label{eq:action_xd}
\end{align}
Adding this dissipator to the von Neumann equation~\eqref{eq:von_neumann} thus accounts for the rates $\gamma_\mathrm{xd}$ and $\gamma=\gamma_\mathrm{xd}/2$ in the Bloch equations with dissipation~\eqref{eq:bloch_eq_diss}. Such a dissipator obviously impacts the probabilities $P_i$ in Eq.~\eqref{eq:def_rho}. If we start in a completely mixed state with $\rho=\frac{1}{2}\ket{g}\bra{g}+\frac{1}{2}\ket{x}\bra{x}$, i.e., $P_x=1/2$ and $P_g=1/2$, we will eventually end up in the ground state with $\rho=\ket{g}\bra{g}$ in the absence of optical driving, i.e., a pure state with $P_x=0$ and $P_g=1$. The dynamics induced by this dissipator are therefore non-unitary, i.e., they cannot be represented in a way analog to Eq.~\eqref{eq:U_rho_U}. We will visualize the impact of the decay dissipator in the following Sec.~\ref{sec:theory_dissipation_vis}.

While the excited state decay dissipator can only describe situations where the dephasing is induced by decay, i.e., $\gamma=\gamma_\mathrm{xd}/2$, we can also include a second dissipator to describe pure dephasing with the rate $\gamma_\mathrm{pd}$. This is done by choosing the transition operator in Eq.~\eqref{eq:D_L} as $L=\ket{x}\bra{x}$, i.e., the occupation of the TLS is not changed by pure dephasing, but we will see that this still impacts the polarization. The pure dephasing dissipator with the rate $\gamma_\mathrm{pd}$ is then given by~\cite{roy2011influence,groll2021controlling,preuss2022resonant}
\begin{equation}
	\mathcal{D}_\mathrm{pd}(\rho)=\gamma_\mathrm{pd}\left(\ket{x}\bra{x}\rho\ket{x}\bra{x}-\frac{1}{2}\left\lbrace\ket{x}\bra{x},\rho\right\rbrace\right)\,.
\end{equation}
Its action on the density matrix elements can be calculated analog to Eq.~\eqref{eq:action_xd} as
\begin{align}
	\left.\dvt \rho(t)\right|_\mathrm{pd}\hat{=}	\left.\left(\begin{matrix}\dvt n_g&\dvt p^*\\\dvt p&\dvt n_x\end{matrix}\right)\right|_\mathrm{pd}&=\left(\begin{matrix}\bra{g}\mathcal{D}_\mathrm{pd}(\rho)\ket{g}&\bra{g}\mathcal{D}_\mathrm{pd}(\rho)\ket{x}\\\bra{x}\mathcal{D}_\mathrm{pd}(\rho)\ket{g}&\bra{x}\mathcal{D}_\mathrm{pd}(\rho)\ket{x}\end{matrix}\right)\notag\\
	&=\gamma_\mathrm{pd}\left(\begin{matrix}0&-p^*/2\\-p/2&0\end{matrix}\right)\hat{=}\mathcal{D}_\mathrm{pd}(\rho)\,.\label{eq:action_pd}
\end{align}
The occupations of excited state and ground state do not change, while the polarization decays with an additional rate of $\gamma_\mathrm{pd}/2$. Depending on the physical realization of the 2LS, pure dephasing can sometimes dominate over excited state decay~\cite{allen1987optical,webb1991stimulated, batalov2008temporal,mermillod2016dynamics,wigger2019phonon}. We will visualize this dissipation process in the following Sec.~\ref{sec:theory_dissipation_vis}.

Adding both dissipators to Eq.~\eqref{eq:von_neumann} yields the Lindblad equation of the optically driven 2LS, subject to pure dephasing and decay
\begin{equation}\label{eq:lindblad}
	\dvt \rho(t)=-\frac{i}{\hbar}\left[H_\mathrm{2LS}(t),\rho(t)\right]+\mathcal{D}_\mathrm{xd}\left[\rho(t)\right]+\mathcal{D}_\mathrm{pd}\left[\rho(t)\right]\,,
\end{equation}
which is equivalent to the Bloch equations with dissipation~\eqref{eq:bloch_eq_diss} with the dephasing rate $\gamma$ in Eq.~\eqref{eq:def_gamma} containing the appropriate contributions from both excited state decay and pure dephasing. Note that both of these dissipators preserve the general density matrix properties from Eqs.~\eqref{eq:prop_rho}, which is generally true for Lindblad equations~\cite{lindblad1976generators,breuer2002theory}.
\subsubsection{Impact of dissipation on Bloch vector dynamics}\label{sec:theory_dissipation_vis}
\begin{figure}[t]
	\centering
	\includegraphics[width = 0.29\textwidth]{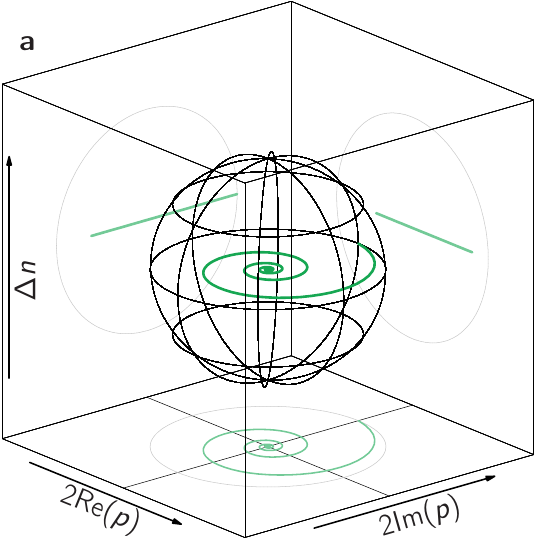}
	\includegraphics[width = 0.29\textwidth]{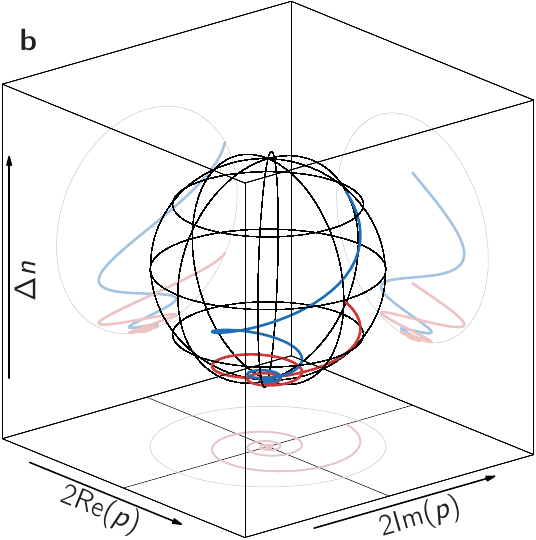}
	\includegraphics[width = 0.39\textwidth]{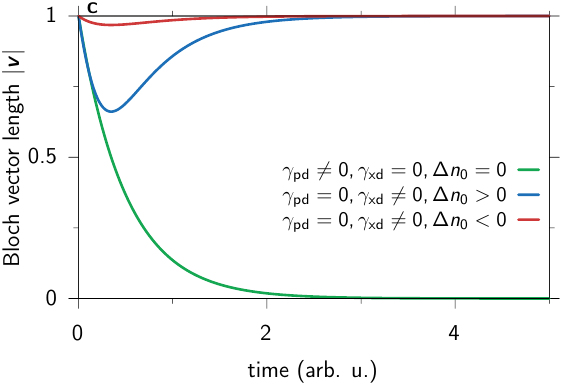}
	\caption{(a) Bloch vector dynamics for pure dephasing. (b) Bloch vector dynamics for excited state decay with two different initial inversions. (c) Bloch vector lengths according to (a) and (b).}
	\label{fig:Bloch_dissipation}
\end{figure}
Now we want to investigate how excited state decay and pure dephasing in, e.g., Eqs.~\eqref{eq:bloch_eq_diss}, modify the dynamics of the Bloch vector. To this aim, Fig.~\ref{fig:Bloch_dissipation} now shows the free dynamics ($\mathcal{A}=0$) of the Bloch vector. 

In (a) we start with an equal superposition of $\ket{g}$ and $\ket{x}$, i.e., a pure state on the equator of the Bloch sphere with maximum polarization $|p|=1/2$. There we consider vanishing excited state decay $\gamma_\mathrm{xd}=0$ and some non-vanishing pure dephasing rate $\gamma_\mathrm{pd}>0$. The dynamics are shown in the frame not rotating with the 2LS frequency $\omega_x$, i.e., the Bloch vector still rotates around the center as in Fig.~\ref{fig:Bloch_rot}, but due to the dephasing it spirals inwards until it reaches the point $n_x=1/2$, $p=0$. Note that, while the Bloch vector had unit length in the pure state case from Fig.~\ref{fig:Bloch_rot}, its length can now decrease. In Fig.~\ref{fig:Bloch_dissipation}~(b) we have a similar situation, but with vanishing pure dephasing rate $\gamma_\mathrm{pd}=0$ and non-vanishing decay rate $\gamma_\mathrm{xd}>0$ for two initial pure states above (blue) and below (red) the equator, respectively. The Bloch vector now falls towards the south pole (ground state $\ket{g}$), which together with the rotation of the Bloch sphere with frequency~$\omega_x$ leads to a downwards spiral. In (c) the Bloch vector length is shown for both cases. While in the situation from (a) (green) the length simply decreases until it reaches zero, in (b) (red, blue) it initially decreases but then increases until it reaches unit length again. In other words: pure dephasing generally decreases the purity of the quantum state, i.e., it tends to produce strongly mixed states like the 50/50 mixture of $\ket{g}$ and $\ket{x}$ in (a). Excited state decay on the other hand can also increase the purity, as it acts to bring the system to the pure state $\ket{g}$ at the south pole of the Bloch sphere. How strong the purity is decreased by pure dephasing depends on the initial state however. If we had started in the north or south pole in (a), the system would remain in that pure state and pure dephasing would have no impact. 
\subsubsection{Impact of dissipation on optical driving}
To study the impact of dissipation on the optical driving of a 2LS, we first consider cw excitation as in Sec.~\ref{sec:cw_nodiss} with
\begin{equation}
	\mathcal{A}(t)=\hbar\Omega_\mathrm{R} e^{-i\omega_lt}\,.
\end{equation}
Going to the frame rotating with the laser frequency, analog to Eqs.~\eqref{eq:bloch_rot_laser}, we transform the Bloch equations~\eqref{eq:bloch_eq_diss} with dissipation to
\begin{subequations}\label{eq:bloch_diss_cw}
	\begin{align}
		\dvt \bar{n}^{(l)}_x(t)&=-\Omega_\mathrm{R}\text{Im}\left[\bar{p}^{(l)}(t)\right]-\gamma_\mathrm{xd}\bar{n}^{(l)}_x(t)\,,\\
		\dvt \bar{p}^{(l)}(t)&=i\delta \bar{p}^{(l)}(t)-\frac{i}{2}\Omega_\mathrm{R}\left[1-2\bar{n}^{(l)}_x(t)\right]-\frac{\gamma_\mathrm{xd}+\gamma_\mathrm{pd}}{2}\bar{p}^{(l)}(t)\,,
	\end{align}
\end{subequations}
which we can use to derive a stationary state focusing on the case of resonant excitation with $\delta=\omega_l-\omega_x=0$. This stationary state is reached when the derivatives on the left hand side vanish, such that
\begin{equation}
	\text{stationary state:}\quad\left[2\bar{n}^{(l)}_x-1\right]=\frac{\gamma_\mathrm{xd}+\gamma_\mathrm{pd}}{i\Omega_\mathrm{R}}\bar{p}^{(l)}\,,\qquad \text{Im}\left[\bar{p}^{(l)}\right]=-\frac{\gamma_\mathrm{xd}}{\Omega_\mathrm{R}}\bar{n}^{(l)}_x\,. \label{eq:rabi_stationary_1}
\end{equation}
The first condition implies that the polarization is purely imaginary, leading to
\begin{equation}
	\text{stationary state:}\quad\left[2\bar{n}^{(l)}_x-1\right]=-\frac{\gamma_\mathrm{xd}+\gamma_\mathrm{pd}}{\Omega_\mathrm{R}}\frac{\gamma_\mathrm{xd}}{\Omega_\mathrm{R}}\bar{n}^{(l)}_x\Rightarrow n_x^{(l)}=\left(2+\frac{\gamma_\mathrm{xd}+\gamma_\mathrm{pd}}{\Omega_\mathrm{R}}\frac{\gamma_\mathrm{xd}}{\Omega_\mathrm{R}}\right)^{-1}\,. \label{eq:rabi_stationary}
\end{equation}

\begin{figure}[t]
	\centering
	\includegraphics[width=0.75\linewidth]{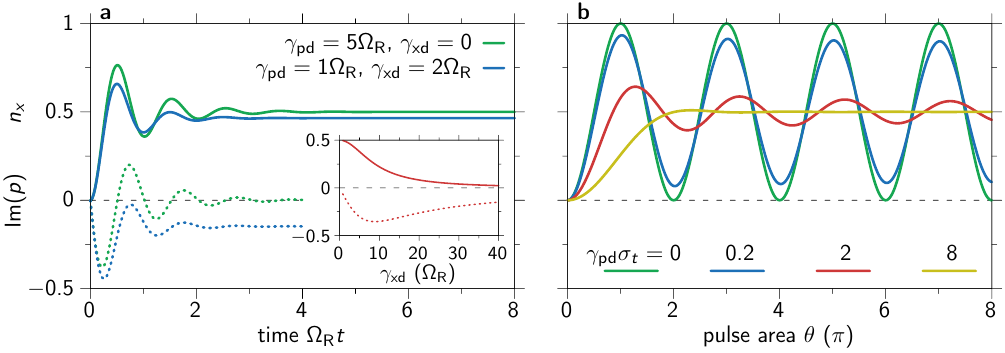}
	\caption{Impact of dissipation on different forms of resonant excitation ($\delta=0$) in the rotating frame from Eqs.~\eqref{eq:bloch_diss_cw}. (a) Rabi oscillations of the occupation (solid) and polarization (dashed) under cw excitation. Inset: Stationary state [Eqs.~\eqref{eq:rabi_stationary_1} and \eqref{eq:rabi_stationary}] for vanishing pure dephasing $\gamma_\mathrm{pd}=0$ as a function of the excited state decay rate $\gamma_\mathrm{xd}$.  (b) Rabi rotations under pulsed excitation with a Gaussian pulse of duration $\sigma_t$ [Eq.~\eqref{eq:gauss_pulse}] for different values of the pure dephasing rate $\gamma_\mathrm{pd}$ and vanishing excited state decay $\gamma_\mathrm{xd}=0$.}
	\label{fig:rabidamping}
\end{figure}

Figure~\ref{fig:rabidamping}~(a) shows Rabi oscillations for different values of the excited state decay rate $\gamma_\mathrm{xd}$ and pure dephasing rate~$\gamma_\mathrm{pd}$. We focus on resonant excitation with $\delta=0$, such that damped oscillations can be seen which tend towards the stationary values described in Eqs.~\eqref{eq:rabi_stationary_1} and \eqref{eq:rabi_stationary}. In addition, the inset shows the properties of this stationary state as a function of the excited state decay rate $\gamma_\mathrm{xd}$ for vanishing pure dephasing $\gamma_\mathrm{pd}=0$.

For vanishing excited state decay (green), the stationary occupation is given by $\bar{n}_x^{(l)}=n_x=1/2$ and therefore $\bar{p}^{(l)}=0=p$, which describes the center of the Bloch sphere. Pure dephasing in the presence of resonant driving forces the Bloch vector towards the center of the Bloch sphere similar to Fig.~\ref{fig:Bloch_dissipation}~(a). This situation describes the completely transparent state, where no external light field can impact the state of the 2LS [the terms $\sim\mathcal{A}$ in the Bloch equations~\eqref{eq:bloch_eq_diss} vanish then]. For finite excited state decay $\gamma_\mathrm{xd}>0$ (blue) we end up with $n_x<1/2$, as can also be seen in Eq.~\eqref{eq:rabi_stationary}. This is of course due to the fact that the excited state decay forces the Bloch vector towards the south pole as in Fig.~\ref{fig:Bloch_dissipation}~(b). Thus, when excited state decay plays a role, we do not reach the fully transparent state, as there is always the possibility of a transition to the ground state $\ket{g}$, e.g., due to spontaneous light emission~\cite{breuer2002theory}, after which excitation by the laser is again possible. 

Next we consider pulsed excitations and investigate the impact of pure dephasing on pulse-excited Rabi rotations, setting $\gamma_\mathrm{xd}=0$. We focus on resonant excitation ($\omega_l=\omega_x$) with a Gaussian pulse 
\begin{equation}
	\mathcal{A}(t)=\frac{\hbar\theta}{\sqrt{2\pi\sigma_t^2}}e^{-\frac{t^2}{2\sigma_t^2}}e^{-i\omega_l t}\,,\label{eq:gauss_pulse}
\end{equation}
where the width $\sigma_t$ quantifies the duration and $\theta$ is the pulse area from Eq.~\eqref{eq:pulse_area}. We start with an initial ground state $\rho(-\infty)=\ket{g}\bra{g}$ and calculate the excited state occupation $n_x(\infty)$ numerically by solving the Bloch equations~\eqref{eq:bloch_eq_diss} in the frame rotating with the laser frequency $\omega_l=\omega_x$. Figure~\ref{fig:rabidamping}~(b) shows the resulting Rabi rotations, analog to the inset in Fig.~\ref{fig:Rabi_pulses}~(b), as a function of the pulse area $\theta$. We distinguish between three cases: (i) long pulses with $\sigma_t\gg 2\gamma_\mathrm{pd}^{-1}$, i.e., much longer than the pure dephasing time (yellow), (ii) pulse lengths on the same timescale as the pure dephasing $\sigma_t=2\gamma_\mathrm{pd}^{-1}$ (red) and (iii) pulses that are much shorter than this timescale $\sigma_t\ll 2\gamma_\mathrm{pd}^{-1}$ (blue). As a reference we also show the dissipation-less case which has already been displayed in the inset of Fig.~\ref{fig:Rabi_pulses}~(b) (green). 

Focusing first on case (i), i.e., long pulses compared to the pure dephasing time (yellow), we see that the Rabi rotations are now damped due to pure dephasing instead of being purely sinusoidal as in the dissipation-less case (green)~\cite{zrenner2002coherent,forstner2003phonon,kruegel2005role,ramsay2010damping}. The reason for this has already been discussed in the current section in the context of cw excitation. There, pure dephasing in combination with continuous driving brought the 2LS into a transparent state, where the external light field cannot impact it anymore. This transparent state lies in the middle of the Bloch sphere with $p=0$, $n_x=1/2$. Since the pulse is now long compared to the dephasing time, pure dephasing can drive the system into this transparent state during the interaction with the pulse. And indeed we see that the Rabi rotations tend towards $n_x=1/2$ for sufficiently large pulse areas. Moreover we see that a full inversion of the 2LS is not possible anymore for sufficiently strong pure dephasing. Note that $n_x=1/2$ is only reached if the pulse area is at least $\pi/2$.

If we now decrease the duration of the pulse to be comparable with the pure dephasing time (red, ii), we see that the Rabi rotations are not as damped anymore, i.e., the impact of pure dephasing during the pulse interaction has been reduced. Decreasing this temporal width even further (blue, iii) we can reach a regime where there is no strong impact on the Rabi rotations due to pure dephasing, when compared to the dissipation-less situation (green).

We have already discussed the ultrashort pulse limit in Sec.~\ref{sec:delta_pulses} and argued that we can replace ultrashort pulses by idealized $\delta$-pulses, if the pulse duration is much shorter than any internal timescale of the 2LS. This is also true in the presence of dissipation, with the relevant internal timescale being $\gamma_\mathrm{pd}^{-1}$ in Fig.~\ref{fig:rabidamping}~(b). The strict limit $\sigma_t\rightarrow 0$ of the Gaussian pulse in Eq.~\eqref{eq:gauss_pulse} yields $\mathcal{A}(t)=\hbar\theta\delta(t)$. This pulse only impacts the 2LS at $t=0$ and has no impact before or after, while the free time evolution, now including dissipation, has no impact during the ultrashort pulse interaction in this limit. That is why we approach the dissipation-less regime of Rabi rotations (green), when the duration of the exciting pulse is sufficiently short (blue). If we consider multiple idealized $\delta$-pulses, the full time evolution is therefore a sequence of pulse transformations via the unitary operator $U_P(\theta,\phi)$ in Eq.~\eqref{eq:U_P} and free time evolution according to Eq.~\eqref{eq:bloch_eq_diss} with $\mathcal{A}=0$ after each pulse. This makes it possible to obtain analytical expressions for the polarization and occupation even when exciting with multiple pulses, as long as they can be idealized as $\delta$-pulses~\cite{vagov2002electron}.
\subsubsection{Impact of dissipation on the linear absorption spectrum}
We will now investigate the impact of pure dephasing and excited state decay on the linear absorption spectrum. If we integrate the linearized version of the polarization equation~\eqref{eq:p_diss} analog to  Eq.~\eqref{eq:2LS_p_linear} we again obtain the linear susceptibility~\cite{krummheuer2002theory,wigger2019phonon}
\begin{equation}
	\chi(t)=\frac{i}{2\hbar}e^{-i\omega_x t}e^{-\gamma t}\Theta(t)\,,\qquad\gamma=(\gamma_\mathrm{xd}+\gamma_\mathrm{pd})/2\,,
\end{equation}
which is still proportional to the polarization after a $\delta$-pulse excitation. In contrast to Eq.~\eqref{eq:chi_nodiss} it now decays exponentially due to both excited state decay and pure dephasing. The corresponding linear absorption spectrum reads
\begin{align}
	\text{Im}\left[\tilde{\chi}(\omega)\right]&=\text{Im}\left[\int\limits_{-\infty}^{\infty}\text{d}t\,\chi(t)e^{i\omega t}\right]=\frac{1}{2\hbar}\text{Re}\left[\int\limits_0^{\infty}\text{d}t\,e^{i(\omega-\omega_x) t}e^{-\gamma t}\right]=\frac{1}{2\hbar}\text{Re}\left[\frac{1}{i(\omega_x-\omega)+\gamma }\right]\notag\\
	&=\frac{1}{2\hbar}\frac{\gamma }{(\omega-\omega_x)^2+\gamma ^2}\,.\label{eq:lin_abs_diss}
\end{align}
Instead of a $\delta$-peak at the resonance of the 2LS $\omega=\omega_x$, we now obtain a Lorentzian curve, whose width $\gamma=(\gamma_\mathrm{xd}+\gamma_\mathrm{pd})/2$ stems from the exponential decay of the polarization due to pure dephasing and decay. In the limit $\gamma\rightarrow 0$ we obtain the result without dissipation from Eq.~\eqref{eq:lin_abs_nodiss}.
\subsection{Homogeneous and inhomogeneous broadening}\label{sec:inhom}
\begin{figure}[t]
	\centering
	\includegraphics[width = 0.6\textwidth]{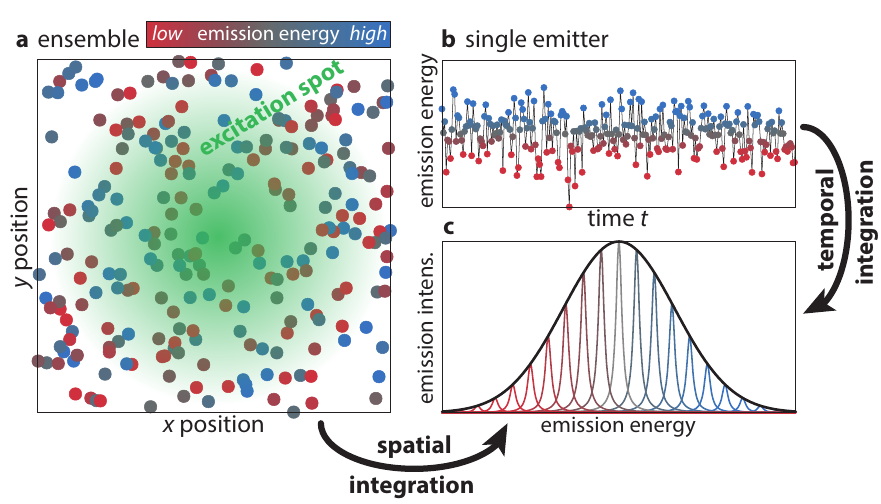}
	\caption{(a) Schematic picture of an ensemble of emitters. The green shaded area symbolizes an optical excitation spot. (b)~Exemplary spectral jitter of a single emitter. (c) Inhomogeneously broadened emission spectrum either generated by spatial integration of an ensemble or by temporal integration of a jittering single emitter.}
	\label{fig:inhom}
\end{figure}

So far we considered a single 2LS and its optical properties. However, it is often simpler or unavoidable to perform measurements on an ensemble of 2LSs.
In two-dimensional excitonic systems like quantum wells the growth process leads to fluctuations of material parameters, e.g., the width of the well~\cite{natali2005inhomogeneous}, while monolayers of transition metal dichalcogenides have internal strain distributions, e.g., from a rough substrate, or a spatially varying dielectric environment~\cite{boule2020coherent}. These factors will directly affect the local transition energies of the investigated optically active excitons in these systems. For self-assembled quantum dot ensembles the growth process results in variations of the dot sizes, strain distributions, and compositions which also lead to different transition energies throughout the full ensemble~\cite{langbein2010coherent,wigger2017systematic,bayer2019bridging,grisard2022multiple}. These situations are schematically depicted in Fig.~\ref{fig:inhom}(a), where red dots symbolize lower transition energies and blue dots higher ones. Measuring now the spatially integrated optical response, e.g., the photoluminescence (PL) spectrum, we effectively produce a histogram of the internal exciton transition energies as depicted in Fig.~\ref{fig:inhom}(c). As typical statistical processes result in normal distributions, one usually finds the depicted Gaussian spectral broadening.

To model such an ensemble, we now consider $N$ 2LSs with in general different values for the transition frequencies $\omega_{x}^{(i=1,...,N)}$. For simplicity, we keep the dissipation rates identical within the ensemble. If the frequencies are all the same, the ensemble is homogeneous, in all other cases it is inhomogeneous. We can of course describe the ensemble using a probability density $P(\omega_x)$ which yields the relative frequency of finding a certain $\omega_x$ in the ensemble.

For most measurements in quantum mechanics, we have to perform a large number of identical repetitions of the same experiment, i.e., exciting the ensemble of 2LSs a large number of times under the same conditions (same initial state, same laser pulse). As long as we are in the weak excitation regime, e.g., we investigate the linear absorption spectrum, we are interested in situations, where only one 2LS from the ensemble is excited in each repetition. The total absorption spectrum is therefore given by a weighted average of the absorption spectrum in Eq.~\eqref{eq:lin_abs_diss} with respect to the probability density $P(\omega_x)$
\begin{equation}
	\text{Im}\left[\tilde{\chi}(\omega)\right]=\int\text{d}\omega_x\,\frac{P(\omega_x)}{2\hbar}\frac{\gamma}{(\omega-\omega_x)^2+\gamma^2}\,.\label{eq:abs_inhom}
\end{equation}
This is a convolution between the probability distribution of the resonance frequencies in the ensemble and the Lorentzian absorption spectrum of each 2LS. The ensemble can thus constitute an additional source of broadening in optical spectra, which is called \textit{inhomogeneous broadening}~\cite{schultheis1985photon,langbein2010coherent,rodek2021local,grisard2022multiple,preuss2022resonant}. In contrast to this, pure dephasing and decay constitute a source of \textit{homogeneous broadening} $\gamma=(\gamma_\mathrm{xd}+\gamma_\mathrm{pd})/2$, i.e., a broadening of spectra which is still present for a completely homogeneous ensemble or even a single 2LS with $P(\omega_x)=\delta\left(\omega_x-\overline{\omega}_{x}\right)$~\cite{schultheis1985photon,langbein2010coherent,boule2020coherent,grisard2022multiple,preuss2022resonant}.

\begin{figure}[t]
	\centering
	\includegraphics[width =0.45\textwidth]{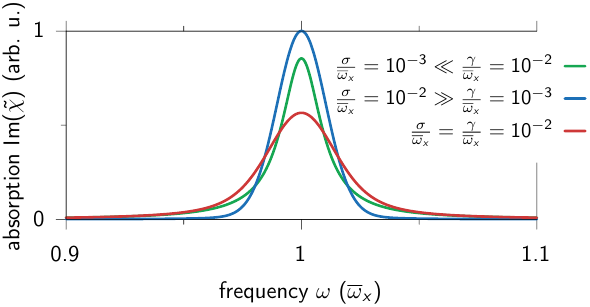}
	\caption{Linear absorption spectra including homogeneous broadening $\gamma=(\gamma_\mathrm{xd}+\gamma_\mathrm{pd})/2$ and inhomogeneous broadening $\sigma$ [Eq.~\eqref{eq:abs_inhom} together with Eq.~\eqref{eq:prob_inhom}]. The three displayed cases describe dominant homogeneous broadening $\gamma\gg\sigma$ (green), dominant inhomogeneous broadening $\gamma\ll\sigma$ (blue), comparable homogeneous and inhomogeneous broadening $\gamma=\sigma$ (red).}
	\label{fig:Im_chi}
\end{figure}

Because it originates from statistical processes such as growth, often one will encounter situations where the ensemble distribution $P(\omega_x$) can be described by a Gaussian of width~$\sigma$, the inhomogeneous broadening, centered around some mean frequency $\overline{\omega}_x$, such that
\begin{equation}
	P(\omega_x)=\frac{1}{\sqrt{2\pi\sigma^2}}\exp\left[-\frac{\left(\omega_x-\overline{\omega}_x\right)^2}{2\sigma^2}\right]\,.\label{eq:prob_inhom}
\end{equation}
Figure~\ref{fig:Im_chi} shows linear absorption spectra for three different situations: (i) dominant homogeneous broadening $\gamma\gg \sigma$ (green), (ii) dominant inhomogeneous broadening $\gamma\ll \sigma$ (blue), and (iii) equal broadenings $\gamma= \sigma$ (red). In case (i), which has a Lorentzian shape, we cannot give a good estimate for the inhomogeneous broadening, while in case (ii), which has a dominant Gaussian shape, we cannot give a good estimate for the homogeneous broadening. The general shape is that of a Voigt profile~\cite{voigt1912uber}, i.e., a convolution of a Lorentzian with a Gaussian. It is difficult to obtain good estimates for both homogeneous and inhomogeneous broadenings separately with linear optical techniques. This is one of the issues that can be solved with wave mixing spectroscopy, as will be discussed in the following Chap.~\ref{sec:theory_2}~\cite{yajima1979spatial,cho1994fifth,langbein2010coherent,cundiff2012optical,boule2020coherent,kasprzak2022coherent}.

While this discussion focused on an ensemble of multiple 2LSs, a single 2LS can show the same features, as schematically depicted in Fig.~\ref{fig:inhom}(b). If the transition frequency $\omega_x$ varies slowly on a very long timescale compared to the repetition frequency of the experiment, each repetition sees a different frequency with a certain probability $P(\omega_x)$ due to some underlying stochastic process modifying the transition frequency. This changing of the frequency is often called spectral jitter and can stem from external charge carriers impacting the 2LS, e.g., electrons moving around in the vicinity of an emitter. Slow spectral jitter (slower than the experiment repetition) therefore also leads to inhomogeneous broadening [Fig.~\ref{fig:inhom}(c)] even for single quantum systems~\cite{hogele2004voltage,delmonte2017coherent,spokoyny2020effect,preuss2022resonant}. This will be discussed in more detail in Sec.~\ref{sec:jitter}.


%% file: 3_Theory_2.tex
\section{Multi-pulse time-resolved spectroscopy}\label{sec:theory_2}
Having introduced the concept of inhomogeneous broadening in the previous Sec.~\ref{sec:inhom}, we can now investigate the dynamics of ensembles of 2LSs. In preparation of Sec.~\ref{sec:FWM_2LS}, where we finally discuss FWM spectroscopy of 2LSs, we consider here specifically the polarization, i.e., coherence, dynamics after excitation with multiple ultrashort pulses. As will be motivated in more detail in the following sections and especially in Chap.~\ref{sec:exp_theo} together with App.~\ref{sec:dipole_emission}, the polarization $p$ is the relevant observable for calculating the FWM signal of 2LSs~\cite{wegener1990line,mukamel1995principles,rossi2002theory,wigger2018rabi,groll2020four}.
\subsection{Bloch vector illustrations of ensembles and echo dynamics}\label{sec:echo}
\subsubsection{Ensemble dynamics after a single $\pi/2$-pulse}
Before investigating the polarization dynamics after excitation with multiple pulses, we start by considering the impact of different types of ensembles on the average coherence
\begin{equation}
	\overline{p}=\int\text{d}\omega_x\, P(\omega_x)p(\omega_x)\label{eq:def_p_avg}
\end{equation}
after a single ultrashort $\pi/2$-pulse, where $P(\omega_x)$ is again the probability density for the transition frequencies $\omega_x$ in the ensemble and $p(\omega_x)$ is the coherence of any 2LS with such a frequency. Note that we assumed here for simplicity that every 2LS from the ensemble contributes with the same strength, i.e., dipole moment. According to Eqs.~\eqref{eq:delta_pulse_transform}, a single $\pi/2$-pulse, i.e., with pulse area $\theta=\pi/2$ and assuming a vanishing phase $\phi=0$, creates the polarization $p=-i/2$ for all 2LSs in the ensemble if they were all in the ground state $\ket{g}$ before. The free evolution of these initial coherences now depends on the specific transition frequency and is described by the Bloch equations~\eqref{eq:bloch_eq_diss} with $\mathcal{A}=0$, yielding
\begin{equation}
	p(\tau;\omega_x)=-\frac{i}{2}e^{-\gamma\tau}e^{-i\omega_x\tau}\,.\label{eq:p_pi_2}
\end{equation}
Here, $\tau>0$ denotes the time that has passed since the pulsed excitation and $\gamma=(\gamma_\mathrm{xd}+\gamma_\mathrm{pd})/2$ is again the total dephasing rate.

\begin{figure}[t]
	\centering
	\includegraphics[width = 0.75\textwidth]{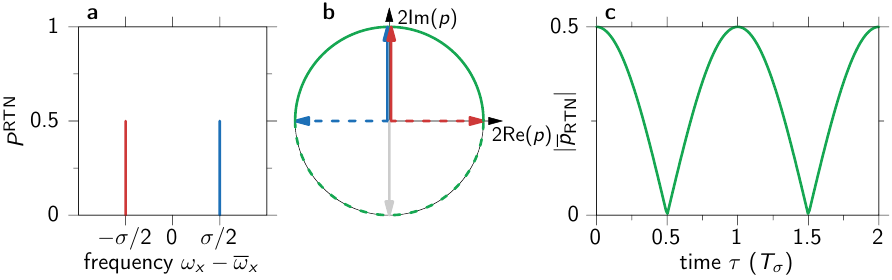}
	\caption{(a) Emitter ensemble of width $\sigma$ containing two frequencies with colors denoting the positive (blue) and negative (red) detuning $\pm\sigma/2$ with respect to the mean frequency $\overline{\omega}_x$ [Eq.~\eqref{eq:ensemble_RTN}]. (b) Coherence dynamics of the ensemble after a single $\pi/2$-pulse, projected on the Re$(p)$-Im$(p)$-plane of the Bloch sphere (equator cut), in the frame rotating with the mean frequency $\overline{\omega}_x$. Blue and red colors correspond to the frequencies in (a). Dashed and solid lines and arrows denote the times $\tau=T_\sigma/2$ and $\tau=T_\sigma$ after the pulse, respectively, with $T_\sigma=2\pi/\sigma$ (c) Absolute value of the average polarization after the $\pi/2$-pulse [Eq.~\eqref{eq:p_avg_RTN}]. The dephasing rate is set to $\gamma=0$ in (b) and (c).}
	\label{fig:rotation_2E}
\end{figure}

We consider now the simplest non-trivial ensemble, which is represented by two uncoupled emitters with unequal transition energies. Such a scenario can not only occur for two distinct emitters but also for a single emitter that experiences random telegraph noise (RTN) which switches between two possible values for the transition frequency~\cite{tran2019suppression,preuss2022resonant}. Note, that the switching rate needs to be sufficiently slow to be able to treat noise as an ensemble, as discussed previously in Sec.~\ref{sec:inhom}. This will also be investigated in detail in Sec.~\ref{sec:jitter}. The ensemble is characterized by the probability density of finding a certain transition frequency $\omega_x$
\begin{align}
	P^\mathrm{RTN}(\omega_x)=\frac12\delta\left(\omega_x- \overline{\omega}_x-\frac{\sigma}{2}\right) + \frac12\delta\left(\omega_x- \overline{\omega}_x+\frac{\sigma}{2}\right)\,,\label{eq:ensemble_RTN}
\end{align}
or simply by the two discrete probabilities $P_{\sigma/2}^\mathrm{RTN}=P_{-\sigma/2}^\mathrm{RTN}=0.5$. Figure~\ref{fig:rotation_2E}(a) shows this ensemble, characterized by the mean transition frequency $\overline{\omega}_{x}$ and the emitter splitting $\sigma$ (ensemble width). The colors denote the red-shifted frequency and the blue-shifted frequency with respect to the average $\overline{\omega}_x$.

To visualize the coherence dynamics of this ensemble we move into a frame rotating with the mean transition frequency $\overline{\omega}_{x}$ and consider the projection of the Bloch vector onto the Re$(p)$-Im$(p)$-plane. For this specific ensemble we can represent each of the two transition frequencies by one Bloch vector. The dynamics of these two Bloch vectors after excitation with a $\pi/2$-pulse are shown in Fig.~\ref{fig:rotation_2E}(b), where we consider a vanishing dephasing rate $\gamma=0$ for simplicity. Directly after the pulsed excitation the two Bloch vectors point to the same spot on the equator (gray arrow), corresponding to $p(\tau=0; \omega_x)=-i/2$, independent of the specific transition frequency.

As the 2LSs have detunings $\pm\sigma/2$ with respect to the rotating frame ($\overline{\omega}_{x}$), the two Bloch vectors rotate in opposite directions on the equator with the same frequency $\sigma/2$. The effect of these rotations is depicted by red and blue arrows in Fig.~\ref{fig:rotation_2E}(b), corresponding to the two frequencies from (a). After $\tau=T_\sigma/2$ with $T_\sigma=2\pi/\sigma$, which is represented by the dashed lines and arrows, the two Bloch vectors point exactly in opposite directions, i.e., they are fully dispersed or dephased. In this situation the average coherence of the ensemble vanishes $\overline{p}=\sum_{j=\pm\sigma/2} P_jp_j=0 $, where $p_j$ are the individual coherences. In the following half-period, i.e., until $\tau=T_\sigma$, represented by the solid lines and arrows, the two Bloch vectors merge and the coherence of the ensemble reaches a maximum again. Therefore, during this second half the ensemble rephases. While time moves on this process of dephasing and rephasing continues periodically with the frequency $\sigma$ and the resulting average polarization, i.e., coherence, reads (including the impact of the dephasing rate~$\gamma$)
\begin{align}
	\overline{p}_\mathrm{RTN}(\tau)&=\int\text{d}\omega_x\, P^\mathrm{RTN}(\omega_x)p(\tau;\omega_x)=-\frac{i}{2}e^{-\gamma\tau}\cos(\sigma\tau/2)e^{-i\overline{\omega}_x\tau}\notag\\
	&=-\frac{i}{2}e^{-\gamma\tau}\cos(\pi\tau/T_\sigma)e^{-i\overline{\omega}_x\tau}\,.\label{eq:p_avg_RTN}
\end{align}
Note that the exponential containing the average frequency $\overline{\omega}_x$ is removed in the rotating frame description in Fig.~\ref{fig:rotation_2E}(b). The absolute value of the average coherence $|\overline{p}_\mathrm{RTN}(\tau)|$ is plotted in Fig.~\ref{fig:rotation_2E}(c) for the case $\gamma=0$ corresponding to the situation in (b). Indeed we see dynamics that are periodic with $T_\sigma$ as was already apparent from the Bloch sphere picture.

\begin{figure}[t]
	\centering
	\includegraphics[width = 0.75\textwidth]{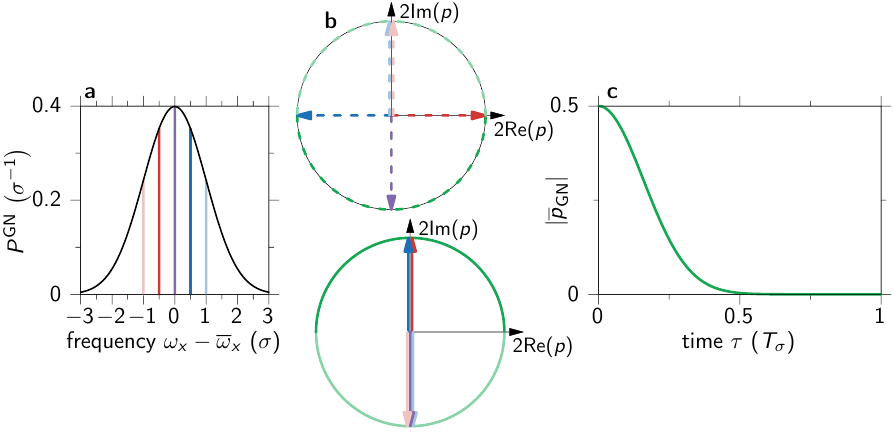}
	\caption{(a) Gaussian ensemble of width $\sigma$ [Eq.~\eqref{eq:ensemble_WN}]. Vertical colored lines mark specific realizations whose dynamics are shown in (b) after a single $\pi/2$-pulse in the Re$(p)$-Im$(p)$-projection of the Bloch sphere and in the frame rotating with the mean frequency $\overline{\omega}_x$. Dashed (top) and solid (bottom) lines and arrows denote the times $\tau=T_\sigma/2$ and $\tau=T_\sigma$ after the pulse, respectively, with $T_\sigma=2\pi/\sigma$ (c) Absolute value of the average polarization after the $\pi/2$-pulse [Eq.~\eqref{eq:p_avg_WN}]. The dephasing rate is set to $\gamma=0$ in (b) and (c).}
	\label{fig:rotation_gauss}
\end{figure}

Next we deal with another common ensemble type, which is given by a Gaussian probability density~\cite{langbein2010coherent,boule2020coherent,preuss2022resonant} of the form from Eq.~\eqref{eq:prob_inhom}
\begin{align}
	P^\mathrm{GN}(\omega_x)=\frac{1}{\sqrt{2\pi\sigma^2}}\exp\left[-\frac{\left(\omega_x-\overline{\omega}_x\right)^2}{2\sigma^2}\right]\,,\label{eq:ensemble_WN}
\end{align}
which can also originate for a single emitter subject to a slow Gaussian noise (GN) process for the transition frequency. Although we are now dealing with a continuous distribution of frequencies, i.e., an infinite amount of relevant coherences  $p(\omega_x)$, we can still visualize the Bloch vector dynamics by selecting a few representative transition frequencies [colored lines in Fig.~\ref{fig:rotation_gauss}(a)] and plot the corresponding Bloch vectors in Fig.~\ref{fig:rotation_gauss}(b) in the frame rotating with the mean frequency~$\overline{\omega}_x$. We consider again the projection of the Bloch vectors onto the Re$(p)$-Im$(p)$-plane and the case $\gamma=0$. The $\pi/2$-pulse with $\phi=0$  again brings all Bloch vectors onto the same point on the equator with $p(\tau=0;\omega_x)=-i/2$. After the time $\tau=T_\sigma/2$ (top circle), the vectors are clearly dispersed (dashed arrows). We see that the collection disperses symmetrically because the different detunings with respect to the mean frequency $\overline{\omega}_x$, i.e., the rotation frequencies, are symmetrically distributed in the ensemble. As all possible Bloch vectors fan out on the equator the average coherence shrinks. In contrast to the RTN ensemble however, there is no rephasing process, as can be seen when looking at a later time $\tau=T_\sigma$ (bottom circle). It is not possible for all arrows to ever meet again at a common point and the ensemble keeps on dephasing. This can also be seen when explicitly calculating the average coherence
\begin{align}
	\overline{p}_\mathrm{GN}(\tau)&=\int\text{d}\omega_x\, P^\mathrm{GN}(\omega_x)p(\tau;\omega_x)=-\frac{i}{2}e^{-\gamma\tau}e^{-\sigma^2\tau^2/2}e^{-i\overline{\omega}_x\tau}\notag\\
	&=-\frac{i}{2}e^{-\gamma\tau}e^{-2\pi^2 \tau^2/T_\sigma^2}e^{-i\overline{\omega}_x\tau}\,,\label{eq:p_avg_WN}
\end{align}
which decays as a Gaussian function on a timescale given by $\sigma^{-1}$, i.e., the inverse of the spectral width of the Gaussian probability distribution describing the ensemble. The exponential containing the average frequency $\overline{\omega}_x$ is again removed in the rotating frame picture of Fig.~\ref{fig:rotation_gauss}(b). In (c) we show the absolute value of the average coherence $|\overline{p}_\mathrm{GN}(\tau)|$ for $\gamma=0$, which is in stark contrast to Fig.~\ref{fig:rotation_2E}(c). For a Gaussian ensemble there is no rephasing after a single pulse. Note that both in Fig.~\ref{fig:rotation_2E} and Fig.~\ref{fig:rotation_gauss}, respectively, the probability densities in (a) and the polarization dynamics in (c) are Fourier transform pairs due to the form of the polarization $p(\tau;\omega_x)$ in Eq.~\eqref{eq:p_pi_2} for $\gamma=0$. This relation between the frequency distribution of the ensemble and the impact on the collective polarization dynamics via Fourier transformation is worth to keep in mind. 

The dephasing rate $\gamma$, which we ignored in Figs.~\ref{fig:rotation_2E} and \ref{fig:rotation_gauss}, simply leads to an additional exponential damping in both cases. For the Gaussian ensemble we thus get two distinct decays of the polarization: an exponential decay with the rate $\gamma$ and a Gaussian decay with the rate $\sigma$ given by the width of the ensemble. As we already discussed in the context of Fig.~\ref{fig:Im_chi}, both of these decays lead to separate types of broadening of the resonance lines, i.e., homogeneous and inhomogeneous, respectively, that cannot be well separated in linear absorption spectra.

\subsubsection{Ensemble dynamics after a $\pi/2$-$\pi$ pulse sequence}
Now we investigate the polarization dynamics after a specific pulse sequence that allows us to control the rephasing dynamics in the case of the RTN ensemble and even enables us to achieve rephasing in the case of the Gaussian ensemble. To this aim the 2LSs are excited by another ultrashort $\pi$-pulse after the $\pi/2$-pulse, i.e., with $\theta=\pi$ and we choose the phase $\phi=0$ for simplicity. The time of the $\pi$-pulse excitation is set to $t=0$ and according to Eqs.~\eqref{eq:delta_pulse_transform} the polarization transforms as 
\begin{equation}
	p_+(t=0)=p_-^*(t=0)\,,\label{eq:action_pi}
\end{equation}
where the indices $\pm$ denote the polarization after/before the ultrashort $\delta$-pulse. In the considered $\pi/2$-$\pi$ pulse sequence, the system has already been excited by a $\pi/2$-pulse at time $t=-\tau$ with $\tau>0$, such that the polarization before the $\pi$-pulse $p_-(t=0)$ is given by Eq.~\eqref{eq:p_pi_2} for any 2LS with the transition frequency $\omega_x$. The $\pi$-pulse together with the subsequent free time evolution of the polarization  yields the overall polarization
\begin{equation}
	p(t,\tau;\omega_x)=\frac{i}{2}e^{-\gamma(t+\tau)}e^{-i\omega_x(t-\tau)}\label{eq:p_pi_2_pi}
\end{equation}
for a single spectral component from an arbitrary ensemble. The average coherence is again determined by Eq.~\eqref{eq:def_p_avg}.

\begin{figure}[t]
	\centering
	\includegraphics[width = 0.7\textwidth]{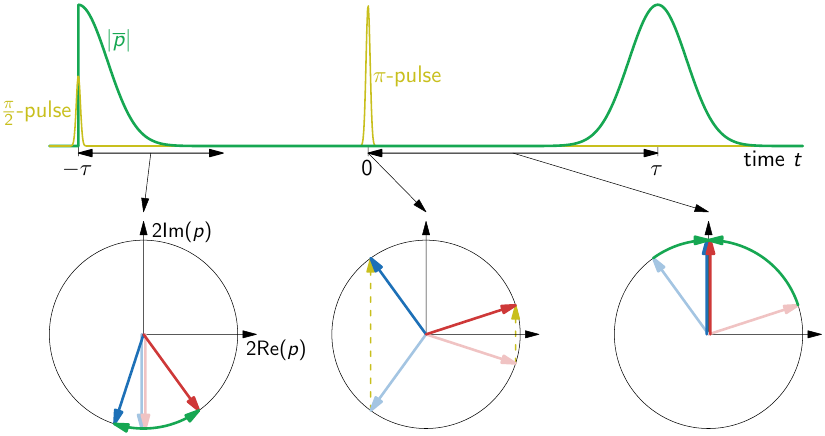}
	\caption{Top: $\pi/2$-$\pi$ pulse sequence (yellow) and resulting coherence dynamics (green) for a Gaussian ensemble. Bottom: Bloch vector dynamics in the Re$(p)$-Im$(p)$-projection of the Bloch sphere and in the frame rotating with the mean frequency $\overline{\omega}_x$ with colors denoting the detuning with respect to this frequency (blue: positive, red: negative). Light arrows denote Bloch vectors at earlier times, dark arrows at later times. Left circle: dynamics after $\pi/2$-pulse. Middle circle: transformation via $\pi$-pulse. Right circle: dynamics after $\pi$-pulse from $t=0$ to $t=\tau$.}
	\label{fig:Echo_nobox}
\end{figure}

Figure~\ref{fig:Echo_nobox} shows this pulse sequence (top, yellow) and its impact on the coherences in the Re$(p)$-Im$(p)$ projection of the Bloch vectors for the case $\gamma=0$ in the frame rotating with the mean frequency $\overline{\omega}_x$ of an arbitrary ensemble (bottom, circles). For simplicity we show the dynamics of only two spectral components with different detuning $\delta_x=\omega_x-\overline{\omega}_x$ with respect to the mean frequency similar to Fig.~\ref{fig:rotation_2E}(b). Directly after the $\pi/2$-pulse (left circle, light arrows) both realizations have the polarization $p(\tau=0;\omega_x)=-i/2$ according to Eq.~\eqref{eq:p_pi_2} and rotate in opposite direction (dark arrows) corresponding to their respective sign of the detuning $\delta_x$ (blue: positive, red: negative). These rotations continue with their respective frequency $|\delta_x|$ in opposite directions until $t=0$ right before the action of the $\pi$-pulse (middle circle, light arrows), which flips the arrows along the Re$(p)$-line according to Eq.~\eqref{eq:action_pi} (dark arrows). After this pulse transformation, the rotations continue in the same direction as before (right circle). Note that now the two arrows move towards each other with the same frequency as of their movement away from each other just before the $\pi$-pulse. This implies that at $t=\tau$ (right circle, dark arrows), both realizations have the same polarization again and are completely rephased. This is true independent of the specific value of their detunings $\delta_x$ since each Bloch vector needs the same time, that they were given ($\tau$) to move away from each other and from the vertical line Re$(p)=0$, to move towards each other and towards this line after being flipped by the $\pi$-pulse. This means that every ensemble rephases at $t=\tau$ for such a pulse sequence. This can already be seen in Eq.~\eqref{eq:p_pi_2_pi}, which yields the same polarization for $t=\tau$, irrespective of the frequency $\omega_x$. The appearance of a complete rephasing exactly at $t=\tau$ is called a photon echo~\cite{kurnit1964observation,abella1966photon,schultheis1985photon,poltavtsev2018photon} (in analogy to the spin or Hahn echo~\cite{hahn1950spin}), since the $\pi$-pulse creates an echo, i.e., a copy, at $t=\tau$ of the initial coherence dynamics from $t=-\tau$. The initial coherence dynamics together with the echo is schematically shown in the top of Fig.~\ref{fig:Echo_nobox} in green for a Gaussian ensemble.

\begin{figure}[t]
	\centering
	\includegraphics[width = 0.6\textwidth]{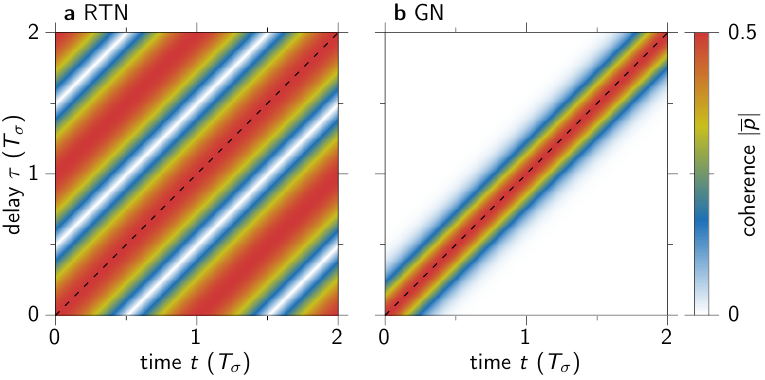}
	\caption{Echo dynamics in time $t$ and delay $\tau$, i.e., absolute values of the average coherences from Eq.~\eqref{eq:p_avg_echo_rtn} in (a) and Eq.~\eqref{eq:p_avg_echo_wn} in (b) for the case of a vanishing dephasing rate $\gamma=0$.}
	\label{fig:echo_dyn}
\end{figure}

We can now investigate the echo dynamics, i.e., the coherence dynamics after this specific two-pulse sequence with varying delay $\tau$, quantitatively for the RTN and Gaussian ensembles. The average coherence dynamics can be calculated using Eq.~\eqref{eq:def_p_avg} together with Eq.~\eqref{eq:p_pi_2_pi} for both ensembles analog to Eqs.~\eqref{eq:p_avg_RTN} and \eqref{eq:p_avg_WN}, yielding
\begin{align}
	\overline{p}_\mathrm{RTN}(t,\tau)&=\frac{i}{2}e^{-\gamma(t+\tau)}\cos[\pi(t-\tau)/T_\sigma]e^{-i\overline{\omega}_x(t-\tau)}\label{eq:p_avg_echo_rtn}\,,\\
	\overline{p}_\mathrm{GN}(t,\tau)&=\frac{i}{2}e^{-\gamma(t+\tau)}e^{-\frac{2\pi^2(t-\tau)^2}{T_\sigma^2}}e^{-i\overline{\omega}_x(t-\tau)}\,.\label{eq:p_avg_echo_wn}
\end{align}
Figure~\ref{fig:echo_dyn} shows the absolute value of both average coherences as a function of the pulse delay $\tau$ and the time after the last pulse $t$ for a vanishing dephasing rate $\gamma=0$. In the case of the RTN ensemble in (a) there is always maximum coherence at the diagonal $t=\tau$, i.e., the photon echo. This maximum is repeated with the period $T_\sigma$ analog to the discussion in the context of Fig.~\ref{fig:rotation_2E}. The $\pi/2$-$\pi$ pulse sequence thus allows to tune the position of the maxima by changing the delay $\tau$. In the case of the Gaussian ensemble in (b) only a single maximum at $t=\tau$ appears. This maximum was already shown schematically in the top of Fig.~\ref{fig:Echo_nobox}. The initial Gaussian dephasing after the $\pi/2$-pulse from Fig.~\ref{fig:rotation_gauss}(c) is inverted by the $\pi$-pulse to yield a rephasing followed by another dephasing leading to an echo with Gaussian temporal shape of width $\sigma^{-1}$. The spectral distribution of the ensemble can therefore be determined by the duration of the echo. In total, the additional $\pi$-pulse after a delay $\tau$ allows for temporal control of rephasing dynamics of ensembles of 2LSs.

\subsubsection{Separation of linewidths}
\begin{figure}[t]
	\centering
	\includegraphics[width = 0.8\textwidth]{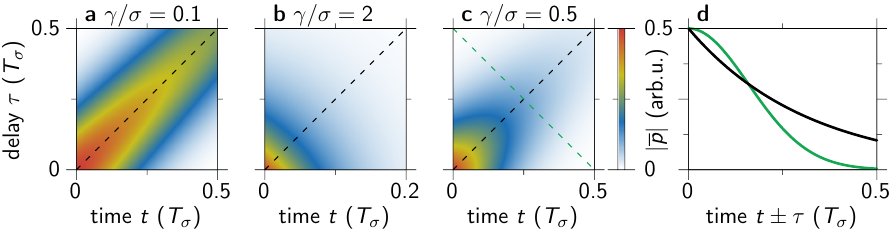}
	\caption{(a-c) Photon echo dynamics in a Gaussian ensemble [Eq.~\eqref{eq:p_avg_echo_wn}] for different ratios of homogeneous $\gamma$ and inhomogeneous broadening $\sigma$. (d) Cuts in (c) along the diagonal $t-\tau=$ const. (black) and cross-diagonal $t+\tau=$ const. (green).}
	\label{fig:echo_with_deph}
\end{figure}

So far we mainly ignored the impact of the dephasing rate $\gamma$ on the echo dynamics. However in Eq.~\eqref{eq:p_avg_echo_wn} we can see something interesting: The width of the Gaussian ensemble $\sigma$ is directly connected to the width measured perpendicular to the diagonal $t=\tau$ from Fig.~\ref{fig:echo_dyn}(b), while the dephasing rate $\gamma$ damps the coherence along the diagonal. Both parameters therefore influence the echo dynamics along two orthogonal directions $t=\tau$ and $t=-\tau$ allowing for separate determination using such an echo spectroscopy technique~\cite{abella1966photon,schultheis1985photon,langbein2010coherent,cundiff2012optical}. Figures~\ref{fig:echo_with_deph}(a-c) show examples of the absolute value of the coherence using a $\pi/2$-$\pi$ pulse sequence for the case of $\gamma>0$. Compared with Fig.~\ref{fig:echo_dyn}(b) the echos are now damped along the diagonal. We consider three different cases: (a) $\gamma\ll\sigma$, yielding a slightly damped echo, (b) $\gamma\gg\sigma$, yielding dominant exponential damping along the diagonal $t=\tau$, and (c) $\gamma\sim\sigma$ showing a mixture of echo and exponential damping. 

Figure~\ref{fig:echo_with_deph}(d) additionally shows cuts in (c) along the diagonal direction with $t-\tau=$ const. (black) and along the cross-diagonal direction with $t+\tau=$ const. (green). The diagonal cut clearly exhibits an exponential damping and from Eq.~\eqref{eq:p_avg_echo_wn} we know that this allows us to extract the dephasing rate $\gamma$, i.e., the homogeneous broadening. The cross-diagonal cut (green) on the other hand yields the Gaussian ensemble width $\sigma$, i.e., the inhomogeneous broadening. While this separation of the two types of broadening, i.e., linewidths, is not easily possible in single pulse experiments like linear absorption in Fig.~\ref{fig:Im_chi}, multi-pulse experiments are an excellent tool to distinguish between them. In the following Sec.~\ref{sec:FWM_2LS} we will finally discuss FWM spectroscopy, which belongs to the class of multi-pulse experiments and therefore also allows for the separation of these different types of linewidths~\cite{schultheis1985photon,langbein2010coherent,cundiff2012optical}.

So far we have called this method echo \textit{spectroscopy}, but only considered time- and delay-resolved signals. These two temporal variables give rise to two frequency variables in Fourier space, i.e., 2D spectroscopy as discussed in more detail in Sec.~\ref{sec:2d_spectra}

\subsubsection{Homogeneous and inhomogeneous broadening from noise}\label{sec:jitter}
At this point it is worth noting that both homogeneous and inhomogeneous broadening of a single emitter can be understood as originating from underlying stochastical processes that impact the transition frequency $\omega_x$~\cite{weiss2012quantum}. To this aim we consider a simple 2LS without any dissipation but with a time-dependent transition frequency $\omega_x(t)$, such that the free Hamiltonian is given by 
\begin{equation}
	H_0(t)=\hbar\omega_x(t)\ket{x}\bra{x}\,.
\end{equation}
This modification does not impact the interaction with $\delta$-pulses and since $H_0$ commutes with itself at different times, the corresponding free time evolution operator is simply given by
\begin{equation}
	U_0(t,t_0)=\exp\left[-i\ket{x}\bra{x}\int\limits_{t_0}^t\text{d}t'\,\omega_x(t')\right]\,.
\end{equation}
The free evolution of the polarization in Eq.~\eqref{eq:2LS_free_dyn} is thus modified to
\begin{equation}
	p(t)=p(t_0)\exp\left[-i\int\limits_{t_0}^t\text{d}t'\,\omega_x(t')\right]\,,
\end{equation}
such that the polarization after the $\pi/2$-$\pi$ pulse sequence in Eq.~\eqref{eq:p_pi_2_pi} is given by
\begin{equation}
	p(t,\tau)[\omega_x]=\frac{i}{2}\exp\left[i\int\limits_{-\tau}^0\text{d}t'\omega_x(t')-i\int\limits_0^t\text{d}t'\,\omega_x(t')\right]\,,
\end{equation}
which becomes a functional of the function $\omega_x(t)$, as denoted by the square brackets on the left-hand side. We now assume that the transition frequency is impacted by some form of noise, i.e., $\omega_x(t)$ becomes a stochastic process. We assume that the fluctuations around some mean value $\overline{\omega}_x$ are due to a stationary process, such that~\cite{van1992stochastic}
\begin{equation}
	\delta_x(t)=\omega_x(t)-\overline{\omega}_x\,,\quad \braket{\delta_x(t)}=0\,,\quad\braket{\delta_x(t)\delta_x(t')}=\braket{\delta_x^2}f(t-t')\,,\quad f(0)=1\,,\label{eq:gauss_cov}
\end{equation}
where $\braket{...}$ denotes the stochastic average. The function $f$ describes the covariance of the stochastic process for different times. We furthermore assume that the process $\omega_x(t)$, and therefore also $\delta_x(t)$, obeys Gaussian statistics. Defining the phase
\begin{equation}
	\varphi(t)=\int\limits_0^t\text{d}t'\delta_x(t')\label{eq:phi_stoch}
\end{equation}
we can write
\begin{equation}
	p(t,\tau)[\omega_x]=\frac{i}{2}e^{-i\overline{\omega}_x(t-\tau)}e^{-i[\varphi(t)+\varphi(-\tau)]}\label{eq:p_stoch}
\end{equation}
for a specific realization of the stochastic process. Since $\delta_x(t)$ is described by a Gaussian stochastic process with vanishing mean value, this also holds for its integral $\varphi(t)$, as well as $\varphi(t)+\varphi(-\tau)$. Therefore, $\left<e^{-i[\varphi(t)+\varphi(-\tau)]}\right>$ is the characteristic function of a Gaussian stochastic process with vanishing mean, i.e., it is completely characterized by the corresponding variance and given by~\cite{weiss2012quantum}
\begin{equation}
	\left<e^{-i[\varphi(t)+\varphi(-\tau)]}\right>=e^{-\frac{1}{2}\braket{[\varphi(t)+\varphi(-\tau)]^2}}=e^{-\frac{1}{2}\left[\braket{\varphi^2(t)}+\braket{\varphi^2(-\tau)}+2\braket{\varphi(t)\varphi(-\tau)}\right]}\,.
\end{equation}
Using Eqs.~\eqref{eq:gauss_cov} and \eqref{eq:phi_stoch} we can calculate
\begin{equation}
	\left<\varphi^2(t)\right>=\int\limits_0^t\text{d}t'\,\int\limits_0^t\text{d}t''\left<\delta_x(t')\delta_x(t'')\right>=\left<\delta_x^2\right>\int\limits_0^t\text{d}t'\,\int\limits_0^t\text{d}t''f(t'-t'')\label{eq:phi_2_t}
\end{equation}
and analog for $\varphi(-\tau)$, as well as 
\begin{equation}
	\left<\varphi(t)\varphi(-\tau)\right>=\left<\delta_x^2\right>\int\limits_0^t\text{d}t'\,\int\limits_0^{-\tau}\text{d}t'' f(t'-t'')\,.
\end{equation}
As a specific model for the covariance function we choose
\begin{equation}
	f(t'-t'')=\exp(-|t'-t''|/\Delta t)=f(|t'-t''|)\,,\label{eq:f_stoch}
\end{equation}
which describes an Ornstein-Uhlenbeck process with correlation time $\Delta t$~\cite{doob1942brownian}. To calculate $\left<\varphi^2(t)\right>$ in Eq.~\eqref{eq:phi_2_t}, we can time-order the integration by separating the $t''$-integral
\begin{align}
	\left<\varphi^2(t)\right>&=\left<\delta_x^2\right>\int\limits_0^t\text{d}t'\,\int\limits_0^{t'}\text{d}t''f(|t'-t''|)+\left<\delta_x^2\right>\int\limits_0^t\text{d}t'\,\int\limits_{t'}^t\text{d}t''f(|t'-t''|)\notag\\
	&=\left<\delta_x^2\right>\int\limits_0^t\text{d}t'\,\int\limits_0^{t'}\text{d}t''f(|t'-t''|)+\left<\delta_x^2\right>\int\limits_0^t\text{d}t''\,\int\limits_{0}^{t''}\text{d}t'f(|t'-t''|)\notag\\
	&=2\left<\delta_x^2\right>\int\limits_0^t\text{d}t'\,\int\limits_0^{t'}\text{d}t''f(|t'-t''|)\,,
\end{align}
where the second integral in the first and second line integrates over the two-dimensional area $[0,t]\times[0,t]$ under the constraint $t''>t'$ and in the last line we used the substitution $t'\leftrightarrow t''$ in this integral. The time-ordering with $t'\geq t''$ now allows us to simply integrate over the exponential $f(t'-t'')$ without the need to consider different cases due to the absolute value therein, which finally yields
\begin{align}
	\left<\varphi^2(t)\right>&=2\left<\delta_x^2\right>\left[t (\Delta t)+ (\Delta t)^2 \big(e^{-t/\Delta t}-1\big)\right]\,.
\end{align}
In an analog fashion we obtain
\begin{align}
	\left<\varphi^2(-\tau)\right>&=2\left<\delta_x^2\right>\left[\tau (\Delta t)+ (\Delta t)^2 \big(e^{-\tau/\Delta t}-1\big)\right]\,.
\end{align}
In the case of $\left<\varphi(t)\varphi(-\tau)\right>$ the integral is already time-ordered, yielding
\begin{equation}
	\left<\varphi(t)\varphi(-\tau)\right>=-\left<\delta_x^2\right>\int\limits_0^t\text{d}t'e^{-t'/\Delta t}\,\int\limits_{-\tau}^{0}\text{d}t'' e^{t''/\Delta t}=-(\Delta t)^2\left<\delta_x^2\right>\big(e^{-t/\Delta t}-1\big)\big(1-e^{-\tau/\Delta t}\big)
\end{equation}

We can now differentiate between two physically distinct cases: (i) fast spectral jitter, i.e., the covariance function $f(t)$ decays rapidly when moving away from $t=0$ on the observation timescale, i.e., for relevant values of $t$ and $\tau$, and (ii) slow spectral jitter, i.e., the covariance function decays very slowly on that timescale. We can realize (i) fast spectral jitter via $\Delta t\ll t,\tau$, which yields in leading order 
\begin{equation}
	\left<e^{-i[\varphi(t)+\varphi(-\tau)]}\right>_\mathrm{fast}=\exp\left[-\Delta t\left<\delta_x^2\right>(t+\tau)\right]\,.
\end{equation}
If we average the polarization in Eq.~\eqref{eq:p_stoch} with respect to such a fast Gaussian jitter process we thus obtain exponential damping of the polarization, i.e., the same effect as homogeneous broadening with the rate $\gamma$ in Eq.~\eqref{eq:p_pi_2_pi}. We can therefore identify $\gamma=\Delta t\left<\delta_x^2\right>$ for a fast Gaussian jitter process.

For (ii) slow spectral jitter we have $\Delta t\gg t,\tau$, which yields in leading order
\begin{equation}
	\left<\varphi^2(t)\right>_\mathrm{slow}\approx 2\left<\delta_x^2\right>\left\{ t (\Delta t)+ (\Delta t)^2 \left[1-\frac{t}{\Delta t}+\frac{t^2}{2(\Delta t)^2}-1\right]\right\}=\left<\delta_x^2\right>t^2
\end{equation}
and analog $\left<\varphi^2(-\tau)\right>_\mathrm{slow}=\left<\delta_x^2\right>\tau^2$, $\braket{\varphi(t)\varphi(-\tau)}_\mathrm{slow}=\left<\delta_x^2\right>t\tau$ using Taylor expansion of the exponentials. In total this yields
\begin{equation}
	\left<e^{-i[\varphi(t)+\varphi(-\tau)]}\right>_\mathrm{slow}=\exp\left[-\frac{\left<\delta_x^2\right>(t+\tau)^2}{2}\right]\,.
\end{equation}
If we average the polarization in Eq.~\eqref{eq:p_stoch} with respect to such a slow Gaussian jitter process we thus obtain Gaussian damping of the polarization, i.e., the same effect as inhomogeneous broadening with a Gaussian ensemble of width $\sigma$ in Eq.~\eqref{eq:p_avg_WN}. We can thus identify $\sigma^2=\left<\delta_x^2\right>$ for a slow Gaussian jitter process. We can of course also consider a sum of two stochastic processes, i.e., fast and slow jitter, which then fully recreates the result in Eq.~\eqref{eq:p_avg_WN}. 

\begin{figure}[t]
	\centering
	\includegraphics[width = \textwidth]{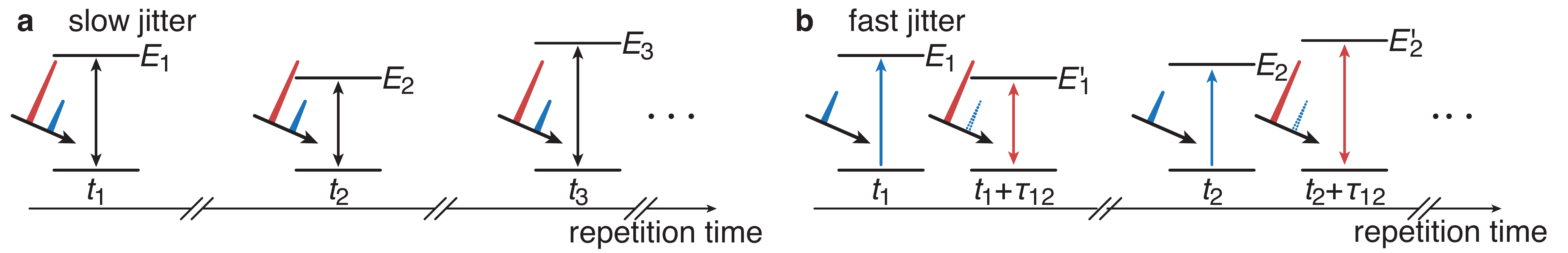}
	\caption{(a) Schematic picture of the influence of slow spectral jitter in the two-pulse experiment. The transition energy remains constant during each two pulse sequence but changes randomly for the next repetition. (b) Fast spectral jitter, where the energy changes randomly between the two pulses in each sequence.}
	\label{fig:jitter}
\end{figure}

The time arguments $t$ and $-\tau$ in Eq.~\eqref{eq:p_stoch} stem from a derivation using two $\delta$-pulses. They relate to a single realization of the stochastic process. In the experiment such two-pulse sequences usually have to be repeated a large number of times, which then results in the stochastical averaging $\braket{...}$. Slow jitter is then given by the case where each repetition of the two-pulse sequence sees a different frequency $\overline{\omega}_x+\delta_x$ which however does not change during the two-pulse sequence, as already discussed in Sec.~\ref{sec:inhom} and schematically shown in Fig.~\ref{fig:jitter}(a). Fast jitter on the other hand is characterized by flucuations of the frequency that are much faster than relevant timescales $t$ and $\tau$ that can be resolved in the experiment. Note that in every experimental realization the time resolution is dictated by the pulse durations. Therefore, fast jitter is characterized by fluctuations also between the pulses in each sequence [see Fig.~\ref{fig:jitter}(b)]. By using a concrete realization of the stochastic process as given in Eq.~\eqref{eq:f_stoch}, we could in principle investigate the transition regime between these two limiting cases.

\subsubsection{Echo spectroscopy for arbitrary pulse areas}\label{sec:echo_arb}
Previously we have seen that echo spectroscopy is very useful for determining homogeneous and inhomogeneous linewidths of emitters independently from each other. There we calculated the echo signal for a specific $\pi/2$-$\pi$ pulse sequence. However this is not the only sequence that can create echos. If we had any arbitrary first pulse instead of the $\pi/2$-pulse creating a non-vanishing coherence $p_0$, i.e., using any pulse area except for multiples of $\pi$, the coherence right before the second pulse would simply be
\begin{equation}
	p_-(t=0)=p_0 e^{-\gamma\tau}e^{-i\omega_x\tau}\,.
\end{equation}
Only the absolute value of the coherence and thus the amplitude of the echo depends on the specific pulse area of the first pulse. Any following $\pi$-pulse with arbitrary phase $\phi$ transforms the polarization according to Eq.~\eqref{eq:action_pi}
\begin{equation}
	p_+(t=0)\sim p_-^*(t=0)\,.\label{eq:action_for_echo}
\end{equation}
This always leads to a polarization signal of the form in Eq.~\eqref{eq:p_pi_2_pi}
\begin{equation}
	p(t,\tau;\omega_x)\sim e^{-\gamma(t+\tau)}e^{-i\omega_x(t-\tau)}\,.\label{eq:p_two_pulse_phases}
\end{equation} 
Whenever the second pulse transforms the polarization according to Eq.~\eqref{eq:action_for_echo} the complex conjugate of the free time evolution before this pulse $\sim\exp(i\omega_x\tau)$ and the free time evolution after this pulse $\sim\exp(-i\omega_x t)$ can interfere constructively at $t=\tau$ for every arbitrary ensemble, enabling a photon echo. Thus, whenever we have a two-pulse sequence where the second pulse is a $\pi$-pulse, we can perform echo spectroscopy. 

In the experiment however it can be challenging to set the pulse area of the second pulse to exactly $\theta=\pi$. Possible reasons include: (i) The required intensity can be too high, such that the laser pulse might irreversibly damage the considered emitter, making repeated measurements impossible~\cite{qin2017photoluminescence}. (ii) The pulse area is not only impacted by the field of the laser itself but also by the dipole of the emitter, which might fluctuate due to influences from its environment or different quantum states~\cite{das2014single}. (iii) Any slight variation in the beam path might change the focusing and therefore the pulse area calibration.

It would therefore be of great benefit to enable the pulse transformation in Eq.~\eqref{eq:action_for_echo} for arbitrary pulse areas $\theta$ of the second pulse. The general polarization transformation rule for ultrashort pulses from Eqs.~\eqref{eq:delta_pulse_transform} reads
\begin{equation}
	p_+=\cos^2(\theta/2) p_-+\frac{i}{2}e^{i\phi}\sin(\theta) (2n_{x-}-1)+\sin^2(\theta/2)e^{2i\phi}p_-^*\,,
\end{equation}
where $\phi$ is the phase of the pulse and $n_{x-}$ is the occupation of the 2LS before this pulse. The three terms on the right hand side have three distinct dependencies on the phase of the pulse. If we could isolate the term $\sim\exp(2i\phi)$ while rejecting all other contributions, we could fulfill the echo condition in Eq.~\eqref{eq:action_for_echo} for arbitrary pulse areas $\theta$. As explained in the following Sec.~\ref{sec:FWM_2LS} and in detail in Chaps.~\ref{sec:exp} and \ref{sec:exp_theo}, heterodyne FWM does exactly that, allowing for echo spectroscopy with arbitrary pulse areas instead of $\pi$-pulses~\cite{abella1966photon,schultheis1985photon,langbein2010coherent}. This fact alone makes FWM spectroscopy a versatile tool for investigating the spectroscopic properties of emitters. Note that in the traditional FWM setup the phases are directly related to the excitation directions $e^{i\bm{k}\bm{r}}$ and the isolation of the correct signal contribution happens via choosing a specific detection direction $\bm{k}_\mathrm{FWM}$ (see Fig.~\ref{fig:WM_scheme}).

\subsection{FWM signals of optically driven 2LSs}\label{sec:FWM_2LS}
\subsubsection{General considerations on FWM}
Four-wave mixing belongs to a general class of multi-pulse experiments, i.e., the system under investigation is excited by multiple laser pulses~\cite{schultheis1985photon,wegener1990line,mukamel1995principles,langbein2010coherent}. If they can be treated as $\delta$-pulses we can write
\begin{equation}
	\mathcal{A}(t)=\sum_j\mathcal{A}_j(t)=\hbar \sum_j \theta_j e^{i\phi_j}\delta(t-t_j)\,,\label{eq:pulse_sequence}
\end{equation}
where $\theta_j$, $\phi_j$, and $t_j$ are the pulse area, phase, and time of excitation of the $j$-th pulse, respectively. Note that we can remove any global phase, as discussed in Sec.~\ref{sec:cw_nodiss}, but not relative phases $\phi_i-\phi_j$ between different pulses. Multiple pulses allow for multi-dimensional investigation of spectroscopy signals using the various relative delays $\tau_{ij}=t_j-t_i$ and the time after the last excitation analog to echo spectroscopy in Fig.~\ref{fig:echo_with_deph}. Being able to vary these delays directly implies that we can perform time-resolved measurements, i.e., investigate dynamical properties~\cite{schultheis1985photon,wegener1990line,langbein2010coherent,wigger2018rabi,groll2020four}. Apart from the possibility to determine different classes of broadenings separately from each other, multi-dimensional spectroscopy allows for the investigation of coupling mechanisms between different states of the investigated system~\cite{patton2006time,langbein2010coherent,kasprzak2011coherent,mermillod2016dynamics,wigger2023controlled}, which will be discussed in more detail in Chap.~\ref{sec:nls} and is another important advantage of FWM over single-pulse spectroscopy techniques.

In the specific heterodyne FWM setup that will be discussed in Chap.~\ref{sec:exp} we directly measure the electric field~\cite{langbein2006heterodyne,langbein2010coherent}, not just the intensity, of the light emitted from the sample that is excited by the laser pulses. In Sec.~\ref{sec:abs_spec} we have discussed that the polarization of a 2LS is connected to the electric field of the exciting laser via its linear response~$\chi$ for sufficiently weak excitation [Eq.~\eqref{eq:2LS_p_linear}]. The reverse statement is also true: the induced electric field in the vicinity of an emitter is proportional to its polarization $p$ for sufficiently weak coupling between the two. This implies that in FWM of 2LSs the polarization $p$ is usually the relevant observable, as already hinted at in Sec.~\ref{sec:echo}. This connection between the emitted light field and the polarization of the emitter is already well known from dipole radiation in classical electrodynamics~\cite{griffiths2005introduction} and can also be derived thoroughly when treating the electromagnetic field quantum mechanically (see App.~\ref{sec:dipole_emission})~\cite{mollow1969power,nazir2016modelling}.

For sufficiently weak exciting laser light the polarization is a linear function of the corresponding field as shown in Eq.~\eqref{eq:2LS_p_linear}. Polarization and field are then connected via the linear response $\chi$, which we call $\chi^{(1)}$ from now on
\begin{equation}
	p\sim \chi^{(1)}\mathcal{A}\,,\qquad\text{linear regime}\,.
\end{equation} 
Here we suppressed the convolution of field and linear response function for notational convenience. In absence of any field the polarization vanishes since the 2LS remains in its ground state. We can understand the above relation as the lowest order of a power series~\cite{shen1984principles,mukamel1995principles}
\begin{equation}
	p\sim\chi^{(1)}\mathcal{A}+\chi^{(2)}\mathcal{A}^2+\chi^{(3)}\mathcal{A}^3+...\,,\label{eq:p_power}
\end{equation}
where we suppress any convolutions and possible complex conjugations of the field in the notation. The higher order terms $\chi^{(n>1)}$ are called nonlinear response functions analog to $\chi^{(1)}$ being the linear response function and they determine how the 2LS is impacted by stronger exciting laser light (or multiple pulses). For many systems the second order response $\chi^{(2)}$ vanishes because they are invariant with respect to spatial inversion, such that the lowest relevant nonlinear response is usually given by $\chi^{(3)}$~\cite{liu2016principles}.

If we are now interested in the impact that a series of multiple pulses has on the polarization of a 2LS we naturally go beyond the linear regime. We can easily see this by looking at the $\delta$-pulse transformations in Eqs.~\eqref{eq:delta_pulse_transform}
\begin{subequations}\label{eq:delta_pulse_transform_linear}
	\begin{align}
		p_+&=\cos^2(\theta/2) p_-+\frac{i}{2}e^{i\phi}\sin(\theta) (2n_{x-}-1)+\sin^2(\theta/2)e^{2i\phi}p_-^* \notag\\
			&\approx p_-+\frac{i}{2}e^{i\phi}\theta (2n_{x-}-1)+\mathcal{O}(\theta^2)\,,\\
		n_{x+}&=n_{x-}+\sin^2(\theta/2)(1-2n_{x-})-\sin(\theta) \text{Im}\left(e^{-i\phi}p_-\right)\approx n_{x-}-\theta\text{Im}\left(e^{-i\phi}p_-\right)+\mathcal{O}(\theta^2) \,.
	\end{align}
\end{subequations}
A single pulse thus impacts the polarization and occupation at least linearly with respect to its pulse area. We can now ask how the polarization behaves depending on the pulse areas and/or phases of $n$ individual pulses. This question can only be meaningfully answered by investigating at least the order $\sim\chi^{(n)}$ in Eq.~\eqref{eq:p_power}. In FWM we consider in general three pulses, such that the measured polarization response is a nonlinear spectroscopy signal of at least third order in the exciting laser pulses. We here usually refer to the excitation with sufficiently weak pulses as $\chi^{(3)}$-regime, where higher orders in Eq.~\eqref{eq:p_power} can be neglected. Note that the term \textit{four}-wave mixing stems from the fact that the three excitation fields are mixed by the sample to create the signal, i.e., the fourth field~\cite{shen1984principles,mukamel1995principles}.

\subsubsection{Rules for calculating heterodyne FWM signals}\label{sec:FWM_rules}
After this preliminary general discussion in the following we formulate how to calculate FWM signals of 2LSs generated by ultrashort pulses. Keep in mind that the ultrashort pulse limit (see Sec.~\ref{sec:delta_pulses}) is an idealization which allows us to perform simple analytical calculations of the signals. In the case of longer pulses in general numerical simulations are required, whose interpretation can often be supported by the results from the ultrashort pulse limit~\cite{hahn2021influence}.
\begin{itemize}
	\item[1.] The system is excited by three, sometimes two, laser pulses with arbitrary independent phases $\phi_j$ and the total field is given by Eq.~\eqref{eq:pulse_sequence}, if the pulses can be idealized as $\delta$-pulses. In this case the pulses transform the system according to Eqs.~\eqref{eq:delta_pulse_transform_linear} (left of the $\approx$). The free time evolution before and after each pulse is described by the Bloch equations~\eqref{eq:bloch_eq_diss} for $\mathcal{A}=0$.
 	\item[2.] The relevant observable is the complex electric field emitted from the sample under investigation, which is proportional to the polarization $p$ in the case of a 2LS (see App.~\ref{sec:dipole_emission}). Since this polarization is created by the exciting pulses, it is in general a function of the pulse phases $p(\phi_1,\phi_2,\phi_3)$.
 	\item[3.] In heterodyne FWM spectroscopy we usually select the specific phase combination $\phi_\mathrm{FWM}=\phi_3+\phi_2-\phi_1$ by integrating the detected signal over the phases of the exciting pulses~\cite{shen1984principles,langbein2006heterodyne,langbein2010coherent}
 	\begin{equation}
 		p_\mathrm{FWM}=\frac{1}{(2\pi)^3}\iiint\limits_{\!\!0}^{\ 2\pi}\text{d}\phi_1\,\text{d}\phi_2\,\text{d}\phi_3\, p(\phi_1,\phi_2,\phi_3)e^{-i\phi_\mathrm{FWM}}	\label{eq:phase_sel}
 	\end{equation}
 	In the case of two pulses the FWM phase is $\phi_\mathrm{FWM}=2\phi_2-\phi_1$. The experimental implementation of this integration will be discussed in Chap.~\ref{sec:exp}. Due to this method we naturally remove the first (and second) order contribution from the polarization in Eq.~\eqref{eq:p_power} since it only contains terms $\sim \exp(\pm i\phi_i)$ (and products of two of these). Only the third order contribution $\sim\chi^{(3)}$ and possibly higher orders contain the correct phase $e^{i(\phi_3+\phi_2-\phi_1)}$. Note that we also achieve background free detection due to the phase selection (more details in Chaps.~\ref{sec:exp} and \ref{sec:exp_theo}), an important benefit of FWM compared to other spectroscopy methods. The laser pulses themselves carry simply the phase $e^{i\phi_j}$ and never the combination $e^{i(\phi_3+\phi_2-\phi_1)}$, such that both resonant reflection and stray light from the laser are filtered out.
\end{itemize}
Before proceeding with straightforward calculations using these rules, it is a good idea to consider the mathematical structure of the problem. To do so, we return to the density matrix representation of the 2LS, as it allows for a very efficient calculation of FWM signals. First of all we can rephrase the three rules from above as
\begin{itemize}
	\item[1.a.] Before and after each pulse the density matrix $\rho$ of the system evolves according to the \textit{linear} Lindblad equation~\eqref{eq:lindblad}
	\begin{equation}\label{eq:Lindblad_FWM}
			\dvt \rho(t)=-i\omega_x\left[\ket{x}\bra{x},\rho(t)\right]+\mathcal{D}_\mathrm{xd}\left[\rho(t)\right]+\mathcal{D}_\mathrm{pd}\left[\rho(t)\right]\,,
	\end{equation}
	where the 2LS Hamiltonian in the absence of optical driving is simply given by $\left.H_\mathrm{2LS}\right|_{\mathcal{A}=0}=\hbar\omega_x\ket{x}\bra{x}$.
	\item[1.b.] A $\delta$-pulse at time $t_j$ with pulse area $\theta_j$ and phase $\phi_j$ transforms the density matrix \textit{linearly} as [see Eq.~\eqref{eq:U_P}]
	\begin{subequations}\label{eq:U_P_FWM}
		\begin{align}
			\rho_+&=U_P(\theta_j,\phi_j)\rho_-U_P^{\dagger}(\theta_j,\phi_j)\,,\\
			U_P(\theta_j,\phi_j)&=\cos(\theta_j/2)-i\left(\ket{x}\bra{g}e^{i\phi_j}+\ket{g}\bra{x}e^{-i\phi_j}\right)\sin(\theta_j/2)\,.
		\end{align}
	\end{subequations}
	\item[2.] The polarization $p$ is the relevant observable and can be calculated from the density matrix via the \textit{linear} operation $p=\text{Tr}(\ket{g}\bra{x}\rho)$. Since the density matrix is impacted by the exciting pulses, it is in general a function of the pulse phases $\rho(\phi_1,\phi_2,\phi_3)$.
	\item[3.] The relevant FWM contribution to the polarization is given by Eq.~\eqref{eq:phase_sel}. There we have $p(\phi_1,\phi_2,\phi_3)= \text{Tr}[\ket{g}\bra{x}\rho(\phi_1,\phi_2,\phi_3)]$. Since the relation between $p$ and $\rho$ is a linear one, we can instead perform the phase selection directly in the density matrix
	\begin{equation}
	\rho_\mathrm{FWM}=\frac{1}{(2\pi)^3}\iiint\limits_{\!\!0}^{\ 2\pi}\text{d}\phi_1\,\text{d}\phi_2\,\text{d}\phi_3\, \rho(\phi_1,\phi_2,\phi_3)e^{-i\phi_\mathrm{FWM}}	\label{eq:phase_sel_rho}
\end{equation}
	and calculate the FWM polarization with it afterwards.
\end{itemize}
According to rule 1.b the $j$-th pulse creates terms in the density matrix $\rho(\phi_j)$ with phase dependencies $\sim 1, e^{\pm i\phi_j}, e^{\pm 2i\phi_j}$. Since all subsequent transformations of the density matrix, i.e., free time evolution and additional pulse transformations, are \textit{linear} operations, the terms with different dependencies on the phase $\phi_j$ are at most shuffled around and added together, but each term retains its specific phase dependence. In the end, using the phase selection in Eq.~\eqref{eq:phase_sel_rho}, we discard most of the terms and only keep a few specific terms, e.g., $\sim e^{-i\phi_1}$. Since all density matrix transformations are linear however and the dependence on $\phi_j$ does not change after the $j$-th $\delta$-pulse, we can simply perform the phase selection
\begin{equation}
	\rho(\phi_j)\rightarrow \frac{1}{2\pi}\int\limits_0^{2\pi}\text{d}\phi_j\, \rho(\phi_j)e^{-il_j\phi_j}\label{eq:phase_sel_rho_single}=\overline{\rho}_{(j)}
\end{equation}
directly after the $j$-th pulse without losing information relevant to the FWM signal~\cite{groll2020four,hahn2021influence,hahn2022destructive}. Here, $l_j$ is a non-zero integer, e.g., $l_1=-1$ in FWM and the index $(j)$ denotes that $\overline{\rho}$ has been phase-selected after the $j$-th pulse. This greatly simplifies the calculation of FWM signals. Note that this is generally not possible when dealing with extended pulses since then there is no properly defined "after" the pulse~\cite{schulze1995pulse,wigger2018rabi}. A very important property of the phase-selected density matrix, courtesy of $\text{Tr}[\rho(\phi_j)]=1$, is given by 
\begin{equation}
\text{Tr}[\overline{\rho}_{(j)}]=0\,,	\label{eq:trace_zero_FWM}
\end{equation}
i.e., the phase-selected density matrix is not a true density matrix in the sense of Eqs.~\eqref{eq:def_rho} anymore. We will still use the name density matrix for notational convenience however. The property in Eq.~\eqref{eq:trace_zero_FWM} implies that whenever the diagonal entries of the phase-selected density matrix in the basis $\lbrace\ket{g},\ket{x}\rbrace$ are non-vanishing, we only need to keep track of the entry $\sim\ket{x}\bra{x}$, i.e., the occupation $n_x$ of the 2LS. The term $\sim\ket{g}\bra{g}$ simply has the opposite sign. We can also understand this from the general relation $n_g=1-n_x$ for any unfiltered density matrix. Since the $1$ does not carry any phase, we are left with $n_g=-n_x$ after the phase selection. Since the Lindblad equation~\eqref{eq:Lindblad_FWM} and the pulse transformation with Eqs.~\eqref{eq:U_P_FWM} are trace-preserving~\cite{breuer2002theory}, the phase-selected density matrix keeps the property $\text{Tr}(\overline{\rho})=0$ also during the free time evolution and subsequent pulse transformations.

To make clear why we choose to investigate the specific phase combinations $\phi_\mathrm{FWM}=2\phi_2-\phi_1$ (two-pulse FWM) or $\phi_\mathrm{FWM}=\phi_3+\phi_2-\phi_1$ (three-pulse FWM) to obtain the signal $p_\mathrm{FWM}$, it is best to perform example calculations, which we do in the following.
\subsubsection{Two-pulse FWM: coherence dynamics}\label{sec:p_FWM_2}
\begin{figure}[h]
	\centering
	\includegraphics[width=0.6\linewidth]{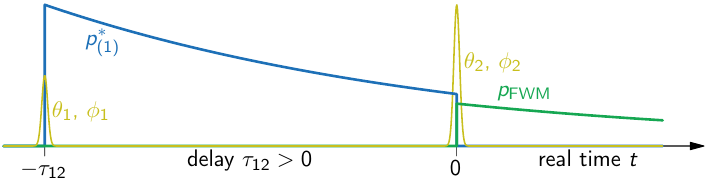}
	\caption{Two-pulse FWM for a single realization from a given ensemble at positive delay $\tau_{12}=t_2-t_1>0$. Pulse 1 excites the system at $t=-\tau_{12}$, creating a conjugated polarization $p^*$ (blue) after the phase selection of terms $\sim e^{-i\phi_1}$. Pulse 2 transforms this conjugated polarization into the FWM polarization (green), whose dynamics are recorded as a function of the real time $t\geq 0$.}
	\label{fig:2pulsesequence}
\end{figure}
As a concrete example we consider two-pulse FWM~\cite{langbein2010coherent} with a relative delay $\tau_{12}=t_2-t_1>0$ chosen positive, as depicted in Fig.~\ref{fig:2pulsesequence} (yellow peaks). This means that pulse 1 excites the system first, it then evolves freely for the duration $\tau_{12}$ followed by pulse 2. The timescale after the last pulse $t$, also called real time in contrast to the delay time $\tau_{12}$, is defined such that the last pulse excites the system at $t=0$. This definition of the real time is common~\cite{langbein2010coherent,groll2020four,li2023optical} and would also be chosen in the alternative case of negative delay $\tau_{12}$, which we discuss shortly in the following, such that pulse 1 would be the last pulse exciting at $t=0$. It is important to note that the labels 1 and 2 do not necessarily describe the temporal order of the pulses. The relevant FWM phase in two-pulse FWM is $\phi_\mathrm{FWM}=2\phi_2-\phi_1$, i.e., $l_1=-1$, $l_2=2$ in Eq.~\eqref{eq:phase_sel_rho_single}.

We assume that the 2LS under investigation is in the ground state $\ket{g}$ prior to the excitation with pulse 1, i.e., $\rho_-(t=-\tau_{12})=\ket{g}\bra{g}$, such that using Eqs.~\eqref{eq:U_P_FWM} the density matrix after pulse 1 is given by
\begin{equation}
	\rho_+(t=-\tau_{12})=\left(\begin{matrix}1-\sin^2(\theta_1/2)&\frac{i}{2}e^{-i\phi_1}\sin(\theta_1)\\-\frac{i}{2}e^{i\phi_1}\sin(\theta_1)&\sin^2(\theta_1/2)\end{matrix}\right)\,.\label{eq:rho_pulse_1}
\end{equation}
According to Eq.~\eqref{eq:phase_sel_rho_single} with $l_1=-1$ we only need to retain
\begin{equation}
	\overline{\rho}_{(1)+}(t=-\tau_{12})=\left(\begin{matrix}0&\frac{i}{2}\sin(\theta_1)\\0&0\end{matrix}\right)\label{eq:rho_pulse_1_phasesel}
\end{equation}
for the calculation of the FWM polarization. We thus discard all entries in the density matrix, except for the conjugate polarization $p^*$ (Fig.~\ref{fig:2pulsesequence}, blue). Note that in the case $\tau_{12}<0$ we would not obtain any FWM signal~\cite{wegener1990line}. As can be seen in Eq.~\eqref{eq:rho_pulse_1}, the first pulse exciting the system from the ground state cannot imprint a phase $\sim e^{\pm 2i\phi}$, which would be needed in that case.

During the following free time evolution according to Eq.~\eqref{eq:Lindblad_FWM} the phase-selected density matrix evolves and reaches at $t=0$ the value 
\begin{equation}
	\overline{\rho}_{(1)-}(t=0)=\left(\begin{matrix}0&\frac{i}{2}\sin(\theta_1)e^{-(\gamma-i\omega_x)\tau_{12}}\\0&0\end{matrix}\right)\,,
\end{equation}
since Eq.~\eqref{eq:Lindblad_FWM} yields
\begin{align}
	\dvt c(t)\ket{g}\bra{x}&=-i\omega_xc(t)\left[\ket{x}\bra{x},\ket{g}\bra{x}\right]+c(t)\mathcal{D}_\mathrm{xd}\left(\ket{g}\bra{x}\right)+c(t)\mathcal{D}_\mathrm{pd}\left(\ket{g}\bra{x}\right)\notag\\
	&=-(\gamma-i\omega_x) c(t) \ket{g}\bra{x}\label{eq:p*_evolve}
\end{align}
when acting on any phase-selected density matrix of the form $\rho(t)=c(t)\ket{g}\bra{x}$ with $c\in\mathbb{C}$, i.e., containing only a conjugate polarization $p^*$ entry. Using again Eqs.~\eqref{eq:U_P_FWM}, the second pulse transforms the density matrix into
\begin{align}\label{eq:rho_pulse_2}
	&\overline{\rho}_{(1)+}(t=0)\\
	&=\left(\begin{matrix}-\frac{1}{2}\sin(\theta_1)e^{-(\gamma-i\omega_x)\tau_{12}}e^{i\phi_2}\sin\left(\frac{\theta_2}{2}\right)\cos\left(\frac{\theta_2}{2}\right)&\frac{i}{2}\sin(\theta_1)e^{-(\gamma-i\omega_x)\tau_{12}}\cos^2\left(\frac{\theta_2}{2}\right)\\\frac{i}{2}\sin(\theta_1)e^{-(\gamma-i\omega_x)\tau_{12}}e^{2i\phi_2}\sin^2\left(\frac{\theta_2}{2}\right)&\frac{1}{2}\sin(\theta_1)e^{-(\gamma-i\omega_x)\tau_{12}}e^{i\phi_2}\sin\left(\frac{\theta_2}{2}\right)\cos\left(\frac{\theta_2}{2}\right)\end{matrix}\right)\,.\notag
\end{align}
From this density matrix we only need to keep 
\begin{equation}
	\overline{\rho}_{(1,2)+}(t=0)=\left(\begin{matrix}0&0\\\frac{i}{2}\sin(\theta_1)\sin^2\left(\frac{\theta_2}{2}\right)e^{-(\gamma-i\omega_x)\tau_{12}}&0\end{matrix}\right)
\end{equation}
using Eq.~\eqref{eq:phase_sel_rho_single} with $l_2=2$. During the subsequent free evolution to the variable real time $t>0$ we obtain [analog to Eq.~\eqref{eq:p*_evolve} with $\rho(t)=c(t)\ket{x}\bra{g}$]
\begin{equation}
	\overline{\rho}_{(1,2)+}(t)=\left(\begin{matrix}0&0\\\frac{i}{2}\sin(\theta_1)\sin^2\left(\frac{\theta_2}{2}\right)e^{-(\gamma-i\omega_x)\tau_{12}}e^{-(\gamma+i\omega_x)t}&0\end{matrix}\right)\label{eq:rho_1_2_t}
\end{equation}
yielding the final FWM polarization (Fig.~\ref{fig:2pulsesequence}, green)
\begin{equation}\label{eq:p_FWM_2}
	p_\mathrm{FWM}(t>0,\tau_{12}>0)=\frac{i}{2}\sin(\theta_1)\sin^2\left(\frac{\theta_2}{2}\right)e^{-\gamma(t+\tau_{12})}e^{-i\omega_x(t-\tau_{12})}\,.
\end{equation}
Note that this is exactly the polarization in Eq.~\eqref{eq:p_pi_2_pi} required to obtain an echo when considering an ensemble of 2LSs, as announced in Sec.~\ref{sec:echo_arb}. Since the ensemble average in Eq.~\eqref{eq:def_p_avg} constitutes another linear transformation, we can apply it also after calculating the FWM polarization for the considered ensemble, assuming that a single run of the FWM experiment only acts on a single frequency component from the ensemble. The $\chi^{(3)}$-regime of this polarization is given by the lowest order in $\theta_1$ and $\theta_2$ and reads
\begin{equation}\label{eq:p_FWM_2_chi3}
	p_\mathrm{FWM}(t>0,\tau_{12}>0)=\frac{i}{8}\theta_1\theta_2^2e^{-\gamma(t+\tau_{12})}e^{-i\omega_x(t-\tau_{12})}\,\qquad\chi^{(3)}\text{-regime}\,.
\end{equation}
To conclude this section, two-pulse FWM allows for the measurement of the dephasing rate $\gamma$, i.e., the homogeneous broadening, as well as separate measurement of any inhomogeneous broadening analog to Fig.~\ref{fig:echo_with_deph}, even for arbitrary small pulse areas. This provides a clear advantage over usual echo spectroscopy~\cite{abella1966photon,schultheis1985photon,langbein2010coherent,boule2020coherent}.

\subsubsection{Three-pulse FWM: occupation dynamics}\label{sec:p_FWM_3}
\begin{figure}[h]
	\centering
	\includegraphics[width=0.6\linewidth]{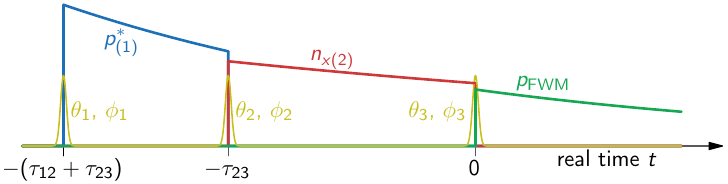}
	\caption{Three-pulse FWM for a single realization from a given ensemble at positive delays $\tau_{12}=t_2-t_1>0$ and $\tau_{23}=t_3-t_2>0$. Pulse 1 excites the system at $t=-\tau_{12}-\tau_{23}$, creating a conjugated polarization $p^*$ (blue) after the phase selection of terms $\sim e^{-i\phi_1}$. Pulse 2, acting at time $t=-\tau_{23}$ transforms this conjugated polarization into an occupation $n_x$ (red) after the phase selection of terms $\sim e^{i\phi_2}$. Pulse~3 transforms this occupation into the FWM polarization (green), whose dynamics are recorded as a function of the real time $t\geq 0$.}
	\label{fig:3pulsesequence}
\end{figure}
We will now consider three pulses and select for the FWM phase $\phi_\mathrm{FWM}=\phi_3+\phi_2-\phi_1$, i.e., $l_3=l_2=-l_1=1$. Specifically we consider a positive delay $\tau_{12}>0$ between pulses 1 and 2 and a positive delay $\tau_{23}>0$, such that pulse 1 and 2 excite the 2LS first, followed by pulse 3, as sketched in Fig.~\ref{fig:3pulsesequence} (yellow peaks). We already know from Eq.~\eqref{eq:rho_pulse_1_phasesel} how pulse~1 including phase-selection transforms the 2LS, inducing a conjugate polarization $p^*$ (Fig.~\ref{fig:3pulsesequence}, blue). From Eq.~\eqref{eq:rho_pulse_2} with $\tau_{12}>0$ we then know the action of pulse 2. Retaining only terms $\sim e^{i\phi_2}$, we get the phase-selected density matrix after pulse 2 as
\begin{align}
	&\overline{\rho}_{(1,2)+}(t=-\tau_{23})\\
	&=\frac{1}{2}\sin(\theta_1)e^{-(\gamma-i\omega_x)\tau_{12}}\left(\begin{matrix}-\sin\left(\frac{\theta_2}{2}\right)\cos\left(\frac{\theta_2}{2}\right)&0\\0&\sin\left(\frac{\theta_2}{2}\right)\cos\left(\frac{\theta_2}{2}\right)\end{matrix}\right)\,.\notag
\end{align}
As discussed previously in Eq.~\eqref{eq:trace_zero_FWM}, the trace of this phase-selected density matrix vanishes such that the excited state occupation $n_x$ (Fig.~\ref{fig:3pulsesequence}, red), i.e., the term $\sim\ket{x}\bra{x}$, fully determines the term $\sim\ket{g}\bra{g}$. The free time evolution can be derived analog to Eq.~\eqref{eq:p*_evolve} with $\rho(t)=c(t)(\ket{x}\bra{x}-\ket{g}\bra{g})$, leading to
\begin{align}
	&\overline{\rho}_{(1,2)-}(t=0)\\
	&=\frac{1}{2}\sin(\theta_1)e^{-(\gamma-i\omega_x)\tau_{12}}\left(\begin{matrix}-\sin\left(\frac{\theta_2}{2}\right)\cos\left(\frac{\theta_2}{2}\right)e^{-\gamma_\mathrm{xd}\tau_{23}}&0\\0&\sin\left(\frac{\theta_2}{2}\right)\cos\left(\frac{\theta_2}{2}\right)e^{-\gamma_\mathrm{xd}\tau_{23}}\end{matrix}\right)\,.\notag
\end{align}
The transformation due to the last pulse, i.e., pulse 3, including the selection of terms $\sim e^{i\phi_3}$ yields
\begin{equation}
	\overline{\rho}_{(1,2,3)+}(t=0)=\left(\begin{matrix}0&0\\i\sin(\theta_1)e^{-(\gamma-i\omega_x)\tau_{12}}\sin\left(\frac{\theta_2}{2}\right)\cos\left(\frac{\theta_2}{2}\right)e^{-\gamma_\mathrm{xd}\tau_{23}}\sin\left(\frac{\theta_3}{2}\right)\cos\left(\frac{\theta_3}{2}\right)&0\end{matrix}\right)\,.
\end{equation}
After the free evolution to the real time $t>0$ analog to Eq.~\eqref{eq:rho_1_2_t} we obtain the FWM polarization (Fig.~\ref{fig:3pulsesequence}, green)
\begin{align}
	&p_\mathrm{FWM}(t>0,\tau_{12}>0,\tau_{23}>0)\notag\\
	&=i\sin(\theta_1)e^{-(\gamma-i\omega_x)\tau_{12}}\sin\left(\frac{\theta_2}{2}\right)\cos\left(\frac{\theta_2}{2}\right)e^{-\gamma_\mathrm{xd}\tau_{23}}\sin\left(\frac{\theta_3}{2}\right)\cos\left(\frac{\theta_3}{2}\right)e^{-(\gamma+i\omega_x)t}\notag\\
	&=\frac{i}{4}\sin(\theta_1)\sin(\theta_2)\sin(\theta_3)e^{-i\omega_x(t-\tau_{12})}e^{-\gamma(t+\tau_{12})}e^{-\gamma_\mathrm{xd}\tau_{23}}\,.\label{eq:p_FWM_3}
\end{align}
Note that for $\tau_{23}=0$ and properly chosen pulse areas $\theta_2$ and $\theta_3$ we recover the result in Eq.~\eqref{eq:p_FWM_2} obtained for two-pulse FWM. While in that case we were only able to determine the dephasing rate $\gamma$, i.e., we could investigate the coherence dynamics, three-pulse FWM also allows for the measurement of the decay rate $\gamma_\mathrm{xd}$ by varying the delay $\tau_{23}$~\cite{mermillod2016dynamics,jakubczyk2016radiatively,wigger2018rabi}. This means that we can fully characterize any 2LS with respect to dephasing, excited state decay and inhomogeneous broadening using heterodyne FWM. In principle we can determine all three quantities in a single experiment when recording the signal as a function of the three time arguments $t-\tau_{12}$ (echo/inhomogeneous coherence dynamics), $t+\tau_{12}$ (dephasing/homogeneous coherence dynamics), and $\tau_{23}$ (occupation/decay dynamics), as demonstrated in Eq.~\eqref{eq:p_FWM_3}. If we compare this to the discussion on linear absorption spectra in Sec.~\ref{sec:inhom}, there we could not separately determine the decay rate or distinguish between homogeneous and inhomogeneous broadening. This impressively demonstrates that FWM constitutes a versatile and powerful spectroscopy method.

%% file: 4_Experiment.tex
\section{Experimental realization of heterodyne wave mixing spectroscopy}\label{sec:exp}
In this chapter we describe the experimental setup for performing heterodyne FWM spectroscopy (Sec.~\ref{sec:setup}). We then discuss the data acquisition and analysis which is required to retrieve complete information on the field emitted by the sample, including amplitude and phase (Sec.~\ref{sec:FWM_data}). We conclude this chapter with a discussion on two-dimensional FWM spectra (Sec.~\ref{sec:2d_spectra}). There we provide typical examples for these spectra measured on different excitonic 2LSs and compare them with the theoretical model from Chaps.~\ref{sec:theory} and \ref{sec:theory_2}.
\subsection{The wave mixing setup}\label{sec:setup}
Setting up a heterodyne wave mixing micro-spectroscopy experiment requires a dedicated optical laboratory. As outlined in Sec.~\ref{sec:FWM_rules} we need the following key ingredients:
\begin{itemize}
	\item Excitation with (typically) three laser pulses with tunable delays.
	\item Assigning phases $\phi_i$ to the excitation pulses.
	\item Isolation of the FWM part of the signal with the correct phase combination of the $\phi_i$'s.
	\item Heterodyne interference to detect the sample's polarization field (not just the intensity).
\end{itemize}
In this chapter we describe in detail how these ingredients are realized in the experimental setup, which needs to meet various conditions, especially when working with individual nanosystems. The experiment requires the combination of knowledge and practical skills in operating ultrafast lasers, nonlinear spectroscopy, cryogenics, radio-frequency (RF) electronics, optics, and in hardware programming. The layout implemented here originates from the heterodyne spectroscopy experiment developed over the years in the group of Wolfgang Langbein and Paola Borri\,\cite{patton2004non, langbein2010coherent, naeem2015giant}, with further specific developments and solutions having been introduced in the setup operating in Grenoble at Institut N\'eel. It is worth mentioning that commercial solutions for coherent spectroscopy recently became available from {\it MONSTR Sense Technologies}~\cite{purz2022rapid,li2023optical}. 

\begin{figure}[t]
	\centering
	\includegraphics[width = 0.9\textwidth]{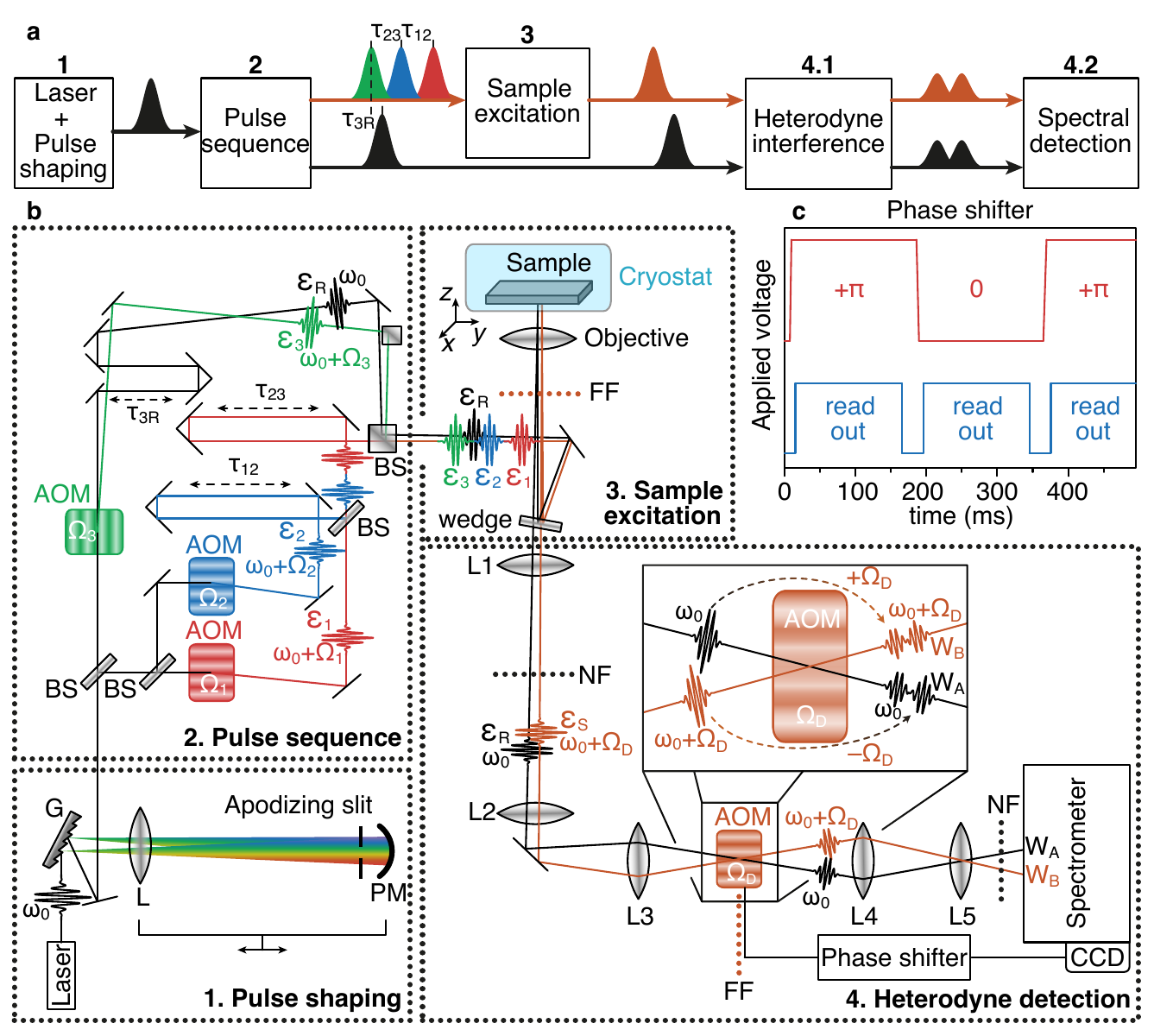}
	\caption{Schematic of the heterodyne wave mixing setup. (a) Flow chart of the different processing stages in the setup. The numbers correspond to the detailed image in (b). The abbreviations are: Grating (G), lens (L), parabolic mirror (PM), beam splitter (BS), 96\% transmission wedged window (wedge), acousto-optic modulator (AOM), near field (NF) and far field (FF) positions. (c) Schematic picture of an oscilloscope image measuring the voltage at the phase shifter in box 4.}
	\label{fig:setup}
\end{figure}

\begin{figure}[t]
	\centering
	\includegraphics[width = 0.7\textwidth]{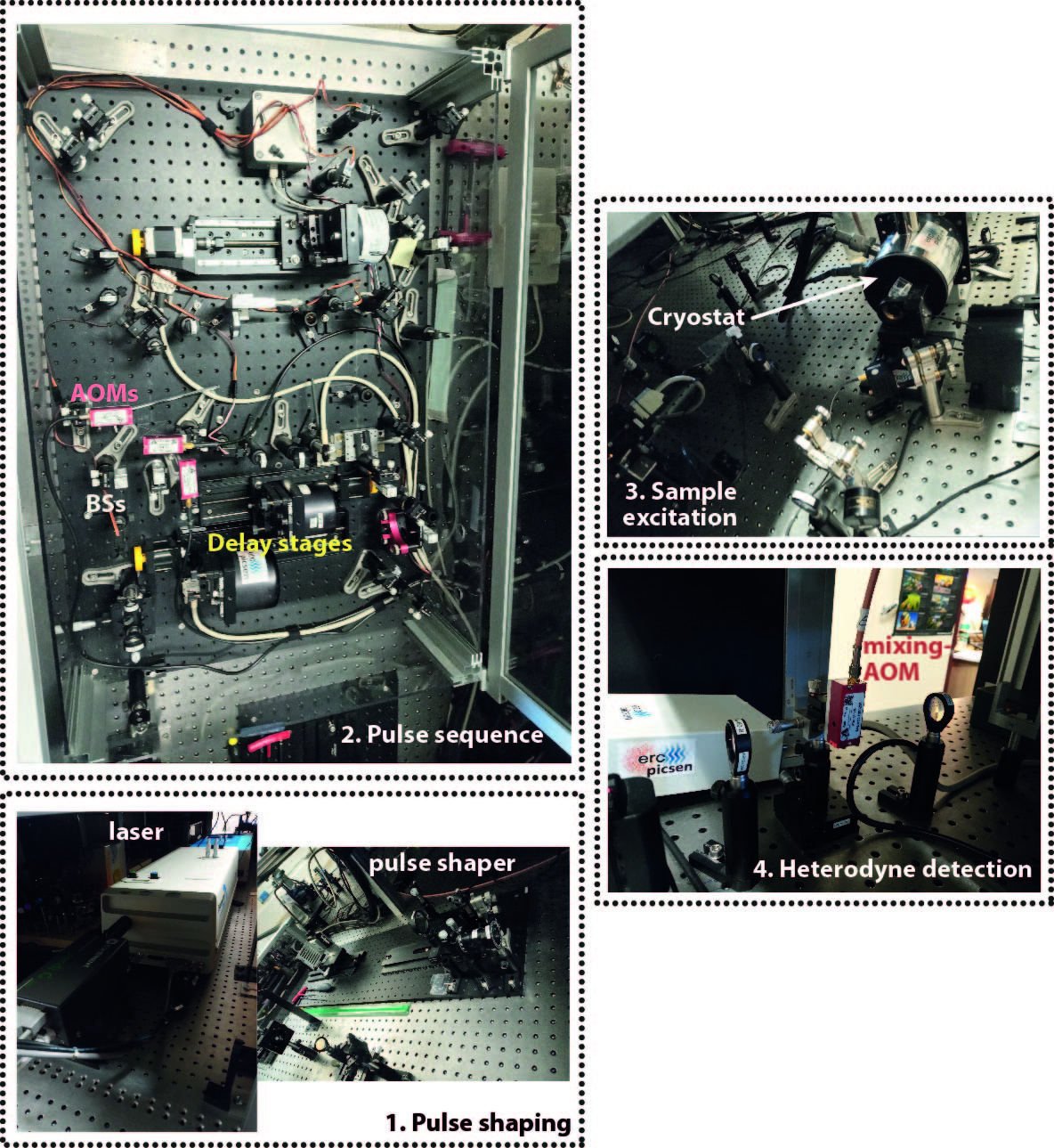}
        \caption{Photos of the FWM setup at Institut N\'eel, Grenoble. The different components are depicted according to the boxes in Fig.~\ref{fig:setup}(b). The red elements in box 2 are the three AOMs for RF labelling, the grey cubes are beam splitters (BSs) and the three delay stages have yellow screws attached at one end.}
        \label{fig:photos}
\end{figure}

In the following we will refer to the current status of the setup at Institut N\'eel. The laboratory is air-conditioned and maintained at a constant temperature of $22^\circ$C and humidity of 50\%, especially enabling a long term stability of mode-locked laser systems. The experiment is constructed on a 9\,m$^2$ large, L-shaped optical table, actively stabilized by air-spring supports from {\it OPTA GmbH}. The optical table is entirely covered by a home-made and adaptable enclosure to minimize air flow instabilities and turbulence which could affect the optical phase of the beams and thus decrease the contrast and phase-stability of the measured spectral interferences. The setup is depicted in Fig.~\ref{fig:setup} and illustrative photos are shown in Fig.~\ref{fig:photos}.

\subsubsection{Pulse shaping}
Optical heterodyne detection~\cite{collett1987quantum, langbein2010coherent, naeem2015giant} relies on a fixed repetition rate within the optical pulse train, in practice produced by a mode-locked laser [box 1 in Fig.~\ref{fig:setup}(b)]. Our working-horse is a Ti:Sapphire femto-second oscillator ({\it Tsunami} by {\it Spectra-Physics}). This laser provides Fourier-limited pulses of $T_{\rm P}=100$~fs full width at half maximum (FWHM) duration with an adjustable repetition rate of $f_{\rm rep}\approx76$~MHz, which corresponds to a temporal pulse spacing of $\tau_{\rm rep}\approx13$~ns, set by the cavity's optical length. The active medium in the {\it Tsunami} is a Ti$_2$O$_3$/Al$_2$O$_3$ crystalline rod with Ti$^{3+}$ ions, which has a peaked absorption at 515~nm (2.41~eV) between a pair of vibrationally broadened levels ${^2}$T$_{2{\rm g}}$\,--\,${^2}$E${_{\rm g}}$. The \textit{Tsunami} is pumped with a 532~nm (2.33~eV) {\it VerdiG} from \textit{Coherent}, offering a cw power of up to 20~W. The gain medium in such optically pumped semiconductor lasers is vanadate (Nd:YVO$_4$) with a characteristic laser transition at 1064~nm (1.165~eV), which is then intracavity doubled via a lithium triborate (LBO) crystal maintained at a rather high temperature of 140$^\circ$C~\footnote{Spectra-Physics, Millennia V: User's Manual, \href{https://archive.org/details/millennia-v-manual/page/n1/mode/2up}{link}, Accessed: 2025-03-19}. According to the Ti:Sapphire gain curve versus wavelength, we inject between 7~W and 11~W into the cavity (see manual for \textit{Tsunami}~\cite{tsunami}\footnote{Tsunami, Mode-locked Ti:sapphire Laser: User's Manual, \href{https://neurophysics.ucsd.edu/courses/physics_173_273/232A_Rev_D_Tsunami_User_Manual.pdf}{link}, Accessed: 2025-03-19}). The optimal operation is achieved by adjusting the pump-beam power, the end mirror of the laser cavity, and the output coupler, as well as the group velocity dispersion compensation employing a 4-prism setup. When the mode-locking sets in properly, a stable Fourier-limited pulse train with no amplitude modulation is generated. A minute portion of the laser can be sent onto a fast-photodiode to confirm the quality of the pulse-train on an oscilloscope. The mode-locking should be robust against mechanical perturbations. A harsh yet effective way to verify this robustness is to tap on the laser coverage which should not disturb the pulsing. We require the mode-locking to be stable over a scale of 100~hours, which calls for keeping temperature and humidity constant in the lab.

\begin{figure}[t]
	\centering
	\includegraphics[width = 0.45\textwidth]{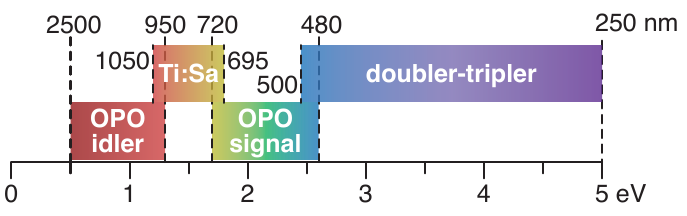}
        \caption{Illustration of the spectral ranges of the different laser sources covering the electromagnetic spectrum from the infrared to the ultra-violet.}
        \label{fig:spec_lasers}
\end{figure}

The Ti:Sapphire laser permits to produce pulses with carrier frequencies spanning from 695~nm (1.78~eV) to 1050~nm (1.18~eV). Hence, to cover more spectral areas than this near-infrared (NIR), other lasers are required. To reach higher energies one can use femto-second pulses from the {\it Tsunami}, centered at 820~nm (1.51~eV) with an average power of 2.5~W, to drive an optical parametric oscillator (OPO) (we use the OPO {\it Inspire-50} from {\it Radiantis}). The pump beam is first frequency-doubled by a nonlinear crystal to 410~nm (3.02~eV) reaching around 1~W of average power, which is then focused onto another built-in nonlinear crystal. Depending on the impinging angle, the crystal generates a signal beam in the visible spectral range, which can be tuned from 480~nm (2.58~eV) to 720~nm (1.72~eV). One can also produce an infrared (IR) idler beam between 950~nm (1.305~eV) and 2500~nm (0.496~eV). To work in the ultraviolet (UV) range we employ a doubler-tripler laser (from \textit{Spectra-Physics}) permitting us to generate pulses from 250~nm (4.96~eV) to 500~nm (2.48~eV) from the beam produced by the {\it Tsunami}. In this way, we can generate ultrafast pulse trains with no spectral gaps from 250~nm (4.96~eV) up to 2500~nm (0.496~eV) as illustrated in Fig.~\ref{fig:spec_lasers}.

The pulse duration $T_{\rm P}$ is an important parameter in time-resolved experiments as it on the one hand determines the addressed spectral window and on the other hand sets a lower limit on the time resolution for investigating dynamical processes. To directly measure $T_{\rm P}$, we use the Auto-Correlator {\it Pulse Check} from {\it APE}. The pulse duration is limited by the Ti:Sapphire laser source ($T_\mathrm{P}=100$~fs), but can be increased by an order of magnitude, up to around $T_{\rm P}=1$~ps by using a passive pulse-shaper~\footnote{Studied in Christian Mann, {\it Vierwellen-Mischen an II-VI-Halbleiter-Nanostrukturen}, Diploma thesis, TU Dortmund, Germany (2000)} [box 1 in Fig.~\ref{fig:setup}(b)]. In case longer pulses are required, up to $T_{\rm P}=10$~ps, we can also use the picosecond oscillator  of the {\it Tsunami}. It should be noted though, that measuring spectral interferences of picosecond laser pulses becomes unhandy because the interference fringes (see Fig.~\ref{fig:process}) for pulses with spectra of envelope widths on the order of $\hbar/T_\mathrm{P}=\hbar/(10\,{\rm ps})\approx 66$~\textmu eV might fall below the spectral resolution limit of standard spectrometers.

\begin{figure}[t]
	\centering
	\includegraphics[width = 0.6\textwidth]{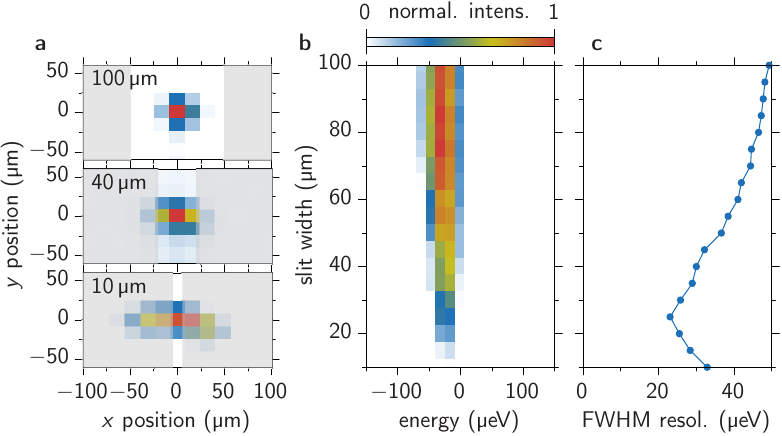}
        \caption{Impact of the spectrometer entrance slit width on the spectral resolution. (a) Examples of the detected intensity on the CCD with growing slit width from bottom to top in zeroth diffraction order of the grating. (b) Detected spectrum in first diffraction order as a function of slit width. (c) Spectral width (FWHM) of (b).}
        \label{fig:slit}
\end{figure}

To demonstrate the resolution of the spectrometer, in Fig.~\ref{fig:slit} we study how the detected spectra of a basic semiconductor diode laser emitting around 750~nm depend on the entry slit width. In Fig.~\ref{fig:slit}(a) we show the detection of the zeroth diffraction order of the grating on the CCD for three different slit openings increasing from 10~\textmu m (bottom) to 100~\textmu m (top). Due to the diffraction of the around 40~\textmu m wide beam at the entry slit, the detection spot on the CCD widens for small openings. For the largest opening, the light beam reaches the CCD without diffraction. In Fig.~\ref{fig:slit}(b) we plot the detected spectrum, i.e., first diffraction order of the grating, as a function of the slit width. We find two features: (i) The detected intensity increases with the slit width as more light enters the spectrometer. (ii) The width of the spectrum depends on the slit opening. To quantify (ii) in Fig.~\ref{fig:slit}(c) we plot the FWHM of the spectrum against the slit width and find a clear minimum around 20~\textmu m with a resolution of 20~\textmu eV due to the competition between diffraction from the slit and the decrease of the illumination spot. For larger slit widths the resolution saturates around 50~\textmu eV, where we have no diffraction from the entry slit but detection of the entire laser spot.

When passing a laser pulse through optical elements, especially thick and high refractive index ones, like microscope objectives or acousto-optic modulators (AOMs), a chirp is accumulated, i.e., different spectral components propagate at different group velocities and result in a frequency modulation within the pulse. Even though no spectral dispersion of the pulse is measured, the pulse duration $T_{\rm P}$ is increased considerably throughout the setup due to this chirp. As a rule of thumb, it typically holds, that the shorter the initial pulse, the more chirp is acquired~\footnote{Studied in Manuel Joffre, {\it Optique non-lin{\'e}aire en r{\'e}gimes continu et femtoseconde}, Master course, Ecole Normale Sup{\'e}rieure - Ecole Polytechnique - Universit{\'e} Pierre et Marie Curie - Universit{\'e} Paris Sud, France (2014)}. If we removed the chirp compensation from our setup, the pulse duration would increase from the initial $T_{\rm P}\approx 100$~fs up to $T_{\rm P}\simeq 1$~ps. A passive pulse-shaper is therefore used to compensate the first-order chirp, as depicted in box 1 in Fig.\,\ref{fig:setup}(b). By changing the G-L and L-PM distances [see labels in Fig.~\ref{fig:setup}(b), box 1], we can vary the optical path of different spectral components. In that manner we introduce a negative first order chirp, such that the pulses arriving at the sample in box 3 have a duration of around $T_{\rm P}=120$~fs for an initial pulse width of $T_\mathrm{P}=100$~fs.

\subsubsection{Pulse sequence}

Now, moving to box 2 in Fig.~\ref{fig:setup}(b), from each laser pulse we generate a sequence of four pulses to perform multi-dimensional time-resolved pump-probe type experiments (see Sec.~\ref{sec:FWM_2LS}). To this aim, the spatially filtered and pre-chirped laser pulse train enters the compact opto-mechanical setup in box 2. By using two non-polarizing beam splitters~(BSs) the beam is divided into three parts $\E_{1,2,3}$ with desired power ratios. Using a telescope configuration with focus ratio 3:1, the beams are focused into AOMs employing Te$_2$O crystals ({\it MT80} by {\it AA Opto Electronic}). The function of the AOMs in our setup is twofold: (I) They introduce frequency shifts and thereby phase labeling to the pulse trains as explained in the following. (II) They act as beam splitters (green AOM with $\Omega_3$).

\begin{figure}[b]
	\centering
	\includegraphics[width = 0.75\textwidth]{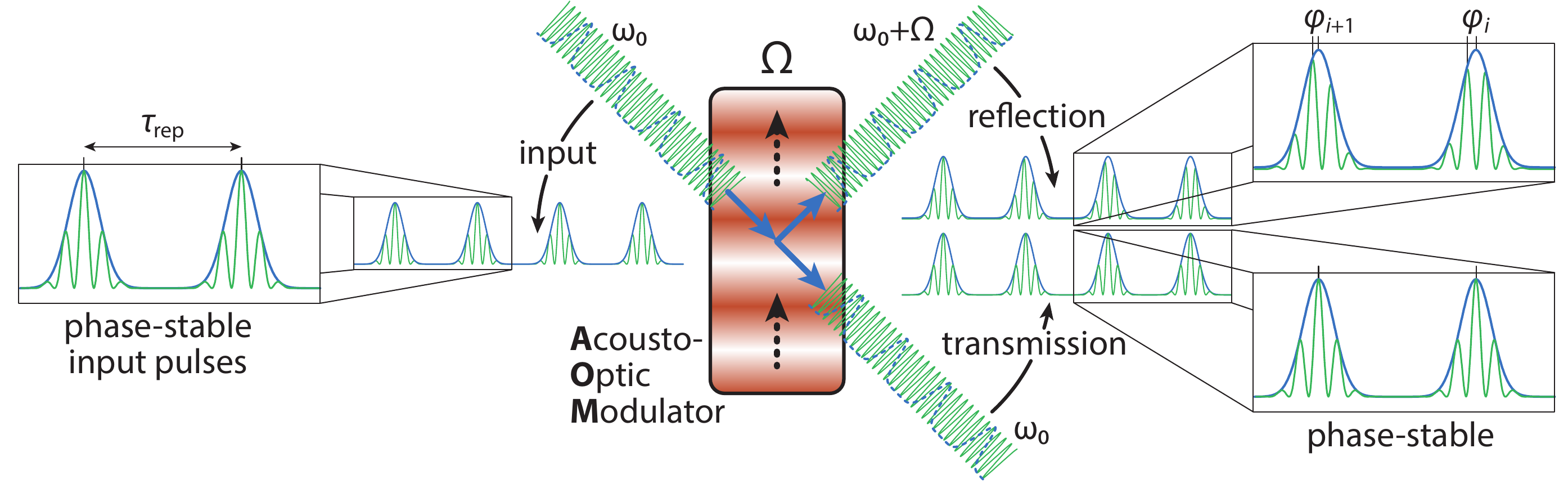}
        \caption{Impact of an AOM on a phase stable pulse train. Carrier fields in green and pulse envelopes in blue. All pulses of the input field have the same vanishing carrier-envelope phase. The transmitted field is still phase stable while the deflected field is shifted to higher frequencies and successive pulses get different phases $\varphi_{i+1}\neq \varphi_{i}$.}
        \label{fig:AOM}
\end{figure}

The impact of an AOM is schematically depicted in Fig.~\ref{fig:AOM}, illustrating that its functionality relies on Brillouin scattering of the injected light beam with acoustic phonons of the set frequency~$\Omega$~\cite{gordon1966review,donley2005double,wan2022highly}. Considering first an injected cw field of frequency $\omega_0$, the light can either be transmitted through the AOM and keep its frequency, or it can be deflected, in the depicted case absorb a phonon, and leave the AOM with the up-converted frequency $\omega_0+\Omega$ (also higher order phonon processes are generally possible leading to larger energy shifts and deflection angles). Note that the frequency in the deflected field will be down-converted if the acoustic field travels in the opposite direction. In the experiment we are dealing with a train of phase-stable laser pulses (constant carrier-envelope phase), which means that the relative phase between the carrier wave (green) and the envelope (blue) is the same in each pulse. The transmitted pulse train keeps its carrier frequency and pulse repetition time $\tau_{\rm rep}$ and therefore remains phase-stable. The deflected pulses however have a larger carrier frequency $\omega_0+\Omega$, which results in a shift of the carrier-envelope phase $\varphi_i$ depending on the number of the pulse $i$ in the pulse train. The phase difference between two successive pulses is given by $\varphi_{i+1}-\varphi_i = \Omega \tau_{\rm rep}$. Consequently, we can use the deflected pulse train and the selected AOM frequency $\Omega$ to phase-label the pulse train (see Sec.~\ref{sec:phase_selection} for more details). The AOM frequency $\Omega$ is typically in the RF range and represents the time-domain equivalent of the excitation direction ${\bm k}$ in angle-resolved, i.e., space-domain, wave mixing experiments~\cite{gobel1990quantum,wegener1990line,kim1992giant,langbein2006heterodyne,langbein2010coherent} (see Fig.~\ref{fig:WM_scheme}). By choosing different frequencies for the three AOMs ($\Omega_{1,2,3}$) used in our pulse sequence generation (colored beams in box 2 of Fig.~\ref{fig:setup}) we have at the same time phase-labeled the beams and brought them into their individual frequency channels: $\E_1 \to \omega_0+\Omega_1$ (red), $\E_2 \to \omega_0+\Omega_2$ (blue), and $\E_3 \to \omega_0+\Omega_3$ (green). In addition we use one transmitted beam (black) as a reference $\E_\mathrm{R}$ for homo- and heterodyne detection in box 4, as discussed later. This beam is not frequency-shifted by the AOMs ($\omega_0$) and is still phase-stable. The choice of the AOM frequencies $\Omega_{1,2,3}$ is discussed in Sec.~\ref{sec:exp_heterodyne} and a detailed explanation of this choice is given in Sec.~\ref{sec:phase_selection}.

Using now telescope configurations with focus ratios of 3:1 the beams are expanded to a beam size of around 6~mm to reduce their divergence during propagation through the setup towards the microscope objective, depicted in box 3 in Fig.~\ref{fig:setup}(b). Typically, we overfill the aperture of the objective, aiming at the tightest spatial focus. The overfilling also helps to maintain a constant beam intensity, form and position on the sample while introducing delays, especially in a ns-range. The different pulses are delayed with respect to each other using mechanical linear stages ({\it VT-80} from {\it Micos}, currently {\it Physik Instrumente}) managed by four-axes {\it Corvus} controllers from {\it Micos}. We typically employ 150\,-\,200~mm stages permitting us to apply delays up to a ns-timescale. In case longer delays are required we can install delay stages up to around 1~m and additionally run multiple paths within the stages. In principle we can thus generate delays of hundreds of ns. When studying very long-lasting signals, that require delays reaching the natural range of the laser repetition time $\tau_{\rm rep}\simeq13$~ns, an additional pulse picker would have to be introduced in the setup in order to decrease the repetition rate of the laser accordingly. After passing the AOMs, the beams are recombined into a common spatial mode [end of box 2, Fig.~\ref{fig:setup}(b)].
\subsubsection{Sample excitation}\label{sec:exp_excitation}
\begin{figure}[h]
	\centering
	\includegraphics[width = 0.67\textwidth]{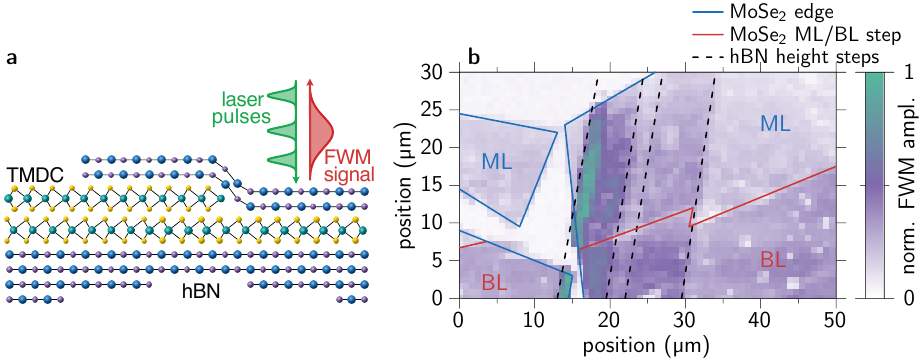}
        \caption{FWM measurement performed on a hBN/MoSe$_2$/hBN heterostructure demonstrating the ability to scan large areas while detecting FWM signals. (a) Schematic picture of the experiment. (b) FWM amplitude map. The lines mark flake edges (blue) and steps between distinct regions (red). The sample was produced in the group of Marek Potemski at LNCMI, CNRS Grenoble, employing hBN from NIMS Tsukuba in Japan.}
        \label{fig:TMDC_map}
\end{figure}

Entering box 3 in Fig.~\ref{fig:setup}(b), all pulses are projected onto the microscope with a $4f$-imaging setup. They are focused on the sample surface using the microscope objective {\it LCPLN50XIR/0.65} from {\it Olympus}, installed on a XYZ closed-loop piezo scanner from {\it Physik Instrumente}. By raster scanning the microscope we can perform imaging of the optical response while simultaneously spectrally resolving the signal, i.e., we can perform spectral imaging. To give a first impression, Fig.~\ref{fig:TMDC_map} shows the FWM amplitude map of a transition metal dichalcogenide (TMDC) MoSe$_2$ sample at room temperature, encapsulated in hexagonal boron nitride (hBN) layers of different thickness. Just from this spatial map, we can distinguish different areas of the sample. The MoSe$_2$ bilayer (BL) region has an overall stronger FWM signal than the MoSe$_2$ monolayer (ML). In addition, the FWM response depends on the thickness of the hBN encapsulation, where height steps are marked by the dashed black lines. Details are discussed in Ref.~\cite{boule2020coherent}. In principle at each pixel an entire multi-parameter scan can be taken (see Sec.~\ref{sec:FWM_scans}), one could for example scan the delay $\tau_{12}$ between pulse 1 and 2 while recording spectra, which could then be processed into two-dimensional (2D) spectra, as explained in detail in Sec.~\ref{sec:2d_spectra}.

The samples are glued with silver paste onto the cold finger of the He-flow cryostat (from {\it Cryo-Vac}) operable between 4.2~K and room temperature. The distance between the sample surface and the objective, i.e., the working distance, is 4.5~mm with a 0.5, 0.2, or 0.1~mm thick, anti-reflex coated, optical window in between. The height of the sample holder can be adjusted, permitting us to study samples of different thicknesses. The cold part of the cryostat is fixed on a XY translation stage used for a coarse alignment of the sample position, with a range of 10~mm by 10~mm. Electrical contacting of the sample can be realized by using feed-through connectors to four BNC outputs which can be connected to voltage/current sources, typically from {\it Keithley}. To improve the mechanical stability of the experiment, the turbo-vacuum pump is disconnected during the measurement. We also use a flexible He-transfer and a heating hose for the He-recovery line, helping to suppress the vibration transfer from the surrounding (building etc.) to the cryostat.
\subsubsection{Heterodyne detection}\label{sec:exp_heterodyne}
\begin{figure}[t]
	\centering
	\includegraphics[width = 0.41\textwidth]{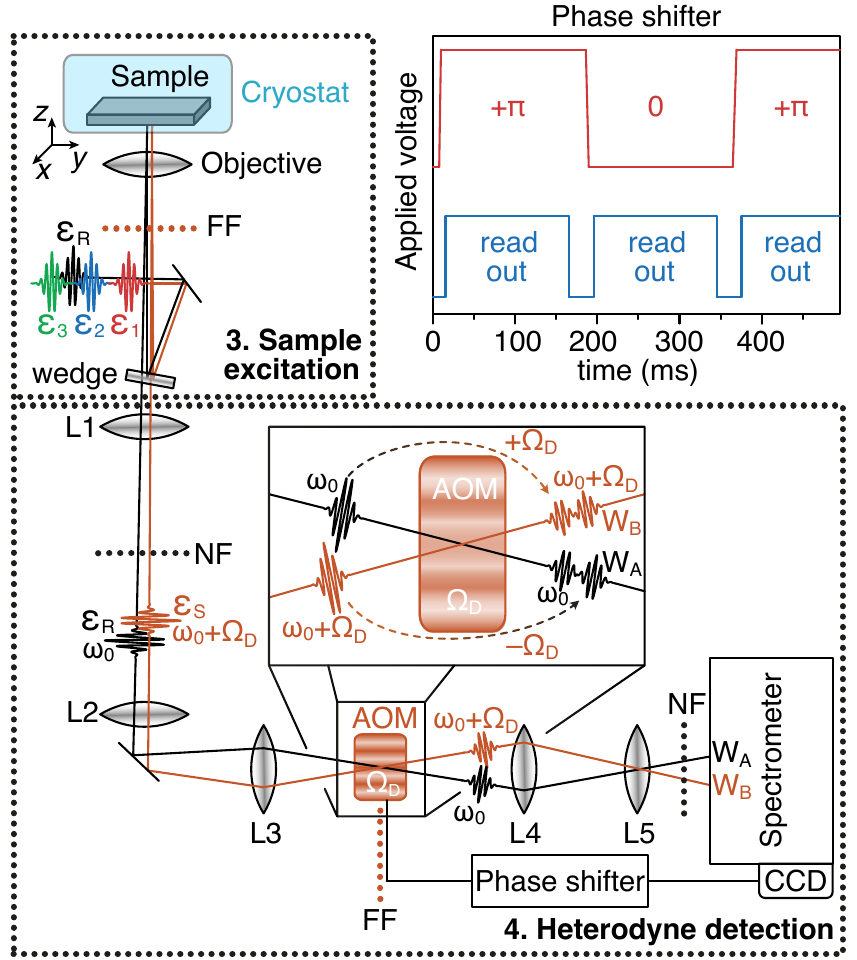}
	\caption{Boxes 3 and 4 from Fig.~\ref{fig:setup}(b) and (c).}
	\label{fig:setup_3_4}
\end{figure}
As depicted in box 4 in Figs.~\ref{fig:setup}(b)/\ref{fig:setup_3_4}, the reflected light from the sample, containing the reference beam $\E_{\rm R}$ and various optical nonlinearities generated by the sample including the desired nonlinear signal $\E_{\rm S}$, is collected by the same objective and imaged onto the entrance slit of an imaging spectrometer of 750~mm focal length provided by {\it Princeton Instruments}. The dispersed light is detected by a CCD camera ({\it Pixis eXcelon}) offering a high quantum detection efficiency above $\geq90$\% between 500~nm (2.48~eV) to 900~nm (1.38~eV)~\cite{CCD}\footnote{Teledyne Vision Solutions, PIXIS 400 eXcelon, \href{https://www.teledynevisionsolutions.com/products/pixis/?vertical=tvs-princeton-instruments&segment=tvs}{link}, Accessed: 2025-03-18}. In order to evaluate the quantum yield of the investigated samples, it is important to specify the efficiency of the setup $\eta$, i.e., the ratio between the number of photons detected by the CCD camera with respect to the number of emitted photons from the sample. The entire quantum efficiency of the setup (boxes 3 and 4) consists of four ingredients: (i) the CCD itself with $\eta_{\rm CCD}\approx 0.475$, (ii) the spectrometer (one grating and two mirrors) with $\eta_{\rm spec}\approx 0.567$, (iii) the transmission through the optical elements of the setup (cryostat window, objective, 96\% transmission wedged window, one mirror, five lenses, one AOM) with $\eta_{\rm setup}\approx 0.45$, and (iv) the numerical aperture (NA) of the objective of $\eta_{\rm NA}\lesssim 0.125$. These contributions multiply to $\eta\lesssim 1.51\%$.

The optical imaging from the sample towards the spectrometer entrance slit takes into account several constraints to isolate the desired nonlinear signal (a detailed theoretical description of the steps is given in Chap.~\ref{sec:exp_theo}):
\begin{enumerate}
    \item The signal $\E_{\rm S}$ is homodyned (for alignment) or heterodyned (for measurement) with the reference pulse $\E_{\rm R}$ via the mixing AOM in Figs.~\ref{fig:setup}(b)/\ref{fig:setup_3_4} (box 4) driven at the radio frequency $\Omega_{\rm D}$. The signal is then detected via their spectral interference. Here, homodyne means the detection of an excitation beam labeled with the frequency $\Omega_i$ via the same mixing AOM frequency $\Omega_{\rm D}=\Omega_i$, while for heterodyning of the signal we use $\Omega_{\rm D}\neq\Omega_i$. As described before, the excitation beams are labeled by distinct radio frequencies $(\Omega_1,\,\Omega_2,\,\Omega_3)/(2\pi)=(79.77,\,79,\,81)$~MHz, generated by fixed frequency drivers, amplified to 2~W as required by the AOMs' transductors. As discussed in detail in Sec.~\ref{sec:phase_selection}, it is crucial to choose the excitation beam AOM frequencies in box 2 of Fig.~\ref{fig:setup}(b) such that (i) they are all slightly different from each other, and (ii) they all lie close to but slightly above the repetition frequency $\Omega_i/2\pi\gtrsim\tau_{\mathrm{rep}}^{-1}\approx76$~MHz (\textit{all} below is also possible). Otherwise one might get unwanted contributions to the signal, different from the desired wave mixing.
    \item The reference $\E_{\rm R}$ propagates in the vicinity of $\E_3$ and impinges on the back focal plane of the objective with a slightly different angle than the excitation pulses $\E_{1,2,3}$ [box 3 in Figs.~\ref{fig:setup}(b)/\ref{fig:setup_3_4}]. $\E_{\rm R}$ therefore focuses on the sample at a different location compared to $\E_{1,2,3}$. After interaction with the sample, the reflected signal $\E_{\rm S}[\omega_0+f(\Omega_{1,2,3})]$ (orange, box 3 and 4), where $f$ is for now an arbitrary function of the AOM frequencies, and reference $\E_{\rm R}(\omega_0)$ (black) are mixed and heterodyned at the fourth AOM, the so-called mixing-AOM, in front of the spectrometer (box 4). A detailed theoretical explanation of this process is given in Secs.~\ref{sec:balanced_heterodyne} and \ref{sec:phase_selection}. To reach a perfect spatial overlap of the two fields in the transmitted and deflected direction, $\E_{\rm S}$ and $\E_{\rm R}$ enter the mixing-AOM with detection frequency $\Omega_{\rm D}$ in a Bragg configuration~\cite{donley2005double} (inset in box 4). The Bragg condition is met when all diffraction orders vanish except the one we are interested in, which happens for a Bragg incidence angle of $\alpha_{\rm B}\approx \sin(\alpha_{\rm B})=\lambda/(2\Lambda)=\lambda \Omega_{\rm D}/(2 n v)$, with the optical wavelength $\lambda$, the acoustic wavelength $\Lambda$, the AOM driving frequency $\Omega_{\rm D}$, refractive index $n$, and speed of sound in the AOM crystal $v$, which in practice yields a few milli-radians~\cite{donley2005double}. In this specific situation, the beams' intensities are split in half between the two output directions. In this way, the AOM-deflected $\E_{\rm R}$ is frequency upshifted to $\omega_0+\Omega_{\rm D}$ and propagates in the same direction as the transmitted $\E_{\rm S}[\omega_0+f(\Omega_{1,2,3})]$, while the AOM-deflected $\E_{\rm S}$ is frequency down-shifted to $\omega_0+f(\Omega_{1,2,3})-\Omega_{\rm D}$ and travels together with the transmitted $\E_{\rm R}(\omega_0)$. The mixing-AOM acts therefore as a 50:50 beam splitter, and brings both $\E_{\rm S}$ and $\E_{\rm R}$ into the same spatial modes $W_{\rm A}$ and $W_{\rm B}$, permitting them to interfere. Energy conservation requires the sum of the intensities to be equal on both sides of the mixing-AOM~\cite{campos1989quantum}. In addition, to achieve interference and to filter a specific combination of $\Omega_{1,2,3}$ from $\E_{\rm S}$, the mixing-AOM frequency is chosen as linear combination $\Omega_{\rm D}=l_1\Omega_1+l_2\Omega_2+l_3\Omega_3$. Standard three-pulse FWM (see Sec.~\ref{sec:FWM_2LS}) is obtained for $l_3=l_2=-l_1=1$. Other possibilities and consequences for the choice of the $l_i$ are discussed in Sec.~\ref{sec:nwm}.
    \item To inject $\E_{\rm S}$ and $\E_{\rm R}$ under the Bragg angle $\alpha_{\rm B}$ into the mixing-AOM, we use a confocal setting. We image the sample surface into the intermediate image plane with a magnification of $\approx 60$, such that $\approx 10$~\textmu m separation on the sample yields $\approx 0.6$~mm separation in the intermediate image plane. This spatial separation in the near-field (image plane of the sample) can now be straightforwardly converted into the far-field image (image plane of the back focal plane of the objective) at the mixing-AOM plane, fulfilling the Bragg angle condition $\alpha_{\rm B}=\lambda \Omega_{\rm D}/(2 n v)$. It is also convenient to construct the intermediate near-field plane and image it onto a camera permitting us to monitor the sample surface during the alignment and between measurements.

    \item To generate a background-free spectral interference, the far-field located at the mixing-AOM is imaged back onto the near-field located at the entrance slit of the spectrometer, dispersed and imaged once again onto the CCD camera. The two areas, separated vertically by around 0.35~mm, detecting $W_{\rm A}$ and $W_{\rm B}$ containing the signal are binned and then subtracted to achieve shot-noise limited detection of the spectral interferences. The entire data acquisition and processing is described in the following Sec.~\ref{sec:FWM_data}.
\end{enumerate}

\subsection{Alignment \& data acquisition}\label{sec:FWM_data}
In the following we discuss proper beam alignment, important details of the heterodyne detection scheme, the importance of the spectrometer response, as well as typical parameter scans that can be performed with the setup. These discussions are supported by the theoretical modeling of the setup in Chap.~\ref{sec:exp_theo}.
\subsubsection{Beam alignment in homodyne detection}\label{sec:data_alignment}
\begin{figure}[b]
	\centering
	\includegraphics[width = 0.63\textwidth]{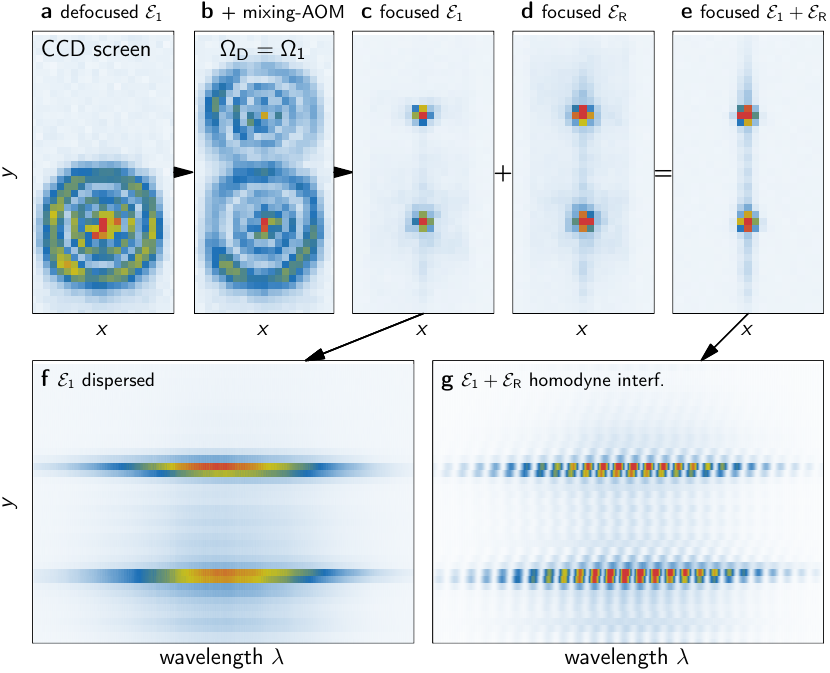}
	\caption{Alignment steps detected on the CCD. (a) Defocused single pulse $\E_1$ showing Airy rings detected in the spatial mode $W_{\rm B}$. (b) Reflection in the mixing-AOM is switched on, deflecting half of the beam to a higher position in the spatial mode $W_{\rm A}$. (c, d, e) Focused beams resulting in single spots on the CCD. (f) Pulse $\E_1$ after dispersed by the spectrometer grating. (g) Homodyne interference between $\E_1$ and $\E_{\rm R}$ showing clear spectral oscillations in the two areas $W_{\rm A}$ (top) and $W_{\rm B}$ (bottom). These data are processed further in Fig.~\ref{fig:process}.}
	\label{fig:CCD}
\end{figure}
\begin{figure}[b]
	\centering
	\includegraphics[width = 0.31\textwidth]{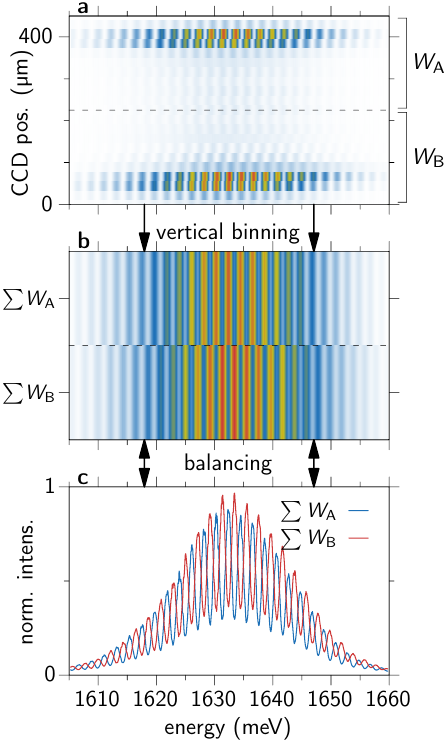}
	\caption{Processing of the detected data. (a) Detected spectra on the CCD with the two areas corresponding to the interference beams in spatial modes $W_{\rm A}$ and $W_{\rm B}$ from Figs.~\ref{fig:setup}(b)/\ref{fig:setup_3_4}. Same data as in Fig.~\ref{fig:CCD}(g). (b) Intensity spectrum after integrating over the marked areas of the CCD in (a). (c) Same as (b) as line plots for easier comparison of the curves. The data from (b) are further processed in Fig.~\ref{fig:process_2}.}
	\label{fig:process}
\end{figure}

To align the setup described in Sec.~\ref{sec:setup} (Fig.~\ref{fig:setup}), we start by properly focussing the beam $\E_1$ on the sample surface, which can be monitored with the CCD camera and the spectrometer set to the 0th diffraction order. Next, we defocus the beams with the condition that the microscope objective aperture is overfilled. We then observe a symmetric pattern of Airy rings as shown in Fig.~\ref{fig:CCD}(a). This procedure is performed for each individual beam identically. In the three-beam configuration, we initially set the same impinging power of around 0.1~\textmu W for each pulse $\E_{1,2,3}$ and 0.5~\textmu W for the reference $\E_{\rm R}$. In our case, going beyond 1~\textmu W leads to saturation of the CCD. By beam-walking $\E_2$ and $\E_3$ we set the spatial and angular overlap with $\E_1$. We then supply RF-power with the frequency $\Omega_1$ to the mixing-AOM [box 4 in Figs.~\ref{fig:setup}(b)/\ref{fig:setup_3_4}] which activates the beam splitter properties, resulting in the image in Fig.~\ref{fig:CCD}(b), and optimize the deflection angle, reaching the Bragg configuration introduced above~\cite{donley2005double}, see Fig.~\ref{fig:CCD}(c-e). After turning the grating of the spectrometer to project the 1st diffraction order onto the CCD [Fig.~\ref{fig:CCD}(f)] we can detect interference patterns in two areas as depicted in Fig.~\ref{fig:CCD}(g). The two areas correspond to the two output directions of the mixing-AOM $W_{\rm A}$ and $W_{\rm B}$ [see Figs.~\ref{fig:setup}(b)/\ref{fig:setup_3_4}, box 4]. After binning the pixels of both areas of the CCD [Figs.~\ref{fig:CCD}(g) and \ref{fig:process}(a)] we arrive at the situation in Fig.~\ref{fig:process}(b), where we have two interference patterns, one for each output channel. By using an attenuator for the RF-power, we equally balance both vertically shifted spectra on the CCD, which can best be checked via line plots as in Fig.~\ref{fig:process}(c). We then perform the same procedure for $\E_{\rm 2}$ and $\E_{\rm 3}$ to guarantee spectral interference between all beams.

To fine-tune the spectral interferences, we go back to the unbinned spectra [see Fig.~\ref{fig:process}(a)], permitting us to monitor stability/verticality of the fringes, and thus to optimize the contrast. We start with the interference between $\E_3$ and $\E_{\rm R}$. By moving the reference linear delay stage, we find the vanishing delay between the two pulses $\tau_{3\rm R}=0$. For a perfect temporal overlap the spectral fringes vanish. We then advance $\E_{\rm R}$ with respect to $\E_3$ by a few ps, such that $t_{\rm R}<0$ when $\E_3$ defines $t=0$. The condition $t_{\rm R}<0$ is required to properly retrieve the signal field in the data post-processing (see Fig.~\ref{fig:amp_phase} and Sec.~\ref{sec:exp_phase}).

In the same way we then find zero delays for the pulses $\E_1$ and $\E_2$ with $\E_3$. Then, $\E_1$ and $\E_2$ have perfect temporal and spatial overlap within the diffraction limit and pulse duration $T_{\rm P}$. By moving the {\it VT-80} linear delay stages over their full range, we align the beam injection angle/position into the delay line, to maintain a perfect beam overlap when scanning the delay. An inaccurate alignment would generate artifacts in the delay dynamics, which are then more or less easy to identify as exemplarily shown in App.~\ref{sec:alignment}.

\subsubsection{Heterodyne data acquisition}\label{sec:exp_heterodyne_data}
At this point we are ready to switch from homodyne to heterodyne detection. We typically choose the heterodyne downshift frequency in the mixing-AOM to $\Omega_\mathrm{D}=\Omega_3+\Omega_2-\Omega_1$, which corresponds to the standard FWM signal (see Secs.~\ref{sec:FWM_2LS} and \ref{sec:phase_selection}) that is in the lowest order proportional to $\E_1^*\E_2^{}\E_3^{}$. The heterodyne frequencies are produced by a home-built three-channel analog mixer with standard components purchased from {\it Mini-Circuits}, yielding a spectral purity of around $-50$~dBm. The frequencies are subsequently re-generated by a phase-locked loop (PLL) of a {\it Stanford Research Systems SR 844} 200~MHz lock-in amplifier. This improves the spectral purity to around $-80$~dBm (see App.~\ref{sec:frequ_filter}). The such prepared RF signal is amplified to 2~W and sent onto the mixing-AOM. Alternatively, optical heterodyning can be performed in a digital fashion, employing a four-channel direct digital synthesizer (for example from {\it Analog Devices} or {\it AA Opto Electronic}) or with an arbitrary wave-form generator ({\it Tektronix}, sampling rate 1.2~G/s), where we found that the retrieved FWM signals from single quantum dots yield a similar signal-to-noise ratio as with analog mixing.

In the heterodyne data acquisition, we typically use short exposure times between 1~ms and 10~ms (corresponding to around $10^5-10^6$ pulse sequence repetitions), permitting us to handle a high photon flux on the CCD pixels, which is additionally set in a high capacity mode and lowest gain. The sought FWM signal is hugely overshadowed by the resonant background of $\E_{1,2,3}$, i.e., scattering of the laser pulses from the sample. The FWM field amplitude (intensity) generated by a single quantum dot is typically six (twelve) orders of magnitude weaker than the resonant laser, illustrating the challenging aspect of this nonlinear experiment. Even though these specifications are more relaxed when working with a larger number of excitons, for example in layered semiconductors, the FWM signal remains weak compared to the driving laser fields. Therefore, the detection scheme always involves numerous acquisitions. At the shot-noise limit, with increasing number of acquisitions $N$, the FWM signal-to-noise ratio increases with $\sqrt{N}$~\cite{schottky1918spontane}.

\begin{figure}[b]
	\centering
	\includegraphics[width = 0.61\textwidth]{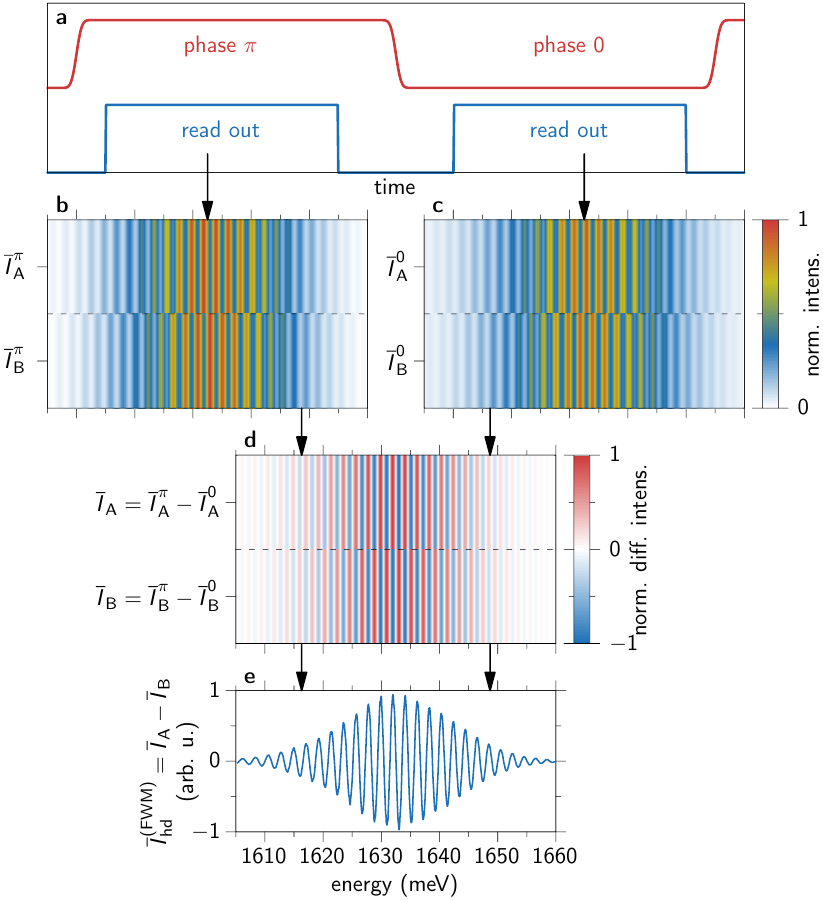}
	\caption{(a) Phase shift and data readout schematic as in Fig.~\ref{fig:setup}(c)/\ref{fig:setup_3_4}. (b, c) Double detection sequence including the phase shift by $\pi$ from (a). (d) Intensity difference spectrum resulting in the CCD-noise--free intensities $\overline{I}_{\rm A}$ and $\overline{I}_{\rm B}$. (e) Final Interference pattern $\overline{I}_{\rm A}-\overline{I}_{\rm B}$.}
	\label{fig:process_2}
\end{figure}
Hence, it is crucial to reach shot-noise limited detection and to minimize the idle time of the CCD. To achieve that, a doubly balanced detection is applied. We double the number of acquisitions by flipping the phase of the RF signal driving the mixing-AOM between 0 and $\pi$, as shown in Fig.~\ref{fig:process_2}(a-c). These phase flips are synchronized with the exposures using the trigger output of the CCD, as schematically depicted in Figs.~\ref{fig:setup}(c)/\ref{fig:setup_3_4} or Fig.~\ref{fig:process_2}(a). Note that each readout section of around 150~ms corresponds to roughly $10^7$ laser repetitions, meaning that the detected signals have collected sufficient phase integration to reach a proper FWM signal (see Sec.~\ref{sec:phase_selection}). By subtracting each consecutive pair to $\overline{I}_\mathrm{A}=\overline{I}_\mathrm{A}^\pi-\overline{I}_\mathrm{A}^0$ and $\overline{I}_\mathrm{B}=\overline{I}_\mathrm{B}^\pi-\overline{I}_\mathrm{B}^0$, the classical noise of the CCD cancels out as seen in Fig.~\ref{fig:process_2}(d).  Next, pairs from the two areas of the CCD, which are also out of phase, are subtracted to $\overline{I}_\text{hd}^{\mathrm{(FWM)}}=\overline{I}_\mathrm{A}-\overline{I}_\mathrm{B}$. This pair-difference routine yields the background free spectral interference.

Note that at this point the exposures are partially binned. While this reduces the stored data size, the remaining data sets permit us to assess and to monitor the phase stability of the spectral interferences during the entire acquisition time (typically between 5~s and 300~s). After that, all exposures are binned into the final interferogram, as shown in Fig.~\ref{fig:process_2}(e). This efficient detection scheme can be extended to more areas of the CCD if required. An example can be a dual-polarization detection scheme~\cite{kasprzak2008vectorial} employing four areas, corresponding to two pairs of orthogonal linear polarizations, which can be produced by placing a highly birefringent calcite plate in front of the spectrometer's slit. Through this method we collect more information on the signal in a single run of the measurement.

An important development was to incorporate a phase referencing routine~\cite{delmonte2017coherent}. By measuring the heterodyne interferences of $\E_1$ for each delay, one can independently reconstruct and track the absolute FWM phase when varying the delay. This piece of information is measured independently from the FWM signal and therefore permits to carry out phase corrections in the data post-procession. This is necessary to accurately perform the Fourier transform with respect to delays and thus to construct two-dimensional FWM spectra that will be introduced in Sec.~\ref{sec:2d_spectra}. Thereby, it is possible to correct the FWM signal phase in the post-processing~\cite{albert2013microcavity} if needed.

\begin{figure}[b]
	\centering
	\includegraphics[width = 0.75\textwidth]{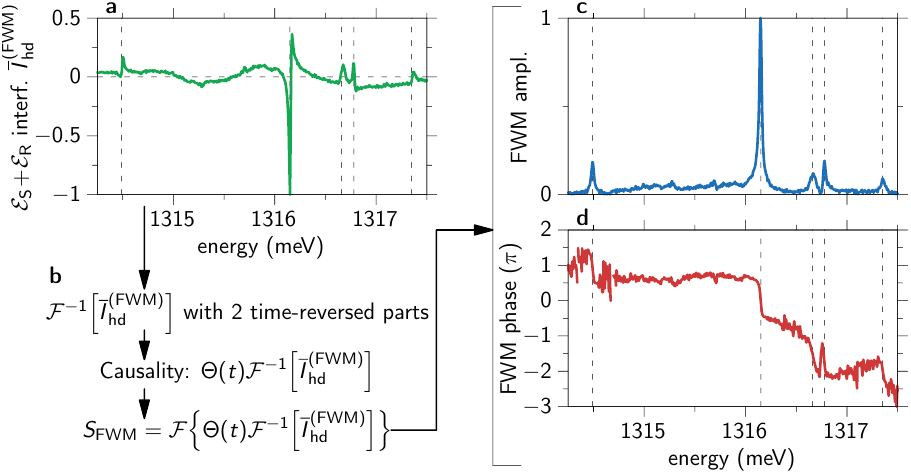}
        \caption{Flow chart depicting the extraction of the FWM spectrum $S_\mathrm{FWM}$ [Eq.~\eqref{eq:I_FWM_1} in Sec.~\ref{sec:exp_phase}]. (a) Detected interference $\overline{I}_\mathrm{hd}^\mathrm{(FWM)}$ [Eq.~\eqref{eq:I_hd_4} in Sec,~\ref{sec:exp_phase}] measured on an individual InAs quantum dot in a low-Q microcavity. (b) Processing steps in accordance to the derivation in Sec.~\ref{sec:exp_phase} and the resulting Eq.~\eqref{eq:I_FWM_1}. (c, d) Amplitude and phase of the FWM spectrum $S_\mathrm{FWM}$, respectively. The measurement has been performed on a single InAs quantum dot in a low-Q microcavity, provided by the University of W{\"u}rzburg.}
        \label{fig:amp_phase}
\end{figure}

The final FWM signal $S_\mathrm{FWM}$ is retrieved in amplitude and phase by applying a Fourier-transform spectral interferometry algorithm~\cite{patton2004non,langbein2006heterodyne,langbein2010coherent} depicted in Fig.~\ref{fig:amp_phase}, and thus can be presented in time or spectral domains, as needed. It is important to note that this algorithm relies on a causality condition between signal and reference: \textit{the reference pulse needs to precede the nonlinear signal}, i.e., $t_\mathrm{R}<0$ where the time origin in each repetition of the pulse trains is given by the arrival of the signal at the spectrometer, i.e., roughly the time of the last excitation of the sample using $\E_{1,2,3}$. More details on the spectral interferometry algorithm are given in Sec.~\ref{sec:exp_phase}.

\subsubsection{Impact of the spectrometer response}\label{sec:exp_spectrometer}
\begin{figure}[h]
	\centering
	\includegraphics[width = 0.55\textwidth]{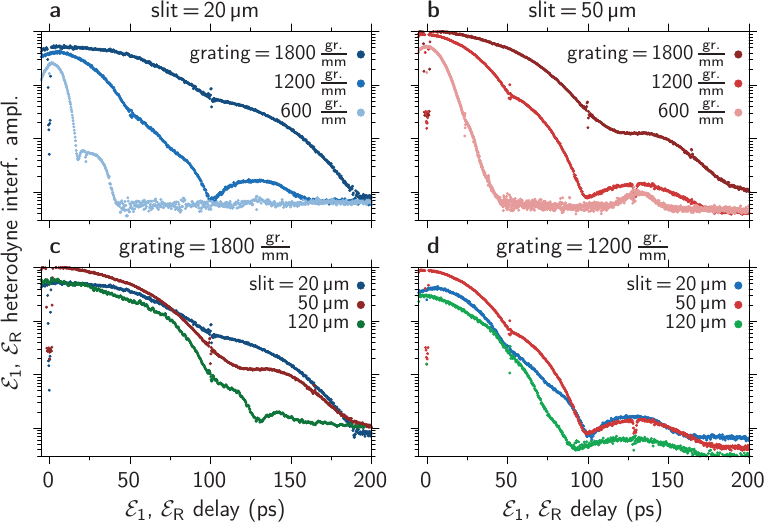}
        \caption{Characterization of the spectrometer response by homodyne interference between $\E_1$ and $\E_{\rm R}$ [see Eq.~\eqref{eq:I_hd_ref} in Sec.~\ref{sec:exp_phase}]. All curves as function of the delay between $\E_1$ and $\E_{\rm R}$. (a) Different spectrometer grating with more groves (dark) and less groves (bright) for a slit width of 20~\textmu m (blue). (b) Same as (a) but for a slit width of 50~\textmu m (red). (c) Comparing different slit widths for a grating with 1800~groves/mm. Slit widths: 20~\textmu m as in (a) (blue), 50~\textmu m as in (b) (red), and 120~\textmu m (green). (d) Same as (c) but for a grating with 1200~groves/mm.}
        \label{fig:spec_exp}
\end{figure}

The FWM signal detection relies on heterodyne interference between an ultrashort reference pulse and the signal from the sample. As stated at the end of the previous section it is crucial that the ultrashort reference pulse precedes the signal such that there is no direct temporal overlap between them. Interference is still possible since the reference pulse is stretched in the spectrometer by the spectrometer response, which therefore provides a time window for the interference and whose role is discussed in detail in Chap.~\ref{sec:exp_theo}. The FWM signal dynamics are thus weighted by the spectrometer response in the time domain and we can obtain this response by homodyning the reference $\mathcal{E}_\mathrm{R}$ and one of the laser pulses $\mathcal{E}_1$ [see Eq.~\eqref{eq:I_hd_ref} in Sec.~\ref{sec:exp_phase} for a theoretical derivation]. To characterize the responses for different configurations of the spectrometer, Fig.~\ref{fig:spec_exp} presents such an interference between these pulses for different slit widths and gratings, i.e., different spectrometer response functions. In Fig.~\ref{fig:spec_exp}(a, b) we increase the grating density from 600~groves/mm (bright colors) to 1800~groves/mm (dark colors) and clearly see that the signal lasts longer. This directly corresponds to an increase in spectral resolution for finer gratings. When comparing the different slit widths 20~\textmu m (blue), 50~\textmu m (red), and 120~\textmu m (green) in Fig.~\ref{fig:spec_exp}(c, d), we find a minor trend that the signal lasts longer for smaller widths. This finding agrees with the one in Fig.~\ref{fig:slit}. We can conclude, that the spectrometer response strongly suppresses the transients spanning beyond 100~ps (see also Fig.~\ref{fig:spec} in Sec.~\ref{sec:exp_phase}). In other words, spectral fringes cannot be well measured below the spectral resolution of the spectrometer, which at best reaches 20~\textmu eV for the finest grating. The impact of the spectrometer response and reference pulse timing is further discussed for time-integrated FWM signals in App.~\ref{sec:ref_timing}.

\subsubsection{Parameter scans}\label{sec:FWM_scans}
The wave mixing experiments allow us to perform parameter scans with different levels of complexity that might require a lot of time accordingly  (total acquisitions times might reach up to 36~hours). We are often predominantly interested in measuring the $\tau_{12}$- and $\tau_{23}$-delay dependence of the three-pulse FWM signal (or the $\tau_{12}$-delay dependence of the two-pulse FWM signal), reflecting the system's coherence and occupation dynamics, respectively, as derived for the example of a 2LS in Sec.~\ref{sec:FWM_2LS}. The delay scans will be used in the following Sec.~\ref{sec:2d_spectra} to construct 2D spectra.

However, sometimes it is also interesting to characterize the dependence of the FWM signal on the applied pulse intensities, which gives insight into the applied pulse areas and regime of nonlinearity (see Sec.~\ref{sec:FWM_2LS})~\cite{wigger2017exploring,groll2020four}. We can also independently scan other parameters, like an applied bias voltage, or light polarizations.

Finally, as already discussed in Sec.~\ref{sec:setup}, in the case of extended samples, all observables can be probed at different positions by raster scanning the microscope, which allows us to perform FWM imaging. It is also possible to position the individual pulses on the sample giving us access to propagation effects. Note that the excitation pulses $\E_{1,2,3}$ and the reference $\E_{\rm R}$ are always reflected from the sample at different positions. A remaining spatial overlap between the reference and the other pulses could here lead to errors in the FWM signal detection as discussed in App.~\ref{sec:overlap}.

\subsection{Comparison between experiment and theory: 2D spectra of 2LSs}\label{sec:2d_spectra}
A useful tool for the investigation of FWM signals, especially beyond the 2LS case (see Chap.~\ref{sec:nls}), is provided by 2D FWM spectra. Let us recall that the FWM polarization of a 2LS generally depends on multiple time variables, as discussed in Sec.~\ref{sec:FWM_2LS}: (i) The real time $t$ after all excitations and (ii) the various delays $\tau_{ij}$ between the pulses. In the experiment we actually detect spectra $S_\mathrm{FWM}(\omega; \lbrace\tau_{ij}\rbrace)$ that depend parametrically on the delays $\tau_{ij}$ as a function of the detection frequency $\omega$, which is related to the signal as a function of the real time $t$ via a Fourier transform.

When discussing dynamical properties of systems, as we did in Chap.~\ref{sec:theory_2}, it seems natural to consider the signal as a function of $t$ and the delays $\tau_{ij}$ instead of $\omega$ and $\tau_{ij}$. Often we are however also interested in the various energy levels and their interactions in the case of a general $N$-level system (see Chap.~\ref{sec:nls}). Then it can be much more useful to discuss the signal completely in frequency space, i.e., in addition to viewing it as a function of $\omega$ (and not $t$), we also perform a Fourier transformation in one of the delays $\tau = \tau_{ij}$, while keeping the others fixed, to plot the signal as a function of the detection frequency $\omega$ and the delay frequency $\omega_\tau$. The resulting 2D spectrum is given by
\begin{align}
	S^{(\text{2D})}_\mathrm{FWM}(\omega,\omega_\tau)=\int\text{d}\tau\,S_\mathrm{FWM}(\omega;\tau)e^{-i\omega_\tau\tau}\,.\label{eq:I_FWM_2D_simple}
\end{align}
Note that the FWM signal as a function of the delay $\tau$ is often discontinuous (in theory) between positive and negative values. In the case of a 2LS for example, the two-pulse FWM signal for negative delays $\tau_{12}$ vanishes as already discussed in Sec.~\ref{sec:p_FWM_2}, while for positive delays it is given by Eq.~\eqref{eq:p_FWM_2}. It is therefore a good idea to discuss such cases separately and integrate $\tau$ in Eq.~\eqref{eq:I_FWM_2D_simple} either over positive or negative values and not the whole range of real numbers. The integration range for the delay $\tau$ is therefore dictated by the choices in the experiment.

As a concrete example for 2D FWM spectra, we consider two-pulse FWM with positive delay $\tau=\tau_{12}>0$ of an optically driven 2LS, including homogeneous~$(\gamma)$ and inhomogeneous~$(\sigma)$ broadenings, such that the signal reads
\begin{equation}
	\mathcal{E}_\mathrm{FWM}(t>0;\tau>0)\sim e^{-\gamma(t+\tau)}e^{-i\overline{\omega}_x(t-\tau)}e^{-\frac{\sigma^2}{2}(t-\tau)^2}
\end{equation}
for a Gaussian ensemble of width $\sigma$ centered around the average transition frequency $\overline{\omega}_x$ [Eq.~\eqref{eq:p_FWM_2_chi3} averaged over Eq.~\eqref{eq:ensemble_WN}]. From this we calculate the normal FWM spectrum $S_\mathrm{FWM}(\omega;\tau>0)$ as a function of the delay via Fourier transformation along $t$. Note that the signal is causal and vanishes for $t\leq 0$. While an accurate modeling of the experimental setup shows that we need to include the spectrometer response when going from the time signal $\mathcal{E}_\mathrm{FWM}(t)$ to the detected spectrum $S_\mathrm{FWM}(\omega)$ [see Eq.~\eqref{eq:I_FWM_1} in Sec.~\ref{sec:exp_phase}], we will ignore these details in this section and assume ideal detection conditions for simplicity. Note that the spectrometer response generally impacts the form of the 2D spectrum as a function of the detection frequency $\omega$, but not of the delay frequency $\omega_\tau$, as the latter simply emerges from a Fourier transformation in Eq.~\eqref{eq:I_FWM_2D_simple}.
\begin{figure}[b]
	\centering
	\includegraphics[width = 0.6\textwidth]{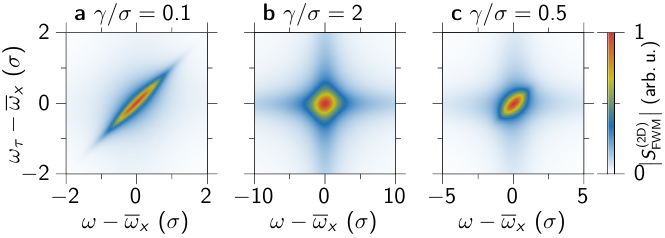}
	\caption{2D FWM spectra corresponding to the echo dynamics from Fig.~\ref{fig:echo_with_deph} for different ratios of homogeneous $\gamma$ and inhomogeneous broadening $\sigma$.}
	\label{fig:echo_with_deph_spec}
\end{figure}

Figure~\ref{fig:echo_with_deph_spec} shows the 2D FWM spectra (absolute values) for three different cases: (a) dominant inhomogeneous broadening $\gamma\ll\sigma$, (b) dominant homogeneous broadening $\gamma\gg\sigma$, and (c) similar broadenings $\gamma\approx\sigma$. This is completely analog to the discussion in the context of Fig.~\ref{fig:echo_with_deph}, however now in frequency instead of time domain. The depicted examples show that we can also use the spectral features to distinguish between dominantly inhomogeneous or homogeneous broadened samples. In the former case [see (a)] the 2D spectrum is significantly stretched along the diagonal which directly reflects the original ensemble shape in this direction~\cite{siemens2010resonance}, i.e., a Gaussian of width $\sigma$. In the case of a more pronounced homogeneous character in (b), the 2D spectrum has a symmetric cross shape where we can read the broadening in a vertical or horizontal cut. In the case of a balanced situation (c) we find a mixed appearance in the 2D spectrum~\cite{siemens2010resonance}. As already shown in the time domain in the context of Fig.~\ref{fig:echo_with_deph} we can in principle separate homogeneous and inhomogeneous broadenings in FWM quantitatively.

These spectra are conventionally interpreted in terms of absorption and emission processes: The free time evolution of the polarization after the pulse sequence in real time $t$ determines the detected spectrum, i.e., the properties of the light \textit{emitted} from the 2LS, while the dynamics with respect to the delay $\tau$ describe how the system evolves after \textit{absorption} of a pulse before the interaction with the next one. In this sense we tentatively interpret the $\omega$-dependence as showing the possible emission frequencies, while the $\omega_\tau$-dependence quantifies the possible absorption frequencies. Furthermore we can identify optical pathways in the system, which will be discussed more in-depth in the following Chap.~\ref{sec:nls}.

In the case of a 2LS the situation is quite trivial. Note however that in the case of dominant inhomogeneous broadening in Fig.~\ref{fig:echo_with_deph_spec}(a) we can use the interpretation in terms of absorption and emission to understand why the spectrum is elongated along the diagonal. Each repetition of the experiment sees a certain fixed frequency $\omega_x$, which does not change during the pulse sequence [see also Fig.~\ref{fig:jitter}], such that absorption $(\omega_\tau)$ and emission $(\omega)$ is at the same frequency for the 2LS (peak on the diagonal). This frequency then varies from repetition to repetition with a Gaussian probability distribution, leading to the elongation of the signal along the diagonal when sampling this distribution.

Note that some authors depict the $\omega_\tau$-axis flipped, i.e., with negative signs, which converts the diagonal into the cross-diagonal~\cite{li2023optical}. While this is simply a matter of convention and does not impact the underlying physics, one should always check which convention is used before interpreting 2D spectra, as it changes the meaning of the (cross-)diagonal.

\begin{figure}[t]
	\centering
	\includegraphics[width = 0.8\textwidth]{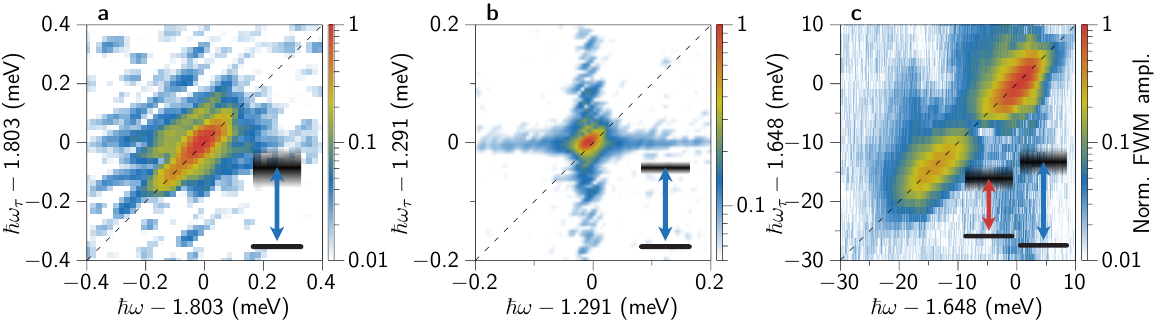}
	\caption{Collection of 2D FWM spectra from different excitonic 2LSs. (a) CdTe quantum dot exciton in a sample produced at University of Warsaw. 2D spectrum corresponding to the photon echo measurement in Ref.~\cite{kasprzak2022coherent}. (b) InAs quantum dot exciton in a photonic wire produced at CEA Grenoble. Same sample and similar data as in Ref.~\cite{mermillod2016harvesting}. (c) Exciton and charged exciton transitions in a MoSe$_2$ monolayer exfoliated at LNCMI, CNRS Grenoble.}
	\label{fig:2D_spec_collection_1}
\end{figure}

For a final discussion on actual experimental data and comparison to the theory presented so far, in Fig.~\ref{fig:2D_spec_collection_1} we collect three exemplary 2D FWM spectra stemming from 2LS-like systems. Figure~~\ref{fig:2D_spec_collection_1}(a) shows the 2D FWM spectrum of a single CdTe quantum dot, which resembles the situation from Fig.~\ref{fig:echo_with_deph_spec}(a) with an elongation along the diagonal. Therefore, we can conclude that the spectral linewidth is dominated by inhomogeneous broadening in this sample. The opposite case is found for the InAs quantum dot shown in Fig.~\ref{fig:2D_spec_collection_1}(b), where we find the characteristic cross shaped pattern similar to Fig.~\ref{fig:echo_with_deph_spec}(b), i.e., the sample is dominantly homogeneously broadened.

Another interesting situation can be seen in Fig.~\ref{fig:2D_spec_collection_1}(c), which shows the 2D FWM spectrum of a MoSe$_2$ monolayer. It consists of a pair of inhomogeneously broadened peaks on the diagonal. From this we can conclude a level structure as depicted in the inset consisting of two 2LSs with separate ground states. The physical origin of this structure is differently charged excitons: For the energetically higher lying transition the ground state is the electronic vacuum and the excited state is a neutral exciton. The energetically lower lying transition has a ground state with charge carriers (electrons or holes) present, e.g., due to doping, and the excited state is a respectively charged exciton, often called trion~\cite{druppel2017diversity}.

This last example already shows that the level structure in relevant samples can, and often does, go beyond the case of a simple 2LS. While the 2LS provides a great model for the introduction of the general features of FWM spectroscopy, as discussed so far from both the theoretical and experimental point of view, in Chap.~\ref{sec:nls} we conclude this tutorial by extending the theoretical modeling to arbitrary $N$-level systems. This will finally allow us to analyze measured 2D FWM spectra of such $N$-level systems in Sec.~\ref{sec:exp_theo_2} analog to Fig.~\ref{fig:2D_spec_collection_1}.


%% file: 5_Experiment_Theory.tex
\section{Theoretical modeling of the experimental setup}\label{sec:exp_theo}
While the previous chapter described the setup in Fig.~\ref{fig:setup}(b) step by step following the beam path, in the following we present how the individual building blocks can be treated in suitable theoretical models. We will move in the opposite direction in the setup, starting with the detection in the spectrometer. By using a general time-resolved spectroscopy theory with quantized light fields~\cite{eberly1977time} we provide a blueprint that can be adapted to the derivation of several different spectroscopy signals, e.g., time-dependent photoluminescence~\cite{groll2021controlling}, resonance fluorescence~\cite{wigger2021resonance,groll2023read}, coherent control~\cite{preuss2022resonant} and of course heterodyne FWM as presented in the following. We highlight the role of the spectrometer response, which should be considered in any quantitative theoretical modeling of FWM signals.
\subsection{Modeling the spectrometer}
We start at the end of the setup from Figs.~\ref{fig:setup}(b)/\ref{fig:setup_3_4} in box 4 with the CCD. Each pixel of the CCD performs an intensity detection of the light impinging on it. Focusing on a single pixel $p$, we denote the electric field operator of the optical signal at the position of the considered pixel (in the Heisenberg picture) by $\hat{E}_p(t)=\hat{E}_p^{(+)}(t)+\hat{E}_p^{(-)}(t)$~\cite{glauber1963coherent,cohen2022photons}. Here the superscripts $(\pm)$ denote the positive (negative) frequency components of the field, which are related by the operator adjoint $\hat{E}_p^{(+)\dagger}(t)=\hat{E}_p^{(-)}(t)$. The positive frequency component is a superposition of the photon annihilation operators $a_j$ for each mode with quantum numbers~$j$, while the negative frequency component contains the creation operators $a_j^\dagger$ (see App.~\ref{sec:dipole_emission} for more details). We suppress the vector character of the electric field here, assuming that the light impinging on the spectrometer has a certain fixed polarization. Keeping in mind that we deal with operators of the light field in general, we drop the hat on the electric field operators from now on for notational simplicity. The intensity, as detected by pixel $p$ at time $t$ is then given by the Glauber formula~\cite{glauber1965optical}
\begin{equation}\label{eq:I_p}
	I_p(t)=\left< E^{(-)}_p E^{(+)}_p\right>(t)\,,
\end{equation}
where the average is taken with respect to the state of the light field, which is described by a density matrix in general. The physical interpretation of this formula becomes clear, if we deal only with a single light mode with annihilation (creation) operator $a^{(\dagger)}$. In this case $E_p^{(+)}$ is proportional to the annihilation operator $a$ and we obtain $I_p(t)\sim\left<a^{\dagger}a\right>(t)$, i.e., the intensity corresponds to the average number of photons in that mode at time $t$. Equation~\eqref{eq:I_p} generalizes this intuitive result to the case of an arbitrary number of light modes.

The light field $E_p(t)$ reaching pixel $p$ of the CCD is directly related to the light $E_{\mathrm{in}}(t)$ that enters the spectrometer (suppressing the positional dependence assuming homogeneous illumination of the spectrometer). This relation has to obey the three properties: (i) causality, i.e., no light is detected by the CCD before entering and passing through the spectrometer, (ii) time-translation invariance, i.e., the properties of the spectrometer are assumed the same throughout the measurement, and (iii) linearity, i.e., the spectrometer does not produce any additional nonlinear response. This leads us to the relation~\cite{eberly1977time}
\begin{equation}\label{eq:def_spec_response}
	E_p^{(+)}(t)=\int\limits_{-\infty}^{\infty}\text{d} t'\,r_p(t-t')E_{\mathrm{in}}^{(+)}(t')
\end{equation}
for the positive frequency component of the field. Here, $r_p(t)$ is the spectrometer response function for the output produced at pixel $p$ of the CCD. If the input is a classical coherent source, e.g., from a laser, of infinitely small temporal width $E_\mathrm{in}^{(+)}(t)\sim\delta(t)$, the spectrometer stretches this $\delta$-input in time, producing a field proportional to the spectrometer response $E_p^{(+)}(t)\sim r_p(t)$, as expected for linear time-invariant systems, i.e., aspect (ii) and (iii) from above~\cite{mcgillem1986probabilistic}. Causality [aspect (i)] is guaranteed by requiring that the spectrometer response vanishes $r_p(t)=0$ for all times $t<t_\mathrm{d}$, where $t_\mathrm{d}$ is the delay time required for the light to pass through the spectrometer. We can however always choose $t_\mathrm{d}=0$ assuming that the time arguments of $E_p^{(+)}(t)$ and $E_\mathrm{in}^{(+)}(t)$ in Eq.~\eqref{eq:def_spec_response} are shifted with respect to each other by this physical delay. We will do so in the following, such that the causality requirement can be written as $r_p(t)=\Theta(t)r_p(t)$, where $\Theta$ is the Heaviside function.

We can now combine the action of the spectrometer [Eq.~\eqref{eq:def_spec_response}] and the intensity detection via pixel $p$ in the CCD [Eq.~\eqref{eq:I_p}] to write 
\begin{equation}
	I_p(t)=\int\limits_{-\infty}^{\infty} {\rm d} t_1\int\limits_{-\infty}^{\infty} {\rm d} t_2\, r_p^*(t-t_1) r_p(t-t_2)\left< {E}^{(-)}_\mathrm{in}(t_1) {E}^{(+)}_\mathrm{in}(t_2) \right>\,.
\end{equation}
The spectrometer grating diffracts the incoming light in such a way that pixel $p$ only detects in an interval around a certain frequency $\omega$ (or wavelength). We assume this interval to be sufficiently narrow and replace $p\rightarrow\omega$, yielding the time-dependent physical spectrum of the light detected by the spectrometer~\cite{eberly1977time} 
\begin{equation}\label{eq:spec_def_general}
	I(t,\omega)=\int\limits_{-\infty}^{\infty} {\rm d} t_1\int\limits_{-\infty}^{\infty} {\rm d} t_2\, r_\omega^*(t-t_1) r_\omega(t-t_2)\left< {E}^{(-)}(t_1) {E}^{(+)}(t_2) \right>\,.
\end{equation}
We drop the label "in" of the incoming light for notational simplicity from now on. This formula can be seen as the backbone for the modeling of all sorts of spectroscopy methods, e.g., photoluminescence~\cite{groll2021controlling,preuss2022resonant}, resonance fluorescence~\cite{wigger2021resonance,groll2023read}, or in this case four-wave mixing.

From this time-dependent spectrum we can calculate the time-integrated spectrum. As an important example we consider a situation, where the correlation function factorizes
\begin{equation}
\left< {E}^{(-)}(t_1) {E}^{(+)}(t_2) \right>=\left< {E}^{(-)}(t_1)\right>\left< {E}^{(+)}(t_2) \right>\equiv\mathcal{E}^*(t_1)\mathcal{E}(t_2)\,.\label{eq:coherent_source_factor}
\end{equation}
This is the case, when the incoming field is derived from a coherent light source, e.g., a laser~\cite{glauber1963coherent,cohen1998atom}. In such situations, the positive frequency component of the field, which contains only photon annihilation operators, yields the corresponding coherent amplitudes when acting to the right and analog for the negative frequency component acting to the left\footnote{For coherent state $\ket{\alpha}$ of photon mode with annihilation (creation) operator $a^{(\dagger)}$: $a\ket{\alpha}=\alpha\ket{\alpha}$, $\bra{\alpha}a^{\dagger}=\bra{\alpha}\alpha^*$.}. With this the time-integrated spectrum can be written as
\begin{align}\label{eq:spec_def_integrated}
\overline{I}(\omega)&=\int\limits_{-\infty}^{\infty}\text{d} t\,I(t,\omega)=\int\limits_{-\infty}^{\infty}\text{d} t\,\left|\,\,\int\limits_{-\infty}^{\infty}\text{d}\tau\,r_\omega(t-\tau)\mathcal{E}(\tau)\right|^2=\int\limits_{-\infty}^{\infty}\text{d}\omega_0 R(\omega,\omega_0)|\widetilde{\mathcal{E}}(\omega_0)|^2\,.
\end{align}
The absolute square of the Fourier transform of the coherent field
\begin{equation}
\widetilde{\mathcal{E}}(\omega)=\int\limits_{-\infty}^{\infty}\text{d}t\,\mathcal{E}(t)e^{i\omega t}
\end{equation}
is here broadened according to the spectral response of the spectrometer
\begin{equation}
R(\omega,\omega_0)=\frac{1}{2\pi}\left|\,\,\int\limits_{-\infty}^{\infty}\text{d}t\,r_\omega(t)e^{i\omega_0 t}\right|^2\label{eq:def_spectral_response}
\end{equation}
to constitute the detected intensity spectrum $\overline{I}(\omega)$. An ideal spectrometer is obviously described by a very sharp spectral response $R(\omega,\omega_0)\sim \delta(\omega-\omega_0)$, i.e., complete loss of any temporal information in Eq.~\eqref{eq:spec_def_general}, yielding the familiar result $\overline{I}(\omega)\sim |\widetilde{\mathcal{E}}(\omega)|^2$. This ideal spectrum of the coherent light source however is in general broadened by the spectral response $R(\omega,\omega_0)$ due to a finite spectrometer resolution~\cite{anderson1975errors,eberly1977time,preuss2022resonant}. Assuming a homogeneous response $R(\omega,\omega_0)=R(\omega-\omega_0)$, i.e., assuming that the spectrometer has the same properties independent of the detection frequency $\omega$, we see that the measured time-integrated spectrum $\overline{I}(\omega)$ is given by a convolution between the ideal spectrum $|\widetilde{\mathcal{E}}(\omega)|^2$ and the spectral response $R(\omega)$. From Eq.~\eqref{eq:def_spectral_response} we see that such a homogeneous response of the spectrometer implies
\begin{equation}
r_\omega(t)= r_{\omega=0}(t)e^{-i\omega t}\Leftrightarrow R(\omega,\omega_0)=R(\omega-\omega_0)\,.\label{eq:homog_response}
\end{equation}
In this case the temporal response $r_\omega(t)$ can be written as an envelope $r_0(t)$ times a carrier wave $e^{-i\omega t}$. In the following we assume that the spectrometer in the experiment fulfills this property sufficiently well in the relevant spectral range. The resolution of the spectrometer is determined by the width of the function $R(\omega)$. The slower the temporal envelope $r_0(t)$ varies, the better the spectral resolution. Since we consider detection in the range of optical frequencies, the relevant frequencies of the carrier wave fulfill $\hbar\omega\sim 1$~eV. At the same time the resolution of the spectrometer is on the order of 20~\textmu eV (see Fig.~\ref{fig:slit}), such that the envelope $r_0(t)$ changes very slowly compared to the carrier wave $e^{-i\omega t}$.

\begin{figure}[t]
	\centering
	\includegraphics[width = 0.35\textwidth]{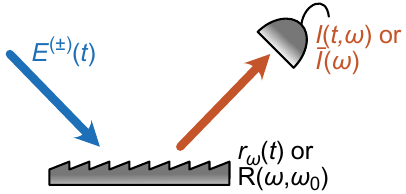}
	\caption{Schematic picture of the performance of a spectrometer modeled in Eq.~\eqref{eq:spec_def_general} for time-dependent and Eq.~\eqref{eq:spec_def_integrated} for time-integrated detection.}
	\label{fig:theory_spectroscopy}
\end{figure}

The role of the spectrometer in the signal detection and the most important quantities introduced in this section are summarized schematically in Fig.~\ref{fig:theory_spectroscopy}.

\subsection{Balanced heterodyne detection}\label{sec:balanced_heterodyne}
\begin{figure}[b]
	\centering
	\includegraphics[width = 0.3\textwidth]{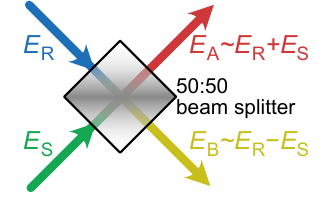}
	\caption{Schematic of the beam mixing by a lossless 50:50 beam splitter from Eq.~\eqref{eq:beam-splitter}. In our case the incoming beams are the signal $E_{\rm S}$ and the reference $E_{\rm R}$ and the outgoing beams are $E_{\rm A}$ and $E_{\rm B}$.}
	\label{fig:beam_splitter}
\end{figure}

Following the beam path in the setup in Figs.~\ref{fig:setup}(b)/\ref{fig:setup_3_4} further backwards, we next reach the mixing-AOM in box 4. As mentioned previously in the context of the inset in box 4, as well as Fig.~\ref{fig:AOM}, this AOM performs two tasks: (I) it acts as a 50:50 beam splitter and (II) it provides the phase shift necessary for the selection of the FWM component of the total signal. Considering first the action as a beam splitter (I, see Fig.~\ref{fig:beam_splitter}), we assume that the two input fields are the signal $E_\mathrm{S}$ from the investigated sample and the reference beam $E_\mathrm{R}$. The two outputs in a 50:50 beam splitter can be written as~\cite{collett1987quantum}
\begin{equation}\label{eq:beam-splitter}
	E_{\mathrm{A/B}}=\frac{1}{\sqrt{2}}\left(E_\mathrm{R}\pm E_\mathrm{S}\right)\,,
\end{equation}
\\
where the two possible signs stem from a relative phase shift of $\pi$ between the two reflection processes (either $E_\mathrm{R}$ or $E_\mathrm{S}$ is deflected). This relative phase shift is a general property of any loss-less beam splitter, whose transfer matrix has to be unitary due to energy conservation~\cite{campos1989quantum}. The output fields $E_{\mathrm{A/B}}$ of the beam splitter are sent into distinct spatial modes $W_\mathrm{\mathrm{A/B}}$ in the experiment and are separately detected using the same spectrometer and different positions on the CCD (see Fig.~\ref{fig:CCD}). Using Eq.~\eqref{eq:spec_def_general}, the time-integrated spectrum of the light in these two spatial modes can be written as

\begin{align}
	\overline{I}_{\mathrm{A/B}}(\omega)&=\int\limits_{-\infty}^{\infty}\!\!\text{d} t\int\limits_{-\infty}^{\infty}\!\!\text{d} t_1\int\limits_{-\infty}^{\infty}\!\!\text{d} t_2\, r_{\omega}^*(t-t_1)r^{}_{\omega}(t-t_2)\left<E_{\mathrm{A/B}}^{(-)}(t_1)E_{\mathrm{A/B}}^{(+)}(t_2)\right>\notag\\
	&=\frac{1}{2}\int\limits_{-\infty}^{\infty}\!\!\text{d} t\int\limits_{-\infty}^{\infty}\!\!\text{d} t_1\int\limits_{-\infty}^{\infty}\!\!\text{d} t_2\, r_{\omega}^*(t-t_1)r^{}_{\omega}(t-t_2)\left[\left<E_\mathrm{R}^{(-)}(t_1)E_\mathrm{R}^{(+)}(t_2)\right>+\left<E_\mathrm{S}^{(-)}(t_1)E_\mathrm{S}^{(+)}(t_2)\right>\right.\notag\\
	&\qquad\qquad\qquad\qquad\qquad\qquad\qquad\qquad\left.\pm\left<E_\mathrm{R}^{(-)}(t_1)E_\mathrm{S}^{(+)}(t_2)\right>\pm\left<E_\mathrm{S}^{(-)}(t_1)E_\mathrm{R}^{(+)}(t_2)\right>\right]\label{eq:I_AB}\,.
\end{align}
Both spectra contain a pure reference part (R), a pure signal part (S) and two interference terms between signal and reference. In the experiment balanced heterodyne detection (hd) is used to remove the pure reference and signal intensities and isolate the interference terms (see Fig.~\ref{fig:process_2}, which contains an additional phase flip to remove classical detection noise not considered here). This isolation of the interference is done by calculating the difference between the two spectra $\overline{I}_\mathrm{hd}=\overline{I}_\mathrm{A}-\overline{I}_\mathrm{B}$.
The reason for doing so is that we are interested in the signal field, not just its intensity, and this brings us one step closer to actually measuring the field itself. Balanced heterodyne detection now yields the time-integrated spectrum
\begin{align}
	\overline{I}_\mathrm{hd}(\omega)&=\!\!\int\limits_{-\infty}^{\infty}\!\!\text{d} t\int\limits_{-\infty}^{\infty}\!\!\text{d} t_1\!\!\int\limits_{-\infty}^{\infty}\!\!\text{d} t_2\, r_{\omega}^*(t-t_1)r^{}_{\omega}(t-t_2)\left[\left<E_\mathrm{R}^{(-)}(t_1)E_\mathrm{S}^{(+)}(t_2)\right>+\left<E_\mathrm{S}^{(-)}(t_1)E_\mathrm{R}^{(+)}(t_2)\right>\right]\notag\\
	&=2\text{Re}\left[\,\,\int\limits_{-\infty}^{\infty}\!\!\text{d} t\int\limits_{-\infty}^{\infty}\!\!\text{d} t_1\int\limits_{-\infty}^{\infty}\!\!\text{d} t_2\, r_{\omega}^*(t-t_1)r^{}_{\omega}(t-t_2)\left<E_\mathrm{R}^{(-)}(t_1)E_\mathrm{S}^{(+)}(t_2)\right>\right]\,.\label{eq:I_hd_1}
\end{align}
Here we have used that the two terms in the integral are complex conjugates of each other due to
\begin{equation}
\left<E_\mathrm{R}^{(-)}(t_1)E_\mathrm{S}^{(+)}(t_2)\right>^*=\left<E_\mathrm{S}^{(-)}(t_2)E_\mathrm{R}^{(+)}(t_1)\right>\,.
\end{equation}
Finally we use that the reference is derived from a coherent light source (the laser), such that the action of $E_\mathrm{R}^{(+)}(t)$ to the right or $E_\mathrm{R}^{(-)}(t)$ to the left in any expectation value with respect to this initial coherent state of the reference beam mode simply yields a classical number (the coherent amplitude), since these operators contain only photon annihilation or creation operators, respectively. This leads to the factorization of the correlation function analog to Eq.~\eqref{eq:coherent_source_factor}
\begin{equation}
	\left<E_\mathrm{R}^{(-)}(t_1)E_\mathrm{S}^{(+)}(t_2)\right>=\left<E_\mathrm{R}^{(-)}(t_1)\right>\left<E_\mathrm{S}^{(+)}(t_2)\right>\equiv \mathcal{E}_\mathrm{R}^*(t_1)\mathcal{E}_\mathrm{S}(t_2)\,,\label{eq:interference_factor}
\end{equation}
implying that there are no quantum mechanical correlations between the signal and the reference beam~\cite{collett1987quantum}, which is easily understandable by the fact that laser light can typically be described accurately using classical fields instead of quantum fields.
\subsection{Phase selection using AOMs}\label{sec:phase_selection}
\begin{figure}[b]
	\centering
	\includegraphics[width = 0.35\textwidth]{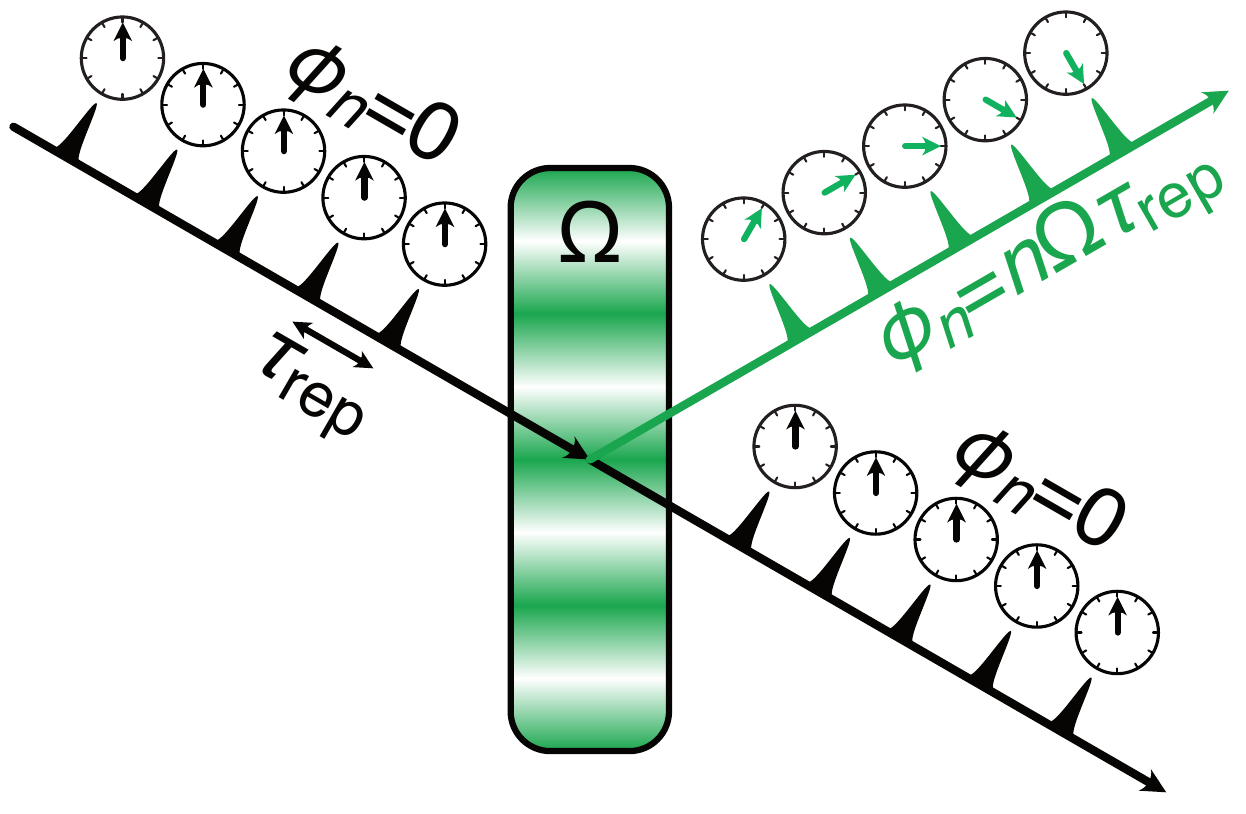}
	\caption{Schematic of the performance of an AOM converting a sequence of phase-stable laser pulses into another phase-stable sequence via transmission (black) and a sequence with continuously progressing phases via phonon scattering according to Eq.~\eqref{eq:r4_pulsetrain_upshift} (green).}
	\label{fig:AOM_clocks_1}
\end{figure}
We continue by investigating aspect (II) of the mixing-AOM, i.e., the frequency and phase shift discussed in the context of Fig.~\ref{fig:AOM} (see also Fig.~\ref{fig:AOM_clocks_1}). Since there is in principle no difference between the mixing-AOM in box 4 of Figs.~\ref{fig:setup}(b)/\ref{fig:setup_3_4} and the AOMs acting on the excitation pulses in box 2 of Fig.~\ref{fig:setup}(b), we consider the general situation of an incoming coherent phase-stable pulse train, whose positive frequency component is described by the classical function
\begin{equation}\label{eq:r4_pulsetrain}
	\mathcal{E}(t)=\sum_n \mathcal{E}_0(t-n\tau_\mathrm{rep})e^{-i\omega_l(t-n\tau_\mathrm{rep})}\,,
\end{equation}
where $\omega_l$ is the carrier frequency and $\mathcal{E}_0$ is the pulse envelope. Single phonon absorption in an AOM driven by an AC voltage with frequency $\Omega$ leads to a frequency upshift $\omega_l\rightarrow\omega_l+\Omega$, as described previously in Fig.~\ref{fig:AOM}. Since we consider mode-locked pulses, i.e., $\omega_l\tau_{\rm rep}=2\pi m$ with $m=1,2,3...$, the upshifted field becomes~\cite{langbein2010coherent}
\begin{equation}\label{eq:r4_pulsetrain_upshift}
	\mathcal{E}_{+\Omega}(t)=\sum_n \mathcal{E}_0(t-n\tau_\mathrm{rep})e^{-i(\omega_l+\Omega)(t-n\tau_\mathrm{rep})}=e^{-i(\omega_l+\Omega)t}\sum_n \mathcal{E}_0(t-n\tau_\mathrm{rep})e^{in\Omega\tau_{\mathrm{rep}}}\,.
\end{equation}
The carrier frequency of the mode-locked pulse train changes to $\omega_l+\Omega$ and each pulse obtains a phase shift with consecutive pulses having the relative phase $(n+1)\Omega\tau_{\rm rep}-n\Omega\tau_{\mathrm{rep}}=\Omega\tau_{\rm rep}$ (see Fig.~\ref{fig:AOM_clocks_1}). The spectral width of the pulses of 100~fs to 1~ps length used in typical experiments is on the order of 1~meV. Compared with this spectral width, the shift of the carrier frequency to $\omega_l+\Omega$ can be neglected since $\hbar\Omega\approx 0.3$~\textmu eV~$\ll1$~meV for the radio frequencies $\Omega/2\pi\simeq 80$~MHz used for the AOMs in the setup [Figs.~\ref{fig:setup}(b)/\ref{fig:setup_3_4}, see also Sec.~\ref{sec:exp_heterodyne}]. Thus the action of an AOM on a mode-locked pulse train via phonon absorption leading to a frequency upshift can be written as
\begin{equation}
		\mathcal{E}_{+\Omega}(t)\approx e^{-i\omega_lt}\sum_n \mathcal{E}_0(t-n\tau_\mathrm{rep})e^{in\Omega\tau_{\mathrm{rep}}}\label{eq:AOM_action}
\end{equation}
and analog for stimulated phonon emission leading to a downshift, described by replacing $+\Omega$ by $-\Omega$ in Eq.~\eqref{eq:AOM_action}. The carrier frequency remains approximately the same, but subsequent pulses obtain a relative phase shift, as already sketched in Fig.~\ref{fig:AOM}.

These phase shifts directly contribute to the interference term in Eq.~\eqref{eq:interference_factor}, which is present in the integrand in Eq.~\eqref{eq:I_hd_1}. We did not consider this when calculating $\overline{I}_\mathrm{hd}=\overline{I}_\mathrm{A}-\overline{I}_\mathrm{B}$ and assumed identical interference terms for both spatial modes $W_\mathrm{A}$ and $W_\mathrm{B}$. This approximation is justified in the following. 

Note that in the experiment mode-locked pulse trains with a repetition time of $\tau_\mathrm{rep}\approx 13$~ns are used to excite the sample and provide the reference beam. Before interacting with the mixing-AOM in Figs.~\ref{fig:setup}(b)/\ref{fig:setup_3_4} (box 4), the reference and signal fields have the form
\begin{subequations}
\begin{align}
	\mathcal{E}_\mathrm{R}(t)&=\sum_n \mathcal{E}_{0,\mathrm{R}}(t-n\tau_{\mathrm{rep}})e^{-i\omega_l t}\\
	\mathcal{E}_\mathrm{S}(t)&=\sum_n \mathcal{E}^{(n)}_\mathrm{S}(t-n\tau_{\mathrm{rep}})\,.
\end{align}
\end{subequations}
\begin{figure}[b]
	\centering
	\includegraphics[width = 0.35\textwidth]{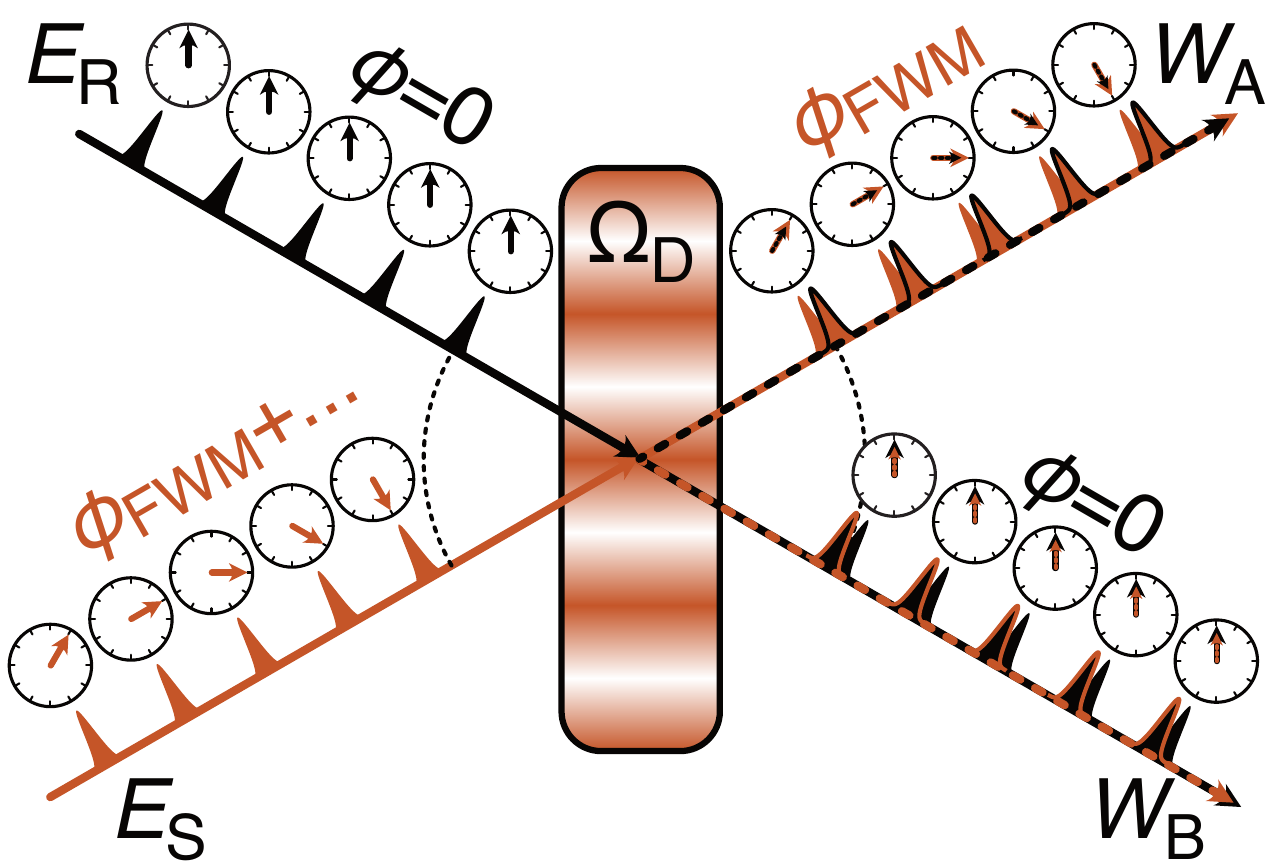}
	\caption{Schematic of the performance of the mixing AOM visualizing Eqs.~\eqref{eq:RS_modes_AB} and \eqref{eq:RS_modes_AB_2}.}
	\label{fig:AOM_clocks_2}
\end{figure}
The label $(n)$ in $\mathcal{E}^{(n)}_\mathrm{S}$ marks that each repetition of the signal is already slightly different due to the action of the other AOMs in box 2 in Fig.~\ref{fig:setup}(b) before excitation of the sample in box 3. Written in this form, we do not explicitly consider a separation into carrier wave and envelope for the signal for notational simplicity. When the mixing-AOM in Fig.~\ref{fig:AOM_clocks_2} [previously: Figs.~\ref{fig:setup}(b)/\ref{fig:setup_3_4} (inset box 4)] interferes signal (orange) and reference (black) beam and converts them into modes $W_\mathrm{A}$ and $W_\mathrm{B}$, respectively, this happens in two ways: (A) the signal simply passes through without phonon scattering, while the reference is upshifted in frequency by $+\Omega_\mathrm{D}$, transforming the interference term in Eq.~\eqref{eq:interference_factor} as
\begin{subequations}\label{eq:RS_modes_AB}
\begin{equation}
	\mathcal{E}_\mathrm{R}^*(t_1)\mathcal{E}^{}_\mathrm{S}(t_2)\rightarrow \sum_{n,n'} \left[\mathcal{E}_{0,\mathrm{R}}(t_1-n\tau_{\mathrm{rep}})e^{-i\omega_l t_1}e^{in(+\Omega_\mathrm{D})\tau_{\mathrm{rep}}}\right]^*\mathcal{E}^{(n')}_\mathrm{S}(t_2-n'\tau_{\mathrm{rep}})\qquad\text{in mode }W_\mathrm{A}\,,
\end{equation}
or (B) the reference passes through, while the signal is downshifted by $\Omega_\mathrm{D}$, i.e., shifted by $-\Omega_\mathrm{D}$
\begin{equation}
	\mathcal{E}_\mathrm{R}^*(t_1)\mathcal{E}^{}_\mathrm{S}(t_2)\rightarrow \sum_{n,n'} \left[\mathcal{E}_{0,\mathrm{R}}(t_1-n\tau_{\mathrm{rep}})e^{-i\omega_l t_1}\right]^*\mathcal{E}^{(n')}_\mathrm{S}(t_2-n'\tau_{\mathrm{rep}})e^{in'(-\Omega_\mathrm{D})\tau_{\mathrm{rep}}}\qquad\text{in mode }W_\mathrm{B}\,.
\end{equation}
\end{subequations}
Since the temporal width of the spectrometer response is much shorter than the repetition time (see also Fig.~\ref{fig:spec_exp}), subsequent repetitions do not overlap in the spectrometer~\cite{langbein2006heterodyne,langbein2010coherent}, i.e., signal and reference from different repetitions do not mix in Eq.~\eqref{eq:I_AB}. Therefore, only those contributions from Eqs.~\eqref{eq:RS_modes_AB} with $n=n'$ effectively contribute. Considering only these overlapping contributions with $n=n'$, the interference terms for both output modes are identical, with the last AOM acting as
\begin{equation}\label{eq:RS_modes_AB_2}
	\mathcal{E}_\mathrm{R}^*(t_1)\mathcal{E}^{}_\mathrm{S}(t_2)\rightarrow \sum_n \left[\mathcal{E}_{0,\mathrm{R}}(t_1-n\tau_{\mathrm{rep}})e^{-i\omega_l t_1}\right]^*\mathcal{E}^{(n)}_\mathrm{S}(t_2-n\tau_{\mathrm{rep}})e^{-in\Omega_\mathrm{D}\tau_{\mathrm{rep}}}\,.
\end{equation}
This justifies that we assumed identical interference terms for $\overline{I}_\mathrm{A}$ and $\overline{I}_\mathrm{B}$ in the calculation of Eq.~\eqref{eq:I_hd_1}. Inserting this result into the heterodyne detected spectrum in Eq.~\eqref{eq:I_hd_1}, we obtain
\begin{align}
	\overline{I}_\mathrm{hd}(\omega)&=\sum_n 2\text{Re}\left[\int\limits_{-\infty}^{\infty}\!\!\text{d} t\int\limits_{-\infty}^{\infty}\!\!\text{d} t_1\int\limits_{-\infty}^{\infty}\!\!\text{d} t_2\, r_{\omega}^*(t-t_1)r^{}_{\omega}(t-t_2)\mathcal{E}^*_{0,\mathrm{R}}(t_1)e^{i\omega_l t_1}\mathcal{E}^{(n)}_\mathrm{S}(t_2)e^{-in\Omega_\mathrm{D}\tau_{\mathrm{rep}}}\right]\,,\label{eq:I_hd_2}
\end{align}
where we performed the substitution $x-n\tau_{\mathrm{rep}}\rightarrow x$ for $x=t,t_1,t_2$ in each term of the sum.

If the $n$-th repetition of the signal $\mathcal{E}_\mathrm{S}^{(n)}(t)$ was independent of $n$, the heterodyne detected spectrum would simply vanish due to the sum over $\exp(-in \Omega_\mathrm{D}\tau_{\mathrm{rep}})$ leading to destructive interference for $\Omega_\mathrm{D}\tau_{\mathrm{rep}}\neq 2\pi m$ with $m=0,1,2..$. However, the signal itself contains a multitude of phase shifts similar to the term $\exp(-in \Omega_\mathrm{D}\tau_{\mathrm{rep}})$, allowing for constructive interference of very specific contributions to the signal. These phase shifts are imprinted on the signal by the exciting laser pulses $\mathcal{E}_j$. Each of these is indeed upshifted by different AOMs, as shown in box 2 of Fig.~\ref{fig:setup}(b). The three excitation pulse trains are therefore of the form [see Eq.~\eqref{eq:AOM_action}]
\begin{equation}
	\mathcal{E}_j(t)=e^{-i\omega_l t}\sum_n \mathcal{E}_{0,j}(t-n\tau_{\mathrm{rep}}) e^{in\Omega_j \tau_{\mathrm{rep}}}\,,\qquad j=1,2,3\,.
\end{equation}
Assuming that the repetition time of the laser is chosen in such a way that the investigated sample returns to the same initial state before excitation by subsequent repetitions of the pulse train, the $n$-th repetition of the detected signal can be written as a general nonlinear power series of the $n$-th repetition of the excitation pulse trains $\mathcal{E}_j$. Apart from the different phase however, each repetition of the excitation is exactly the same, leading to the same dependence of the signal on the excitation pulses plus an additional imprinted phase. We can therefore write the $n$-th repetition of the signal as~\cite{shen1984principles,langbein2006heterodyne,langbein2010coherent}
\begin{equation}
	\mathcal{E}_\mathrm{S}^{(n)}(t)=\sum_{l_1,l_2,l_3=-\infty}^{\infty}\mathcal{E}_\mathrm{S}^{(l_1,l_2,l_3)}(t)e^{in(l_1\Omega_1+l_2\Omega_2+l_3\Omega_3)\tau_{\mathrm{rep}}}\label{eq:E_S_n_sum}\,.
\end{equation}
Here, the function $\mathcal{E}_\mathrm{S}^{(l_1,l_2,l_3)}(t)$ is the (non-)linear response of the investigated sample on the excitation meaning that the linear response is contained in the cases with $|l_1|+|l_2|+|l_3|=1$.

Considering the FWM signals derived in Sec.~\ref{sec:FWM_2LS} the nonlinear FWM response that we want to isolate is $\mathcal{E}_\mathrm{FWM}=\mathcal{E}_\mathrm{S}^{(-1,1,1)}\sim \mathcal{E}_1^*\mathcal{E}^{}_2\mathcal{E}^{}_3$. Note that $\mathcal{E}_\mathrm{S}^{(-1,1,1)}$ also contains higher order contributions like $\sim\mathcal{E}_1^*\mathcal{E}^{}_2\mathcal{E}^{}_3|\mathcal{E}^{}_3|^2$, which carry the same phase combination but stem from higher order nonlinearities that become relevant beyond the $\chi^{(3)}$-regime. To achieve the isolation of $\mathcal{E}_\mathrm{FWM}=\mathcal{E}_\mathrm{S}^{(-1,1,1)}$, slightly different AOM frequencies $\Omega_i$ for the three excitation pulse trains need to be considered (see Sec.~\ref{sec:exp_heterodyne}) and we accordingly choose the mixing-AOM frequency as $\Omega_\mathrm{D}=\Omega_3+\Omega_2-\Omega_1$. When we now insert Eq.~\eqref{eq:E_S_n_sum} into Eq.~\eqref{eq:I_hd_2}, we can evaluate the sum over $n$ to determine which contributions interfere constructively. Omitting all terms that do not depend on $n$ in Eq.~\eqref{eq:I_hd_2}, we obtain the sum
\begin{equation}
	\sum_n e^{in[(l_1+1)\Omega_1+(l_2-1)\Omega_2+(l_3-1)\Omega_3]\tau_{\rm rep}}
\end{equation}
which forces 
\begin{equation}
	[(l_1+1)\Omega_1+(l_2-1)\Omega_2+(l_3-1)\Omega_3]\tau_{\rm rep}=2\pi m\,,\qquad m\in\mathbb{Z}
\end{equation}
for a sufficiently large number of repetitions. This is obviously fulfilled for $l_1=-1$, $l_2=1$, and $l_3=1$. Furthermore, all AOM frequencies $\Omega_i$ are chosen only slightly different (see Sec.~\ref{sec:exp_heterodyne}), such that this matching condition cannot be satisfied for other small values of $l_i\sim\mathcal{O}(1)$. Very large values for the $l_i$ on the other hand are generally suppressed as they require very high nonlinear responses far beyond the third order nonlinearity examined in FWM. Furthermore the repetition time in the experiment is chosen in such a way that $\Omega_i\tau_{\mathrm{rep}}\gtrsim 2\pi$, $\Omega_i\tau_{\mathrm{rep}}\neq2\pi$ lie close to, but slightly above $2\pi$ for all AOMs that label the excitation pulses. The next possible values of the $l_i$ satisfying the matching condition then require large $|l_i|\gg 1$ and can be neglected~\cite{langbein2006heterodyne,langbein2010coherent}.

\begin{figure}[h]
	\centering
	\includegraphics[width = 0.6\textwidth]{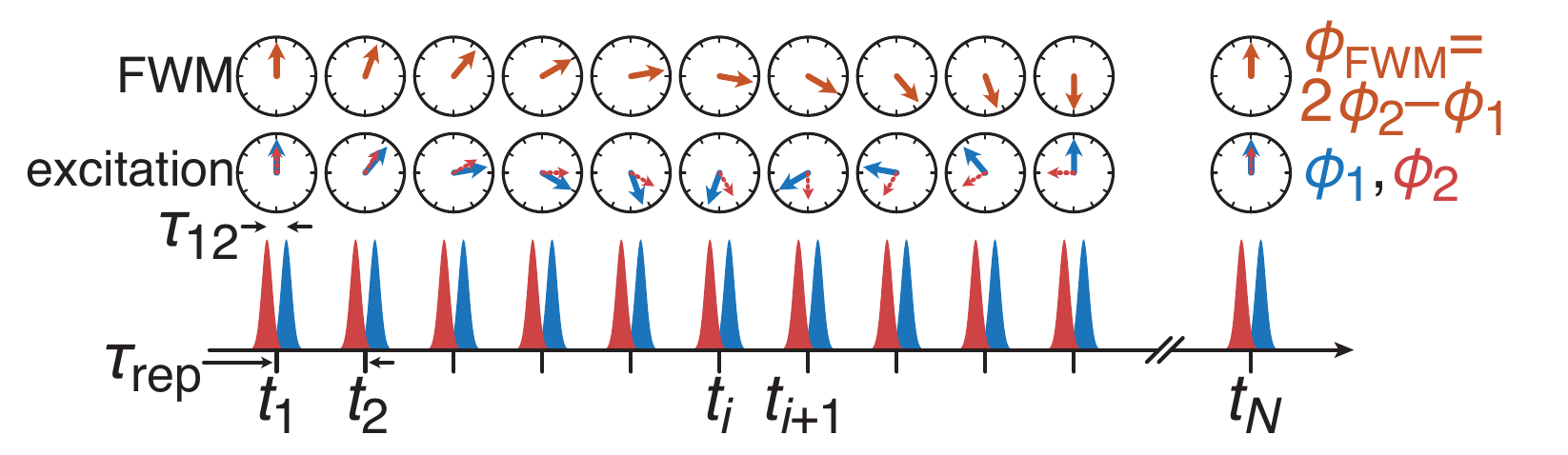}
	\caption{Schematic construction of the FWM phase for two-pulse FWM. The pulse phases are illustrated by the clock symbols above each pulse pair. The corresponding FWM phase, required to filter the desired signal, is shown in the top.}
	\label{fig:clocks}
\end{figure}

To visualize the impact of the AOMs on the pulse and FWM phases, in Fig.~\ref{fig:clocks} we consider two-pulse FWM with $\phi_\mathrm{FWM}=2\phi_2-\phi_1$. The double pulse sequence is repeated several times while the phases of the pulses $\phi_1,\,\phi_2$ change from repetition to repetition, visualized by the excitation clocks above each pulse pair (blue and red). The corresponding FWM phase relations $\phi_{\rm FWM}=2\phi_2-\phi_1$ are depicted in the top clocks. The important aspect is here that the relative phase between the two pulses varies throughout the repetition and covers the entire interval $[0,2\pi)$ with a sufficiently dense discretization. 

If the AOMs are operated in this way, we finally obtain the heterodyne detected (three-pulse) FWM spectrum
\begin{align}
	\overline{I}_\mathrm{hd}^\mathrm{(FWM)}(\omega)&=2\text{Re}\left[\int\limits_{-\infty}^{\infty}\!\!\text{d} t\int\limits_{-\infty}^{\infty}\!\!\text{d} t_1\int\limits_{-\infty}^{\infty}\!\!\text{d} t_2\, r_{0}^*(t-t_1)r^{}_{0}(t-t_2)\mathcal{E}^*_{0,\mathrm{R}}(t_1)e^{i(\omega_l-\omega) t_1}\mathcal{E}_\mathrm{S}^{(-1,1,1)}(t_2)e^{i\omega t_2}\right]\,,\label{eq:I_hd_3}
\end{align}
where we assume a homogeneous spectrometer response $r_\omega(t)=r_0(t)e^{-i\omega t}$ [see Eq.~\eqref{eq:homog_response}]. This phase selection process from the experiment is typically modeled by simulating the dynamics of and emitted field from the investigated system for a single repetition, i.e., excitation with three pulses (not pulse trains) $\mathcal{E}_j(t)$ with $j=1,2,3$. In such a situation, we then calculate $\mathcal{E}_\mathrm{S}(t)$, which can take on different forms, e.g., the polarization of a 2LS~\cite{jakubczyk2016impact,wigger2017exploring,wigger2018rabi} (see also App.~\ref{sec:dipole_emission}) or the electric field inside a microcavity~\cite{kasprzak2013coherence,groll2020four}. To extract the FWM response from the signal, the excitation pulses are given independent phases $\mathcal{E}_j(t)\rightarrow \mathcal{E}_j(t) e^{i\phi_j}$, such that the signal is a function of these phases $\mathcal{E}_\mathrm{S}(t)\rightarrow \mathcal{E}_\mathrm{S}(t;\phi_1,\phi_2,\phi_3)$ analog to Eq.~\eqref{eq:E_S_n_sum}, as already discussed in Sec.~\ref{sec:FWM_2LS}. The simulation is then performed for a large number of different phase values, reflecting the pulses in the pulse trains from the experiment (see Fig.~\ref{fig:clocks}), and the FWM signal is extracted by calculating
\begin{equation}
	\mathcal{E}_{\mathrm{FWM}}(t)=\mathcal{E}_\mathrm{S}^{(-1,1,1)}(t)=\frac{1}{(2\pi)^3}\int\limits_0^{2\pi}\int\limits_0^{2\pi}\int\limits_0^{2\pi}\text{d}\phi_1\,\text{d}\phi_2\,\text{d}\phi_3\, \mathcal{E}_\mathrm{S}(t;\phi_1,\phi_2,\phi_3)e^{-i(\phi_3+\phi_2-\phi_1)}\,.
\end{equation}
This integration corresponds to the constructive interference reached with the mixing-AOM that is set to $\Omega_\mathrm{D}=\Omega_3+\Omega_2-\Omega_1$. It is straightforward to generalize such simulations from FWM to $N$-wave mixing with an arbitrary number of pulses and corresponding phase combinations~\cite{fras2016multi,hahn2022destructive,grisard2023temporal}, discussed in more detail in Sec.~\ref{sec:nwm}.
\subsection{Extracting amplitude and phase of the signal}\label{sec:exp_phase}
Finally, we discuss how to retrieve the FWM signal $\mathcal{E}_\mathrm{FWM}(t)\equiv\mathcal{E}_\mathrm{S}^{(-1,1,1)}(t)$ from the heterodyne detected spectrum in Eq.~\eqref{eq:I_hd_3}, including amplitude and phase. This is done via post-processing of the measured spectral interferograms~\cite{patton2004non,langbein2006heterodyne,langbein2010coherent} (see Fig.~\ref{fig:amp_phase}). We are usually dealing with measurements where a single reference pulse in the pulse train created in box 1 of Fig.~\ref{fig:setup} is much shorter in time than the temporal width of the envelope $r_0(t)$ of the spectrometer response and detection happens close to the carrier frequency of the laser $\omega\approx\omega_l$. Then we can evaluate the $t_1$-integral in Eq.~\eqref{eq:I_hd_3} approximately by replacing $t_1$ with $t_\mathrm{R}$ in $r_0(t-t_1)$ and $e^{i(\omega_l-\omega) t_1}$, where $t_\mathrm{R}$ describes the time at which this single reference pulse arrives in the spectrometer. This approximation implies that the single reference pulse is spectrally much broader than the resolution of the spectrometer, i.e., that the reference pulse can be well resolved in frequency (see Fig.~\ref{fig:CCD} and discussion in the context of Fig.~\ref{fig:slit}).
Note that we performed a time-shift for each iteration of the pulse trains when calculating Eq.~\eqref{eq:I_hd_2}. The reference time $t_\mathrm{R}$ is therefore counted relative to each individual repetition of the experiment, e.g., $t_\mathrm{R}=0$ actually means $t_\mathrm{R}=0+n\tau_{\rm rep}$ for the $n$-th repetition.

After this replacement we can write the $t_1$-integral over the reference pulse envelope in Eq.~\eqref{eq:I_hd_3} as a general complex number with amplitude $A_\mathrm{R}$ and phase $\phi_\mathrm{R}$. Under these conditions the heterodyne detected spectrum reads
\begin{align}
	\overline{I}^\mathrm{(FWM)}_\mathrm{hd}(\omega)&=2A_\mathrm{R}\text{Re}\left\lbrace\,\int\limits_{-\infty}^{\infty}\!\!\text{d} t_2\, (r_{0}\star r_{0})(t_\mathrm{R}-t_2)e^{i[\phi_\mathrm{R}+(\omega_l-\omega)t_\mathrm{R}]}\mathcal{E}_\mathrm{FWM}(t_2)e^{i\omega t_2}\right\rbrace\,,\label{eq:I_hd_4}
\end{align}
where the $\star$-operator denotes the autocorrelation of the spectrometer response envelope $r_0$~\cite{mcgillem1986probabilistic}. To obtain the FWM signal the detected spectrum is now post-processed. The first step is to calculate the inverse Fourier transform
\begin{align}
&\mathcal{F}^{-1}\left[\overline{I}_\mathrm{hd}^\mathrm{(FWM)}(\omega)\right](t-t_\mathrm{R}) \notag\\
	&=\frac{1}{2\pi}\int\limits_{-\infty}^{\infty}\text{d}\omega\, e^{-i\omega (t-t_\mathrm{R})}\overline{I}_\mathrm{hd}^\mathrm{(FWM)}(\omega)\notag\\
	&=A_\mathrm{R} (r_{0}\star r_{0})(t_\mathrm{R}-t)\left[e^{i[\phi_\mathrm{R}+\omega_lt_\mathrm{R}]}\mathcal{E}_\mathrm{FWM}(t) +e^{-i[\phi_\mathrm{R}+\omega_lt_\mathrm{R}]}\mathcal{E}_\mathrm{FWM}^{*}(2t_\mathrm{R}-t)\right]\,,\label{eq:I_hd_FT}
\end{align}
where we have used the general property $(r_0\star r_0)^*(t)=(r_0\star r_0)(-t)$ of the autocorrelation function and shifted the time arguments by $t_\mathrm{R}$ for convenience. Here, the $\omega$-integral produced $\delta$-functions which allowed for the evaluation of the $t_2$-integral. We see that there are two distinct FWM signal terms in the square brackets: a backward time component ($2t_\mathrm{R}-t$) and a forward time component ($t$). To obtain direct information on the function $\mathcal{E}_\mathrm{FWM}(t)$, we have to somehow separate these two contributions. This is done by using a causality condition: Each repetition of the signal can only be non-zero after excitation of the sample with the incoming laser pulses. Let us denote the point in time where the signal from the excited sample reaches the spectrometer by $t=0$. Then the FWM signal has the causality property $\mathcal{E}_\mathrm{FWM}(t)=\mathcal{E}_\mathrm{FWM}(t)\Theta(t)$. The forward time component therefore vanishes at $t<0$, while the backward time component vanishes at $2t_\mathrm{R}-t<0$, i.e., $t>2t_\mathrm{R}$.

\begin{figure}[t]
	\centering
	\includegraphics[width = 0.9\textwidth]{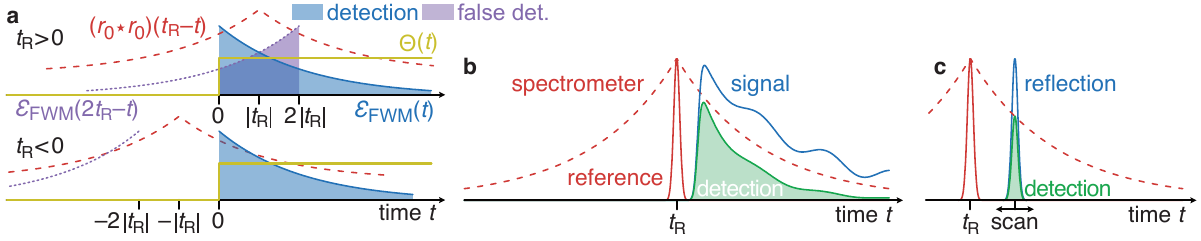}
	\caption{Illustration of the impact of the spectrometer response. (a) Comparison of the two situations $t_\mathrm{R}\gtrless 0$, i.e., reference arriving at the spectrometer after (top) or before (bottom) the signal for Eq.~\eqref{eq:I_hd_FT}. Separation of backward (violet) and forward (blue) component of the signal by multiplication with an appropriate Heaviside function (yellow) is only possible for $t_\mathrm{R}<0$. (b) Impact of the spectrometer response in Eq.~\eqref{eq:I_hd_isol}. In the time domain the detected signal (green) is the product of the emitted signal $\mathcal{E}_\mathrm{FWM}(t)=\mathcal{E}_\mathrm{S}^{(-1,1,1)}(t)$ (blue) and the (autocorrelation of the) spectrometer response $(r_{0}\star r_{0})(t_\mathrm{R}-t)$ (dashed red), centered around the reference pulse arriving at the spectrometer at time $t_\mathrm{R}$ before the signal does (solid red). (c) Same as (b) but with the signal being replaced by one of the laser beams, reflected from the sample, yielding the spectrometer response when scanning the delay between reference and pulse, as shown in Eq.~\eqref{eq:I_hd_ref}.}
	\label{fig:spec}
\end{figure}

Figure~\ref{fig:spec}(a) shows two possible choices for the reference time $t_\mathrm{R}$: $t_\mathrm{R}>0$ (top), i.e., the reference pulse reaches the spectrometer after the signal does, and $t_\mathrm{R}<0$ (bottom), i.e., the reference pulse reaches the spectrometer before the signal does. In the first case of $t_\mathrm{R}>0$ we see that the forward $(t$, blue) and backward $(2t_\mathrm{R}-t$, violet) component of the signal can possibly overlap and are not guaranteed to be separated in time. In the second case of $t_\mathrm{R}<0$ they are always separated which allows us to retrieve the FWM signal via appropriate multiplication by the Heaviside function (yellow)
\begin{equation}
\Theta(t)\mathcal{F}^{-1}\left[\overline{I}^\mathrm{(FWM)}_\mathrm{hd}(\omega)\right](t_\mathrm{R}-t)=A_\mathrm{R} (r_{0}\star r_{0})(t_\mathrm{R}-t)e^{i[\phi_\mathrm{R}+\omega_lt_\mathrm{R}]}\mathcal{E}_\mathrm{FWM}(t)\,,\qquad t_\mathrm{R}<0\,.\label{eq:I_hd_isol}
\end{equation}
The significance of choosing the reference to precede the signal in time $t_{\rm R}<0$ is highlighted by the example of echo dynamics in App.~\ref{sec:ref_timing}.

There are some important things to note here: (i) The post-processed signal scales with the amplitude $A_\mathrm{R}$ of the reference due to the heterodyne detection. The strength of the detected signal can therefore be increased by increasing the amplitude of the reference pulses. (ii) We cannot obtain the FWM signal field  $\mathcal{E}_\mathrm{FWM}(t)\equiv\mathcal{E}_\mathrm{S}^{(-1,1,1)}(t)$ just on its own. The detected signal always contains the autocorrelation $(r_{0}\star r_{0})(t_\mathrm{R}-t)$ of the spectrometer response, which needs to be considered when making quantitative comparisons between theory and experiment~\cite{jakubczyk2016impact}. This is sketched in Fig.~\ref{fig:spec}(b). The autocorrelation yields a time-window for detection (dashed red) and generally suppresses the detected signal (green) for times that are much later than the reference time $t_\mathrm{R}$. (iii) The heterodyne detection measures an interference between the signal from the sample and the reference pulse. In the previous derivation we considered ultrashort reference pulses by replacing $t_1$ with $t_\mathrm{R}$ in Eq.~\eqref{eq:I_hd_4}. Furthermore we assumed that the reference pulse arrives at the spectrometer before the signal does. This prohibits any direct interference between these two fields outside of the spectrometer. However, the reference pulse is stretched in the spectrometer by the spectrometer response, such that the interference happens in the spectrometer with this stretched-out reference.

By Fourier transforming Eq.~\eqref{eq:I_hd_isol} we obtain the so-called FWM spectrum
\begin{align}
	S_\mathrm{FWM}(\omega)&=\mathcal{F}\left\lbrace\Theta(t)\mathcal{F}^{-1}\left[\overline{I}_\mathrm{hd}^\mathrm{(FWM)}\right](t-t_\mathrm{R})\right\rbrace(\omega)=\int\limits_{-\infty}^{\infty}\text{d}t\,e^{i\omega t}\Theta(t)\mathcal{F}^{-1}\left[\overline{I}_\mathrm{hd}^\mathrm{(FWM)}\right](t-t_\mathrm{R})\notag\\
	&=A_\mathrm{R}e^{i[\phi_\mathrm{R}+\omega_lt_\mathrm{R}]}\int\limits_{-\infty}^{\infty}\!\!\text{d} t\, (r_{0}\star r_{0})(t_\mathrm{R}-t)\mathcal{E}_\mathrm{FWM}(t)e^{i\omega t}\,.\label{eq:I_FWM_1}
\end{align}
Comparing this result to the full heterodyne spectrum in Eq.~\eqref{eq:I_hd_4}, we see that we essentially got rid of the real part projection $\text{Re}[...]$ and obtained the full information on amplitude and phase by the application of an inverse Fourier transform $\mathcal{F}^{-1}$, multiplication by an appropriate Heaviside function $\Theta(t)$ using the causality requirements of the signal, and a final application of a Fourier transform back to the frequency domain~\cite{patton2004non,langbein2006heterodyne,langbein2010coherent} (see also Fig.~\ref{fig:amp_phase}). The transformation that is used to obtain the FWM signal is an application of Kramers-Kronig relations, which connect real and imaginary part of spectra of causal time signals~\cite{toll1956causality}.

Lastly we have to somehow determine the spectrometer response, as we have seen that it is important for an accurate theoretical modeling. The total signal from the sample of course does not only contain the desired FWM signal, but also, e.g., the reflection of the exciting laser beams. In terms of the series in Eq.~\eqref{eq:E_S_n_sum}, the reflection is part of the linear response, e.g., contained in $\mathcal{E}_\mathrm{S}^{(1,0,0)}$, and it usually dominates that part of the signal $\mathcal{E}_\mathrm{S}^{(1,0,0)}(t)\sim \mathcal{E}_1(t)=\mathcal{E}_{1,0}(t)e^{-i\omega_l t}$. 

In the same way that we isolated the FWM part of the total signal, we can isolate this laser reflection by choosing the frequency of the last AOM to be $\Omega_\mathrm{D}=\Omega_1$ (homodyne detection, see Sec.~\ref{sec:exp_heterodyne}). As the exciting laser pulse $\mathcal{E}_1$ and the reference $\mathcal{E}_\mathrm{R}$ originate from the same source, we can assume the same approximations that were used in deriving the heterodyne spectrum in Eq.~\eqref{eq:I_hd_4}: the exciting laser pulse is much shorter in time than the spectrometer response and we want to consider detection close to the laser frequency $\omega\approx\omega_l$, such that we can replace all times~$t_2$ in Eq.~\eqref{eq:I_hd_4} by the arrival time of the signal, i.e., the reflection of pulse 1. Without loss of generality, we assume this to be at $t_2=0$, such that we obtain
\begin{align}
	\overline{I}_\mathrm{hd}(\omega)&\sim2A_\mathrm{R}\text{Re}\left[(r_{0}\star r_{0})(t_\mathrm{R})e^{i[\phi_\mathrm{R}+(\omega_l-\omega)t_\mathrm{R}]}\right]\,,\label{eq:I_hd_ref}
\end{align}
under these conditions. By scanning the reference time $t_\mathrm{R}$ with respect to the reflection signal and choosing the mixing-AOM to isolate exactly this reflection component, we can thus obtain information on the autocorrelation of the spectrometer response~\cite{jakubczyk2016impact} (see Fig.~\ref{fig:spec_exp} for experimental demonstration). We can also understand this with the help of Fig.~\ref{fig:spec}(c): The reflection signal (blue) is now a very narrow peak in time, which effectively selects the autocorrelation of the spectrometer response (dashed red) at that specific point in time for detection (green). 

From the previous discussion it also becomes clear that -- under heterodyne FWM detection conditions with $\Omega_\mathrm{D}=\Omega_3+\Omega_2-\Omega_1$ -- the mixing-AOM acts as a filter removing an otherwise dominant background stemming from laser reflection.


%% file: 6_nls.tex
\section{FWM signals and 2D spectra of few-level systems}\label{sec:nls}
After the extensive discussion on the experimental setup in Chap.~\ref{sec:exp} and its theoretical modeling in Chap.~\ref{sec:exp_theo}, we here come back to the description of FWM signals, based on Chaps.~\ref{sec:theory} and \ref{sec:theory_2}, where we focused on 2LSs. In the following we will extend the description to arbitrary optically driven $N$-level systems ($N$LS). Certain applications, e.g., for entangled photon emission~\cite{akopian2006entangled,huber2018semiconductor} actually require multiple levels, where the description as a 2LS is not sufficient anymore. As we will discuss here, in such $N$LSs wave mixing spectroscopy provides a powerful tool to investigate the specific level structure as well as coupling mechanisms between the different levels~\cite{smallwood2018multidimensional}. 
\subsection{The optically driven $N$-level system}\label{sec:optics_NLS}
\begin{figure}[b]
	\centering
	\includegraphics[width=0.25\linewidth]{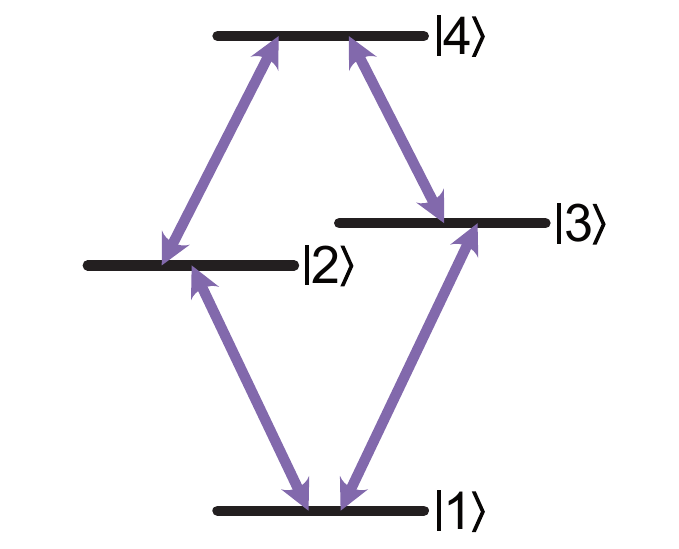}
	\caption{Example of an optically driven 4LS with allowed dipole transitions marked by arrows.}
	\label{fig:4ls}
\end{figure}
We consider a system consisting of $N$ energy levels, described by the Hamiltonian
\begin{equation}
	H_0=\sum_{m=1}^N E_m\ket{m}\bra{m}\,,
\end{equation}
where $\ket{m}$ are the energy eigenstates with energy $E_m$. The 2LS case from Chap.~\ref{sec:theory} is obviously recovered for $N=2$, $E_1=0$, $E_2=\hbar\omega_x$, $\ket{1}\hat{=}\ket{g}$, and $\ket{2}\hat{=}\ket{x}$. We assume that the levels are ordered, such that $n>m$ implies $E_n\geq E_m$ (see Fig.~\ref{fig:4ls} for a 4LS). Optical driving is again described within the dipole approximation with~\cite{cohen1998atom,cohen2022photons} [see also Eq.~\eqref{eq:dipole_approx}]
\begin{equation}
	H_I(t)=-\frac{\bm{d}}{2}\left[\bm{E}(t)+\bm{E}^*(t)\right]=-\sum_{m,n}\ket{m}\bra{n}\frac{\bra{m}\bm{d}\ket{n}}{2}\left[\bm{E}(t)+\bm{E}^*(t)\right]\,,
\end{equation}
where $\bm{E}(t)=\bm{E}_0(t)e^{-i\omega_l t}$ now denotes the complex electric field and $\bm{d}$ is again the dipole operator. By using this approximation we  assume that the wavefunctions belonging to the different energy levels are sufficiently localized in a volume whose dimensions are much smaller than the wavelength of the light field. Applying the RWA~\cite{cohen1998atom}, only those transitions with $E_m-E_n\approx \hbar\omega_l>0$ need to be taken into account (purple arrows in Fig.~\ref{fig:4ls}), leading to
\begin{equation}
	H_I(t)\approx \frac{1}{2}\sum_{m>n}\ket{m}\bra{n}\mathcal{A}_{mn}(t)+h.c.
\end{equation}
analog to Eq.~\eqref{eq:H_2LS_RWA}, where the effective amplitude for the transition between levels $\ket{m}$ and $\ket{n}$ is now given by
\begin{equation}
	\mathcal{A}_{mn}(t)=\left\lbrace\begin{matrix}-\bm{E}_0(t) e^{-i\omega_l t}\bra{m}\bm{d}\ket{n}\,,\quad & E_m-E_n\approx \hbar\omega_l\\
		0\,,\quad & \text{else}\end{matrix}\right.
\end{equation}
analog to Eq.~\eqref{eq:def_eff_field}. By using this definition for the amplitudes $\mathcal{A}_{mn}$ we automatically exclude far off-resonant transitions. 
\subsection{FWM in the optically driven $N$-level system}
We consider again excitation with ultrashort laser pulses, such that the effective fields are modeled via
\begin{equation}
	\mathcal{A}_{mn}(t)=\sum_j \mathcal{A}_{mn,j}(t)=\hbar\sum_j \theta_{mn,j} e^{i\phi_j} \delta(t-t_j)
\end{equation}
analog to Eq.~\eqref{eq:pulse_sequence}~\cite{mukamel1995principles}. The driving consists of $\delta$-pulses acting at times $t_j$ with phases $\phi_j$ and the pulse areas
\begin{equation}
	\theta_{mn,j}=\left\lbrace\begin{matrix}|\bm{E}_{0,j}\bra{m}\bm{d}\ket{n}|/\hbar\,,\quad & E_m-E_n\approx \hbar\omega_l\\
		0\,,\quad & \text{else}\end{matrix}\right.\label{eq:theta_mn_j}
\end{equation}
are now in general different for the different transitions. $\bm{E}_{0,j}$ is here the electric field amplitude of the $j$-th pulse. 

We now have to specify the rules for calculating FWM signals analog to Sec.~\ref{sec:FWM_rules}. Similar to Eq.~\eqref{eq:U_P_deriv} we can construct the action of a single $\delta$-pulse acting at time $t_j$ by calculating the full time evolution operator $U(t_j+\epsilon, t_j-\epsilon)$ due to the Hamiltonian $H(t)=H_0+H_I(t)$ and letting $\epsilon\rightarrow 0$. In an analog fashion we then obtain the pulse transformation operator
\begin{subequations}\label{eq:U_P_nls}
\begin{equation}
	U_{\rm P}(\theta_j,\phi_j)=\exp\left[-\frac{i}{2}\sum_{m>n}\left(\theta_{mn,j}\ket{m}\bra{n}e^{i\phi_j}+\theta_{mn,j}\ket{n}\bra{m}e^{-i\phi_j}\right)\right]\,,
\end{equation}
where the argument $\theta_j$ now denotes the matrix defined by Eq.~\eqref{eq:theta_mn_j}. This operator transforms the density matrix at time $t_j$ according to
\begin{equation}
	\rho_+(t_j)=U^{}_{\rm P}(\theta_j,\phi_j)\rho_-(t_j)U_{\rm P}^{\dagger}(\theta_j,\phi_j)
\end{equation}
\end{subequations}
analog to Eqs.~\eqref{eq:U_P_FWM}. The free time evolution between the pulses is governed by 
\begin{equation}
	U_0(t,t_0)=\exp[-iH_0(t-t_0)]\label{eq:U_0_nls}
\end{equation}
and we ignore dissipative processes in the following discussion for simplicity. However, if required for a more accurate description, decay and decoherence processes can directly be added, either via phenomenological damping rates or via Lindblad dissipators (see Sec.~\ref{sec:theory_dissipation}).

Since we now have multiple dipole transitions, the relevant observable is not simply the microscopic polarization $p=\braket{\ket{g}\bra{x}}$ of a 2LS anymore, but a suitable weighted sum of all relevant microscopic polarizations
\begin{equation}
	p_{mn}=\left\lbrace\begin{matrix}\braket{\ket{n}\bra{m}}=\text{Tr}(\ket{n}\bra{m}\rho)=\bra{m}\rho\ket{n}\,,\quad&E_m-E_n\approx\hbar\omega_l\\0\,,\quad&\text{else}\end{matrix}\right.\,.\label{eq:micro_p_mn}
\end{equation}
Here we excluded polarizations whose frequencies are far away from the laser frequency $\omega_l$, assuming that we are interested in optical detection of the FWM signal in the vicinity of this excitation frequency. From these microscopic polarizations we can construct the complex macroscopic polarization~\cite{mukamel1995principles,koch2001correlation,krugel2007monitoring,haug2009quantum}
\begin{equation}
	\bm{P}=\sum_{m>n}p_{mn}\bra{m}\bm{d}\ket{n}^*\,,
	\label{eq:polarization_vector}
\end{equation}
which constitutes the source for the signal field. Since there can now be multiple possible dipole transitions, all of them contribute to the total polarization weighted by their respective dipole moments $\bra{m}\bm{d}\ket{n}^*$ describing their respective emission strengths. Note that in Sec.~\ref{sec:exp_theo} we suppressed the vector character of the signal field, assuming that it has uniform polarization or is filtered, e.g., with a polarizer, before detection. While such details have to be considered when modeling an actual experiment~\cite{kasprzak2008vectorial,kasprzak2013vectorial,mermillod2016dynamics}, in the following we work with the complex amplitude of the macroscopic polarization as given by $P\sim \sum_{m>n}p_{mn}$ for simplicity, assuming that all microscopic polarizations enter equally. Note that we did not need to consider the macroscopic polarization separately in the 2LS case since there micro- and macroscopic polarization are proportional to each other.

As discussed in Sec.~\ref{sec:FWM_rules} it is a useful strategy to calculate the FWM dynamics in the density matrix itself and obtain the relevant microscopic polarizations afterwards. After each pulse we can directly perform the phase selection obtaining
\begin{equation}
	\rho(\phi_j)\rightarrow \frac{1}{2\pi}\int\limits_0^{2\pi}\text{d}\phi_j\, \rho(\phi_j)e^{-il_j\phi_j}=\overline{\rho}_{(j)}
\end{equation}
as described in the context of Eq.~\eqref{eq:phase_sel_rho_single}. In the following we are interested in weak optical driving, i.e., the $\chi^{(3)}$-regime of the FWM response, allowing us to make analytical progress without lengthy calculations. In the linear order with respect to the pulse area(s) $\theta_j$ the density matrix transformation in Eqs.~\eqref{eq:U_P_nls} can be written as
\begin{equation}
	\rho_+(t_j)=1-\frac{i}{2}\sum_{m>n}\left[\theta_{mn,j}\ket{m}\bra{n}e^{i\phi_j}+\theta_{mn,j}\ket{n}\bra{m}e^{-i\phi_j},\rho_-(t_j)\right]+\mathcal{O}(\theta_j^2)\,.\label{eq:pulse_nls_1st}
\end{equation}
The phase-selected density matrix then reads
\begin{equation}
	\overline{\rho}_{(j)}=\left\lbrace\begin{matrix}-\frac{i}{2}\sum_{m>n}\theta_{mn,j} \left[\ket{m}\bra{n},\rho_-(t_j)\right]\,,\quad & l_j=1\\[2mm] -\frac{i}{2}\sum_{m>n}\theta_{mn,j} \left[\ket{n}\bra{m},\rho_-(t_j)\right]\,,\quad & l_j=-1\end{matrix} \right.\,.\label{eq:pulse_nls_1st_selected}
\end{equation}
Depending on the specific sign of the phase that is selected $l_j=\pm 1$, the density matrix transforms via the commutator with the transition operators $\ket{m}\bra{n}$ or $\ket{n}\bra{m}$, respectively, i.e., transitions upwards or downwards with respect to the ordering of the energy levels $m>n$.

Note that the lowest order of the transformation does not contain any terms $\sim e^{\pm 2i\phi_j}$ which are necessary for two-pulse FWM. To calculate these terms we have to go to the second order with respect to the pulse areas, i.e., we obtain the double commutator
\begin{align}
	\rho_+(t_j)=... -\frac{1}{8}&\Bigg[\sum_{m'>n'}\left(\theta_{m'n',j}\ket{m'}\bra{n'}e^{i\phi_j}+\theta_{m'n',j}\ket{n'}\bra{m'}e^{-i\phi_j}\right),\notag\\
	&\bigg[\sum_{m>n}\left(\theta_{mn,j}\ket{m}\bra{n}e^{i\phi_j}+\theta_{mn,j}\ket{n}\bra{m}e^{-i\phi_j}\right),\rho_-(t_j)\bigg]\Bigg] + \mathcal{O}(\theta_j^3)\,,
\end{align}
where the dots denote the zeroth and first order given in Eq.~\eqref{eq:pulse_nls_1st}. If we phase select the terms with $l_j=\pm 2$, i.e., $\sim e^{\pm 2i\phi_j}$ from this density matrix, the result is identical to twice the action of a pulse that is phase selected for $l_j=\pm1$ from Eq.~\eqref{eq:pulse_nls_1st_selected}, apart from a factor $1/2$. This implies that at lowest order in the pulse areas there is no distinction between two-pulse FWM with the phase $\phi_\mathrm{FWM}=2\phi_2-\phi_1$ and three-pulse FWM with the phase $\phi_\mathrm{FWM}=\phi_3+\phi_2-\phi_1$ and a delay $\tau_{23}=0$ between identical pulses 2 and 3. In the following we only consider the $\chi^{(3)}$-regime such that the transformation rule Eq.~\eqref{eq:pulse_nls_1st_selected} suffices for our calculations. 


\subsection{Double-sided Feynman diagrams for the $\chi^{(3)}$-regime}\label{sec:feynman}
In the previous section we have seen how to obtain the phase-selected density matrix after a weak pulse for the case $l_j=\pm1$. We can summarize this transformation together with the free time evolution in the form of double-sided Feynman diagrams which are a handy graphical representation of the relevant states throughout the FWM pulse sequence~\cite{yee1978diagrammatic,mukamel1995principles}. To construct these diagrams we consider the following rules:
\begin{itemize}
	\item[1.] The state of the system relevant for the FWM signal is represented by a density matrix element $\ket{m}\bra{n}$. Initially the system is usually in its ground state $\ket{1}\bra{1}$.
	\item[2.] A pulse at time $t_j$ transforms the density matrix according to Eq.~\eqref{eq:pulse_nls_1st_selected}, which is represented by a horizontal arrow pointing away from or towards the state. 
	\begin{itemize}
		\item[2.a.] If we select for negative phase factors, i.e., $l_j=-1$, the arrow points to the left $\boxed{\leftarrow -\phi_j}$. In the case of positive phase factors $l_j=+1$ it points to the right $\boxed{\rightarrow +\phi_j}$.
		\item[2.b.] Arrows pointing away from the state lead to downwards transition, i.e., they act as $\frac{i}{2}\sum_{m>n}\theta_{mn,j}\ket{m}\bra{n}$ on the right and as $-\frac{i}{2}\sum_{m>n}\theta_{mn,j}\ket{n}\bra{m}$ on the left.
		In practice: $\boxed{\ket{a}\bra{b}\rightarrow}$ becomes $\boxed{\ket{a}\bra{c},\,c<b}$\quad   or\quad $\boxed{\leftarrow\ket{a}\bra{b}}$ becomes $\boxed{\ket{d}\bra{b},\,d<a}$.
		\item[2.c.] Similarly, arrows pointing towards the state lead to upwards transitions, i.e., they act as $\frac{i}{2}\sum_{m>n}\theta_{mn,j}\ket{n}\bra{m}$ on the right and $-\frac{i}{2}\sum_{m>n}\theta_{mn,j}\ket{m}\bra{n}$ on the left. 
		In practice: $\boxed{\ket{a}\bra{b}\leftarrow}$ becomes $\boxed{\ket{a}\bra{c},\,c>b}$\quad or\quad $\boxed{\rightarrow\ket{a}\bra{b}}$ becomes $\boxed{\ket{d}\bra{b},\,d>a}$.
		\item[2.d.] We have to construct all possible non-vanishing diagrams.
		\item[2.e.] Higher phase factors $l_j=\pm k\ (k>1)$ are obtained by applying the pulse transformation rules $k$ times. This leads to additional numerical prefactors, which do not change the FWM polarization dynamics in the $\chi^{(3)}$-regime and can thus be ignored.
	\end{itemize}
\item[3.] Free time evolution after a pulse $j$ is marked by a vertical downward arrow with the delay $\tau_{jk}$ until the next pulse $k$. This leads to an overall factor of $\exp[-i\tau_{jk} (E_m-E_n)/\hbar]$ for a state $\ket{m}\bra{n}$ due to the action of $U_0$ in Eq.~\eqref{eq:U_0_nls}. The time evolution after the last pulse is labeled accordingly as $t$.
\end{itemize}
If we keep track of all factors, adding up all relevant diagrams yields the total phase-selected FWM density matrix, from which we can calculate the relevant microscopic polarizations in Eq.~\eqref{eq:micro_p_mn}.

To give a concrete example for double-sided Feynman diagrams, we apply these rules to the familiar case of a 2LS which we can compare to our results from Sec.~\ref{sec:FWM_2LS}. We consider three-pulse FWM with the phase combination $\phi_\mathrm{FWM}=\phi_3+\phi_2-\phi_1$ and positive delays $\tau_{12},\tau_{23}\geq 0$, i.e., pulse 1 excites first, followed by pulse 2 and then pulse 3. Following all rules we obtain the two diagrams given in Tab.~\ref{tab:feyn_2ls}.
\begin{table}[h]
\begin{center}
	\begin{tabular}{r r | c | l }
		initial state:&				&$1 \rangle\langle 1$	&$\leftarrow  -\phi_1$		\\
		after pulse 1:&				&$1 \rangle\langle 2$	&						\\
		delay propagation:&			&	$\downarrow$	 		&$\tau_{12}$				\\
		&$+\phi_2 \rightarrow$		&$1 \rangle\langle 2$	& 						\\
		after pulse 2:&				&$2 \rangle\langle 2$	&						\\
		delay propagation:&			&	$\downarrow$	 		&$\tau_{23}$				\\
		&						&$2 \rangle\langle 2$	&$ \rightarrow +\phi_3$		\\
		after pulse 3:&				&$2 \rangle\langle 1$	&						\\
		final propagation:&			&	$\downarrow$	 &$t$					
	\end{tabular}\qquad
	\begin{tabular}{r r | c | l }
		&						&$1 \rangle\langle 1$	&$\leftarrow  -\phi_1$		\\
		&						&$1 \rangle\langle 2$	&						\\
		&						&	$\downarrow$	 		&$\tau_{12}$				\\
		&						&$1 \rangle\langle 2$	&$\rightarrow +\phi_2 $	\\
		&						&$1 \rangle\langle 1$	&						\\
		&						&	$\downarrow$	 		&$\tau_{23}$				\\
		&$+\phi_3 \rightarrow $		&$1 \rangle\langle 1$	& 						\\
		&						&$2 \rangle\langle 1$	&						\\
		&						&	$\downarrow$	 &$t$					
	\end{tabular}
\end{center}
\caption{Double-sided Feynman diagrams for three-pulse FWM of a 2LS with positive delays $\tau_{12}$ and $\tau_{23}$.}
\label{tab:feyn_2ls}
\end{table}

Let us break these diagrams down step by step. Initially the system is in its ground state, described by the density matrix element or projection operator $\ket{1}\bra{1}$. Pulse 1 carries a negative phase, such that it acts with a left-pointing arrow $\leftarrow -\phi_1$. We need to consider both possibilities (i) $-\phi_1\leftarrow\ket{1}\bra{1}$ and (ii) $\ket{1}\bra{1}\leftarrow -\phi_1$. In case (i) the arrow points outwards, i.e., it induces a downwards transition, which is not possible for the ground state as there is no state with less energy. Thus, case (i) does not lead to a separate non-vanishing diagram. In case (ii) we act from the right with $\frac{i}{2}\theta_{21,1} \ket{1}\bra{2}$ leading to the state $\ket{1}\bra{2}$ shown in line 2 of the diagrams. After pulse 1 follows free time evolution for the interval $\tau_{12}$ yielding the factor $\exp[-i\tau_{12}(E_1-E_2)/\hbar]$.

Pulse 2 carries the phase factor $l_2=+1$, yielding a right pointing arrow. This time we indeed get two separate diagrams, as $+\phi_2\rightarrow \ket{1}\bra{2}=-\frac{i}{2}\theta_{21,2}\ket{2}\bra{2}$ (left diagram, upwards transition) and $\ket{1}\bra{2}\rightarrow +\phi_2=\frac{i}{2}\theta_{21,2}\ket{1}\bra{1}$ (right diagram, downwards transition). Since we did not include any dissipation and $E_1-E_1=E_2-E_2=0$, the free time evolution does not yield any additional factor.

Pulse 3 now brings both diagrams into the same final state by inducing a downwards transition in the state $\ket{2}\bra{2}$ towards $\ket{2}\bra{1}$ (left diagram), yielding a factor $\frac{i}{2}\theta_{21,3}$ and similarly in the right diagram, inducing an upwards transition of the state $\ket{1}\bra{1}$ yielding a factor $-\frac{i}{2}\theta_{21,3}$. The final time evolution for time $t$ is identical for both diagrams, yielding $\exp[-i(E_2-E_1)t/\hbar]$. Note that the diagram with $+\phi_3\rightarrow \ket{2}\bra{2}$ vanishes, since there is no state with larger energy than $\ket{2}$.

Collecting all factors properly and adding both diagrams, we obtain the final microscopic polarization
\begin{equation}
	p_{21}(t,\tau_{12}>0,\tau_{23}>0)=\frac{i}{4}\theta_{21,1}\theta_{21,2}\theta_{21,3}e^{-i(E_2-E_1)(t-\tau_{12})/\hbar}\,.\label{eq:p_21}
\end{equation}
%
We can compare this result to the calculations in Sec.~\ref{sec:FWM_2LS} by identifying $\ket{1}=\ket{g}$, $\ket{2}=\ket{x}$, $\theta_{21}=\theta$, and $E_2-E_1=\hbar\omega_x$. We thus recover the result from Eq.~\eqref{eq:p_FWM_3} in the limiting case of vanishing dissipation and for small pulse areas. The calculation of this result however was much more efficient than in Sec.~\ref{sec:FWM_2LS}. Note that the delay $\tau_{23}$ does not impact the FWM polarization here, as we neglected excited state decay. Inclusion of dissipation is in principle easily possible, as it simply requires a slight modification of rule 3 for the free time evolution step~\cite{mukamel1995principles}. 

Finally we note that the double-sided Feynman diagrams in Tab.~\ref{tab:feyn_2ls} are great tools for understanding which quantity is probed by which time variable. Clearly the dynamics in real time $t$ after the last pulse is dictated by the coherence $\ket{2}\bra{1}$, i.e., on this timescale we measure the polarization of the 2LS. When varying the delay $\tau_{12}$, we see that we instead probe the conjugate coherence $\ket{1}\bra{2}$ containing the same information. The delay $\tau_{23}$ probes the occupations of the 2LS, i.e., $\ket{1}\bra{1}$ and $\ket{2}\bra{2}$. This is very helpful for understanding which quantities are measured in a certain experimental configuration of the delays~\cite{dai2012two,wigger2023controlled}.

With the help of these diagrams it is also easy to see that there is no signal for $\tau_{12}<0$ and $\tau_{23}\leq 0$ in the ordinary 2LS, i.e., pulses 2 and 3 excite the 2LS first, followed by pulse 1~\cite{wegener1990line}. In such a configuration, pulse 3 acts as $+\phi_3\rightarrow\ket{1}\bra{1}$, creating a coherence $\ket{2}\bra{1}$. The action of pulse 2 then automatically leads to a vanishing diagram, as $+\phi_2\rightarrow\ket{2}\bra{1}=0$ and $\ket{2}\bra{1}\rightarrow+\phi_2=0$, since the excited state $\ket{2}$ cannot be further excited in this 2LS configuration, while the ground state cannot be de-excited.
\subsection{FWM signals of three-level systems}\label{sec:FWM_3LS}
\begin{figure}[t]
	\centering
	\includegraphics[width = 0.55\textwidth]{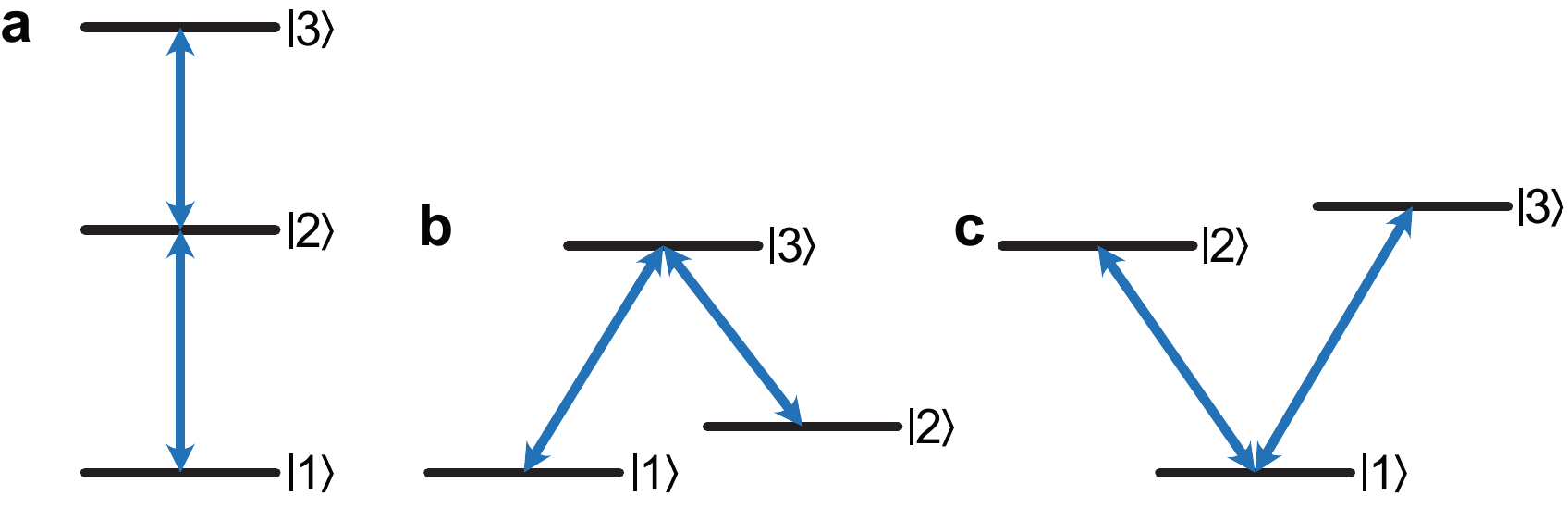}
	\caption{Possible dipole couplings in 3LSs: (a) cascade, (b) $\Lambda$-system, and (c) V-System.}
	\label{fig:3LS}
\end{figure}

In the following we provide a qualitative discussion on FWM signals of 3LSs using Feynman diagrams to discuss typical 2D spectra. To this aim, in Fig.~\ref{fig:3LS} we present three different ways in which an optically driven 3LS can be realized: (a) the cascade system with optically active transitions (blue arrows) between the ground state $\ket{1}$ and the first excited state $\ket{2}$, as well as between the first $\ket{2}$ and the second excited state $\ket{3}$~\cite{fu1995ultrafast,gerardot2009dressed,hohn2023energy}, (b) the $\Lambda$-system, where the ground state $\ket{1}$ does not couple optically to the first excited state $\ket{2}$, only to the second excited state $\ket{3}$, which in turn couples to $\ket{2}$~\cite{fleischhauer2005electromagnetically}, and (c) the V-system, where the ground state $\ket{1}$ couples optically to both $\ket{2}$ and $\ket{3}$, which are themselves however not coupled by relevant dipole transitions~\cite{grynberg1996amplification,ficek2004simulating,wang2005coherent}. These qualitative level schemes are sufficient for determining the Feynman diagrams. For example in the cascade system (a) we have non-vanishing $\theta_{21,j}>0$ and $\theta_{32,j}>0$, while $\theta_{31,j}=0$ [see Eq.~\eqref{eq:theta_mn_j}], as marked by the blue arrows, and analog for the other two cases.

We again consider three-pulse FWM with $\phi_\mathrm{FWM}=\phi_3+\phi_2-\phi_1$ and positive delays $\tau_{12}\geq0$ and $\tau_{23}\geq0$ as in the 2LS case in Tab.~\ref{tab:feyn_2ls}. The resulting non-vanishing diagrams are collected in Tab.~\ref{tab:feyn_3ls}. We already see that they are much richer than in the 2LS case. Furthermore, already at the level of the diagrams, there are huge qualitative differences between the different 3LSs with the V-shaped system in (c) showing the largest set of diagrams. From these diagrams we can now obtain the relevant microscopic polarizations. For simplicity we set $\tau_{23}=0$ and obtain for the cascade (C) system
\begin{subequations}\label{eq:p_3LS}
\begin{align}\label{eq:p_C}
	p^{(\text{C})}_{21}(t>0,\tau>0)&=\frac{i}{4}\theta_{21,1}\theta_{21,2}\theta_{21,3}e^{-i\omega_{21}(t-\tau)}\,,\notag\\
	p^{(\text{C})}_{32}(t>0,\tau>0)&=-\frac{i}{8}\theta_{21,1}\theta_{21,2}\theta_{32,3}e^{i\omega_{21}\tau}e^{-i\omega_{32}t}\,,
\end{align}
for the $\Lambda$-system
\begin{align}\label{eq:p_lambda}
	p^{(\Lambda)}_{31}(t>0,\tau>0)&=\frac{i}{4}\theta_{31,1}\theta_{31,2}\theta_{31,3}e^{-i\omega_{31}(t-\tau)}\,,\notag\\
	p^{(\Lambda)}_{32}(t>0,\tau>0)&=\frac{i}{8}\theta_{31,1}(\theta_{31,2}\theta_{32,3}+\theta_{32,2}\theta_{31,3})e^{i\omega_{31}\tau}e^{-i\omega_{32}t}\,,
\end{align}
and for the V-system
\begin{align}\label{eq:p_V}
	p^{(\text{V})}_{21}(t>0,\tau>0)&=\frac{i}{8}\left[2\theta_{21,1}\theta_{21,2}\theta_{21,3}e^{i\omega_{21}\tau}+\theta_{31,1}(\theta_{21,2}\theta_{31,3}+\theta_{31,2}\theta_{21,3})e^{i\omega_{31}\tau}\right]e^{-i\omega_{21}t}\,,\notag\\
	 p^{(\text{V})}_{31}(t>0,\tau>0)&=\frac{i}{8}\left[\theta_{21,1}(\theta_{31,2}\theta_{21,3}+\theta_{21,2}\theta_{31,3})e^{i\omega_{21}\tau}+2\theta_{31,1}\theta_{31,2}\theta_{31,3}e^{i\omega_{31}\tau}\right]e^{-i\omega_{31}t}\,,
\end{align}
\end{subequations}
where we wrote $\tau_{12}=\tau>0$ and $\omega_{ij}=(E_i-E_j)/\hbar$. As discussed in the context of Eq.~\eqref{eq:polarization_vector}, the various microscopic polarization can contribute with different strength to the overall detected macroscopic polarization, depending on the relative orientation of their corresponding dipole moments, as well as detection conditions. For simplicity we assume here that all microscopic polarizations contribute with equal strength to the signal
\begin{equation}
	\mathcal{E}_\mathrm{FWM}^{(\text{S})}(t>0,\tau>0)\sim\sum_{i>j} p_{ij}^{(\text{S})}(t>0,\tau>0)\,,\qquad\text{S}=\text{C, }\Lambda\text{, \text{V}}\,.\label{eq:E_FWM_3LS}
\end{equation}

\clearpage
\textbf{(a)}\\[-5mm]
\begin{table}[h]\small
~\hspace{-1cm}
		\begin{tabular}{l r c | c | l }
			&& (C1)				&$1 \rangle\langle 1$	&$\leftarrow  -\phi_1$		\\
			&&					&$1 \rangle\langle 2$	&						\\
			&&					&$\downarrow$	 			&$\tau_{12}$				\\
			&&$+\phi_2 \rightarrow$&$1 \rangle\langle 2$	& 						\\
			&&					&$2 \rangle\langle 2$	&						\\
			&&					&$\downarrow$	 			&$\tau_{23}$				\\
			&&					&$2 \rangle\langle 2$	&$ \rightarrow +\phi_3$		\\
			&&					&$2 \rangle\langle 1$	&						\\
			&&					&$\downarrow$	 &$t$					
		\end{tabular}
\hspace{-0.1cm}
		\begin{tabular}{r c | c | l }
			&(C2)				&$1 \rangle\langle 1$	&$\leftarrow  -\phi_1$		\\
			&					&$1 \rangle\langle 2$	&						\\
			&					&$\downarrow$	 			&$\tau_{12}$				\\
			&$+\phi_2\rightarrow$	&$1 \rangle\langle 2$	& 						\\
			&					&$2 \rangle\langle 2$	&						\\
			&					&$\downarrow$	 		&$\tau_{23}$				\\
			&$+\phi_3 \rightarrow$	&$2 \rangle\langle 2$	& 						\\
			&					&$3 \rangle\langle 2$	&						\\
			&					&$\downarrow$	 &$t$					
		\end{tabular}
\hspace{-0.1cm}
		\begin{tabular}{r c | c | l }
			&(C3)				&$1 \rangle\langle 1$	&$\leftarrow  -\phi_1$		\\
			&					&$1 \rangle\langle 2$	&						\\
			&					&$\downarrow$	 			&$\tau_{12}$				\\
			&					&$1 \rangle\langle 2$	&$\rightarrow +\phi_2 $		\\
			&					&$1 \rangle\langle 1$	&						\\
			&					&$\downarrow$	 			&$\tau_{23}$				\\
			&$+\phi_3 \rightarrow$	&$1 \rangle\langle 1$	& 						\\
			&					&$2 \rangle\langle 1$	&						\\
			&					&$\downarrow$	 &$t$					
		\end{tabular}
\end{table}\\[-5mm]
\textbf{(b)}\\[-5mm]
\begin{table}[h]\small
~\hspace{-1cm}
		\begin{tabular}{l r c | c | l }
			&& ($\Lambda$1)		&$1 \rangle\langle 1$	&$\leftarrow  -\phi_1$		\\
			&&					&$1 \rangle\langle 3$	&						\\
			&&					&$\downarrow$	 			&$\tau_{12}$				\\
			&&$+\phi_2 \rightarrow$&$1 \rangle\langle 3$	& 						\\
			&&					&$3 \rangle\langle 3$	&						\\
			&&					&$\downarrow$	 			&$\tau_{23}$				\\
			&&					&$3 \rangle\langle 3$	&$ \rightarrow +\phi_3$		\\
			&&					&$3 \rangle\langle 1$	&						\\
			&&					&$\downarrow$	 &$t$					
		\end{tabular}
\hspace{-0.1cm}
		\begin{tabular}{r c | c | l }
			&($\Lambda$2)			&$1 \rangle\langle 1$	&$\leftarrow  -\phi_1$		\\
			&					&$1 \rangle\langle 3$	&						\\
			&					&$\downarrow$	 			&$\tau_{12}$				\\
			&$+\phi_2\rightarrow$	&$1 \rangle\langle 3$	& 						\\
			&					&$3 \rangle\langle 3$	&						\\
			&					&$\downarrow$	 			&$\tau_{23}$				\\
			&					&$3 \rangle\langle 3$	&$\rightarrow+\phi_3  $		\\
			&					&$3 \rangle\langle 2$	&						\\
			&					&$\downarrow$	 &$t$					
		\end{tabular}
\hspace{-0.1cm}
		\begin{tabular}{r c | c | l }
			&($\Lambda$3)			&$1 \rangle\langle 1$	&$\leftarrow  -\phi_1$		\\
			&					&$1 \rangle\langle 3$	&						\\
			&					&$\downarrow$	 			&$\tau_{12}$				\\
			&					&$1 \rangle\langle 3$	&$\rightarrow +\phi_2 $		\\
			&					&$1 \rangle\langle 1$	&						\\
			&					&$\downarrow$	 			&$\tau_{23}$				\\
			&$+\phi_3 \rightarrow $	&$1 \rangle\langle 1$	& 						\\
			&					&$3 \rangle\langle 1$	&						\\
			&					&$\downarrow$	 &$t$					
		\end{tabular}
\hspace{-0.1cm}
		\begin{tabular}{r c | c | l }
			&($\Lambda$4)			&$1 \rangle\langle 1$	&$\leftarrow  -\phi_1$		\\
			&					&$1 \rangle\langle 3$	&						\\
			&					&$\downarrow$	 			&$\tau_{12}$				\\
			&					&$1 \rangle\langle 3$	&$\rightarrow +\phi_2 $		\\
			&					&$1 \rangle\langle 2$	&						\\
			&					&$\downarrow$	 			&$\tau_{23}$				\\
			&$+\phi_3 \rightarrow $	&$1 \rangle\langle 2$	& 						\\
			&					&$3 \rangle\langle 2$	&						\\
			&					&$\downarrow$	 &$t$					
		\end{tabular}
\end{table}\\[-5mm]
\textbf{(c)}\\[-5mm]
\begin{table}[h]\small
~\hspace{-1cm}
		\begin{tabular}{l r c | c | l }
			& &(V1)				&$1 \rangle\langle 1$	&$\leftarrow  -\phi_1$		\\
			&&					&$1 \rangle\langle 2$	&						\\
			&&					&$\downarrow$	 			&$\tau_{12}$				\\
			&&$+\phi_2 \rightarrow$&$1 \rangle\langle 2$	& 						\\
			&&					&$2 \rangle\langle 2$	&						\\
			&&					&$\downarrow$	 			&$\tau_{23}$				\\
			&&					&$2 \rangle\langle 2$	&$ \rightarrow +\phi_3$		\\
			&&					&$2 \rangle\langle 1$	&						\\
			&&					&$\downarrow$	 &$t$					
		\end{tabular}
\hspace{-0.1cm}
		\begin{tabular}{r c | c | l }
			&(V2)				&$1 \rangle\langle 1$	&$\leftarrow  -\phi_1$		\\
			&					&$1 \rangle\langle 2$	&						\\
			&					&$\downarrow$	 			&$\tau_{12}$				\\
			&$+\phi_2\rightarrow$	&$1 \rangle\langle 2$	& 						\\
			&					&$3 \rangle\langle 2$	&						\\
			&					&$\downarrow$	 			&$\tau_{23}$				\\
			&					&$3 \rangle\langle 2$	&$\rightarrow+\phi_3  $		\\
			&					&$3 \rangle\langle 1$	&						\\
			&					&$\downarrow$	 &$t$					
		\end{tabular}
\hspace{-0.1cm}
		\begin{tabular}{r c | c | l }
			&(V3)				&$1 \rangle\langle 1$	&$\leftarrow  -\phi_1$		\\
			&					&$1 \rangle\langle 2$	&						\\
			&					&$\downarrow$	 			&$\tau_{12}$				\\
			&					&$1 \rangle\langle 2$	&$\rightarrow +\phi_2 $		\\
			&					&$1 \rangle\langle 1$	&						\\
			&					&$\downarrow$	 			&$\tau_{23}$				\\
			&$+\phi_3 \rightarrow$	&$1 \rangle\langle 1$	& 						\\
			&					&$2 \rangle\langle 1$	&						\\
			&					&$\downarrow$	 &$t$					
		\end{tabular}
\hspace{-0.1cm}
		\begin{tabular}{r c | c | l }
			&(V4)				&$1 \rangle\langle 1$	&$\leftarrow  -\phi_1$		\\
			&					&$1 \rangle\langle 2$	&						\\
			&					&$\downarrow$	 			&$\tau_{12}$				\\
			&					&$1 \rangle\langle 2$	&$\rightarrow +\phi_2 $		\\
			&					&$1 \rangle\langle 1$	&						\\
			&					&$\downarrow$	 			&$\tau_{23}$				\\
			&$+\phi_3 \rightarrow$	&$1 \rangle\langle 1$	& 						\\
			&					&$3 \rangle\langle 1$	&						\\
			&					&$\downarrow$	 &$t$					
		\end{tabular}
\end{table}\\[-10mm]
\begin{table}[h!]\small
~\hspace{-1cm}
		\begin{tabular}{l r c | c | l }
			&&(V5)				&$1 \rangle\langle 1$	&$\leftarrow  -\phi_1$		\\
			&&					&$1 \rangle\langle 3$	&						\\
			&&					&$\downarrow$	 			&$\tau_{12}$				\\
			&&$+\phi_2 \rightarrow$&$1 \rangle\langle 3$	& 						\\
			&&					&$2 \rangle\langle 3$	&						\\
			&&					&$\downarrow$	 			&$\tau_{23}$				\\
			&&					&$2 \rangle\langle 3$	&$ \rightarrow +\phi_3$		\\
			&&					&$2 \rangle\langle 1$	&						\\
			&&					&$\downarrow$	 &$t$					
		\end{tabular}
\hspace{-0.1cm}
		\begin{tabular}{r c | c | l }
			&(V6)				&$1 \rangle\langle 1$	&$\leftarrow  -\phi_1$		\\
			&					&$1 \rangle\langle 3$	&						\\
			&					&$\downarrow$	 			&$\tau_{12}$				\\
			&$+\phi_2\rightarrow$	&$1 \rangle\langle 3$	& 						\\
			&					&$3 \rangle\langle 3$	&						\\
			&					&$\downarrow$	 			&$\tau_{23}$				\\
			&					&$3 \rangle\langle 3$	&$\rightarrow+\phi_3  $		\\
			&					&$3 \rangle\langle 1$	&						\\
			&					&$\downarrow$	 &$t$					
		\end{tabular}
\hspace{-0.1cm}
		\begin{tabular}{r c | c | l }
			&(V7)				&$1 \rangle\langle 1$	&$\leftarrow  -\phi_1$		\\
			&					&$1 \rangle\langle 3$	&						\\
			&					&$\downarrow$	 			&$\tau_{12}$				\\
			&					&$1 \rangle\langle 3$	&$\rightarrow +\phi_2 $		\\
			&					&$1 \rangle\langle 1$	&						\\
			&					&$\downarrow$	 			&$\tau_{23}$				\\
			&$+\phi_3 \rightarrow$	&$1 \rangle\langle 1$	& 						\\
			&					&$2 \rangle\langle 1$	&						\\
			&					&$\downarrow$	 &$t$					
		\end{tabular}
\hspace{-0.1cm}
		\begin{tabular}{r c | c | l }
			&(V8)				&$1 \rangle\langle 1$	&$\leftarrow  -\phi_1$		\\
			&					&$1 \rangle\langle 3$	&						\\
			&					&$\downarrow$	 			&$\tau_{12}$				\\
			&					&$1 \rangle\langle 3$	&$\rightarrow +\phi_2 $		\\
			&					&$1 \rangle\langle 1$	&						\\
			&					&$\downarrow$	 			&$\tau_{23}$				\\
			&$+\phi_3 \rightarrow$	&$1 \rangle\langle 1$	& 						\\
			&					&$3 \rangle\langle 1$	&						\\
			&					&$\downarrow$	 &$t$					
		\end{tabular}
	\caption{Double-sided Feynman diagrams for three-pulse FWM of a 3LS with positive delays $\tau_{12}$ and $\tau_{23}$. Orderings of the levels and dipole transitions as in Fig.~\ref{fig:3LS} with (a) cascade system, (b) $\Lambda$-system, and (c) V-system.}
	\label{tab:feyn_3ls}
\end{table}

\begin{figure}[t]
	\centering
	\includegraphics[width = 0.7\textwidth]{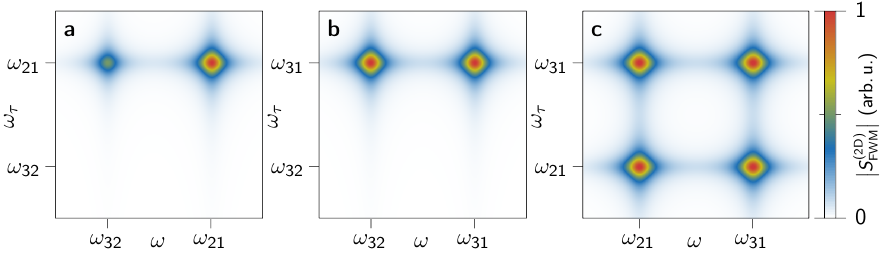}
	\caption{Qualitative 2D spectra [absolute value of Eq.~\eqref{eq:I_FWM_2D_simple} w/ Eq.~\eqref{eq:I_FWM_1} and $r_0\star~r_0 =$~const.] for the possible three level systems in Fig.~\ref{fig:3LS} with the peak amplitudes scaled according to Eqs.~\eqref{eq:p_3LS} and \eqref{eq:E_FWM_3LS} with $\theta_{ij,k}=$const. for all appearing pulse areas. We consider a positive delay $\tau_{12}=\tau>0$ between pulse 1 and 2 and a vanishing delay $\tau_{23}=0$ between pulse 2 and 3. (a) Cascade system, (b) $\Lambda$-system, (c) V-system.}
	\label{fig:2D_spec_3LS}
\end{figure}

Analog to Fig.~\ref{fig:echo_with_deph_spec}, in Fig.~\ref{fig:2D_spec_3LS} we present 2D spectra for all three 3LSs from Fig.~\ref{fig:3LS} using identical pulse areas for all transitions and pulses $\theta_{ij,k}=$~const. for simplicity. Note that only the height but not the position of the peaks is impacted by the pulse areas, such that the qualitative features in Fig.~\ref{fig:2D_spec_3LS} are representative for the respective 3LS. For the same reason we also ignored the impact of the spectrometer response, i.e., $r_0\star r_0=$~const., in Eq.~\eqref{eq:I_FWM_1}, as was also done in Fig.~\ref{fig:echo_with_deph_spec}. We however added a simple Lorentzian broadening to the peaks by multiplying the signal in Eq.~\eqref{eq:E_FWM_3LS} with a damping factor $e^{-\gamma(t+\tau)}$ before Fourier-transforming along $t$ according to Eq.~\eqref{eq:I_FWM_1} and along $\tau$ according to Eq.~\eqref{eq:I_FWM_2D_simple}. Note that this is only necessary for visualization here as we did not include any dissipation explicitly in our model. 

We quickly see that the cascade (a) and $\Lambda$-system (b) have similar 2D spectra, while the V-system (c) shows a qualitatively different peak structure. In the first two cases we only have two horizontally aligned peaks in the spectrum, while the latter system exhibits four peaks arranged in a square. In terms of their interpretation as emission and absorption peaks, we see that the cascade system emits a FWM signal at the transition frequencies $\omega_{21}$ and $\omega_{32}$ after absorbing at $\omega_{21}$ and similar for the $\Lambda$-system with switched $2\leftrightarrow 3$. The presence of only one absorption frequency in these two systems is due to the fact that the ground state is only connected to one other energy level via a dipole transition [blue arrows in Fig.~\ref{fig:3LS}(a,b)]. This level in turn is however connected to two other levels leading to the presence of two possible emission peaks. Note that in this discussion we consider a pure initial (ground) state $\ket{1}\bra{1}$, i.e., low temperatures with $k_BT\ll E_{(n>1)}-E_1$. Especially in the $\Lambda$-system larger temperatures can lead to modifications of the initial state since $\ket{1}$ and $\ket{2}$ are energetically close to each other.

The much richer structure of the V-system (c) can be understood in a similar fashion: there the ground state is connected to two levels [Fig.~\ref{fig:3LS}(c)] such that there are two possible absorption frequencies and two possible emission frequencies yielding the four peaks in Fig.~\ref{fig:2D_spec_3LS}(c). These concrete examples show that FWM signals, especially in the form of 2D spectra, can be used to determine relevant optical transitions within multi-level systems, which is another important benefit of this and other similar multi-dimensional spectroscopy methods~\cite{smallwood2018multidimensional}. The interpretation in terms of absorption and emission spectra allows to identify possible optical pathways, i.e., we can answer which coherence (after absorption) is transformed into which other coherences that are then detected in emission.

While we can easily distinguish between the V-system and the other two systems in Fig.~\ref{fig:2D_spec_3LS}, we lack a simple qualitative distinction between cascade and $\Lambda$-system. However here we considered only one of the possible choices for the delays, i.e., we scanned over positive $\tau_{12}$. In the case of negative $\tau_{12}<0$ with $\tau_{23}=0$, pulse 2 and 3 excite the system at the same time. Similar to the situation in the 2LS, this does not lead to a FWM signal in the $\Lambda$-system, since we cannot excite two times, i.e., $+\phi_3\rightarrow +\phi_2\rightarrow\ket{1}\bra{1}=0$. In the cascade system however we obtain the Feynman diagrams shown in Tab.~\ref{tab:feyn_3ls_tau_neg} and thus a non-vanishing FWM signal for $\tau_{12}<0$ (not contributing to Fig.~\ref{fig:2D_spec_3LS}). By choosing different signs for the delays we generally obtain physically different situations, which yield even more information on the investigated system. Varying the negative delay $\tau_{12}$ then probes the coherence $\ket{3}\bra{1}$, which is not optically active as it is not connected by a simple dipole transition [blue arrows in Fig.~\ref{fig:3LS}(a)]. Indeed it is connected by two downwards transitions and is thus also called a two-photon (or two-quantum) coherence~\cite{kasprzak2010up,smallwood2018multidimensional,groll2020four}. 

\begin{table}[b]\small
		\begin{tabular}{l r r | c | l }
			& (C$^\prime$1)&	$+\phi_2 \rightarrow$	&$1 \rangle\langle 1$	&					\\
			&&									&$2 \rangle\langle 1$	&					\\
			&&									&$\downarrow$	 			&$\tau_{23}=0$		\\
			&&				$+\phi_3 \rightarrow$	&$3 \rangle\langle 1$	& 					\\
			&&									&$3 \rangle\langle 1$	&					\\
			&&									&$\downarrow$	 			&$|\tau_{12}|$		\\
			&&									&$3 \rangle\langle 1$	&$ \leftarrow -\phi_1$	\\
			&&									&$3 \rangle\langle 2$&						\\
			&&									&$\downarrow$	 &$t$					
		\end{tabular}
		\quad
		\begin{tabular}{r r | c | l }
			(C$^\prime$2)&		$+\phi_2 \rightarrow$	&$1 \rangle\langle 1$	&				\\
			&									&$2 \rangle\langle 1$	&				\\
			&									&$\downarrow$	 			&$\tau_{23}=0$	\\
			&				$+\phi_3 \rightarrow$	&$3 \rangle\langle 1$	& 				\\
			&									&$3 \rangle\langle 1$	&				\\
			&									&$\downarrow$	 			&$|\tau_{12}|$	\\
			&				$ -\phi_1 \leftarrow$	&$3 \rangle\langle 1$	& 				\\
			&									&$2 \rangle\langle 1$	&				\\
			&									&$\downarrow$	 &$t$						
		\end{tabular}
	\caption{Double-sided Feynman diagrams for three-pulse FWM with negative delay $\tau_{12}<0$ in the cascade system.}
	\label{tab:feyn_3ls_tau_neg}
\end{table}

We can also access another class of optically inactive transitions using FWM. This can be seen when going back to Tab.~\ref{tab:feyn_3ls}, e.g., diagram ($\Lambda$4). By varying the delay $\tau_{23}$ we probe the dynamics of the coherence $\ket{1}\bra{2}$ between the two low-lying states in the $\Lambda$-system. This transition is again optically inactive, as it is not connected via a single dipole transition [blue arrows in Fig.~\ref{fig:3LS}(b)]. However it is also not a two-photon coherence, since the states are not connected via emission of two photons. This renders an interesting benefit of the FWM method, as it provides a tool to directly measure quantum properties of coherences which are not connected by spontaneous emission of light and are therefore interesting for quantum applications~\cite{fleischhauer2002quantum,yale2013all,neuwirth2021quantum}.

As was found for the 2LS, scanning the delay $\tau_{23}$ while keeping $\tau_{12}$ fixed generally provides information on the occupation dynamics of the involved states, this is also true for the different 3LSs in Tab.~\ref{tab:feyn_3ls}. Note however, that the information on some of the occupations remains elusive in the $\chi^{(3)}$-regime, e.g., the occupation of the highest lying state $\ket{3}\bra{3}$ in the cascade system. We can however go beyond the $\chi^{(3)}$-regime by using sufficiently strong laser pulses. The FWM signal then also contains information on the occupations of such higher lying states~\cite{mermillod2016dynamics}. Alternatively we can investigate higher order wave mixing signals, such as six-wave mixing~\cite{fras2016multi,hahn2022destructive,grisard2023temporal}, as will be discussed in Sec.~\ref{sec:nwm}.

Thus, also in the case of 3LSs, FWM allows us to obtain a near complete characterization of the properties of the system. If we had included dissipation in the model, we would obtain the dephasing and decay rates in a fashion analog to the case of the 2LS in Chap.~\ref{sec:theory}. The complexity of the investigated spectroscopic signals increases with the number of relevant energy levels, which should already be clear from this discussion on the simplest non-trivial systems beyond the 2LS.

\subsection{Comparison between experiment and theory: 3LS and 4LS}\label{sec:exp_theo_2}
\begin{figure}[b]
	\centering
	\includegraphics[width = 0.6\textwidth]{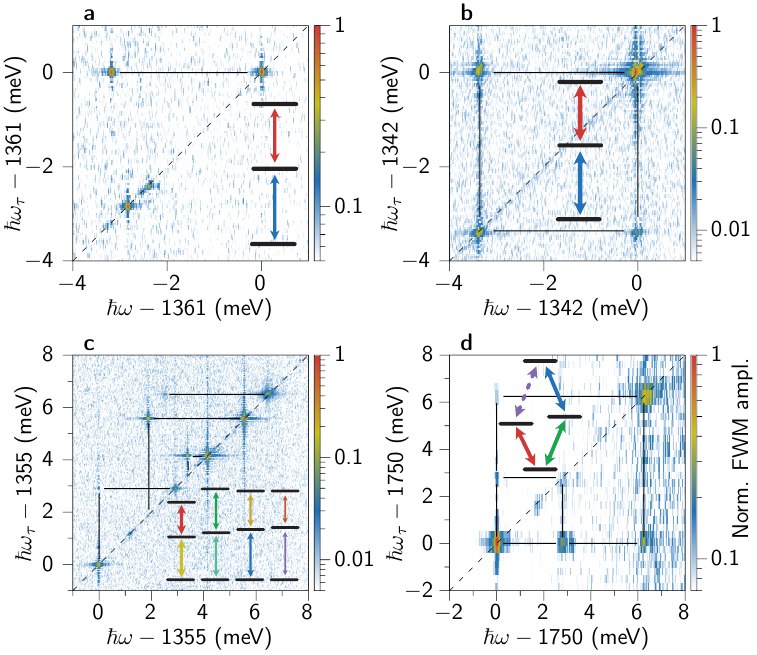}
	\caption{Collection of 2D FWM spectra from different excitonic few-level systems. (a) Cascade-type 3LS [see Figs.~\ref{fig:3LS}(a) and \ref{fig:2D_spec_3LS}(a)] in an InAs quantum dot. The level structure was confirmed by a $\tau_{12}<0$ scan. Same sample as in Ref.~\cite{mermillod2016dynamics}. (b) Cascade-type 3LS driven beyond the $\chi^{(3)}$-regime. The level structure was confirmed by a $\tau_{12}<0$ scan. Same quantum dot molecule sample as in Ref.~\cite{wigger2023controlled} operated far from trion hybridization. (c) Series of uncoupled cascade 3LSs. Same sample as in Ref.~\cite{delmonte2017coherent}. (d) Diamond-shaped 4LS showing a combination of cascade and V-shaped 3LS in a CdTe quantum dot grown by Wojciech Pacuski at the University of Warsaw.}
	\label{fig:2D_spec_collection_2}
\end{figure}

To go beyond the so far purely theoretical discussion of the current chapter, we will use our understanding of 2D FWM spectra of few-level systems here to analyze actual experimental data. To this aim, Fig.~\ref{fig:2D_spec_collection_2} shows a collection of 2D FWM spectra from different few-level systems. 

Starting with Fig.~\ref{fig:2D_spec_collection_2}(a), which shows the FWM spectrum of a single InAs quantum dot (same sample as in Ref.~\cite{mermillod2016dynamics}), we find two horizontally aligned peaks (marked by the black horizontal line). There are also additional peaks on the diagonal (dashed line), which however stem from situations where the quantum dot was differently charged during the measurement, as discussed in Ref.~\cite{mermillod2016dynamics}. The spectrum thus might indicate a cascade level structure as depicted in Fig.~\ref{fig:2D_spec_collection_2}(a) [compare with Fig.~\ref{fig:2D_spec_3LS}(a,b), the possibility of a $\Lambda$-system was ruled out using negative delay scans, see Tab.~\ref{tab:feyn_3ls_tau_neg}]. Indeed a single quantum dot can be described with a cascade three-level structure, which represents an empty dot (ground state), the excitation of a single exciton (first excited state), and a biexciton (second excited state)~\cite{gerardot2009dressed,mermillod2016dynamics}. The biexciton typically has a slightly different energy than twice that of the exciton due to the interaction between the charge carriers~\cite{chen2002biexciton}. In principle there are two first excited states, i.e., two possible single exciton states, in quantum dots. Suitable polarization selection however allows to neglect one of them, leading to the cascade structure shown in Fig.~\ref{fig:2D_spec_collection_2}(a)~\cite{gerardot2009dressed,mermillod2016dynamics}.

Figure~\ref{fig:2D_spec_collection_2}(b) now presents the FWM spectrum of a quantum dot molecule (same sample as in Ref.~\cite{wigger2023controlled}). While the square-shaped pattern similar to Fig.~\ref{fig:2D_spec_3LS}(c) might indicate that the system can be described by a V-shaped 3LS, from excitation intensity scans we know that the two peaks at lower energies $\hbar\omega_\tau$ should disappear at smaller pulse areas~\cite{wigger2023controlled}. Consequently, we are in the sketched cascade-shaped 3LS situation from Fig.~\ref{fig:2D_spec_3LS}(a) but for stronger excitation here (again the $\Lambda$-structure was ruled out using negative delay scans). For higher pulse areas beyond the linear response regime for a single pulse, the highest excited state can already be reached by the first pulse giving rise to the full square pattern of peaks in the 2D spectrum~\cite{mermillod2016dynamics,wigger2017exploring}. 

In such quantum dot molecules couplings and state hybridization between two quantum dots in close proximity are possible. The resulting level structure can be tuned by electric fields and single charges can be injected into the structure which changes the ground state character from an empty dot to a single electron or hole. In such a system, depending on the external fields, we can not only get a cascade system as in Fig.~\ref{fig:2D_spec_collection_2}(b), but also a V-shaped or a $\Lambda$-shaped 3LS as studied in detail in Ref.~\cite{wigger2023controlled}.

%

A more involved situation is found in Fig.~\ref{fig:2D_spec_collection_2}(c) with the spectrum exhibiting now a variety of peaks, mainly along the diagonal. As discussed in Ref.~\cite{delmonte2017coherent} the entire 2D FWM spectrum stems from four different quantum dots, exhibiting each a cascade level scheme. Also here, as in the situation in Fig.~\ref{fig:2D_spec_collection_2}(b) the dots are excited beyond the $\chi^{(3)}$-regime which explains why the spectrum does not simply consist of four distinct pairs of horizontally aligned peaks [see Fig.~\ref{fig:2D_spec_3LS}(a) for $\chi^{(3)}$-regime of a single cascade system].

Finally in Fig.~\ref{fig:2D_spec_collection_2}(d) we find a pattern of two squares sharing a corner. This clearly goes beyond any of the previously discussed 3LSs and a diamond-shaped 4LS like the one shown in the inset is required. This is the typical level structure of a quantum dot which in principle always has two possible single exciton states~\cite{chen2002biexciton,mermillod2016dynamics}. Here it is recorded on a single CdTe quantum dot with a polarization of the exciting laser which does not allow for the isolation of only one of the two single exciton states. The fact that the fourth possible transition energy is missing indicates that this transition was driven less efficiently maybe also due to the selected light polarization in the experiment.

\subsection{Generalized rules for $N$-wave mixing}\label{sec:nwm}
Having introduced the Feynman rules in Sec.~\ref{sec:feynman} we can generalize the spectroscopy method and discuss possible $N$-wave mixing ($N$WM) signals~\cite{mukamel1995principles,bolton2000demonstration,hahn2022destructive,grisard2023temporal}. As in the case of $(N=4)$-wave mixing, i.e., FWM, we have at most $N-1$ excitation pulses with phases $\phi_j$. The $N$WM signal is obtained by filtering for a specific phase combination
\begin{equation}
	\phi_{N\text{WM}}=\sum_j l_j\phi_j \,.
\end{equation}
To efficiently deduce which $N$WM signals are possible we introduce the notion of $n$-photon coherences. We already talked about two-photon coherences in the context of the cascade system in Tab.~\ref{tab:feyn_3ls_tau_neg}. There this term described the coherence $\ket{3}\bra{1}$, i.e., the ket and bra are connected by two optically active dipole transitions, either only upwards or only downwards in the energy level scheme in Fig.~\ref{fig:3LS}(a). To have a well-defined meaning we call $\ket{3}\bra{1}$ a two-photon coherence and $\ket{1}\bra{3}$ a conjugate two-photon coherence. In a similar fashion we call every density matrix element $\ket{m}\bra{m'}$ an $n$-photon coherence, if the state $\ket{m}$ can be reached from the state $\ket{m'}$ by $k$ photon absorption processes and $l$ photon emission processes with $n=k-l>0$. The corresponding conjugate matrix element $\ket{m'}\bra{m}$ is then called conjugate $n$-photon coherence, which is equivalent to a $-|n|$-photon coherence. Note that this notion of $n$-photon coherence is sufficiently well-defined due to the RWA applied in Sec.~\ref{sec:optics_NLS} since dipole transitions are only possible between states of energy separation $\approx \hbar\omega_l$. Thus a one-photon coherence cannot simultaneously be a two-photon coherence. In this sense occupations are "zero-photon" (or "zero-quantum") coherences~\cite{smallwood2018multidimensional}.

From the Feynman rules in Sec.~\ref{sec:feynman} we can deduce that a pulse filtered with respect to the phase $l_j\phi_j$ changes each $n$-photon coherence in the Feynman diagrams into $n+l_j$-photon coherences. To reach an optically active dipole transition, i.e., a one-photon coherence, after all pulses we therefore need 
\begin{equation}
	\sum_j l_j=1
\end{equation}
since we start in the ground state, i.e., with a "zero-photon-coherence". Note that this restriction on the possible $N$WM phases made explicit use of the RWA. If we did not use this approximation we could in principle address higher order emission processes by choosing $\sum_j l_j>1$, e.g., second-harmonic generation for $\sum_j l_j=2$ and high-harmonics in general~\cite{mukamel1995principles,langbein2010coherent}.

In the same way that FWM selects the third order nonlinearity for weak excitation [$\chi^{(3)}$-regime], $N$WM is defined as selecting the $N-1$-th order nonlinearity, which is ensured by
\begin{equation}
	\sum_j |l_j|=N-1\,.
\end{equation}


%% file: 7_Conclusions.tex
\section{Conclusions}\label{sec:conclusions}
We conclude this tutorial by recapitulating the benefits of heterodyne wave mixing spectroscopy, discussed throughout the tutorial mainly for the special case of FWM.
\begin{itemize}
	\item Due to the phase filtering of the detection (equivalent to directional filtering in conventional FWM setups) the FWM signals are background free [see discussion at the end of Sec.~\ref{sec:exp_phase}]. 
	\item Due to the heterodyne wave mixing setup and using suitable analysis of the data we obtain information on the full signal field, including amplitude and phase, and not only the intensity [see Secs.~\ref{sec:exp_heterodyne_data} and \ref{sec:exp_phase}].
	\item As discussed in Chap.~\ref{sec:theory_2}, FWM spectroscopy allows to separate different sources of line broadening. We have explicitly discussed how homogeneous broadening, stemming from pure dephasing or a fast spectral jitter, results in an exponential damping of the FWM signal in time after the last exciting laser pulse. On the contrary, inhomogeneous broadening, due to ensemble averaging or slow spectral jitter, leads to the formation of a photon echo. The temporal dynamics of the echo is determined by the spectral distribution of the source of inhomogeneity. Converted to the spectral domain, in 2D FWM spectra the inhomogeneous broadening results in an elongation along the diagonal [see Sec.~\ref{sec:2d_spectra}].
	\item By varying different delays in FWM and related multi-pulse experiments,  information on the dynamics of optically active quantum coherences or occupations can be obtained. Furthermore also optically dark coherences can be measured which are usually not directly accessible in other spectroscopy experiments. By a systematic combination of FWM experiments with different parameter variations, such as delay, excitation intensities, optical polarizations, it is sometimes possible to determine the dynamics of all  density matrix elements of a given few-level system [see Sec.~\ref{sec:FWM_2LS} and Sec.~\ref{sec:FWM_3LS}].
	\item Connected to the previous point, visualizing FWM signals in multi-dimensional spectra gives direct insight into the coupling mechanisms within few-level systems [see Sec.~\ref{sec:FWM_3LS}]. In this context it is again important to note that we have access to amplitude and phase of the signal field, not just its intensity. It has been shown in Refs.~\cite{kasprzak2011coherent,li2023optical} that the phase-information contained in 2D FWM spectra can give additional insight into the internal coupling mechanisms.
	\item Finally, a huge benefit of the presented FWM setup is the possibility to perform diffraction-limited imaging of all possible signals [see discussion in Sec.~\ref{sec:exp_excitation}]. One obvious example is mapping the ratio between homogeneous and inhomogeneous dephasing rates across extended samples~\cite{jakubczyk2019coherence,boule2020coherent,rodek2023controlled}.
\end{itemize}

%% file: A_Appendix.tex
\section*{Appendix}\label{sec:app}
\section{Light emission from a 2LS}\label{sec:dipole_emission}
The measured quantities in wave-mixing experiments are optical signals, i.e., the electromagnetic field emitted from the sample. In the theory parts of this tutorial, however, we have always calculated polarizations, i.e., material quantities. Therefore, here we will briefly address the connection between the material quantities and the optical field. Such a relation can be established both on the semiclassical level, where the emitted field is treated on a classical level~\cite{haken1984laser,hess1996maxwell,koch2001correlation,kira2011semiconductor} (Maxwell-Bloch equations), and on a fully quantum mechanical level with a quantized electromagnetic field. Here we will discuss the quantum mechanical treatment of the field to derive the light emission properties of a 2LS.

For the following discussion it is helpful to remind ourselves of some basic facts from quantum electrodynamics as presented, e.g., in Refs.~\cite{glauber1963coherent,cohen2022photons}. It is well known, that we can break down the electromagnetic field into normal modes, which are completely decoupled in the absence of any sources. In quantum theory these normal modes become independent harmonic oscillators and the Hamiltonian of the light field reads
\begin{equation}\label{eq:H_L}
	H_L=\sum_{n} \hbar\omega^{}_{n}a_{n}^{\dagger}a_{n}^{}\,,
\end{equation}
where we dropped the vacuum energy. The operators $a_n^{\dagger}$ and $a_n^{}$ are the bosonic creation and annihilation operators for the photon mode with quantum number $n$, respectively. They create or destroy a photon with frequency $\omega_n$. The quantum numbers $n$ are used to characterize these normal modes. For example, in the case of light inside a cubic volume $V$ with periodic boundary conditions, they are given by $n=(\bm{k},s)$, where $\bm{k}$ is the wavevector of the mode and $s=1,2$ denotes the polarization, e.g., right- or left-handed circular polarization. For a finite volume $V$ the set of wavevectors is discrete and becomes continuous in the limit $V\rightarrow\infty$.

In the interaction picture with respect to the light-field Hamiltonian $H_L$ the electric and magnetic field operators obey Maxwell's equations. The time-dependent electric field operator $\bm{E}(\bm{r},t)$ is given by~\cite{glauber1963coherent}
\begin{equation}\label{eq:E(t)}
	\bm{E}(\bm{r},t)=i\sum_n \sqrt{\frac{\hbar\omega_n}{2\epsilon_0}}\left[\bm{u}_n(\bm{r})a_n e^{-i\omega_n t}-\bm{u}^*_n(\bm{r})a_n^{\dagger} e^{i\omega_n t}\right]\,,
\end{equation}
where $\epsilon_0$ is the dielectric constant (to be modified in a material) and $\bm{u}_n(\bm{r})$ are mode functions, which describe the spatial distribution of the electromagnetic modes with quantum numbers $n$. The corresponding operator in the Schrödinger picture is obtained by setting $t=0$, i.e., dropping the time-dependent exponentials. We see that the electric field contains a positive and negative frequency component
\begin{equation}\label{eq:E_parts}
	\bm{E}(\bm{r},t)=\bm{E}^{(+)}(\bm{r},t)+\bm{E}^{(-)}(\bm{r},t)\,.
\end{equation}
The positive frequency component is given by
\begin{equation}\label{eq:E_+(t)}
	\bm{E}^{(+)}(\bm{r},t)=i\sum_n \sqrt{\frac{\hbar\omega_n}{2\epsilon_0}}\bm{u}_n(\bm{r})a_n e^{-i\omega_n t}
\end{equation}
and contains only photon annihilation operators. On the contrary, the negative frequency component contains only creation operators and obeys
\begin{equation}\label{eq:E_-(t)}
	\bm{E}^{(-)\dagger}(\bm{r},t)=\bm{E}^{(+)}(\bm{r},t)\,.
\end{equation}
As discussed in Sec.~\ref{sec:dipole_RWA}, we consider dipole interaction and apply a rotating wave approximation (RWA) to the coupling between the considered 2LS and the light field, such that the interaction with a quantized field in the Schrödinger picture is described by~\cite{cohen2022photons,cohen1998atom} [see Eq.~\eqref{eq:H_I_RWA} for a non-quantized version]
\begin{equation}\label{eq:H_I_RWA_quant}
	H_{I}=-\bra{x}\bm{d}\ket{g}\cdot \bm{E}^{(+)}\ket{x}\bra{g} +h.c.\,
\end{equation}
with $\bm{E}^{(+)}=\bm{E}^{(+)}(\bm{r}_\mathrm{2LS},t=0)$ being the Schrödinger picture field operator at the position of the 2LS. Note that the explicit time-dependence in Eq.~\eqref{eq:H_I_RWA} is removed from this Schrödinger picture interaction Hamiltonian but can be reintroduced when going to the interaction picture with respect to $H_L$, i.e., using Eq.~\eqref{eq:E_+(t)}.

The total Hamiltonian of the 2LS interacting with the surrounding quantized light modes is given by $H=H_0+H_I+H_L$ with $H_0=\hbar\omega_x\ket{x}\bra{x}$ as in Eq.~\eqref{eq:H_0}. Using the coupling between the emitter and the light modes we can describe the impact of light emission on the electromagnetic field. To this aim we consider the complete dynamics of the operator $\bm{E}^{(+)}$ in the Heisenberg picture of the total coupled light-matter system~\cite{nazir2016modelling}. The Heisenberg equations of motion for the annihilation operators $a_n^H(t)$ are then given by
\begin{align}
	\frac{\text{d}}{\text{d}t}a_n^H(t)&=-i\omega_n a_n^H(t) +(\ket{g}\bra{x})^H(t)\underbrace{\bra{g}\bm{d}\ket{x}\cdot \bm{u}_n^*(\bm{r}_\mathrm{2LS})\sqrt{\frac{\omega_{n}}{2\epsilon_0\hbar}}}_{g_n}\,,
\end{align}
where we have used $\left[a_m^{H}(t),a_n^{H\dagger}(t)\right]=\delta_{m,n}$ and $\left[a_m^H(t),a_n^H(t)\right]=0$ for the bosonic operators of the light modes. The label $H$ denotes the operators in the Heisenberg picture. We can formally integrate this equation obtaining
\begin{equation}
	a_n^H(t)=e^{-i\omega_n t}a_n^H(0)+g_n\int\limits_0^t \text{d} t'\, (\ket{g}\bra{x})^H(t') e^{i\omega_n (t'-t)}\,.
\end{equation}
In the case of vanishing coupling $g_n\rightarrow 0$ we recover Eq.~\eqref{eq:E_+(t)}. Therefore, the first term describes the free evolution of the light field, whereas the second term describes the additional influence from the presence of the emitter. Reinserting this result into $\bm{E}^{(+)}$ yields
\begin{equation}
	\bm{E}^{(+)H}(\bm{r}_\mathrm{2LS},t)=\bm{E}^{(+)H}_0(\bm{r}_\mathrm{2LS},t)+i\sum_n \sqrt{\frac{\hbar\omega_n}{2\epsilon_0}}\bm{u}_n(\bm{r}_\mathrm{2LS}) g_n \int\limits_0^t\text{d} t'\,(\ket{g}\bra{x})^H(t')e^{i\omega_n(t'-t)}\,.
\end{equation}
Here, $\bm{E}^{(+)H}_0(\bm{r}_\mathrm{2LS},t)$ is the free field without the influence of the 2LS, i.e., Eq.~\eqref{eq:E_+(t)}. As a final step we introduce the spectral density of the light modes, as defined by
\begin{equation}\label{eq:g_Omega}
	\bm{g}(\omega)=\sum_n \sqrt{\frac{\hbar\omega_n}{2\epsilon_0}}\bm{u}_n(\bm{r}_\mathrm{2LS}) g_n \delta (\omega-\omega_n)=\sum_n \bm{u}_n(\bm{r}_\mathrm{2LS})\left[\bra{g}\bm{d}\ket{x}\cdot \bm{u}_n^*(\bm{r}_\mathrm{2LS})\right] \frac{\omega}{2\epsilon_0} \delta(\omega-\omega_n)\,,
\end{equation}
which describes the effective coupling between the emitter and all modes with frequency $\omega$. With this we can write
\begin{equation}\label{eq:E_w_g}
	\bm{E}^{(+)H}(\bm{r}_\mathrm{2LS},t)-\bm{E}^{(+)H}_0(\bm{r}_\mathrm{2LS},t)= i\int\limits_{0}^{\infty}\text{d}\omega\,\bm{g}(\omega)\int\limits_0^t\text{d} t'\,(\ket{g}\bra{x})^H(t')e^{i\omega(t'-t)}\,.
\end{equation}
In the case of a continuum of optical modes, e.g., for modes in a cubic volume $V$ with periodic boundary conditions in the limit $V\rightarrow \infty$, the sum over $n$ appearing in Eq.~\eqref{eq:g_Omega} leads to an integral over wavevectors $\bm{k}$. This renders $\bm{g}(\omega)$ a well-behaved function.

We assume a sufficiently weak coupling between light-field and emitter, such that
\begin{equation}\label{eq:approx_XH}
	(\ket{g}\bra{x})^H(t')\approx \exp(iH_0t')\ket{g}\bra{x}\exp(-iH_0t')=e^{-i\omega_x t'}\ket{g}\bra{x}
\end{equation}
for $t'\in[0,t]$. The frequencies contained in $(\ket{g}\bra{x})^H(t')$ are thus peaked at $\omega_x$, such that for sufficiently long times $t$, i.e., in the Markov limit~\cite{breuer2002theory}, only contributions to the $\omega$-integral in Eq.~\eqref{eq:E_w_g} with $\omega\approx\omega_x$ are relevant. Assuming that the effective coupling $\bm{g}(\omega)$ to the modes with frequencies $\omega$ is sufficiently slowly varying around the resonance $\omega\approx\omega_x$, we can replace it with $\bm{g}(\omega_x)$. Since only the modes with $\omega\approx\omega_x$ contribute significantly, we can make a crude approximation and extend the integration over the continuum of frequencies even to unphysical negative values, such that the field due to the presence of the emitter becomes~\cite{nazir2016modelling}
\begin{equation}
	\bm{E}^{(+)H}(\bm{r}_\mathrm{2LS},t)-\bm{E}^{(+)H}_0(\bm{r}_\mathrm{2LS},t)\approx i\bm{g}(\omega_x)\int\limits_{-\infty}^{\infty}\text{d}\omega\,\int\limits_0^t\text{d} t'\,(\ket{g}\bra{x})^H(t')e^{i\omega(t'-t)}\,.
\end{equation}
These approximations are well-known from the Weisskopf-Wigner theory of spontaneous emission~\cite{cohen1998atom,weisskopf1930berechnung}. The $\omega$-integral together with the exponential function yields a $\delta$-function, enforcing $t=t'$. After the $t'$-integration we thus obtain~\cite{mollow1969power,nazir2016modelling}
\begin{equation}
	\bm{E}^{(+)H}(\bm{r}_\mathrm{2LS},t)-\bm{E}^{(+)H}_0(\bm{r}_\mathrm{2LS},t)\sim(\ket{g}\bra{x})^H(t)\,.
\end{equation}
The correction to the positive frequency component of the light field due to the presence of a weakly coupled emitter is proportional to the transition operator $\ket{g}\bra{x}$. The average emitted field is therefore proportional to the (microscopic) polarization $p(t)$ of the 2LS
\begin{equation}
	\braket{\bm{E}^{(+)}(\bm{r}_\mathrm{2LS})-\bm{E}^{(+)}_0(\bm{r}_\mathrm{2LS})}(t)\sim p(t)\,,
\end{equation}
which justifies its name given in Sec.~\ref{sec:bloch}.

\section{Inaccurate beam alignment}\label{sec:alignment}
\begin{figure}[h]
	\centering
	\includegraphics[width = 0.55\textwidth]{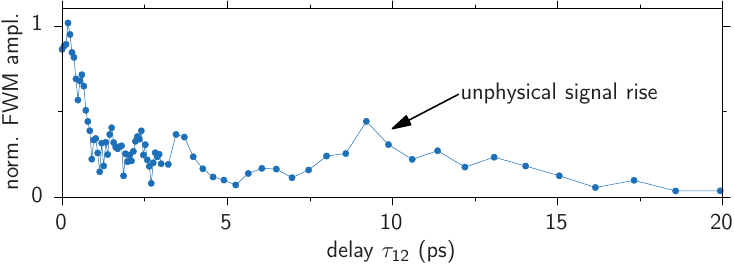}
        \caption{Faulty coherence dynamics measurement with an unphysical rise of the signal after the initial drop due to phonon induced dephasing~\cite{wigger2020acoustic}.}
        \label{fig:error_data}
\end{figure}

An inaccurate beam alignment (see Sec.~\ref{sec:data_alignment}) would generate artifacts in the delay dynamics, which are then more or less easy to identify in data post-processing. This can for example produce unphysical intermediate increases of the detected signal as exemplary shown for a single quantum dot measurement in Fig.~\ref{fig:error_data}. For this reason the beam overlap should always be checked carefully before executing delay scans.

\section{Radio frequency filtering}\label{sec:frequ_filter}
\begin{figure}[h]
	\centering
	\includegraphics[width = 0.4\textwidth]{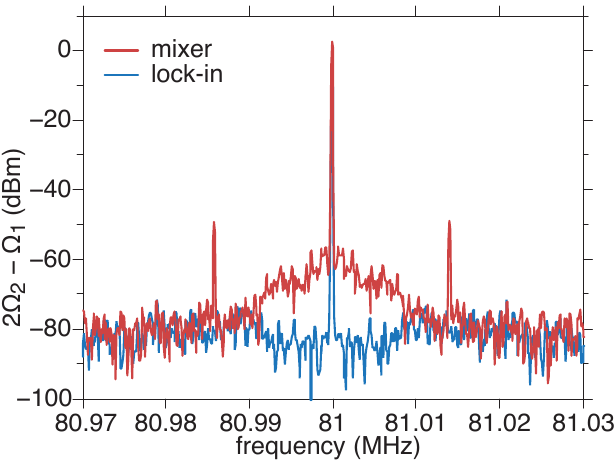}
	\caption{Power spectrum after the mixer (red) and after the lock-in amplifier (blue). Adapted from Ref.~\cite{mermillod2016spectroscopie}.}
	\label{fig:dB}
\end{figure}

The heterodyne frequencies (see Sec.~\ref{sec:exp_heterodyne_data}) are produced by a home-build three-channel analog mixer with standard components purchased from {\it Mini-Circuits}, yielding a spectral purity of around $-50$~dBm, as shown by the red curve in Fig.~\ref{fig:dB} for AOM frequencies of $\Omega_1=80$~MHz and $\Omega_2=79$~MHz and the FWM signal is filtered with respect to $2\Omega_2-\Omega_1$~\cite{mermillod2016spectroscopie}. Note, that the smaller side peaks stem from harmonic frequencies generated by the frequency mixer. The frequencies are subsequently re-generated by a phase-locked loop (PLL) of a {\it Stanford Research Systems SR 844} 200~MHz lock-in amplifier (blue spectrum in Fig.~\ref{fig:dB}). This improves the spectral purity to around $-80$~dBm and suppresses the side peaks. 

\section{Impact of reference timing and spectrometer response on echo detection}\label{sec:ref_timing}
\begin{figure}[h]
	\centering
	\includegraphics[width = 0.45\textwidth]{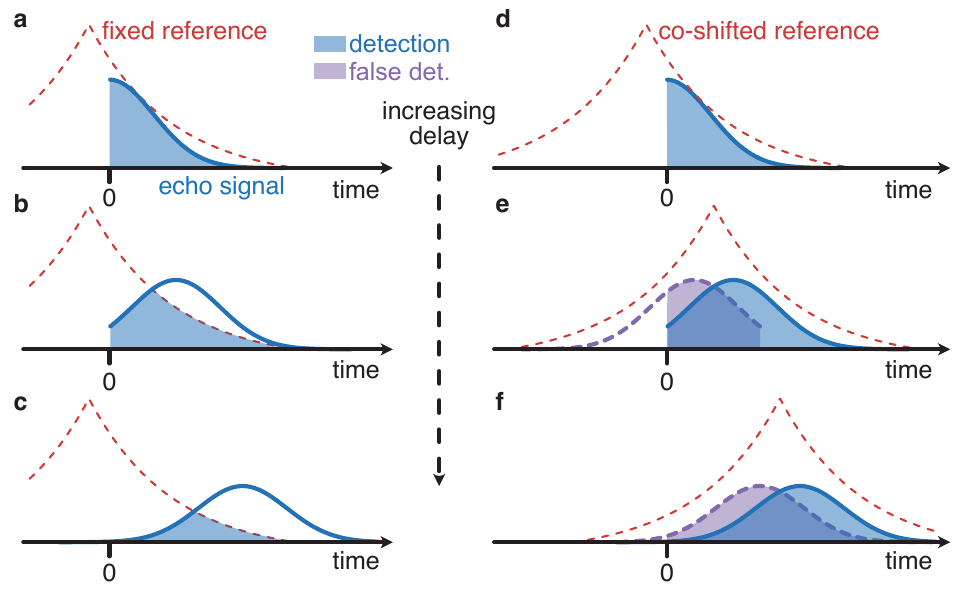}
	\includegraphics[width = 0.42\textwidth]{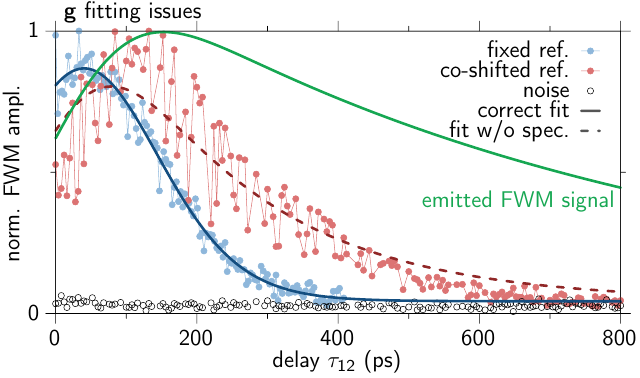}
        \caption{Impact of the spectrometer response on the detection of a photon echo. (a-c) Illustration of the detected signal for a temporally fixed reference pulse at $t_{\rm R}<0$. Detection window around reference as dashed red line, FWM echo signal is blue line, and detected signal as blue shaded area. The delay increases from (a) to (c). (d-f) Illustration of the measurement with a co-shifted reference. The violet curve shows the undesired signal contribution discussed in the context of Fig.~\ref{fig:spec}. (g) Measurements according to (a-c) in blue and (d-f) in red. The experimental data (bright dots) are fitted with Eq.~\eqref{eq:A_FWM} (dark lines) to determine the inhomogeneous $(\sigma)$ and homogeneous $(\gamma)$ broadening as given in the text. The green solid line shows the emitted FWM signal, i.e., the signal that would theoretically be detected in ideal conditions with $\gamma_\mathrm{D}=0$. The measurement was performed on a single InAs quantum dot in an optical microlens as in Ref.~\cite{jakubczyk2016impact}.}
        \label{fig:ref_shift}
\end{figure}

It is common practice to discuss the delay dynamics of FWM signals~\cite{kasprzak2010up,mermillod2016dynamics,jakubczyk2016impact,groll2020four,kasprzak2022coherent}. To this end one may calculate the delay-dependent FWM amplitude from the FWM spectrum $S_\mathrm{FWM}(\omega;\tau)$ in Eq.~\eqref{eq:I_FWM_1} via
\begin{equation}
	A_\mathrm{FWM}(\tau)=\int\text{d}t\, \left|\mathcal{F}^{-1}\left[S_\mathrm{FWM}\right](t;\tau)\right|\,.
\end{equation}
Considering echo dynamics with a Gaussian ensemble as in Eq.~\eqref{eq:p_avg_echo_wn} we thus get 
\begin{align}
	A_\mathrm{FWM}(\tau)&\sim \int\text{d}t\, \left|(r_0\star r_0)(t_\mathrm{R}-t)\mathcal{E}_\mathrm{FWM}(t;\tau)\right|\notag\\
	&\sim \int\limits_0^{\infty}\text{d}t\, \left|(r_0\star r_0)(t_\mathrm{R}-t)e^{-\gamma(t+\tau)}e^{-\frac{\sigma^2(t-\tau)^2}{2}}\right|
\end{align}
with $\gamma$ and $\sigma$ denoting homogeneous and inhomogeneous broadening, respectively. As discussed in Sec.~\ref{sec:exp_phase}, the reference time $t_\mathrm{R}<0$ has to be negative for a signal starting at $t=0$. As a simple model for the autocorrelation of the spectrometer response function we choose an exponential decay
\begin{equation}
(r_0\star r_0)(t_\mathrm{R}-t)\sim e^{-\gamma_\mathrm{D}|t-t_\mathrm{R}|}=e^{-\gamma_\mathrm{D}(t-t_\mathrm{R})}\,,\qquad t>0,\  t_\mathrm{R}<0
\end{equation}
corresponding to a Lorentzian spectral response $R(\omega-\omega_0)$ in Eq.~\eqref{eq:def_spectral_response} with a spectrometer resolution $\sim\gamma_\mathrm{D}$. The time-integrated FWM amplitude can then be evaluated as
\begin{align}
	A_\mathrm{FWM}(\tau)&\sim \int\limits_0^{\infty}\text{d}t\, e^{-\gamma_\mathrm{D} t}e^{-\gamma(t+\tau)}e^{-\frac{\sigma^2(t-\tau)^2}{2}}\notag\\
	&=e^{-\gamma\tau}e^{-\frac{\sigma^2\tau^2}{2}}\int\limits_0^{\infty}\text{d}t\,e^{-\left(\gamma_\mathrm{D}+\gamma-\sigma^2\tau\right)t}e^{-\frac{\sigma^2 t^2}{2}}\notag\\
	&\sim e^{-\left(2\gamma+\gamma_\mathrm{D}\right)\tau}\text{erfc}\left(\frac{\gamma+\gamma_\mathrm{D}-\sigma^2\tau}{\sqrt{2}\sigma}\right)\,.\label{eq:A_FWM}
\end{align}
In the case of vanishing homogeneous broadening $\gamma=0$ and perfect spectrometer resolution $\gamma_\mathrm{D}=0$ the signal is given by
\begin{equation}
	\left.A_\mathrm{FWM}(\tau)\right|_{\gamma,\gamma_\mathrm{D}=0}\sim \text{erfc}\left(-\frac{\sigma\tau}{\sqrt{2}}\right)\sim \int\limits_0^{\infty}\text{d}t\,e^{-\frac{\sigma^2(t-\tau)^2}{2}}\,,\label{eq:A_FWM_simple}
\end{equation}
i.e., it simply integrates the ideal photon echo as shown in Fig.~\ref{fig:ref_shift} (a-c) (blue curve). For large delays $\tau\gg\sigma^{-1}$ in (c) the echo lies almost completely at positive times $t>0$, such that the $t$-integration yields a constant contribution in Eq.~\eqref{eq:A_FWM_simple}. For smaller delays in (a,b) only part of the echo lies at positive times $t>0$, such that the integration only covers part of it with the smallest value for $A_\mathrm{FWM}(\tau)$ given by half of the value for large $\tau$: $A_\mathrm{FWM}(\tau=0)=\frac{1}{2} A_\mathrm{FWM}(\tau\rightarrow\infty)$. Note that in the derivation of the echo signal in Eq.~\eqref{eq:p_avg_echo_wn} we assumed $t>0$ and $\tau>0$. For vanishing homogeneous broadening and perfect spectrometer resolution the $t$-integrated FWM amplitude in Eq.~\eqref{eq:A_FWM_simple} is thus clearly a monotonously growing function in the delay $\tau$.

This situation is changed when considering a non-vanishing homogeneous broadening $\gamma>0$ or spectrometer resolution $\gamma_\mathrm{D}>0$. The signal in Eq.~\eqref{eq:A_FWM} then decays asymptotically in an exponential fashion with the rate $2\gamma+\gamma_\mathrm{D}$, i.e., both the dephasing of the considered emitter and the spectrometer response lead to a suppression of the echo signal at large delays. The impact of dephasing has already been discussed in the context of Fig.~\ref{fig:echo_with_deph}. To understand the contribution from the spectrometer, in Fig.~\ref{fig:ref_shift} (a-c) we show the spectrometer response triggered by the arrival of the reference pulse at $t_\mathrm{R}<0$ (red dashed lines). The overlap between the ideal echo signal (blue lines) and the spectrometer response is marked as a blue shaded area and corresponds to the part of the signal that is detected in the spectrometer. We see that this overlap decreases for increasing delay from (a) to (c). As discussed in Sec.~\ref{sec:exp_phase} the spectrometer response constitutes a time-window for the heterodyne interference between signal and reference pulse. If the signal, i.e., the photon echo, shifts out of this time-window as in (c), the overall detected signal gets suppressed, leading to the detector-induced damping with the rate $\gamma_\mathrm{D}$ in Eq.~\eqref{eq:A_FWM}.

To circumvent this detector-induced damping and increase the signal strength one might get the idea to shift the reference time $t_\mathrm{R}$ in accordance to the delay $\tau$, such that there is a fixed distance between the echo maximum at $t=\tau$ and the reference at $t=t_\mathrm{R}$, i.e., $\tau-t_\mathrm{R}=$~const. As discussed in detail in Sec.~\ref{sec:exp_phase} it is crucial that the reference arrives prior to the (entire) signal. If we consider the dominant contribution to the signal to be the echo, this means we should have $\max(\tau-n\sigma^{-1},0)>t_\mathrm{R}$ for a sufficiently large $n$, i.e., the reference should precede the echo/signal. Otherwise, as already discussed in the context of Fig.~\ref{fig:spec} we get a faulty signal detection, i.e., we do not only detect the FWM field $\mathcal{E}_\mathrm{FWM}(t)$ weighted by the spectrometer response, but also an additional time-reversed component $\mathcal{E}_\mathrm{FWM}(2t_\mathrm{R}-t)$. This is schematically shown in Fig.~\ref{fig:ref_shift} (d-f) for a co-moving reference pulse with a badly chosen value for $\tau-t_\mathrm{R}$, such that the time-reversed component (violet) is detected in addition to the desired signal (blue).

To conclude this discussion on the impact of the spectrometer and the reference timing, Fig.~\ref{fig:ref_shift} (g) shows FWM photon echo data (dots) from a single InAs quantum dot embedded in an optical microlens as in Ref.~\cite{jakubczyk2016impact}. The blue solid line shows a fit to the data with fixed reference timing $t_\mathrm{R}<0$ (blue dots) using Eq.~\eqref{eq:A_FWM}, yielding the spectrometer broadening $\gamma_\mathrm{D}=(61~\text{ps})^{-1}$ (consistent with Ref.~\cite{jakubczyk2016impact}), the homogeneous broadening $\gamma=(1484~\text{ps})^{-1}$ and the inhomogeneous broadening $\sigma=(95~\text{ps})^{-1}$. The green solid line shows the emitted FWM signal $\mathcal{E}_\mathrm{FWM}$, i.e., the signal that would be measured in ideal detection conditions with $\gamma_\mathrm{D}=0$. We thus again see the suppression of the signal due to the spectrometer response when comparing the green and blue lines.

If we consider a co-moving reference for the FWM signal detection (red dots and line), the measured signal (dots) naturally lasts longer. But as discussed previously it can be the case that there are unwanted contributions from the time-reversed signal field. Fitting the data (red line) in the co-moving case naively with Eq.~\eqref{eq:A_FWM} for $\gamma_\mathrm{D}=0$ (since our intention was to circumvent the impact of the spectrometer by co-moving the reference) yields a homogeneous broadening of $\gamma=(425~\text{ps})^{-1}$ and an inhomogeneous broadening of $\sigma=(84~\text{ps})^{-1}$. The discrepancy between the fitted parameters in the fixed reference and the co-moving reference case demonstrates that: (A) It is usually very important to consider the impact of the spectrometer response and the reference time in the theoretical modeling, and (B) It is crucial that the reference precedes the (entire) signal, especially when considering a co-moving reference pulse in photon echo measurements.

\section{Spatial overlap between excitation and reference}\label{sec:overlap}
\begin{figure}[t]
	\centering
	\includegraphics[width = 0.5\textwidth]{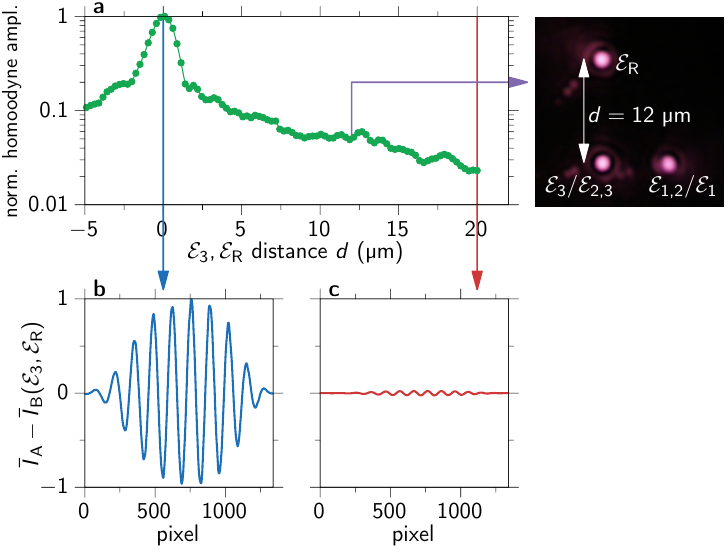}
	\caption{Impact of residual spatial pulse overlap. (a) Amplitude of the $\E_3+\E_{\rm R}$ homodyne amplitude versus their relative distance as illustrated by the CCD image on the right. The CCD image also shows that all pulses can be individually positioned on the sample. (b, c) Interferograms for $d=0$ (blue) and $d=20$~\textmu m (red) as marked by the arrows in (a).}
	\label{fig:distance_ref}
\end{figure}
We can use the independent positioning of the pulses on the sample (see Fig.~\ref{fig:TMDC_map} and Sec.~\ref{sec:FWM_scans}) also to determine a possible source of error, which is the remaining pulse overlap at a significant spatial distance $d$. To quantify this dependence, in Fig.~\ref{fig:distance_ref}(a) we plot the homodyne interference amplitude between the pulses $\E_3$ and $\E_{\rm R}$ ($\Omega_{\rm D}=\Omega_3$) as function of their distance $d$. Although the signal is naturally strongest at perfect overlap $d=0$ [Fig.~\ref{fig:distance_ref}(b)], still at $d=20$~\textmu m we can detect interference fringes [Fig.~\ref{fig:distance_ref}(c)]. This could for example lead to misinterpretation of remaining spatial overlap as propagation effects. \\